\documentclass{aa}  

\usepackage{amssymb}
\usepackage{graphicx}
\usepackage{epsfig}
\usepackage{subfigure}
\usepackage[authoryear]{natbib}
\usepackage[]{rotating}

\usepackage{txfonts}

\newcommand{\zsun}{\ifmmode Z_{\odot} \else Z$_{\odot}$\fi}
\newcommand{\rsun}{\ifmmode R_{\odot} \else R$_{\odot}$\fi}
\newcommand{\msun}{\ifmmode M_{\odot} \else M$_{\odot}$\fi}
\newcommand{\teff}{\ifmmode T_{\rm eff} \else T$_{\mathrm{eff}}$\fi}
\newcommand{\logg}{\ifmmode \log g \else $\log g$\fi}
\newcommand{\lL}{\ifmmode \log \frac{L}{L_{\odot}} \else $\log \frac{L}{L_{\odot}}$\fi}
\newcommand{\mdot}{$\dot{M}$}
\newcommand{\myr}{M$_{\odot}$ yr$^{-1}$}
\newcommand{\vsini}{$V$ sin$i$}
\newcommand{\vinf}{$v_{\infty}$}

\newcommand{\vmac}{v$_{\rm mac}$}

\newcommand{\kms}{km~s$^{-1}$}
\newcommand{\ha}{H$\alpha$}

\begin{document}

\title{Radial dependence of line profile variability in seven O9--B0.5 stars}
\author{F. Martins\inst{1}
\and W. Marcolino\inst{2}
\and D.J. Hillier\inst{3}
\and J.-F. Donati\inst{4}
\and J.-C. Bouret\inst{5}
}
\institute{LUPM, Universit\'e Montpellier 2, CNRS, Place Eug\`ene Bataillon, F-34095 Montpellier, France \\
           \email{fabrice.martins@univ-montp2.fr}
\and
            Observat\'orio do Valongo, Universidade do Rio de Janeiro, Ladeira Pedro Ant\'onio, 43, CEP 20080-090, Brasil\\
\and
            Department of Physics and Astronomy \& Pittsburgh Particle physics, Astrophysics and Cosmology Center (PITT PACC), University of Pittsburgh, 3941 O'Hara Street, Pittsburgh, PA 15260, USA \\
\and
            IRAP, CNRS \& Univ.\ de Toulouse, 14 Av.\ E.~Belin, F--31400 Toulouse, France\\
\and
            LAM--UMR 6110, CNRS \& Universit\'e de Provence, rue Fr\'ed\'eric Joliot-Curie, F-13388, Marseille Cedex 13, France \\ 
}

\date{Received: 25 March 2014 / Accepted}

\abstract
{Massive stars show a variety of spectral variability: presence of discrete absorption components in UV P-Cygni profiles, optical line profile variability, X-ray variability, radial velocity modulations.}
{Our goal is to study the spectral variability of single OB stars to better understand the relation between photospheric and wind variability. For that, we rely on high spectral resolution, high signal-to-noise ratio optical spectra collected with the spectrograph NARVAL on the T\'elescope Bernard Lyot at Pic du Midi.}
{We investigate the variability of twelve spectral lines by means of the Temporal Variance Spectrum (TVS). The selected lines probe the radial structure of the atmosphere, from the photosphere to the outer wind. We also perform a spectroscopic analysis with atmosphere models to derive the stellar and wind properties, and to constrain the formation region of the selected lines.}
{We show that variability is observed in the wind lines of all bright giants and supergiants, on a daily timescale. Lines formed in the photosphere are sometimes variable, sometimes not. The dwarf stars do not show any sign of variability. If variability is observed on a daily timescale, it can also (but not always) be observed on hourly timescales, albeit with lower amplitude. There is a very clear correlation between amplitude of the variability and fraction of the line formed in the wind. Strong anti-correlations between the different part of the temporal variance spectrum are observed.}
{Our results indicate that variability is stronger in lines formed in the wind. A link between photospheric and wind variability is not obvious from our study, since wind variability is observed whatever the level of photospheric variability. Different photospheric lines also show different degrees of variability.}

\keywords{Stars: massive -- Stars: atmospheres -- Stars: fundamental parameters -- Stars: variability -- Stars: individual: $\epsilon$~Ori, HD~167264, HD~188209, HD~207198, HD209975, 10~Lac, AE~Aur}

\authorrunning{Martins et al.}
\titlerunning{Variability of O9-B0.5 stars}

\maketitle


\section{Introduction}
\label{s_intro}

Massive stars are important objects for several fields of astrophysics. Due to their high temperature and large luminosity, they have powerful ionizing spectra that create \ion{H}{II} regions. With their large initial mass, they are able to go beyond the fusion of carbon in their core, thus producing a large fraction of the elements heavier than oxygen and contributing to the chemical evolution of galaxies. They explode as core-collapse supernovae, leading to the formation of neutron stars and stellar black holes. They are also the precursors of long-soft gamma ray bursts \citep{wb06}. 

One of the main characteristics of massive stars is their strong winds. Because of the large luminosity, a large number of photons can be absorbed or scattered through spectral lines in the outer layers of massive stars, leading to a radiative acceleration strong enough to generate an outflow: the radiatively driven stellar wind. A massive star can lose 90$\%$ of its mass through such a wind during its life \citep{mm00}. The rate of mass loss ranges from $10^{-9}$ to $10^{-4}$ \myr, depending on the evolutionary status and luminosity. The velocity at which this stellar wind is driven can reach 4000 \kms\ in the most extreme cases. The amount of mechanical energy released by stellar wind over the lifetime of a massive star is similar to the energy produced by the final supernova. The consequence is that massive stars keep injecting energy in the interstellar medium, participating in the support of turbulence.  

The process of acceleration through spectral line is intrinsically unstable. The resulting acceleration is directly related to the velocity gradient in the atmosphere. Hence, a slight increase of the velocity gradient leads to a stronger acceleration, which in turn increases the velocity gradient \citep{owo84}. This instability produces fluctuations in the velocity structure of a massive star's atmosphere \citep{mg79,owo88,feld95}. Due to mass conservation, these velocity fluctuations induce density variations. Since spectral lines are sensitive to both the velocity and density structure, they are also affected. 

Spectroscopic variability is commonly observed in massive stars in general and O stars in particular. UV resonance lines of O supergiants sometimes show Discrete Absorption Components (DACs) corresponding to zones of over-absorption throughout the blueshifted absorption part of the P-Cygni profile \citep{kaper96,ful12}. They are usually attributed to the presence of large scale structures leading to optical depth enhancement. Such structures may be due to Corotating Interacting Regions (CIRs) born in the photosphere. CIRs are thought to be produced by surface brightness variations, leading to modulations of the launching velocity between adjacent surface regions and ultimately to interaction between slower and faster material. Due to rotation, CIRs may subsequently develop through a spiral pattern, explaining that DACs are observed to move out of P-Cygni profiles with time. Optical observations also show line profile variability in the strongest wind lines of O stars, especially H$\alpha$ \citep[e.g.][]{dejong01}. \citet{morel04} reported H$\alpha$ variability in all their sample stars showing wind dominated lines. \citet{markova05} studied a sample of O supergiants and concluded that the detected H$\alpha$ variability was usually consistent with a model of broken shells for the atmosphere, at least for the stars with the strongest winds. Such a model is a way of representing the density fluctuations caused by the line driving instability. Lines formed closer to the photosphere also vary. \citet{ful96} conducted a survey of 30 O stars and detected variability in HeI~5876 and sometimes CIV~5802-5812 for 77\% of their sample, indicating that spectroscopic variability is widespread among O stars even in the deepest parts of the atmosphere. Subsequent studies have revealed variability in both wind and photospheric lines of O stars \citep{dejong01,kaufer02,prinja04,prinja06}. 

Whether there is a direct causal link between photospheric, surface variability and wind variability is still a matter of debate. It is not clear if the line driving instability can act on its own, at rather large velocities, or if it is an amplifier of deeply seated variations. \citet{dejong01} investigated this question in the case of the O7.5III star $\xi$ Per, using UV and optical spectroscopy. Although variability was observed in photospheric and wind features, a direct link was not obvious because of the different timescales of these variations. \citet{kaufer06} reached similar conclusions for the B0.5Ib supergiant HD~64760, although part of the H$\alpha$ variability has a periodicity comparable to that of lines formed closer to the photosphere. On the theoretical ground, surface variability is expected if the star experiences non-radial pulsations, as is the case of several O stars \citep{smith78,kambe90,ful91,dejong99,degroote10}. Surface inhomogeneities, possibly caused by magnetic spots, could also trigger variability.

To further test this relationship between photospheric and wind variability, more spectroscopic time series of O stars are required. As many photospheric and wind lines as possible should be monitored in order to sample the transition region between the stellar surface and the wind-dominated atmosphere. The classical indicators are usually H$\alpha$ and \ion{He}{I}~5876. In this study, we consider 10 additional lines probing the inner part of the wind, down to the photosphere. 
We report on the spectroscopic variability of seven late-O/early-B stars using high resolution, high signal-to-noise spectra. We quantify the degree of variability by means of the Temporal Variance Spectrum (TVS) analysis developed by \citet{ful96}. We also model the average profiles with the atmosphere code CMFGEN. Based on the models, we constrain the line formation region of each line. We investigate the radial dependence of variability throughout the photosphere-wind transition region of the sample stars. Our spectroscopic analysis also provides the fundamental parameters of the sample stars. 

The paper is organized as follows: we present the observations in Sect.\ \ref{s_obs}; the line profile variability is presented in Sect.\ \ref{s_var}; we describe the modelling of the stars and the determination of the line formation region in Sect.\ \ref{s_spatialvar}; we discuss our results in comparison to previous studies in Sect.\ \ref{s_comp} and present our conclusions in Sect.\ \ref{s_conc}.


\section{Observations and data reduction}
\label{s_obs}

The sample stars were initially selected as part of a program aiming at searching magnetic fields in massive stars. They are bright enough to allow repeated and frequent observations on a two meter telescope. Previous variability in the giants and supergiants was used as a potential indication of magnetic field, based on previous detections \citep[e.g.][]{donati02,donati06}. The dwarf stars were objects with very weak winds which, according to \citet{ww05} could be due to the presence of a magnetic field. We selected the stars with a sufficient number of spectra to allow a variability study. In total, seven O stars with spectral type from O9 to B0.5 and luminosity classes from V to Iab were chosen. None of them is a known binary.

We used the spectropolarimeter NARVAL mounted on the \textit{T\'elescope Bernard Lyot} at the Pic du Midi observatory to obtain high signal-to-noise, high spectral resolution data \citep[see][]{bouret08,martins10}. The observations were obtained between October 16$^{th}$ and October 25$^{th}$ 2007, in June 2008 and between July 20$^{th}$ and August 4$^{th}$ 2009. NARVAL provides \'echelle spectra between 3700 \AA\ and 1.05 $\mu$m. The signal to noise ratio per 2.6\kms\ resolution bin depends on weather conditions and wavelength but is always larger than 200. The resolving power is 65\,000. Table \ref{tab_obs} gives the journal of observations. 

The data were automatically reduced on site by the \textit{Libre Esprit} software, a fully automatic reduction package for NARVAL data. Ample details on the reduction process can be obtained in \citet{donati97} and we refer the reader to this paper for information.

\section{Line profile variability}
\label{s_var}

To study the spectral variability of our sample stars, we have computed the Temporal Variance Spectrum (TVS) first introduced by \citet{ful96} and defined as follows:

\begin{equation}
TVS=\frac{1}{n-1}\sum_{i=1}^{n} \frac{w_{i}}{F_{i}} (F_{i}-F_{av})^{2} 
\end{equation}

\noindent where $F_{i}$ is the rectified flux, $F_{av}$ the mean rectified flux, $n$ the number of spectra available and $w_{i}$ a weighting factor taking into account the variation of the signal-to-noise ratio from spectrum to spectrum \citep[see Eq.\ 11 of][]{ful96}. Given the relatively large exposure time, we have assumed that the photon noise was the main source of noise, implying that the factor $\alpha_{ij}$ of Fullerton et al., encompassing the wavelength dependence of noise, is equal to the flux $F_{i}$. The TVS is computed at all wavelengths and quantifies the degree of variability in a series of spectra. We have checked the effect of normalization uncertainties on the resulting TVS. One of us has normalized the same spectrum five times, and used the five resulting spectra to compute the TVS. The intensity we obtain is usually lower than 0.005, and is seen over the entire spectrum, not just in lines. We present two types of TVS depending on the data available: TVS computed for one night, when at least five spectra are available for the night; or TVS computed for several days. In the latter case, we use the average spectrum of the night (when several spectra are available) for each date.  

We have selected twelve spectral lines to look for spectroscopic variability. In addition to the classical H$\alpha$ and \ion{He}{I}~5876 features, we have added H$\beta$, H$\gamma$ which probe deeper layers than H$\alpha$ due to their reduced oscillator strength. We have also included three other \ion{He}{I} lines  (\ion{He}{I}~4471, \ion{He}{I}~4026 and \ion{He}{I}~4712 in order of decreasing oscillator strength\footnote{\ion{He}{I}~5876 being strongest of all \ion{He}{I} lines.}), three \ion{He}{II} lines (\ion{He}{II}~4542, \ion{He}{II}~4686, \ion{He}{II}~5412), \ion{O}{III}~5592 and \ion{C}{IV}~5802. The high ionization lines (\ion{C}{IV}, \ion{He}{II}, \ion{O}{III}) probe deeper layers of the atmosphere compared to Balmer lines. The \ion{He}{I} lines are triplets and belong to the 2p\,$^3$P$^o$ series. With these lines, we are able to sample variability throughout the atmosphere of the sample stars. 

We used continuum regions adjacent to the lines of interest to evaluate the one and three $\sigma$ levels. The one $\sigma$ level is set to the standard deviation around the mean value of the TVS in this continuum region.

\subsection{Daily variability}
\label{s_var_day}

\begin{figure*}
     \centering
     \subfigure[\ha]{
          \includegraphics[width=.28\textwidth]{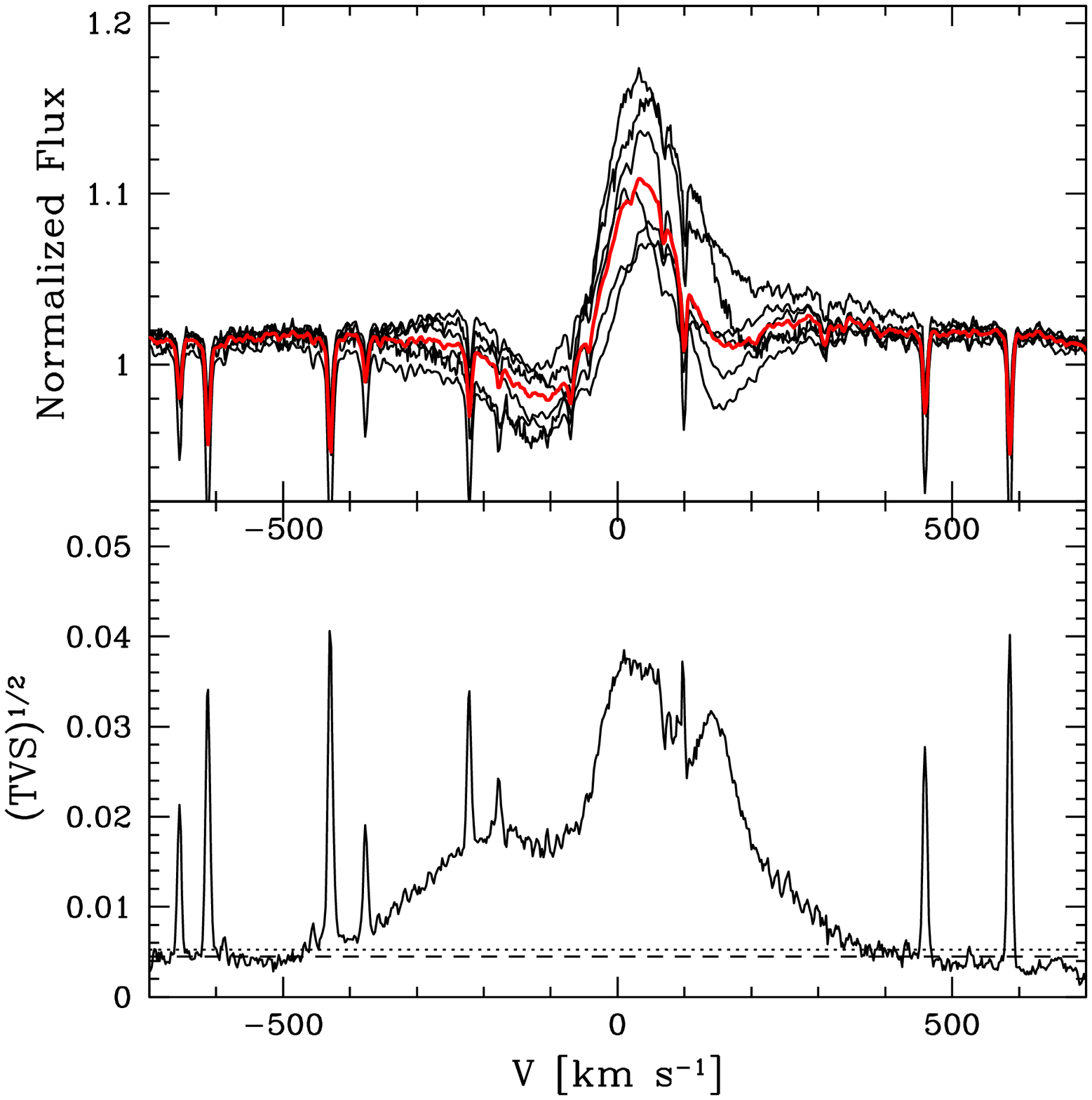}}
     \hspace{0.2cm}
     \subfigure[H$_{\beta}$]{
          \includegraphics[width=.28\textwidth]{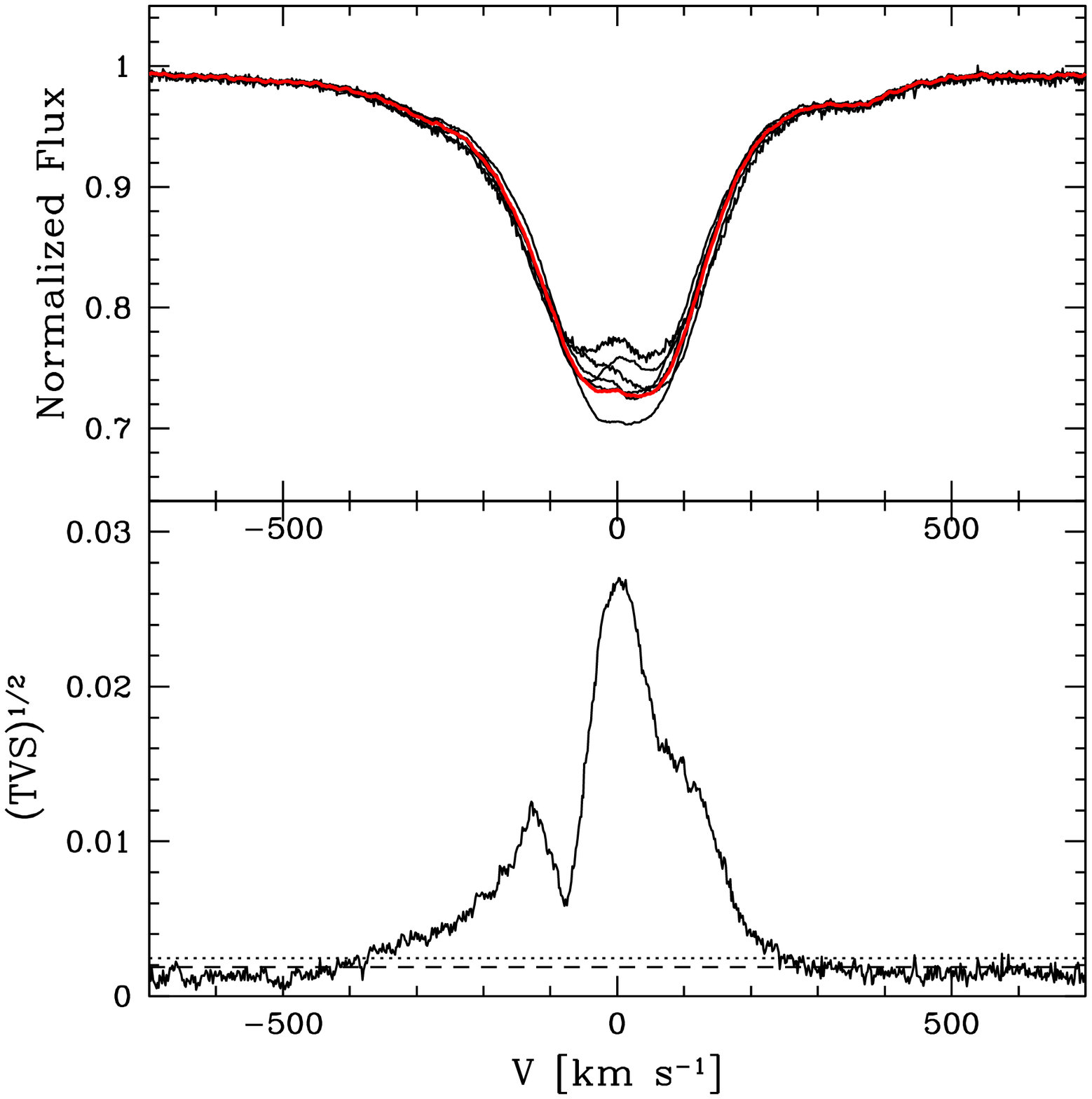}}
     \hspace{0.2cm}
     \subfigure[H$_{\gamma}$]{
          \includegraphics[width=.28\textwidth]{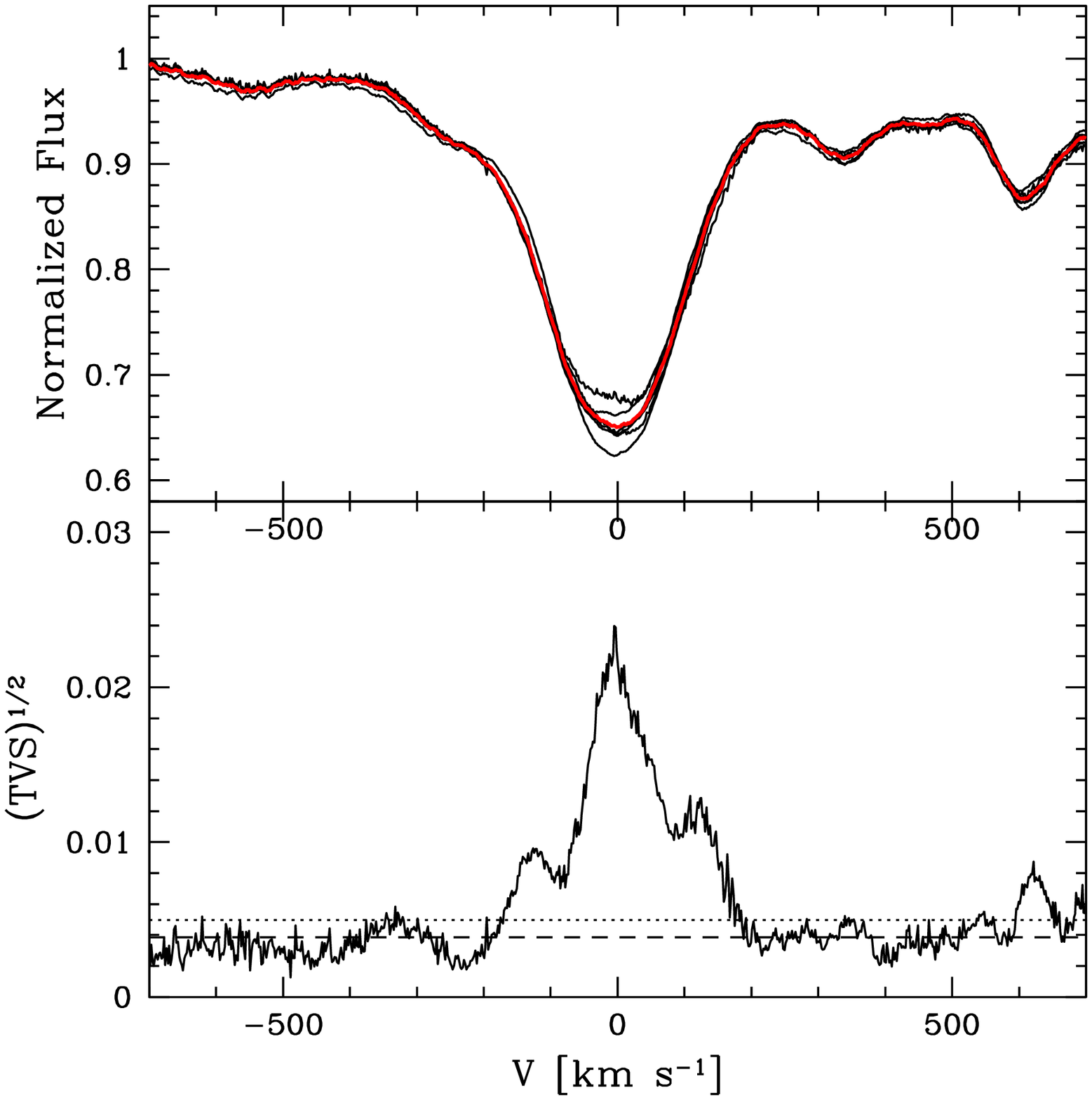}}\\
     \subfigure[\ion{He}{I} 4026]{
          \includegraphics[width=.28\textwidth]{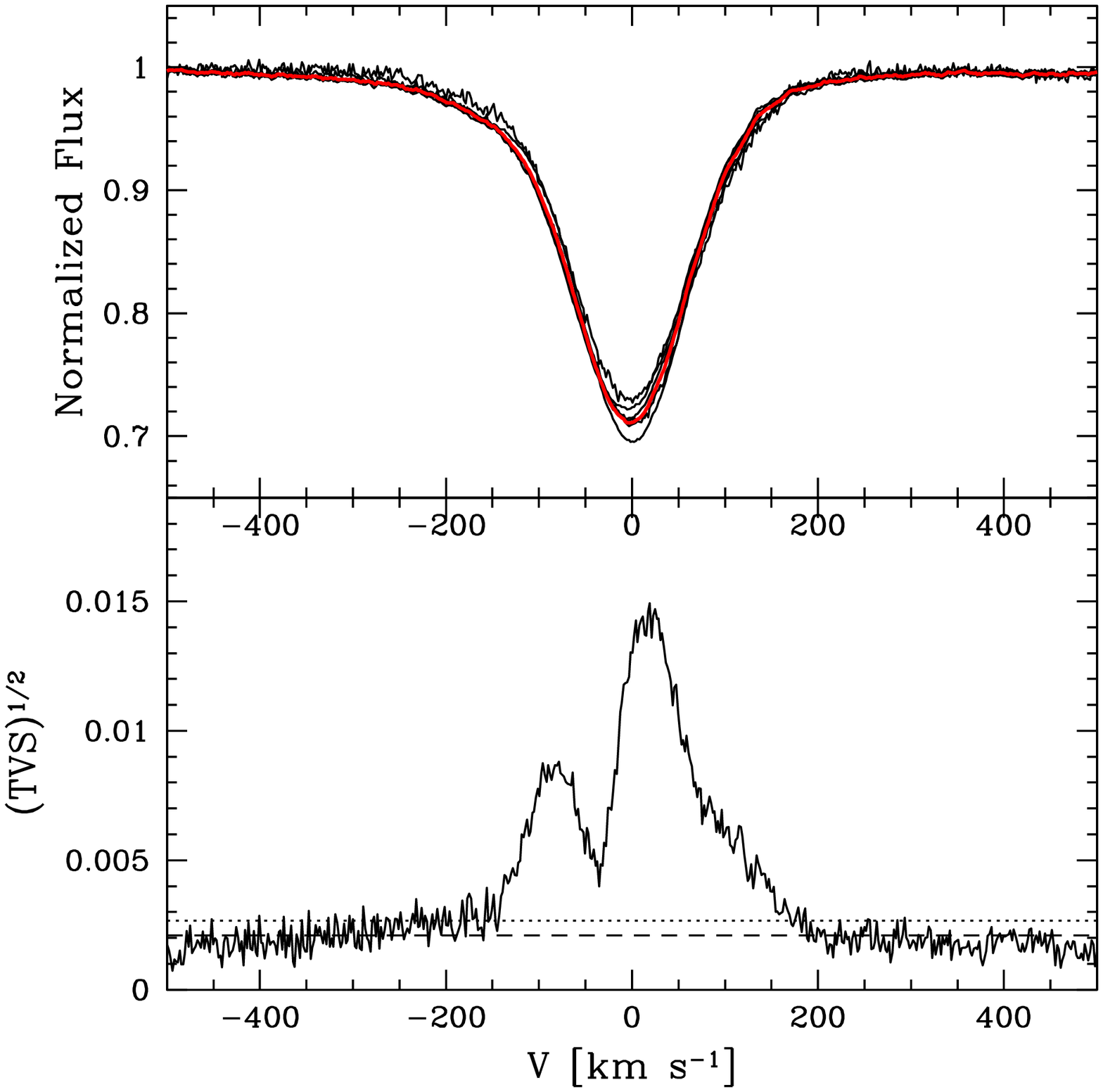}}
     \hspace{0.2cm}
     \subfigure[\ion{He}{I} 4471]{
          \includegraphics[width=.28\textwidth]{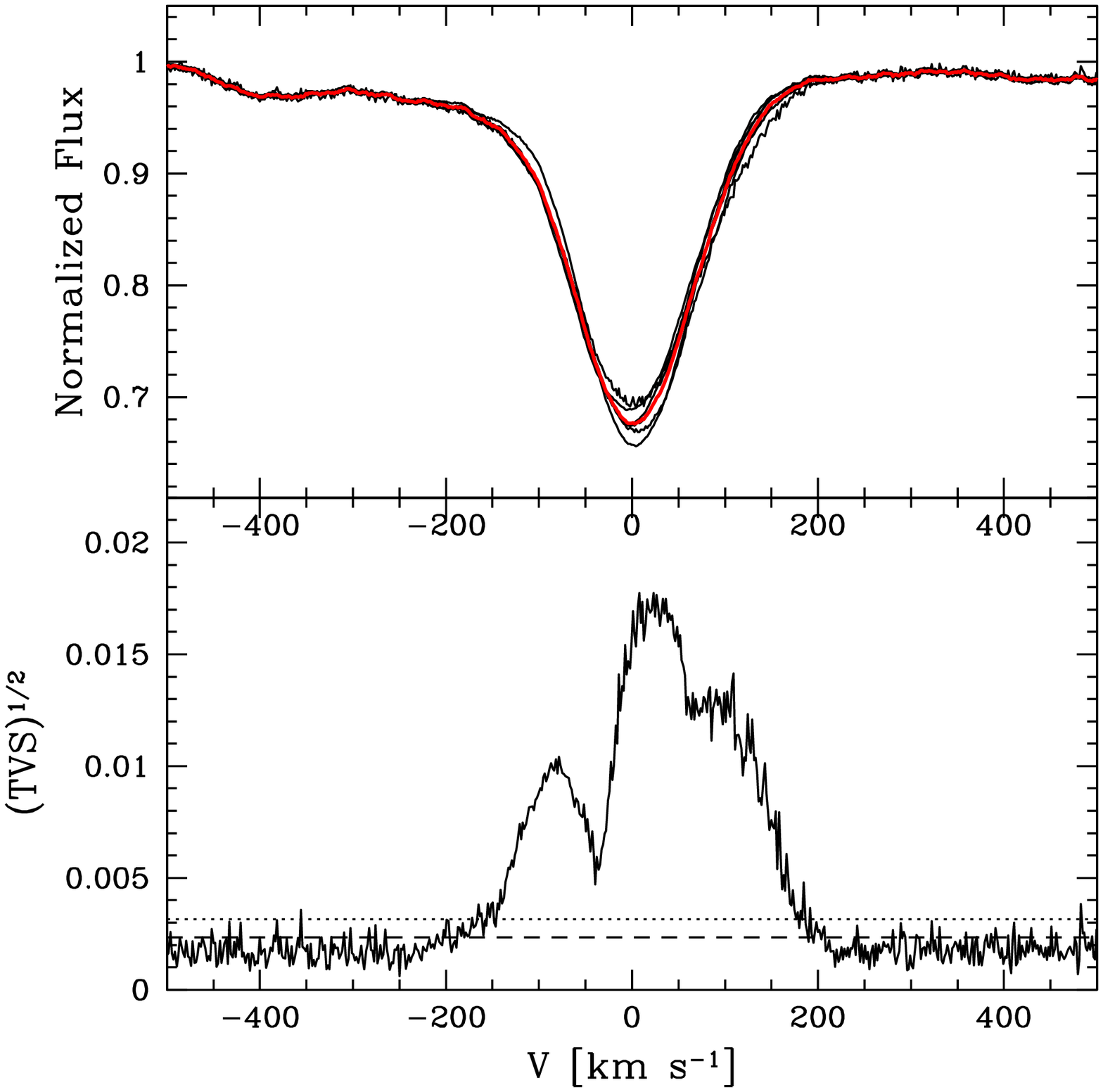}}
     \hspace{0.2cm}
     \subfigure[\ion{He}{I} 4712]{
          \includegraphics[width=.28\textwidth]{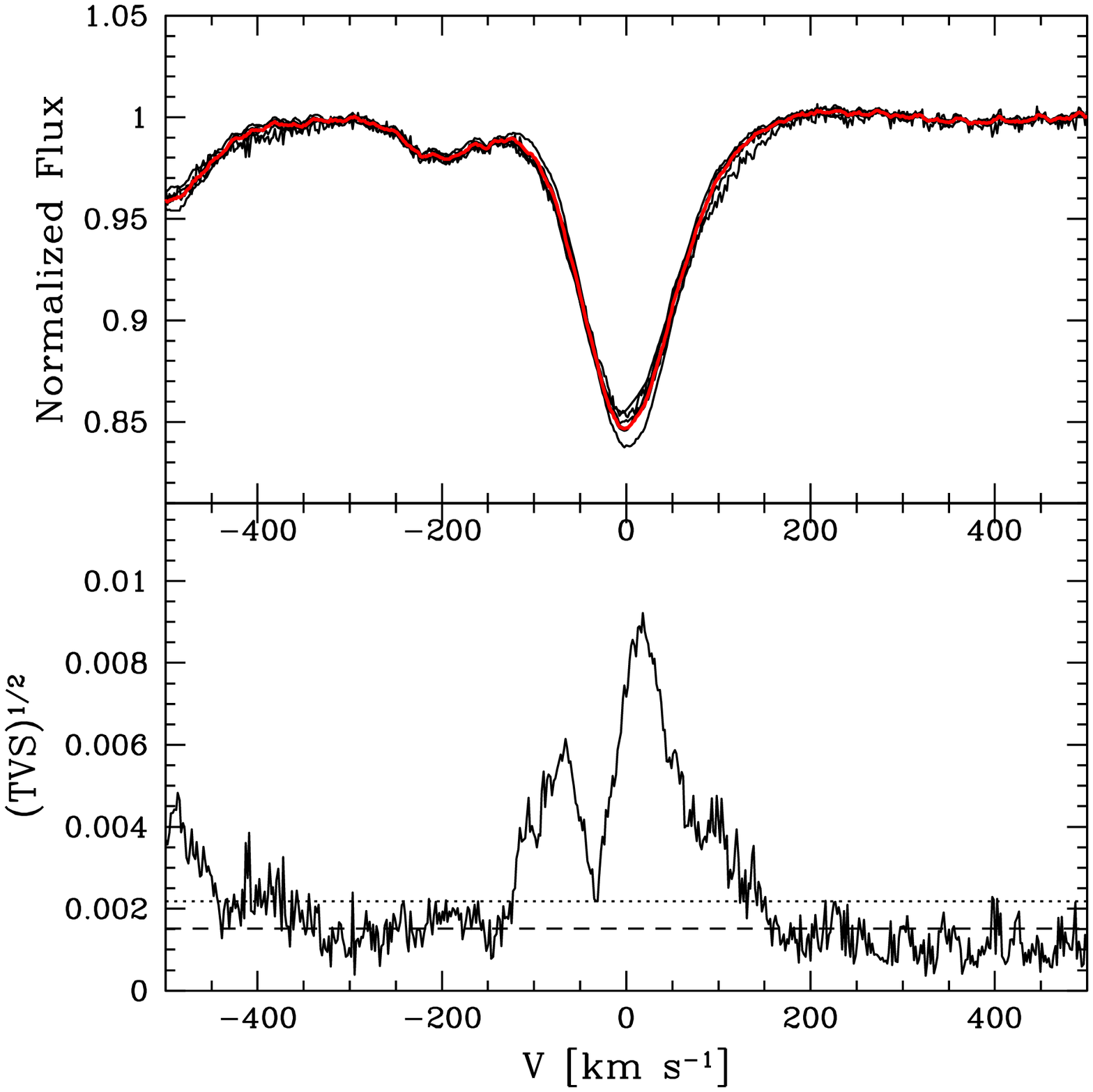}}\\
     \subfigure[\ion{He}{I} 5876]{
          \includegraphics[width=.28\textwidth]{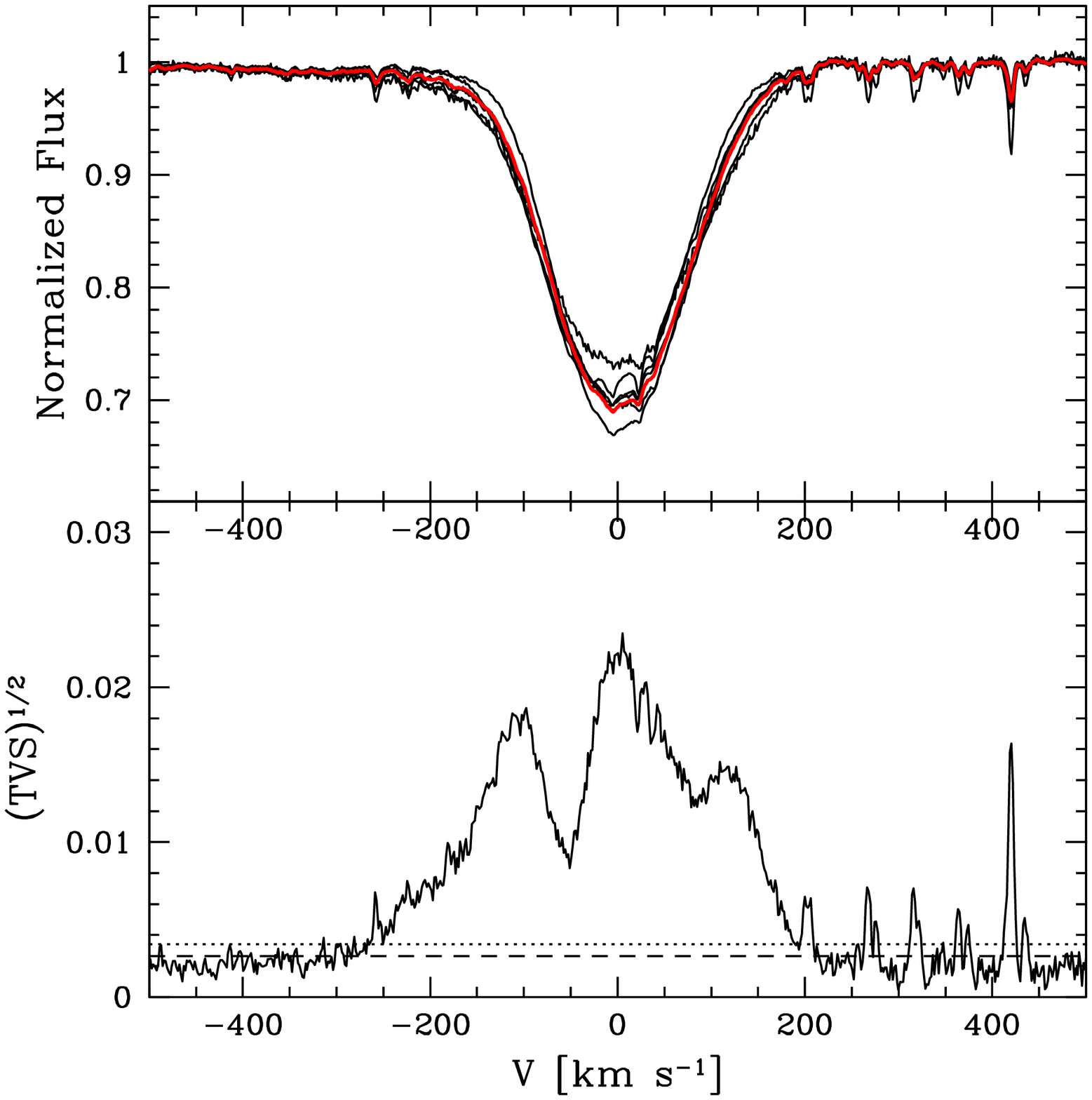}}
     \hspace{0.2cm}
     \subfigure[\ion{He}{II} 4542]{
          \includegraphics[width=.28\textwidth]{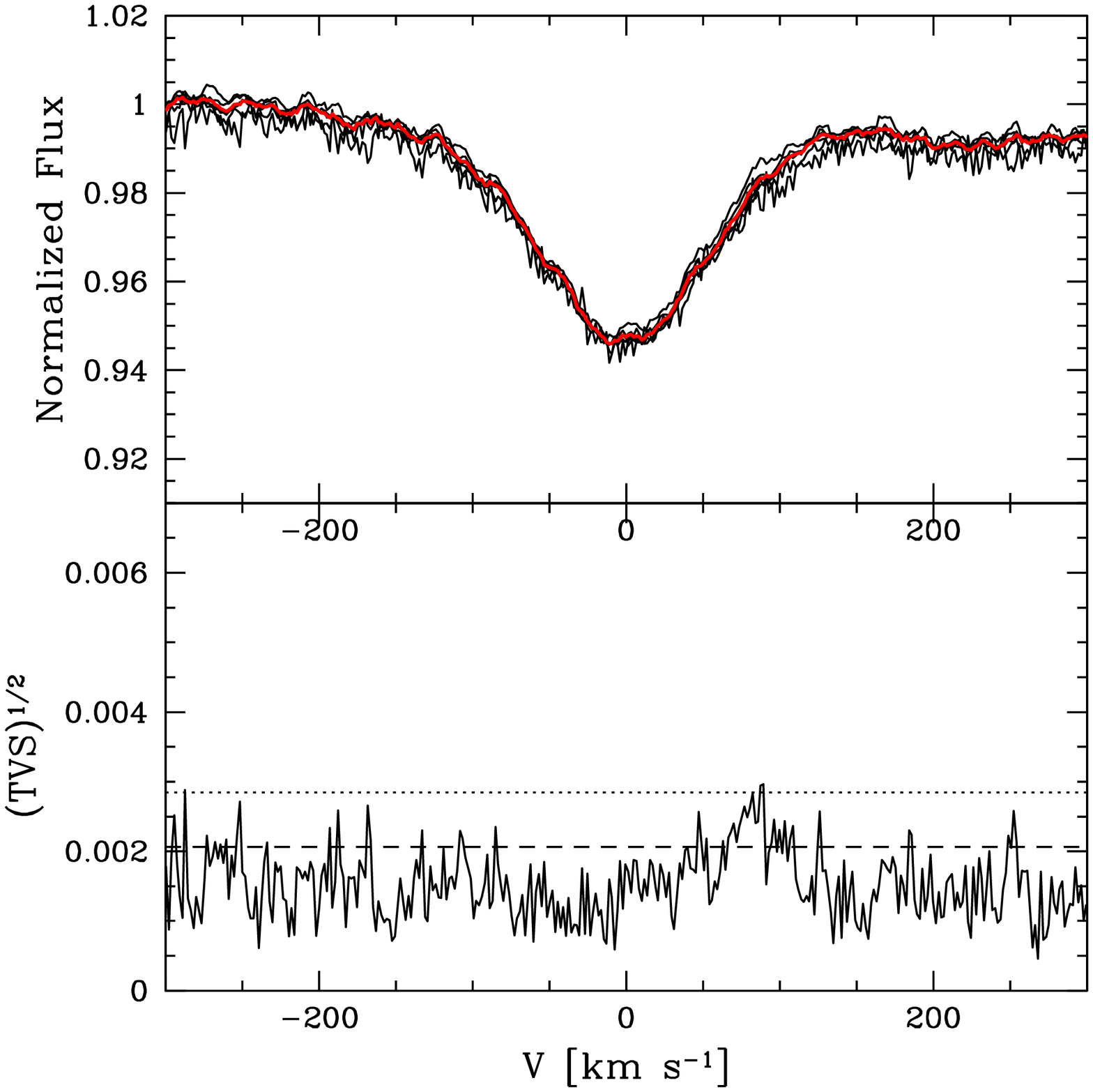}}
     \hspace{0.2cm}
     \subfigure[\ion{He}{II} 4686]{
          \includegraphics[width=.28\textwidth]{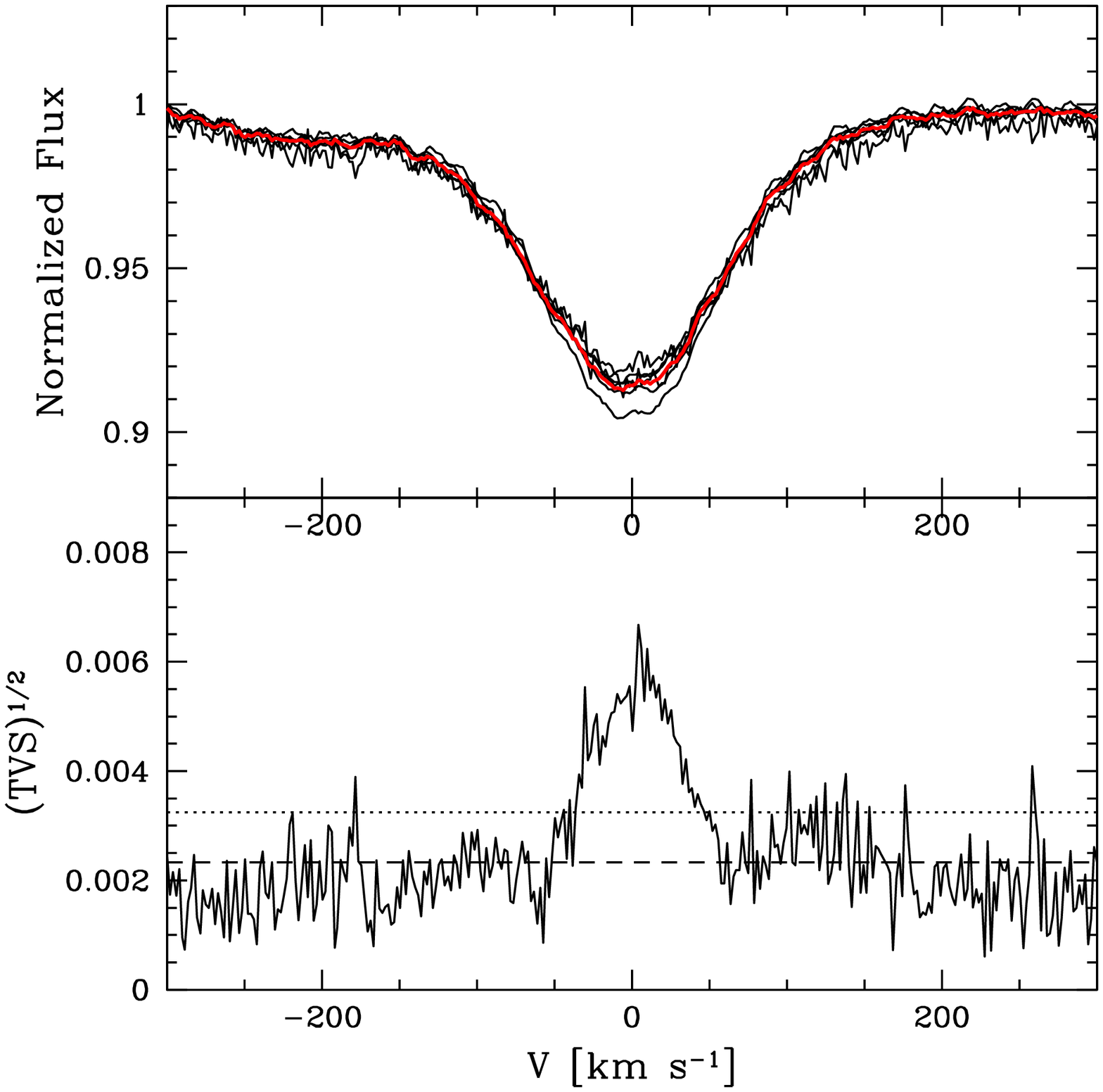}}\\
     \subfigure[\ion{He}{II} 5412]{
          \includegraphics[width=.28\textwidth]{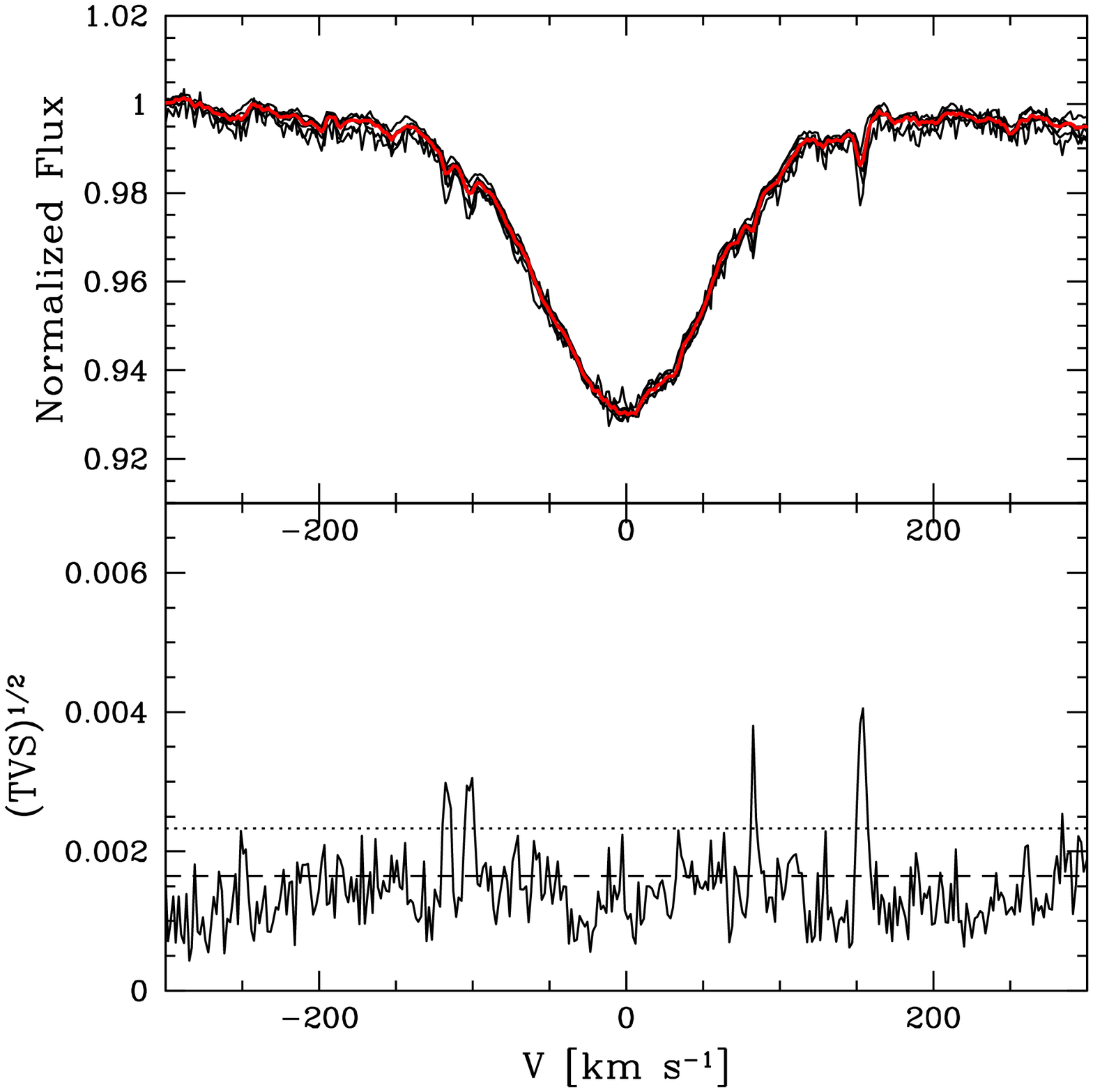}}
     \hspace{0.2cm}
     \subfigure[\ion{O}{III} 5592]{
          \includegraphics[width=.28\textwidth]{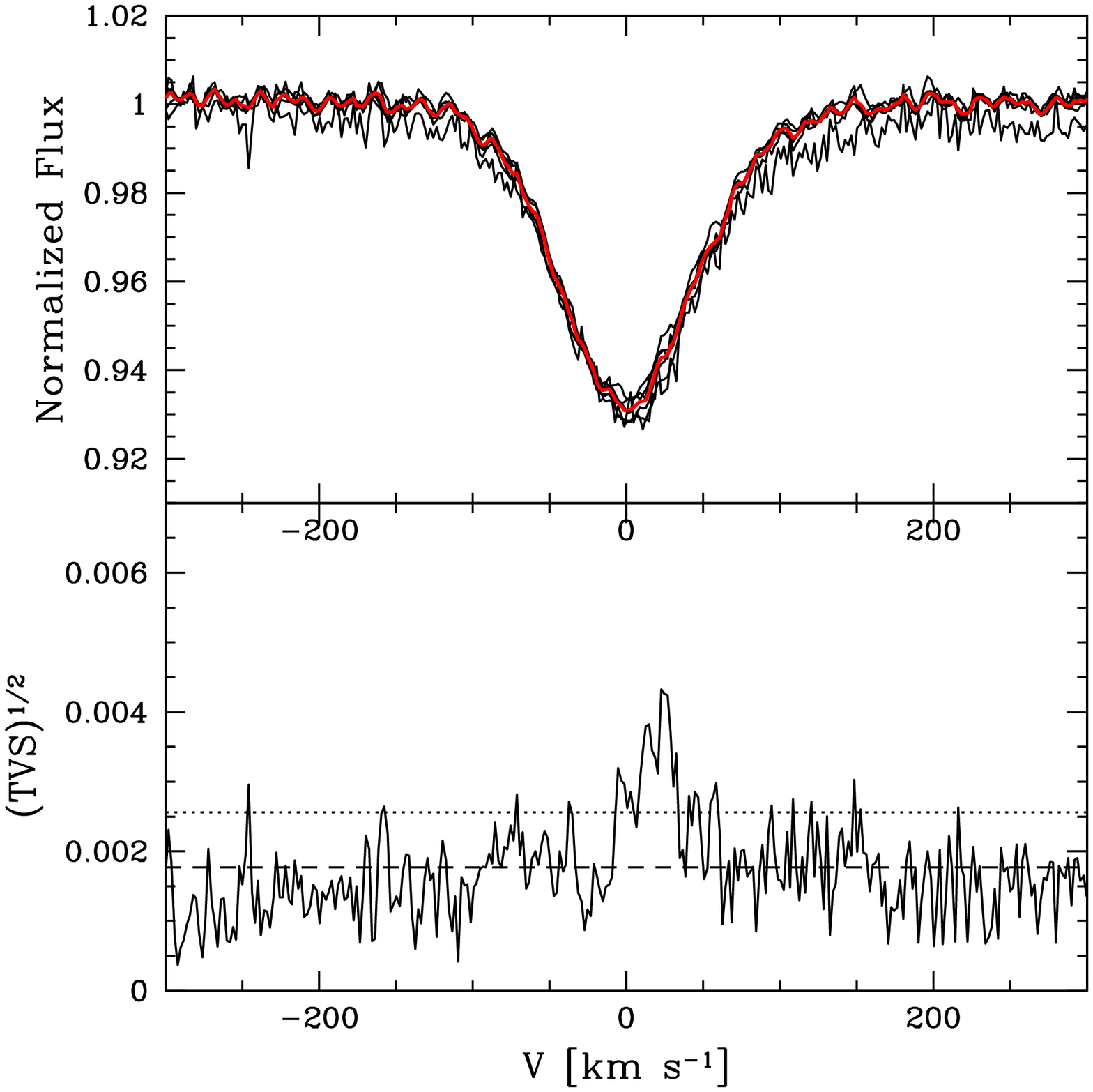}}
     \hspace{0.2cm}
     \subfigure[\ion{C}{IV} 5802]{
          \includegraphics[width=.28\textwidth]{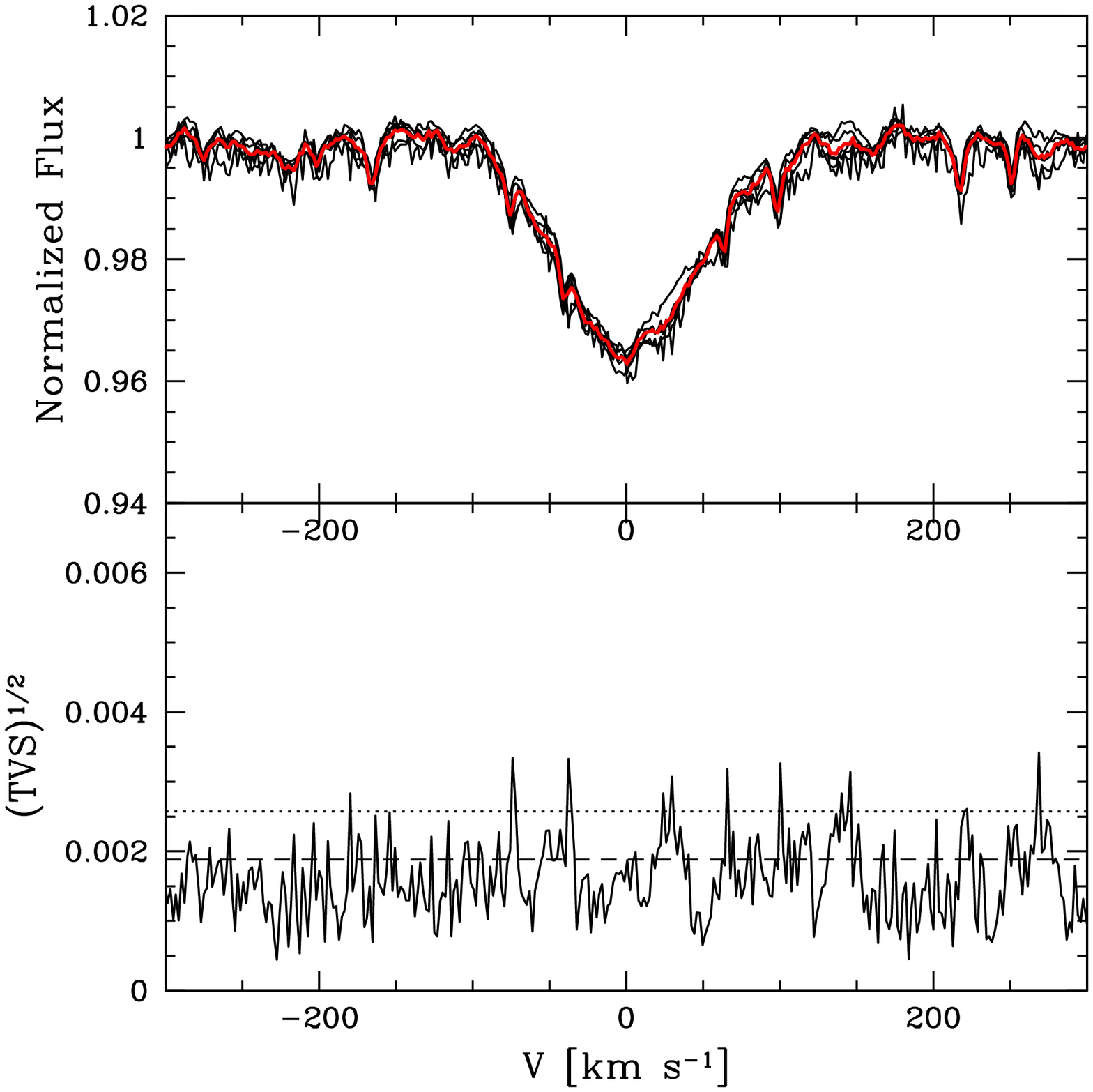}}\\
     \caption{Variability of $\epsilon$ Ori on a daily timescale (from 16$^{th}$ to 25$^{th}$ October 2007). Upper panels: individual spectra  in black (each spectrum corresponds to a night average) and global average spectrum in red. Lower panels: Temporal Variance Spectrum, together with the one sigma (dashed line) and three sigma (dot-dashed line) limits.}
     \label{fig_var_months}
\end{figure*}

In Fig.\ \ref{fig_var_months} we show the TVS and individual spectra of $\epsilon$ Ori between October 16$^{\rm th}$ and October 25$^{\rm th}$ 2007. We detect variability in all H and \ion{He}{I} lines, as well as in \ion{He}{II}~4686. The other \ion{He}{II} lines, as well as \ion{O}{III}~5592 and \ion{C}{IV}~5802 do not show any significant variability. The shape of the TVS is similar in all lines for which variability is detected. There is a triple peak system, with the central peak being the strongest. It corresponds to variability in the line core. The two secondary peaks are located in the line wings, relatively far from the line core, at velocities up to 500 \kms\ in \ha. 
Variability is usually larger and more extended in stronger lines. For example, the TVS reaches 0.042 in H$\alpha$, 0.024 in H$\beta$ and 0.020 in H$\gamma$. Variability extends up to 500 \kms\ in \ha, 200 \kms\ in H$\beta$ and 100 \kms\ in H$\gamma$. The same trend of larger variability in stronger lines is seen in \ion{He}{I} with the largest TVS maximum in \ion{He}{I} 5876, then \ion{He}{I}~4471, \ion{He}{I}~4026 and \ion{He}{I}~4712, corresponding to decreasing oscillator strength (see above).

We gather in Figs.\ \ref{fig_var_15sgr} to \ref{fig_var_10lac} the same figures as Fig.\ \ref{fig_var_months} for the other sample stars. The variability of these stars is summarized as follows:

\begin{itemize}

\item {\it HD~167264}: this star shows variability in the same lines as $\epsilon$~Ori. The amplitude of the TVS is also of the same order (e.g. 0.04 for H$\alpha$). The shape of the TVS is mostly single peaked, except for \ion{He}{I}~4026, \ion{He}{I}~4471 and \ion{He}{I}~5876 for which a double peak structure is observed. 

\item {\it HD~207198}: variability is observed in the H and \ion{He}{I} lines, but also in all photospheric lines (\ion{He}{II}, \ion{O}{III}, \ion{C}{IV}) although marginally in \ion{He}{II}~4542. The shape of the TVS is mostly double peaked, as in HD~167264, with the red peak stronger than the blue one in the most strongly variable lines. In the \ion{He}{I} lines, the amplitude of the blue peak in the TVS decreases with decreasing oscillator strength (it vanishes in \ion{He}{I}~4712). 

\item {\it HD~188209}: it is variable in all lines, photospheric or wind dominated. The Balmer lines are mostly triple peaked, the red component being the strongest. The TVS of the \ion{He}{I} lines is double peaked, with the red peak being usually stronger than the blue one. The high ionization lines (from \ion{C}{IV}, \ion{O}{III}, \ion{He}{II}) are clearly variable.

\item {\it HD~209975}: as in HD~188209, all lines are showing some degree of variability, with rather similar amplitudes. A double peak is seen in the TVS with an asymmetry red/blue, the red peak being stronger at least in the most variable lines. 

\item {\it 10~Lac and AE~Aur}: both late O dwarfs do not show any evidence for variability in any of the lines. The spectra are very stable over several nights. 

\end{itemize}

In general, we observe strong variability in giants/supergiants on a daily timescale. All low ionization lines (\ion{H}{I} and \ion{He}{I}) are variable. In two cases ($\epsilon$ Ori and HD~167264), the high ionization lines (\ion{He}{II}, \ion{O}{III} and \ion{C}{IV}) are stable over days, while in the three other stars (HD~188209, HD~207198, and HD~209975) they also show clear changes from day to day. The former stars are B0-0.5 supergiants, while the latter are O9-9.5 supergiants. The two late-type O dwarfs (AE~Aur and 10~Lac) do not show any variability on a daily timescale. 

\subsection{Hourly variability}
\label{s_var_hour}

If we now look at Figs.\ \ref{fig_var_epsori_day} to \ref{fig_var_209975_day}, we see how the lines vary during a few hours (for the stars for which we have at least five spectra over a night).

\begin{figure}
\centering
\includegraphics[width=9cm]{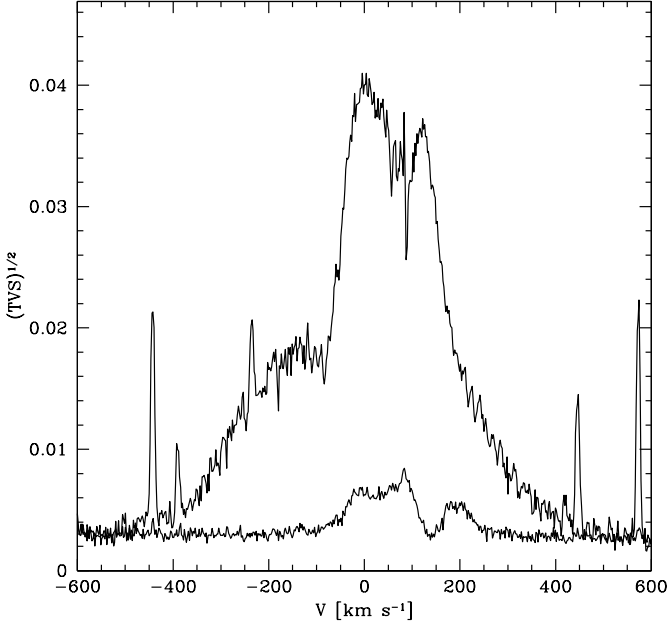}
\caption{Daily and hourly H$\alpha$ TVS of $\epsilon$ Ori. The largest TVS is for the daily timescale. The hourly TVS is observed during the night of October 19$^{th}$ 2007. } \label{comp_var_ha}
\end{figure}

\begin{itemize}

\item {\it $\epsilon$ Ori}: the changes during the night of October 19$^{th}$ 2007 are shown. The amplitude is reduced compared to Fig.\ \ref{fig_var_months} (e.g. 0.008 instead of 0.042 for H$\alpha$) but the same lines show variability (and the same lines remain stable). Fig.\ \ref{comp_var_ha} compares the TVS of the night of October 19$^{th}$ 2007 to the TVS obtained on the night-averaged spectra over the October 2007 month, for the H$\alpha$ line. We see that on a timescale of hours, the variability is smaller than on a daily timescale. In addition, the strong variability observed over days extends up to 500 \kms\ away from the line core. The shorter timescale variability is restricted to the line cores, either because the physical origin of this shorter timescale variability is different, or because the level of variability at high velocities is too small to be detected with our datasets.
We have also computed the TVS during the night of October 22$^{nd}$ and October 25$^{th}$ 2007, but we did not detect any significant variability. The number of spectra available is much larger during the night of October 19$^{th}$, but if we compute the TVS using only a sample of six spectra taken during that night, the variability is still present. Thus, the lower number of spectra available for the night of October 22$^{nd}$ and October 25$^{th}$ does not explain the non detection of variability.   

\item {\it HD~188209}: Fig.\ \ref{fig_var_188209_day} reveals that HD~188209 is variable in all lines also on an hourly timescale during the night of June 25$^{th}$ 2008. The variations in the high ionization lines are marginal. The variability is observed again during the night of June 27$^{th}$ and 28$^{th}$ 2008. 

\item {\it HD~209975}: during the night of June 28$^{th}$ 2008, variability is observed in all lines except \ion{He}{II}~4542 and \ion{C}{IV}~5802. OIII~5592 is also very marginally variable. The variability is smaller than that observed on a daily timescale. The selected lines are also variable on the night of June  21$^{st}$, 22$^{nd}$ and 27$^{th}$. The amplitude of the variations on June 21$^{st}$ are smaller than on June 28$^{th}$, and only the most strongly variable lines are actually changing during that night. The time span of the observations is the same during all nights (about 2 hours).

\end{itemize}

The general conclusion is that stars showing variability on a daily timescale can also vary over a few hours. The variability on this shorter timescale is smaller (in intensity and velocity), and sometimes rescticted to the strongest wind lines. Conversely, if a line does not vary on a daily basis, it does not vary either on an hourly timescale. If the processes responsible for the daily and hourly variability are the same, then our observations indicate that the associated timescale is longer than days. However, if the processes are different, Fig.\ \ref{comp_var_ha} suggests that the one acting on longer timescales leads to stronger variability than that acting on timescales of hours. 

\section{Spatial distribution of variability}
\label{s_spatialvar}

\subsection{Stellar and wind parameters}
\label{s_lineform}

In order to better understand the origin of the observed variability, we have identified the line formation region for all features displayed in Fig.\ \ref{fig_var_months} and Figs \ref{fig_var_15sgr} to \ref{fig_var_209975_day}. For that purpose, we have analyzed the spectroscopic data with atmosphere models computed with the code CMFGEN \citep{hm98}. Our goal was to best reproduce the observed spectra to constrain the atmosphere structure and thus the line formation regions, directly accessible from the models. In practice, we have worked on the grand average of all optical spectra for each star. In order to constrain the wind structure, we also relied on far--UV spectra retrieved from the \textit{IUE} archive. We used averaged spectra to have the highest possible signal-to-noise ratio, and to provide average parameters for the sample stars. 

We used different diagnostics to compute the best fitting model for each star. In practice, we have proceeded as follows to determine those parameters:

\begin{itemize}

\item[$\bullet$] {\it Effective temperature}: we used the ionization ratio method to constraint \teff. The \ion{He}{I} and \ion{He}{II} lines were the main indicators: \ion{He}{I}~4026, \ion{He}{I}~4388, \ion{He}{I}~4471, \ion{He}{I}~4712, \ion{He}{I}~4920, \ion{He}{I}~5876, \ion{He}{II}~4200, \ion{He}{II}~4542, \ion{He}{II}~5412. \ion{He}{II} 4686 is not considered since it also depends on the wind density.

\item[$\bullet$] {\it Gravity}: the wings of the Balmer lines are the main diagnostics of \logg. They are broader at larger gravities. We used, H$_{\beta}$, H$_{\gamma}$,  H$_{\delta}$,  H$_{\epsilon}$, \ion{H}{I}~3889 and \ion{H}{I}~3835. The typical uncertainty on \logg\ is 0.1 dex.

\item[$\bullet$] {\it Luminosity}: we fitted the UV-optical-infrared SED to constrain the luminosity. When reliable (i.e. the uncertainty is lower than 30\%), we used the Hipparcos parallaxes as constraints on the distances. The distances we used are reported in Table \ref{tab_param}. We used the Galactic extinction law of \citet{seaton79} and \citet{howarth83} to constrain the extinction. 

\item[$\bullet$] {\it Surface abundances}: C and N abundances have been derived using the numerous carbon and nitrogen lines present in the optical range. \ion{C}{III}~4070, \ion{C}{III}~4153-56-63, \ion{C}{III}~4326, \ion{C}{III}~5305, \ion{C}{III}~6205 were the main carbon abundance indicator. For nitrogen, we used \ion{N}{II}~3995, \ion{N}{II}~4041, \ion{N}{III}~4196, \ion{N}{III}~4216, NII~4447, \ion{N}{III}~4511, \ion{N}{III}~4515, \ion{N}{III}~4523, \ion{N}{III}~4602, NII~4607, NII~4621, \ion{N}{III}~4907, \ion{N}{II}~5676, \ion{N}{II}~5679. 

\item[$\bullet$] {\it Mass loss rate}: the mass loss rate was determined from the strength of the UV P--Cygni lines (\ion{C}{III}~1176, \ion{N}{V}~1238-1242, \ion{C}{III}~1247, \ion{Si}{IV}~1393--1403, \ion{C}{IV}~1548--1550) and from H$_{\alpha}$. The typical uncertainty on \mdot\ is 0.2 dex.

\item[$\bullet$] {\it Wind terminal velocity}: \vinf\ was directly measured from the blueward extinction of the UV P--Cygni profiles. A typical uncertainty of 100\kms\ is achieved. 

\item[$\bullet$] {\it Clumping}: clumping is treated by means of an exponential law in CMFGEN, where the volume filling factor $f$ has the following dependence on velocity: $f=f_{\infty}+(1-f_{\infty})e^{-\frac{v}{v_{cl}}}$. $f_{\infty}$ is the maximum clumping factor at the top of the atmosphere, and $v_{cl}$ determines the velocity at which clumping becomes non negligible. We adopted $v_{cl}$=30 \kms\ and we derived $f_{\infty}$ from the shape of NIV 1720 line \citep{jc05}. 

\item[$\bullet$] {\it velocity slope}: we constrained the slope of the velocity field - the so-called $\beta$ parameter - from the condition that both H$\alpha$ and the UV lines should be fitted for the same mass loss rate. Increasing the value of $\beta$ allows to reproduce the emission peak sometimes observed at the core of the broader H$\alpha$ absorption. The UV P-Cygni profiles are usually reproduced simultaneously. For the almost pure photospheric H$\alpha$ profiles of AE~Aur end 10~Lac, $\beta$ was adopted.

\item[$\bullet$] {\it rotational velocity / macroturbulence}: we used the Fourier transform method \citep{sergio07} to constrain the projected rotational velocity (\vsini). We relied on the averaged spectrum to benefit from the highest signal-to-noise ratio. We used \ion{He}{I}~4712 and \ion{O}{III}~5592 as diagnostics. Once the value of \vsini\ was estimated, we used synthetic spectra (from our private database) to determine the macroturbulent velocity. We approximated macroturbulence by a Gaussian broadening. We tested several values of \vmac\ to best reproduce the diagnostic lines, for the estimated \vsini. The best fit (selected by eye) provided \vmac.   

\end{itemize}

The internal velocity structure of the atmosphere was iterated: we first converged the populations, then computed the resulting radiative acceleration and included it in a solution of the hydrodynamical equation to obtain a new velocity field. The populations was then re-converged with this hydrodynamical structure and the process was iterated. After a couple of these hydrodynamical iterations, the structure was usually consistent with the radiative transfer solution, except around the sonic point where a discrepancy was still present. Once the model atmosphere was converged, we performed a formal solution of the radiative transfer equation adopting a microturbulent velocity ranging from 10 \kms\ in the photosphere to 0.1 $\times$ \vinf\ in the outer atmosphere. The resulting spectrum was compared to the observed spectra to constrain the stellar and wind parameters. The results of the spectroscopic analysis are gathered in Tab.\ \ref{tab_param}. 

\begin{figure*}
\begin{center}
\includegraphics[width=0.49\textwidth]{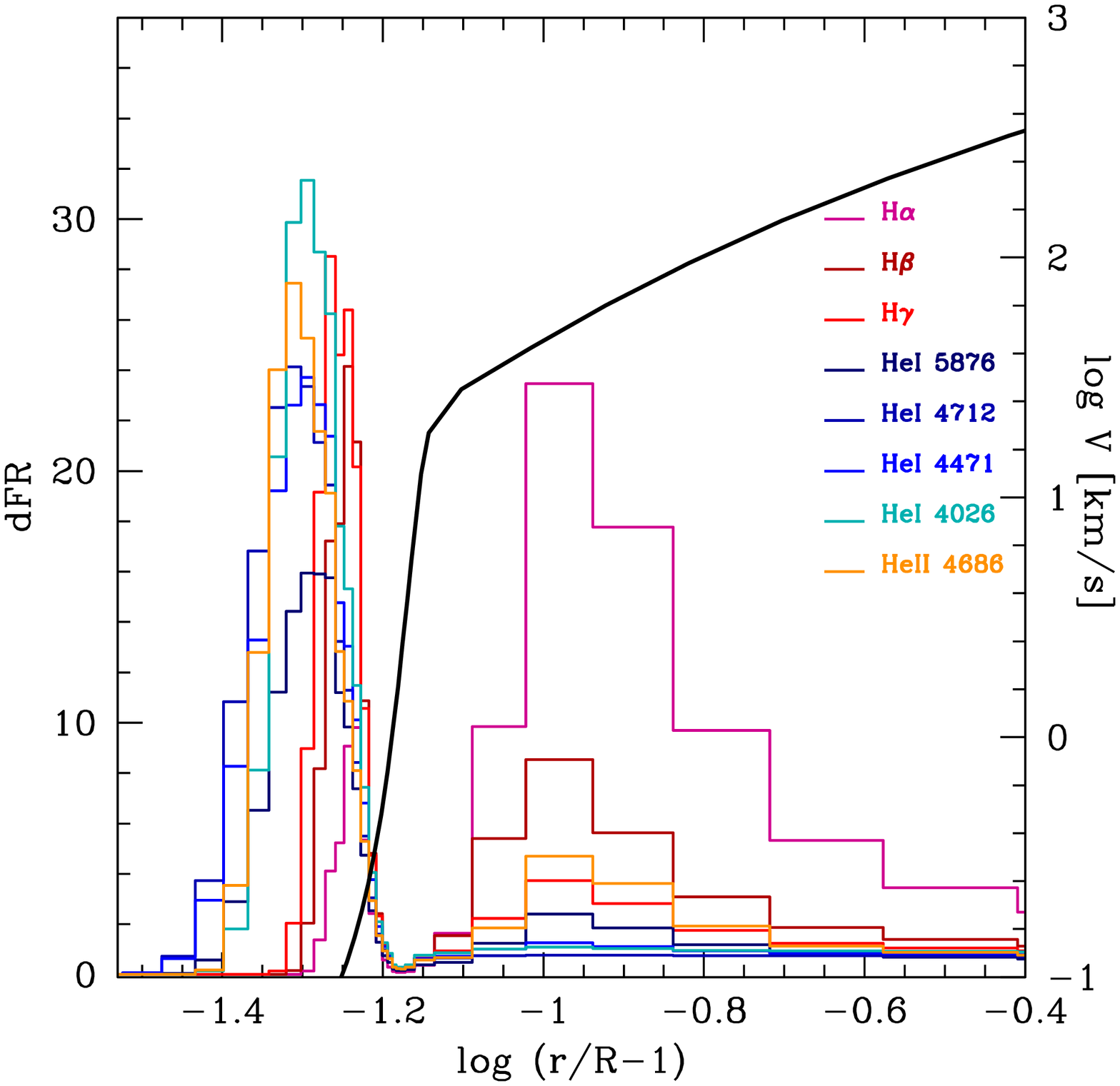}
\includegraphics[width=0.49\textwidth]{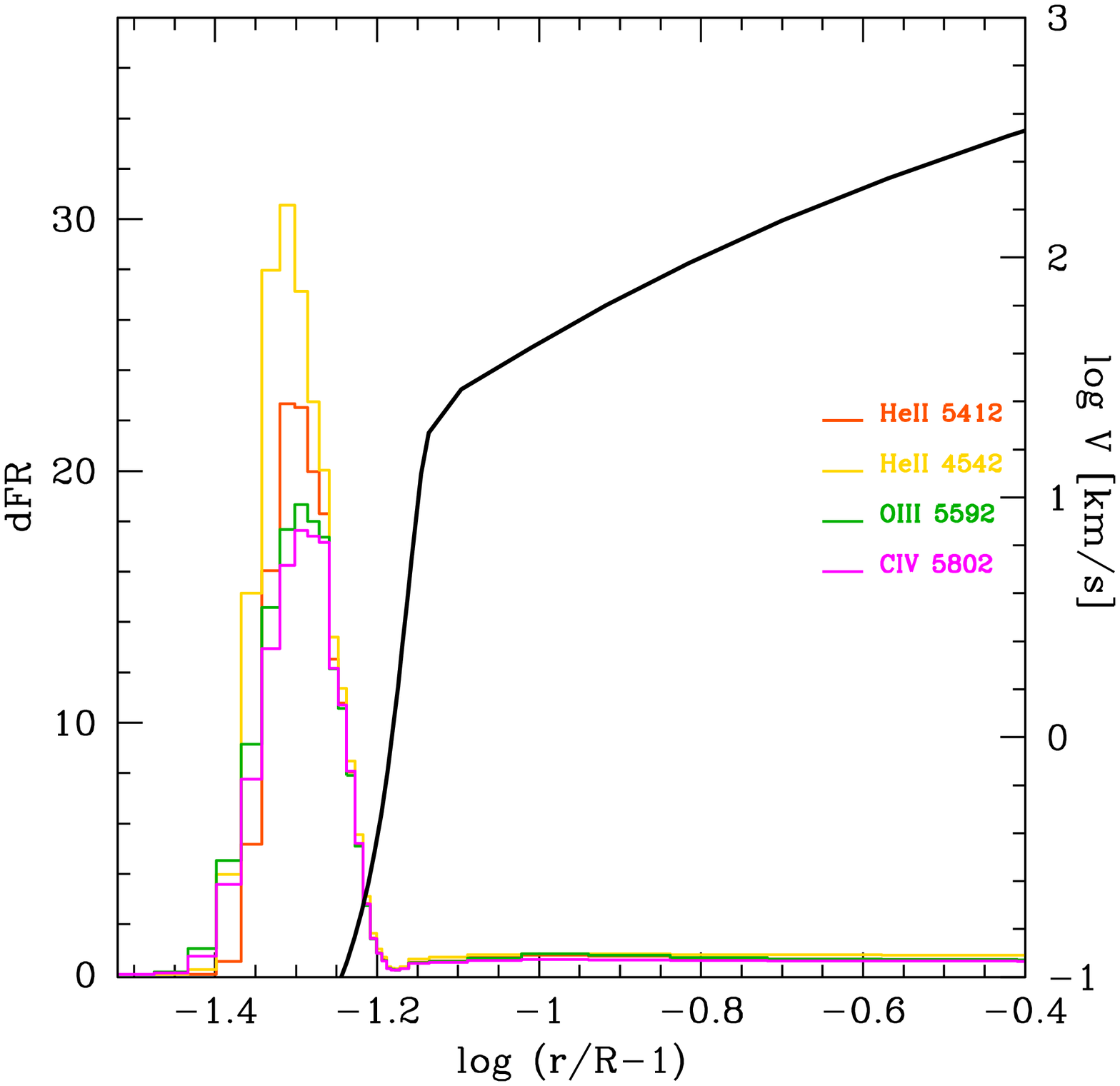}
\caption{Formation region of lines shown in Figs.\ \ref{fig_var_months}. The contribution function for a given line  (in arbitrary units) is shown as a function of height above the photosphere. \textit{Left}: lines showing variability in Fig.\ \ref{fig_var_months}. \textit{Right}: lines not showing variability in Fig.\ \ref{fig_var_months}. The black solid line in both panels is the velocity structure in the best fit model.} \label{form_reg}
\end{center}
\end{figure*}

\subsection{Spatially resolved variability}
\label{s_spatial}

Once a satisfactory fit was obtained, we retrieved the line formation region from the best fit model. The formation region is quantified by the contribution function, i.e. the fraction of the line produced as a function of radius. In Fig.\ \ref{form_reg}, we show the formation region of the lines of $\epsilon$ Ori displayed in Fig.\ \ref{fig_var_months}. The contribution function is evaluated at the line core. No correction for continuum has been performed, but we verified that the continuum just outside the line wings forms closer to the photosphere than the line core. We separate the lines showing variability (left panel) and the lines showing no variability at the 3$\sigma$ level (right panel). All lines are partly formed in the photosphere (i.e. below 10 \kms). The main difference between both sets of lines is that the variable lines have a significant contribution coming from high velocity, i.e. from the wind. This is clearly seen in H$\alpha$: the line is mainly formed between 1.1 and 1.25 R$_{\rm star}$ above the photosphere, which corresponds to velocities between 5 and 200 \kms. We thus conclude that the H$\alpha$ variability observed in Figs.\ \ref{fig_var_months} and \ref{fig_var_epsori_day} is rooted in the wind and not in the photosphere. More specifically, it is the inner part of the wind, at velocities of a few tens to a few hundreds of \kms, which is probed by H$\alpha$.

\begin{figure}
\centering
\includegraphics[width=9cm]{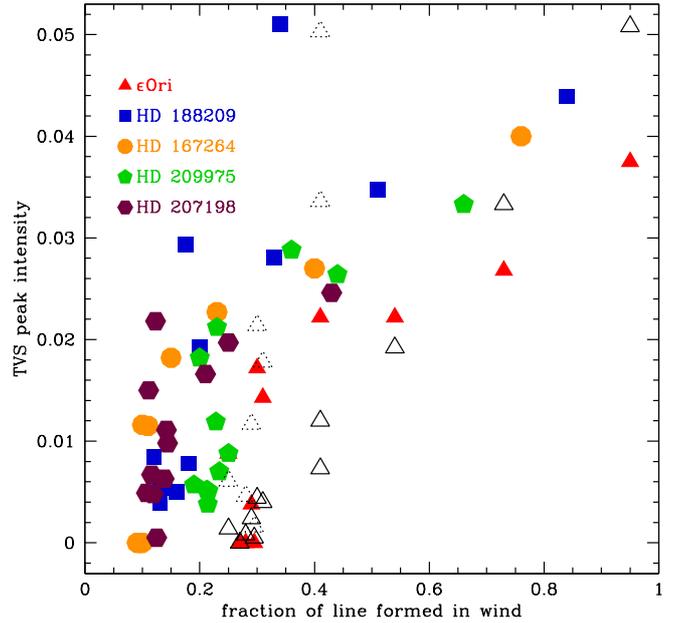}
\caption{TVS peak intensity as a function of fraction of the line formed in the wind. Filled symbols correspond to the sample stars showing variability on a daily timescale. The solid open triangles show the variability of synthetic lines for which the mass loss rate varies by 25\%, starting from the best fit model of $\epsilon$ Ori. The dotted triangles represent the theoretical variability due to changes in $\beta$ (from 0.8 to 1.4) in the best fit model of $\epsilon$ Ori. The points for H$\alpha$, H$\beta$ and H$\gamma$ are out of the plot (amplitude of 0.135, 0.104 and 0.070 respectively).} \label{ampl_tvs_fracwind}
\end{figure}

Using the best fit model for each star, we computed the fraction of the line formed above 10 \kms, i.e. in the wind part of the atmosphere. We did that for all lines. The choice of 10 \kms\ is governed by the shape of the contribution function shown in Fig.\ \ref{form_reg}. It corresponds to the velocity of the radius separating the two main peaks in the contribution function. A velocity of 10 \kms\ is also close to the sonic point considered to be around the transition between the photosphere and the wind. Fig.\ \ref{ampl_tvs_fracwind} shows the intensity of the TVS as a function of the fraction of the line formed in the wind. We quantified the variability by the intensity of the TVS peak.
We see a clear correlation: lines formed mostly in the wind are more variable than lines formed in the photosphere. The correlation is observed for all stars showing variability. 
This correlation between amplitude of variability and fraction of the line formed in the wind is valid for stars both showing and not showing changes in the photospheric lines. More specifically, the same level of H$\alpha$ variability is observed in $\epsilon$ Ori and in HD~188209. The former star show no sign of photospheric changes, while the latter is clearly variable in lines from high ionization states. It seems that the degree of variability in the photosphere is not directly correlated to the variability in the wind. This does not exclude a link between photospheric and wind changes in line profiles. But it suggests that instabilities in the wind can develop even if photospheric changes are absent (or at least weak). 

Variability in lines formed mainly in the photosphere is quite complex. When present, variability is weaker than in wind lines. But lines formed at the bottom of the atmosphere do not behave similarly. For instance, the HeII and OIII lines in $\epsilon$ Ori do not vary while HeI~4026 and HeI~4713 do. All lines have very similar contribution functions (see Fig.\ \ref{form_reg}). A similar behaviour is observed in HD~167264. This shows that the line formation position in the atmosphere is not the only parameter controlling line variability. For instance, \citet{reid93} suggested that for the fast rotator $\zeta$~Oph equator-to-pole temperature gradients lead to formation of \ion{He}{ii} (\ion{He}{i}) lines preferentially at the pole (equator). A latitudinal variation of the variability mechanism could then naturally explain the observed different behaviour of \ion{He}{i} and \ion{He}{ii} lines. 

\begin{figure}
\centering
\includegraphics[width=9cm]{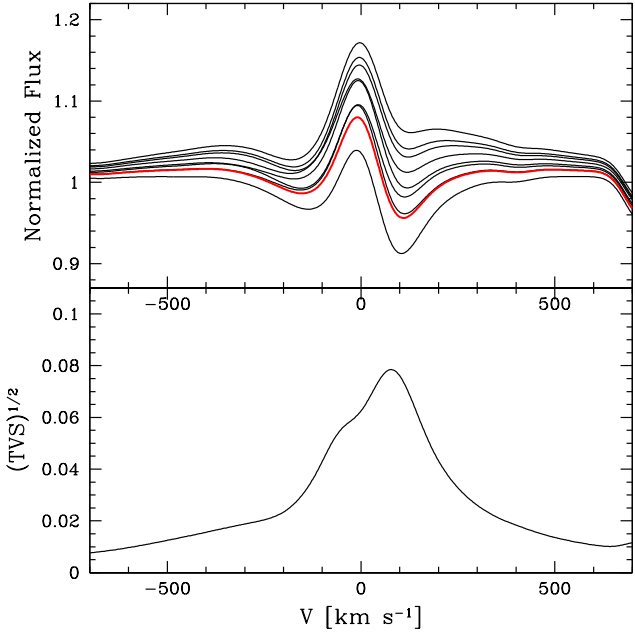}
\caption{TVS of a serie of synthetic spectra computed from atmosphere models with \mdot\ varying from 2.7 to 3.5 10$^{-6}$ \myr. The models are tailored to best represent the observed spectra of $\epsilon$ Ori.} \label{tvs_synth_mdot}
\end{figure}

\begin{figure}
\centering
\includegraphics[width=9cm]{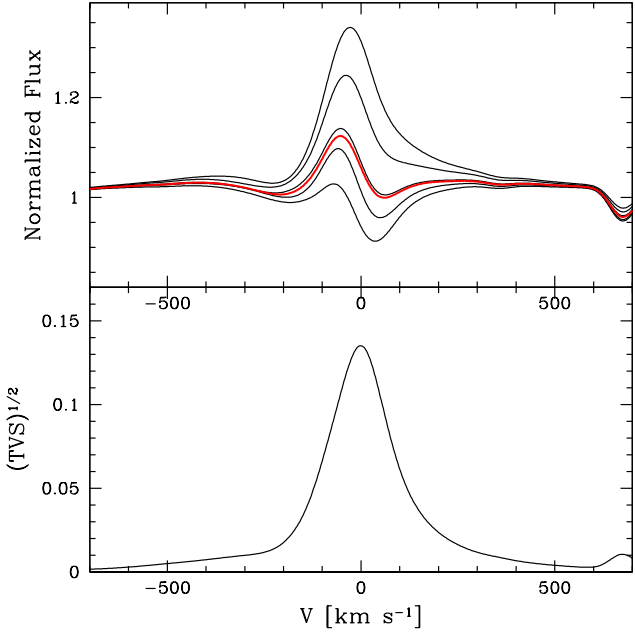}
\caption{TVS of a serie of synthetic spectra computed from atmosphere models with $\beta$ varying from 0.8 to 1.4 for the case of $\epsilon$ Ori.} \label{tvs_synth_beta}
\end{figure}

In Fig.\ \ref{tvs_synth_mdot} we show the effect of a variation of mass loss rate on the H$\alpha$ synthetic spectrum of $\epsilon$ Ori. The best fit model for $\epsilon$ Ori was used as a starting point. The TVS was computed assigning the same weight to all models. The mass loss rate varies by $\pm$25\%. The associated TVS shows a single peak located close to the line core, slightly to its red side. The blue side of the TVS displays a shoulder. The TVS amplitude reaches 0.04, similar to what is observed in $\epsilon$ Ori. However, we do not observed the three peak structure in the TVS. In Fig.\ \ref{tvs_synth_beta}, the influence of a change in the velocity law slope ($\beta$) is shown. Here again, the best fit model for $\epsilon$ Ori is the starting point. $\beta$ is varied between 0.8 and 1.4. The TVS is strongly peaked around the line core, and is much narrower than the TVS resulting from changes in mass loss rates or the observed TVS. Its amplitude reaches a maximum of 0.15, much higher than what is observed for $\epsilon$ Ori.  

The open black triangles in Fig.\ \ref{ampl_tvs_fracwind} correspond to the model shown in Fig.\ \ref{tvs_synth_mdot}. A change in mass loss rate by 25\% produces the same type of correlation between variability and fraction of line formed in the wind as what we observe for $\epsilon$ Ori (filled red triangles). 
The dotted triangles in Fig.\ \ref{ampl_tvs_fracwind} illustrate the effect of a variation of the slope of the velocity field (the $\beta$ parameter). We see that the same correlation is observed: the larger the fraction of the line formed in the wind, the stronger the variability. For the changes adopted ($\beta$ between 0.8 and 1.4) the variability of the lines with the smallest wind contribution is correctly reproduced, but the main wind lines show too strong variability compared to what is observed in $\epsilon$ Ori (field triangles).

\begin{figure}
\centering
\includegraphics[width=9cm]{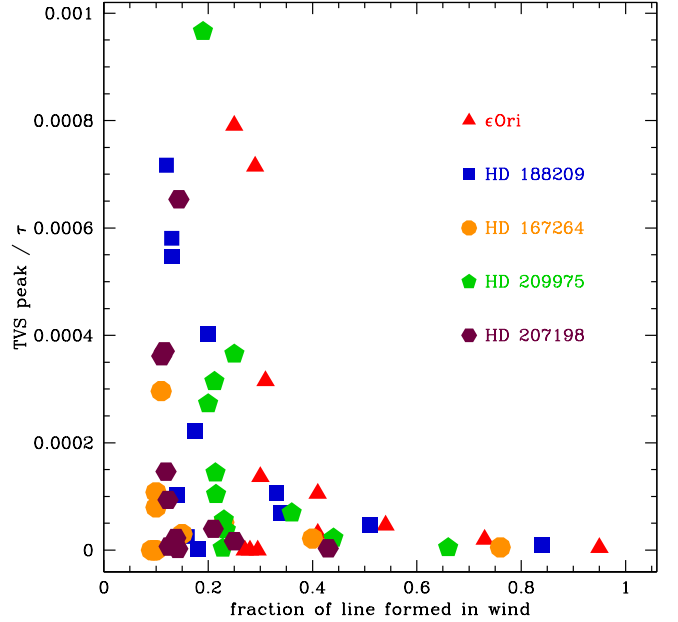}
\caption{Same as Fig.\ \ref{ampl_tvs_fracwind} for the ratio of the TVS peak intensity to the line optical depth at the photosphere.} \label{ampl_normtau}
\end{figure}

The strongest variability in wind lines (quantified by the absolute value of the TVS maximum intensity) does not necessarily imply that the underlying mechanism responsible for the variability is stronger in the wind. Indeed, lines formed in the outer parts of the atmosphere are on average stronger (i.e. they have a larger oscillator strength) and form over wider regions. To first order, one expects a stronger reaction to any instability in stronger lines. To further investigate this issue, we computed the ratio of the maximum amplitude of the TVS divided by the line optical depth at the photosphere ($TVS^{peak}/\tau$), i.e. the depth at which the electron scattering optical depth is equal to 2/3. This is first order approximation to take into account line strength effects. The results are shown in Fig.\ \ref{ampl_normtau}. We see a different behaviour compared to Fig.\ \ref{ampl_tvs_fracwind}: in all stars, there is a large scatter of $TVS^{peak}/\tau$ close to the photosphere and a decrease from the maximum value of $TVS^{peak}/\tau$ towards a minimum as the fraction of the line formed in the wind increases. Does this mean that the underlying variability mechanism is stronger close to the photosphere and that the stronger variability in lines formed mainly in the wind is only due to their higher strength and their formation over wider regions? Not necessarily either. Indeed, lines formed over a wide region may be subject to cancellation effects: variability at a given wavelength and a given position in the wind can be compensated by variability at another wavelength and another position corresponding to the same wavelength in the observer's frame. Such cancellation effects are smaller in photospheric lines which all form in narrow regions.

Our investigations do not allow us to constrain the \textit{nature} of the variability mechanism, nor its variation with depth. Nevertheless, we are able to show very clearly that there is a gradient of variability from the photosphere to the outer parts of the wind. Modulations in the wind parameters (mass loss rate, slope of the velocity field) can qualitatively explain the variability of the wind lines. We do not claim that true changes in the wind parameters are responsible for the observed variability. But the global density fluctuations they imply are \textit{qualitatively} consistent with the observed line profile variations. In the photosphere, variability is quite complex and lines formed in very close regions may show different variability patterns.

\subsection{TVS profile}
\label{s_tvs_prof}

The TVS provides more information that the global level of variability. We illustrate this on the cases of $\epsilon$ Ori and HD~207198 which are typical of the types of variability observed: $\epsilon$ Ori does not show photospheric variability on the timescales we probe, while HD~207198 does. Both stars show wind line variability.

We use the autocorrelation function to extract information on the line profile variability. It is defined as

\begin{equation}
A(j,k) = \frac{\Sigma (s(j,l)\times s(k,l))}{\sqrt{(\Sigma s^2(j,l))\times (\Sigma s^2(k,l))}}
\end{equation}

\noindent where $s(j,l)$ is the l$^{th}$ spectrum at wavelength $j$. The $l$ index in the sum has been dropped for clarity. For a perfect (anti-)~correlation between wavelengths $j$ and $k$, $A(j,k)$ is equal to 1 (-1).

\begin{figure*}[t]
     \centering
     \subfigure[H$\alpha$]{
          \includegraphics[width=.45\textwidth]{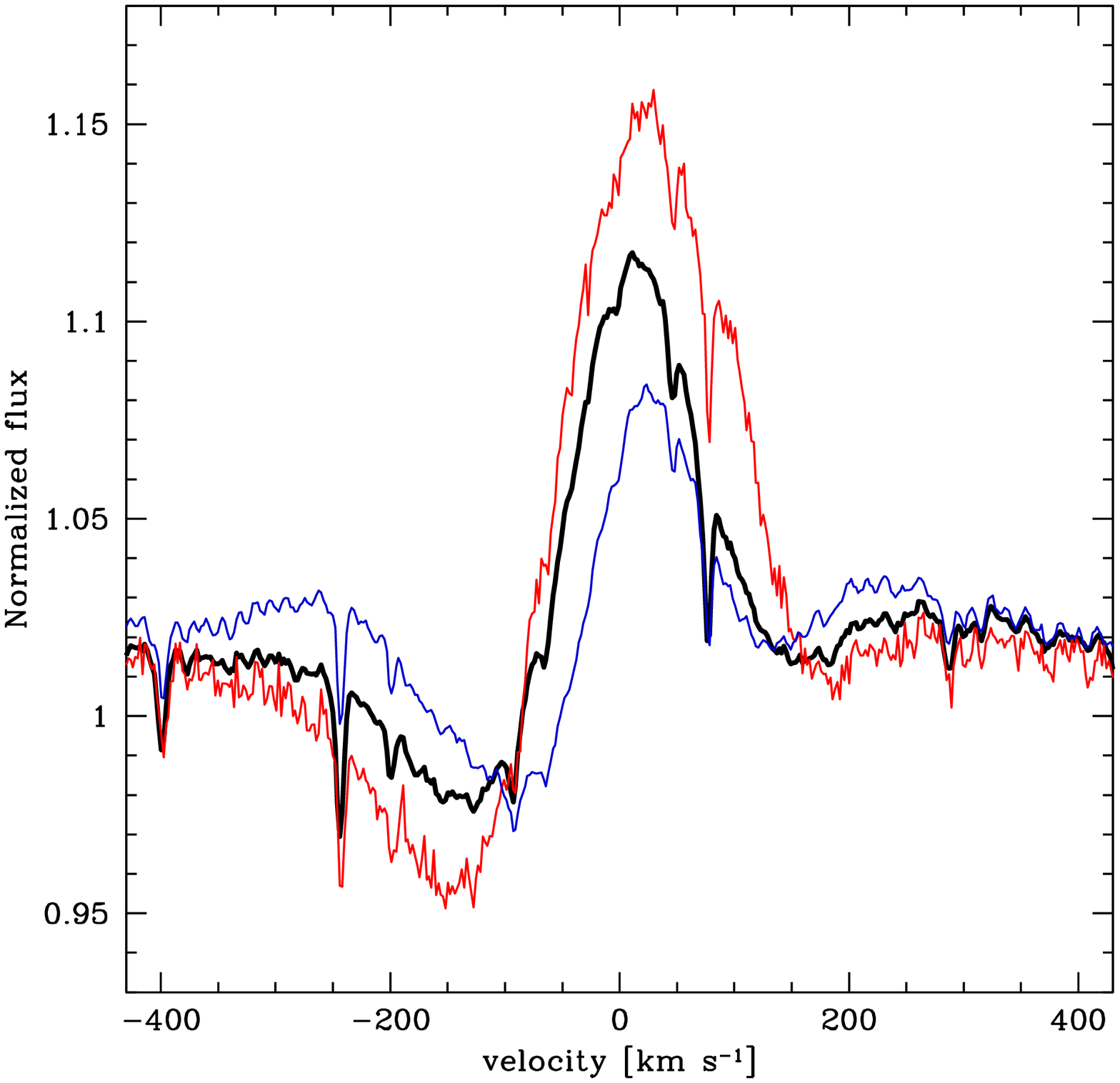}}
     \hspace{0.2cm}
     \subfigure[H$\gamma$]{
          \includegraphics[width=.45\textwidth]{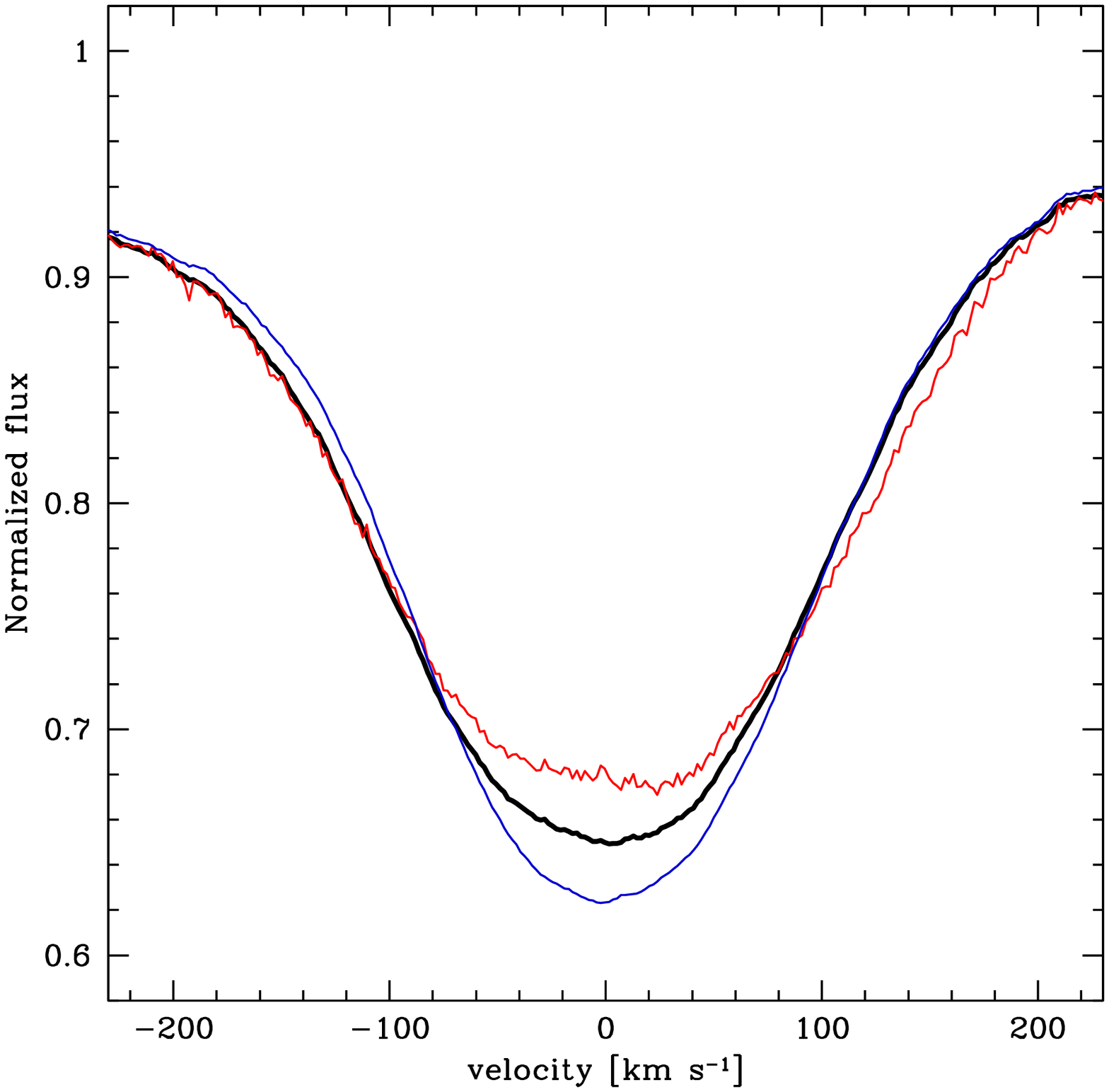}}
     \subfigure[H$\alpha$]{
          \includegraphics[width=.45\textwidth]{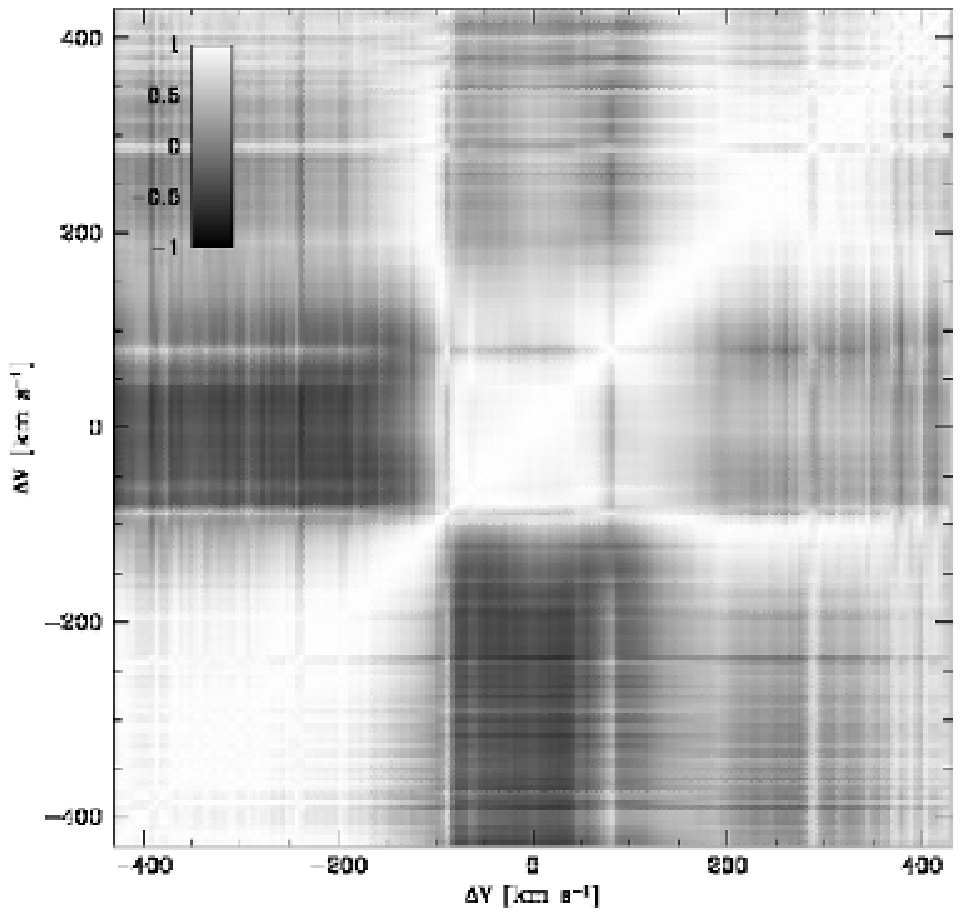}}
     \hspace{0.2cm}
     \subfigure[H$\gamma$]{
          \includegraphics[width=.45\textwidth]{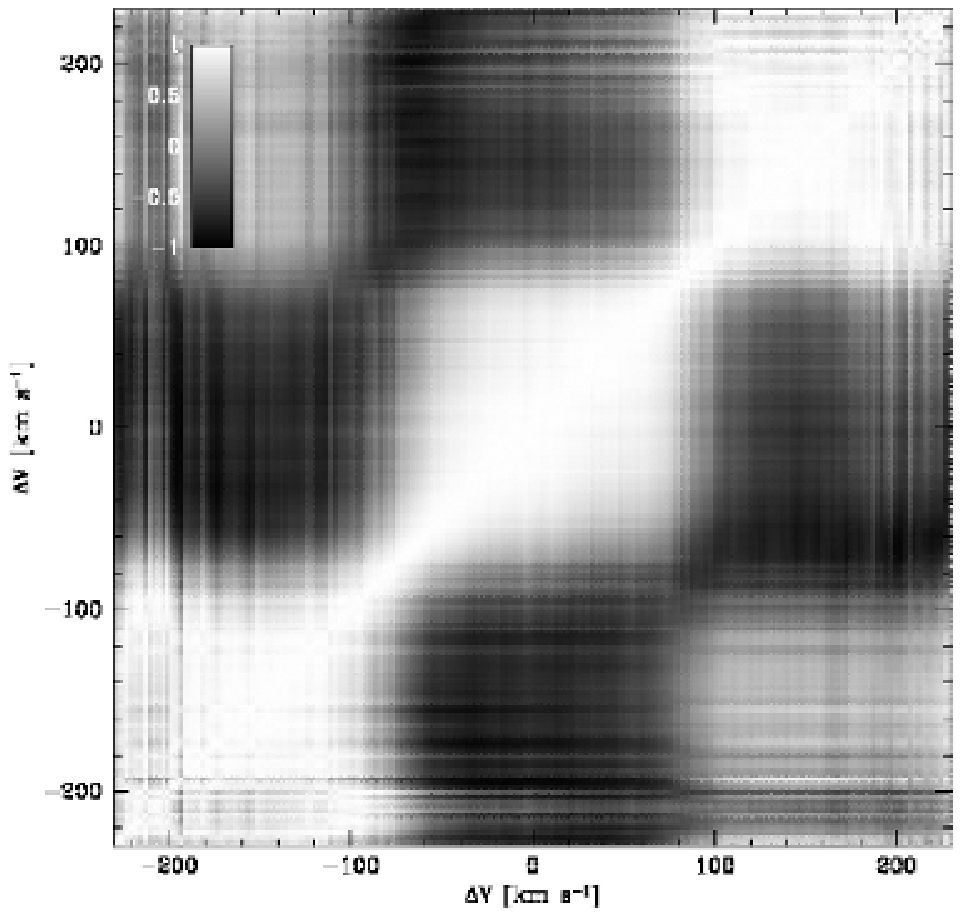}}
     \caption{\textit{Top panels:} Selected H$\alpha$ (left) and H$\gamma$ (right) spectra of $\epsilon$ Ori showing an anticorrelated variation (with the average spectrum in bold line). \textit{Bottom panels:} Autocorrelation matrixes. White (black) corresponds to a full (anti-) correlation. Only the part of the spectrum showing variability on a daily timescale is displayed.}
     \label{fig_acor_epsori}
\end{figure*}

 In the upper panels of Fig.\ \ref{fig_acor_epsori} we show selected H$\alpha$ and H$\gamma$ spectra of $\epsilon$ Ori illustrating the autocorrelation quantified in the matrixes displayed in the bottom panels.
The triple-peak structure observed in Fig.\ \ref{fig_var_months} is also seen in the autocorrelation matrixes. There is a clear anti-correlation between the variability of the line core and of the line wings. The core is stronger when the wings are weaker. This type of anti-correlation is seen in all lines showing the triple peak structure. A profile showing regular core strengthening and line narrowing could account for such a behaviour. For instance, a change in the line broadening caused by a variable macroturbulence could explain such a behaviour. We simulated such a change by varying the macroturbulent velocity between 50 and 75 \kms\ and we could reproduce a triple peaked structure in the TVS as well as the associated anti-correlation. One can also think of an extra emission on top of a photospheric profile. It could be caused by a shell of material initially just above the photosphere. At first, it would contaminate only the line core, making the overall profile less strong than a pure photospheric profile. The shell could subsequently expand, reaching larger velocity. The emission would then be spread over a range of velocities, and would consequently contaminate more the wings and less the line core. In that case the core would be stronger and the wings weaker than when the shell was just above the photosphere. This could reproduce the anti-correlation observed in $\epsilon$ Ori.

\begin{figure*}[t]
     \centering
     \subfigure[H$\alpha$]{
          \includegraphics[width=.45\textwidth]{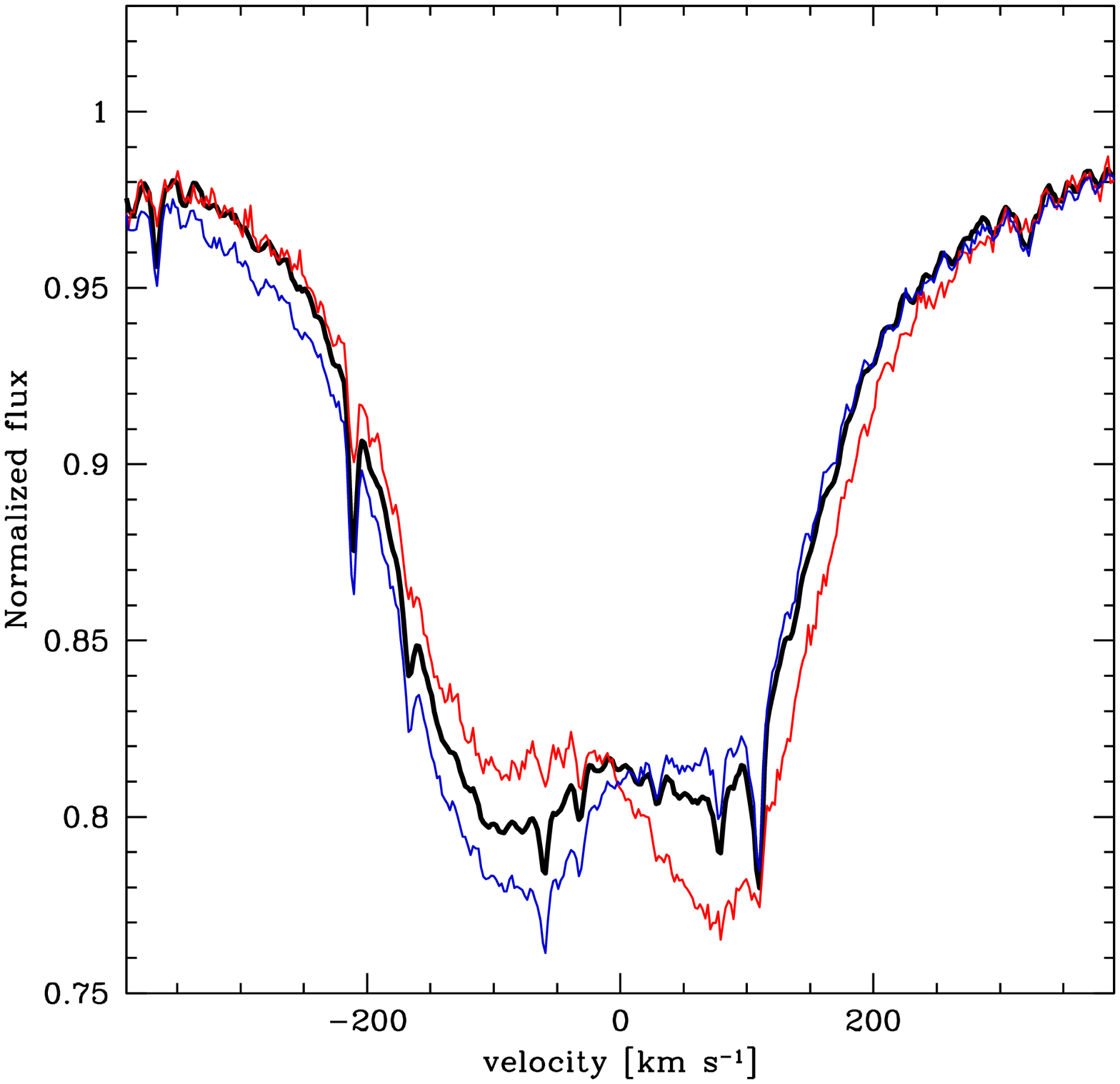}}
     \hspace{0.2cm}
     \subfigure[H$\gamma$]{
          \includegraphics[width=.45\textwidth]{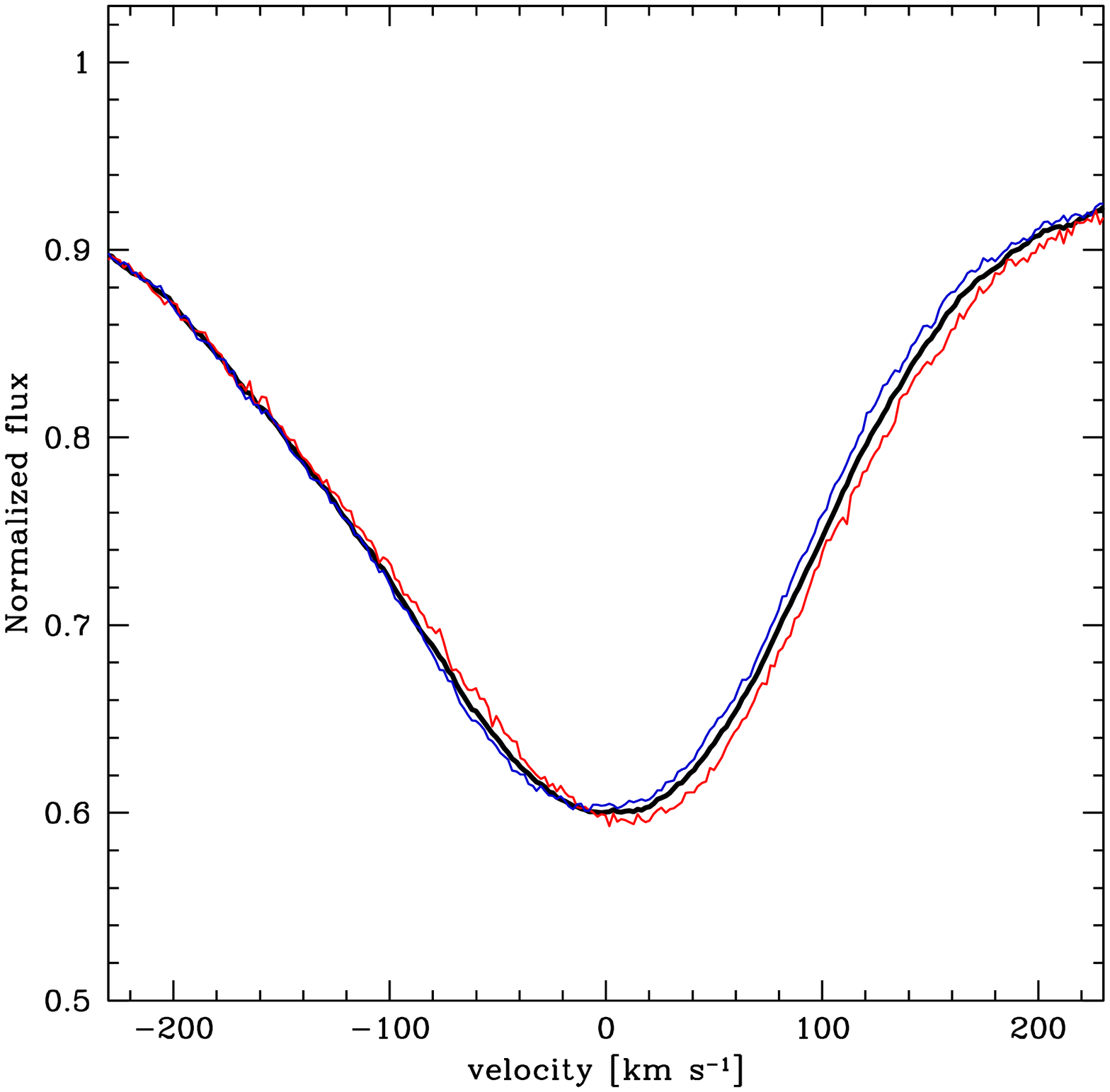}}
     \subfigure[H$\alpha$]{
          \includegraphics[width=.45\textwidth]{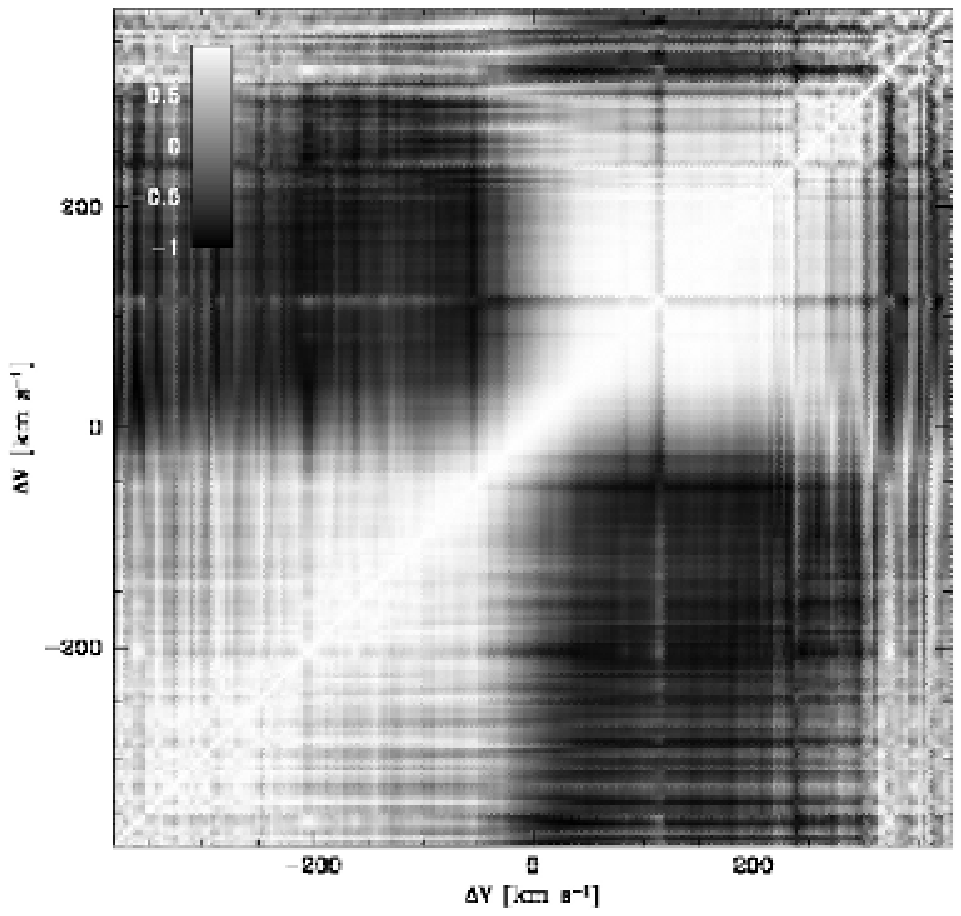}}
     \hspace{0.2cm}
     \subfigure[H$\gamma$]{
          \includegraphics[width=.45\textwidth]{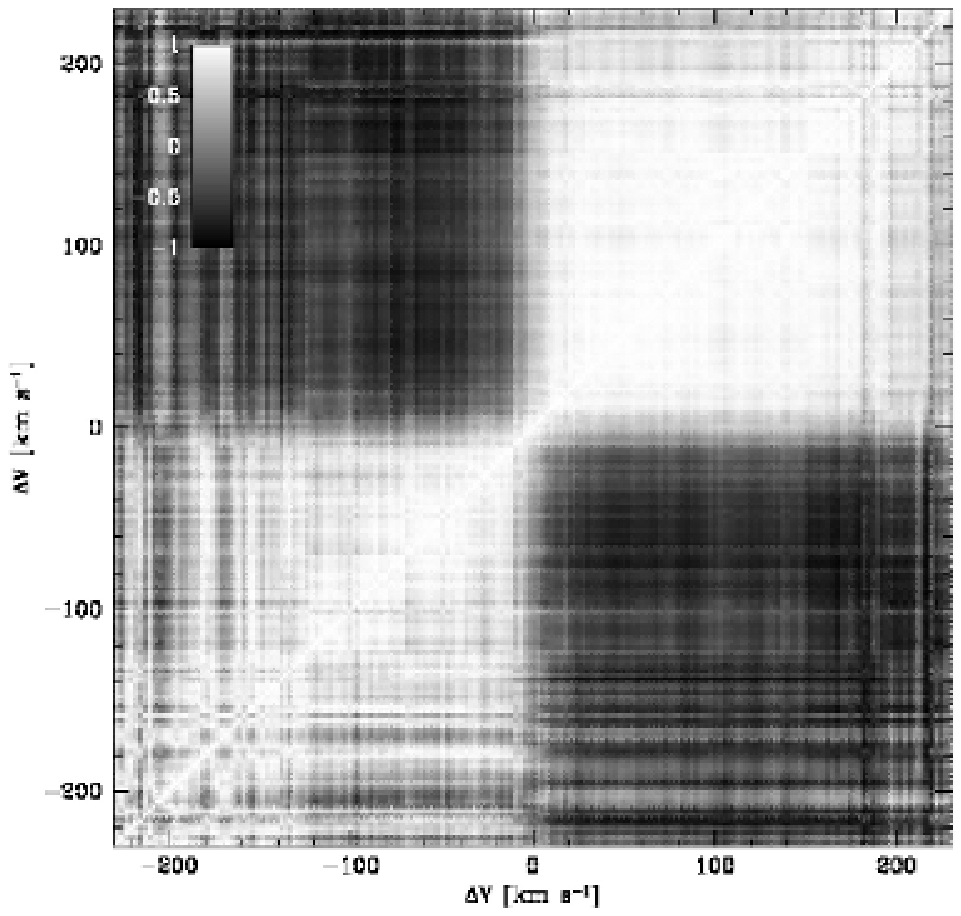}}
     \caption{Same as Fig.\ \ref{fig_acor_epsori} for HD~207198.}
     \label{fig_acor_207198}
\end{figure*}

Fig.\ \ref{fig_acor_207198} shows the correlation matrixes and selected spectra for HD~207198. This star shows mainly double peaked structures in its TVS (see Fig.\ \ref{fig_var_207198}). The H$\alpha$ and H$\gamma$ lines show a clear anti-correlation between their blue and red wings. One possibility to explain such a behaviour is a simple radial velocity shift. But since not all lines show variability, we can exclude a bulk motion of the star. In addition, the two peaks are not always of the same strength, which is inconsistent with a radial velocity modulation. A moving shell does not account for the observed anti-correlation of the red and blue wings (see above). An asymmetry between the blue and red TVS peak can be explained by some types of wind asymmetries. For instance, \citet{markova05} performed simulations of clumpy winds with broken shells (see their Fig.\ 10). The resulting TVS is asymmetric, but with a stronger blue peak, contrary to what we observe. Their simulations of winds with a spiral pattern yield a symmetric TVS. The presence of blue and red variability could be explained in the case of winds confined by a magnetic field. Simulations by \citet{ud13} show that if the star is not rotating too fast, material channelled along the field lines can settle in a disk-like structure and episodically fall back onto the star. Material reaching the equatorial plane at larger distance or flowing away from the surface at the poles can escape the star. This can qualitatively explain blue and red variability. However, since both infall and ejection of material happen at the same time, no anti-correlation is expected. In addition, no magnetic field has been detected on HD~207198. In Fig.\ \ref{fig_var_207198} we see that as we move from lines mainly formed in the wind to photospheric lines (e.g. \ion{He}{II} 5412) the asymmetry of the TVS disappears. This implies that any mechanism leading to this asymmetry is linked to the wind itself, and not to the photosphere. 

In conclusion, the TVS shapes are complex and vary from star to star. But a qualitative interpretation indicates the important role of the wind to produce the observed structures, in agreement with the conclusions regarding the degree of variability (see Sect.\ \ref{s_spatial}). 

\section{Discussion}
\label{s_comp}

Several of our sample stars have been subject of variability study in the past.

\citet{kaper96} reported variability of 10~Lac in the bluest part of the \ion{N}{V}~1238-1242 UV line, but not on all of their time series. No H$\alpha$ variability was detected by \citet{kaper97}. 10~Lac was found to be a non-radial pulsator by \citet{smith78}. Our time series do not show any variability.

HD~188209 was studied by \citet{ful96}. They gathered spectra over one night and detected variability in \ion{He}{I}~5876 and \ion{C}{IV}~5802. The variation they observe span the range [-200;100] \kms\ about the line center. Our results obtained on the same timescale (night of 25-26 June 2008) are very similar, with variations in the range [-70;150] \kms. We also detect variability on June 27 and June 28. \citet{markova05} reported H$\alpha$ variability on a timescale of months. They concluded that changes in the mass loss rate by 17\% could account for this variability. \citet{isr00} reported variability with a period of 6.4 days that they tentatively attribute to binarity, although there are difficulties with this scenario (see their Sect.\ 7).

HD~209975 was also part of the sample of \citet{ful96}. They collected spectra over one day and reported a variability of \ion{He}{I}~5876 with an extension over $\pm$170 \kms. We also detect variability in \ion{He}{I}~5876 on hourly timescale for all nights covered by our observations. The variability can be restricted to the first 100 \kms or extend up to 200 \kms\ depending on the night, consistent with the findings of Fullerton et al. \citet{markova05} observed a monthly variability in H$\alpha$ that they could reproduce by 27\% change in the mass loss rate of the star. \citet{kaper96} reported the presence of DACs in UV spectra of HD~209975, causing variations on a timescale of at leat 5 days. Subsequent observations in the UV and H$\alpha$ by \citet{kaper97} confirmed this timescale. The H$\alpha$ variability was found to extend over the first 200 \kms\ around the line center. In our case, the TVS extends up to $\pm$300 \kms\ on the timescale of the observations (7 nights).

$\epsilon$ Ori was observed by \citet{prinja04}. Variability was detected in several lines, including H$\alpha$ and the weaker \ion{Si}{III}~4553 feature. They highlighted a periodicity of 1.9 d in the variations of both the strong wind line H$\alpha$ and the \ion{Si}{III}~4553, formed closer to the photosphere. They concluded that a direct link between photospheric and wind variability was revealed. \citet{th13} confirmed the presence of periodic signals on timescales of 2 to 7 days in both H$\alpha$ and \ion{He}{I}~5876, which they interpreted as evidence for a connexion between photospheric and wind variability. Our results are consistent in the sense that we detect a clear variability in the wind lines. But we do not see any change in many photospheric lines (the \ion{He}{II}, \ion{O}{III} and \ion{C}{IV} lines). We checked the behaviour of the \ion{Si}{III}~4553 and detected a variability with a peak of the TVS of the order 0.010, in agreement with \citet{prinja04}. As noted previously, some of the \ion{He}{I} lines formed in the photosphere (\ion{He}{I}~4026, \ion{He}{I}~4713) do show variability, making the interpretation of the photospheric variability of $\epsilon$ Ori quite complex.  \\ 
We investigated the presence of periodic signals in the variability we detect for $\epsilon$ Ori. We focused on H$\alpha$ since it is the most variable line, and we relied on its maximum emission as a proxy for our measurement. We performed a period search using the technique of \citet{scargle82}. We only detected a broad peak in the power spectrum at about 2 days. This is consistent with the findings of \citet{prinja04} and \citet{th13}. However, our sampling is too coarse to get a good period identification. 

Previous analysis were based on only a few lines (mainly H$\alpha$, \ion{He}{I}~5876 and \ion{C}{IV}~5802). Our study extends the investigation of the line profile variability to a much larger list of spectral features. We confirm that variability is mainly observed in wind sensitive lines. We show that there is a significant correlation between variability and fraction of the line formed in the wind. 
The question of whether these instabilities can be triggered by a photospheric mechanism or if they can be self-sustained is not settled. Our results indicate that some photospheric lines can be remarkably stable even if wind lines are variable. At the same time, different photospheric lines may be stable or not, as in the case of $\epsilon$ Ori or HD~167264. The effect of sub-surface convection in evolved massive stars \citep{cantiello09} may explain the properties of stars showing photospheric variability, but fail to account for the objects without detectable variability or variability in only selected photospheric lines. 

The presence of small scale magnetic loops could trigger magnetic activity close to the photosphere. This could be the seeds for wind variability \citep[e.g.][]{ful96,morel04}. All our sample stars were initially part of a sample dedicated to the search for magnetic field among O stars. We did not detect any evidence for Zeeman signatures in the analysis of the spectro-polarimetric data, excluding strong  magnetic field (B $>$ few 100 G, Grunhut et al.\ in prep.). In addition, all known O stars with a strong magnetic field show well defined rotational modulations \citep[e.g.][]{donati02,donati06,wade12} due to the presence of equatorial overdensities caused by magnetic wind confinment. In our sample stars, we detect variability on various timescales (hour-to-hour and night-to-night). This is not fully consistent with the presence of large scale magnetic field. 
However we cannot exclude much weaker fields of the order of a few Gauss (as discovered by \cite{lignieres09} on Vega). Such fields could create surface inhomogeneities (temperature, composition) that might trigger photospheric variability subsequently amplified in the wind (e.g. by the line-driving instability). The photospheric variability could be weak so that we do not detect it in some of our sample stars.

Pulsations are also often claimed to be the seeds of spectral variability in OB-type stars. \citet{kaufer02} performed a large survey of the early B supergiant HD~64670 during 10 nights and reported the existence of periodic variability in photospheric lines. They could explain it by a model of stellar oscillations. Variability was also observed in H$\alpha$. Some of the features observed in this line could be assigned a periodicity similar to that of the photospheric lines. Some other patterns escaped this relation. \citet{kaufer02} concluded that there was a clear link between photospheric and wind variability. \citet{prinja06} studied the late-O supergiant $\alpha$ Cam. They reported hourly and daily variability in both H$\alpha$ and \ion{He}{I}~5876. From the daily variability of \ion{He}{I}~5876, they suggested that photospheric variations are present in the time serie analysis. In our analysis, we show that \ion{He}{I}~5876 is one of the most variable line after the Balmer lines. In $\epsilon$ Ori, \ion{He}{I}~5876 has a significant contribution from the wind (about 35\%). Other lines are mainly photospheric (e.g. \ion{He}{II} lines) and show no or weak variability. Hence in our sample stars, pulsations at the photosphere are difficult to relate to variability in lines formed further out in the wind, at least for stars without clear photospheric variability.

Line profile variability in OB stars is a complex process. Our results seem to indicate that a photospheric seed is not mandatory to trigger wind variability. However the behaviour of photospheric variability is difficult to interpret since lines formed over relatively similar regions, in the photosphere, may show different degrees of variability. In addition, the anti-correlations in the TVS are not easy to interpret. Dedicated simulations of specific variability mechanisms coupled to radiative trsnefer calculations are necessary to test possible excitation mechanisms.

\begin{sidewaystable*}
\vspace{5cm}
\begin{center}
\caption{Physical parameters of the sample stars.}
\begin{tabular}{lccccccccccccccccc}
\hline
Star      & ST         & d            & \teff\ & \lL  & \logg\ & R        &   \vsini\   &  \vmac\  &  C/H      &  N/H       & \mdot\   &  \vinf & $\beta$ & f \\
          &            & [pc]         & [kK]   &      &        & [\rsun]  &  [\kms]     &  [\kms]  &  [10$^{-4}$] & [10$^{-5}$]  & [\myr]   & [\kms] \\
\hline
$\epsilon$~Ori  & B0Iab      & 410$\pm$150  & 27.5$\pm$1.0  & 5.60$\pm$0.33 & 3.10   & 28.0$\pm$16.0         &   40        &  38  &   1.4$\pm$1.0 &  10.3$\pm$6.8  & -6.25   & 1800 & 1.2 & 0.05 \\
HD~167264 & B0.5Ia/Iab & 1700$\pm$510 & 28.0$\pm$1.0  & 5.65$\pm$0.27 & 3.10   & 28.6$\pm$12.5         &   70        &  22  &   1.2$\pm$0.5 &  18.0$\pm$8.0  & -6.50  & 2000 & 1.8 & 0.1 \\
HD~207198 & O9II       & 620$\pm$180  & 32.5$\pm$1.0  & 5.05$\pm$0.26 & 3.50   & 10.6$\pm$4.4          &   60        &  27  &   1.6$\pm$0.8 &  20.0$\pm$6.0  & -7.0   & 2000  & 3.0 & 1.0 \\
HD~188209 & O9.5Ib     & 2000$\pm$600 & 29.8$\pm$1.0  & 5.65$\pm$0.26 & 3.20   & 25.2$\pm$10.5         &   45        &  33  &   0.7$\pm$0.5 &  35.0$\pm$18.0 & -6.4   & 2000 & 2.2 & 0.05 \\
HD~209975 & O9Ib       & --           & 30.5$\pm$1.0  & 5.35$\pm$0.30 & 3.35   & 17.0$\pm$8.2          &   48        &  40  &   1.2$\pm$0.9 &  20.5$\pm$11.4 & -6.50  & 2000 & 2.9 & 1.0 \\
HD~34078  & O9.5V      & 445$\pm$145  & 34.0$\pm$1.0  & 4.65$\pm$0.29 & 4.10   & 5.4$\pm$2.6           &   25        &   5  &   2.5$\pm$1.5 &  7.0$\pm$1.1   & -10.0  & 1000 & 1.0\tablefootmark{a} & 1.0\tablefootmark{a}  \\
10~Lac    & O9V        & 325$\pm$65   & 35.0$\pm$1.0  & 4.35$\pm$0.18 & 4.05   & 4.1$\pm$1.1           &   15        &  15  &   2.3$\pm$1.0 &  17.6$\pm$3.9  & -9.70  & 1200 & 1.0\tablefootmark{a}  & 1.0\tablefootmark{a}  \\
\end{tabular}
\end{center}
\tablefoot{Distances are from Hipparcos catalog \citep{hipcat} when the uncertainty on the parallax is lower than 30\%. For the remaining objects, the distances provided by \citet{mason98} where used, with a conservative uncertainty of 30 per cent. The luminosity of HD~209975 is adopted (HD~209975 being a Ib supergiant, we chose a value intermediate between a O9III and a O9I according to Martins, Schaerer \& Hillier 2005). This corresponds to a distance of 1350 pc. The Hipparcos distance (630$\pm$190 pc) implies a very low value (\lL=4.7) for a late O supergiant.} 
\tablefoottext{a}{Adopted}
\label{tab_param}
\end{sidewaystable*}


\section{Conclusion}
\label{s_conc}

We have performed time series analysis of optical spectra of seven late-O early-B stars to investigate line profile variability. We have used spectra collected with the spectropolarimeter NARVAL mounted on the T\'elescope Bernard Lyot at Pic du Midi. The spectra have been obtained in several observing campaign between 2007 and 2009. Depending on the star, spectra have been collected over a single night or over several nights (sometimes with several exposures during individual nights). We have performed the analysis of the variability using the Temporal Variance Spectrum \citep[TVS, ][]{ful96}. We have also computed atmosphere models with the code CMFGEN to determine the fundamental properties of the sample stars, and to constrain the formation regions of the lines for which variability was investigated. Contrary to most previous studies, we did not restrict to lines formed in the wind, but we analyzed twelve spectral lines sampling the radial structure of the atmosphere. Our results can be summarized as follows:

\begin{itemize}

\item[$\bullet$] Variability is detected in the wind lines of all bright giants and supergiants. The dwarfs stars do not show any sign of line profile variability in any line.

\item[$\bullet$] Photospheric variability in lines from high ionization states is sometimes observed, sometimes not, even if wind lines variability is definitely detected.

\item[$\bullet$] In some cases the weakest photospheric \ion{He}{I} lines may be variable even if the higher excitation lines are stable.

\item[$\bullet$] When variability is detected on daily timescales, it is can also (but not always) be observed on hourly timescales, with a much lower amplitude.

\item[$\bullet$] The shape of the temporal variance spectrum is usually double or triple peaked. Most of the double peaked structures show a red/blue asymmetry, the former being the strongest.

\item[$\bullet$] Clear anticorrelations between these variable structures is observed.

\item[$\bullet$] We find a clear correlation between amplitude of the variability and fraction of the line formed in the wind. 

\item[$\bullet$] There is no obvious relation between photospheric variability and stellar/wind parameters.

\end{itemize}

Our results are the first to provide variability information for a large sample of lines probing the entire radial structure of the atmosphere. They show for the first time a very clear correlation between the level of variability and the fraction of the line formed in the wind. At the same time, our results do not indicate that wind variability is always related to photospheric modulations.

\section*{Acknowledgments}

We thank an anonymous referee for a careful reading of the manuscript and a constructive report. FM thanks the Agence Nationale de la Recherche for financial support (grant ANR-11-JS56-0007). DJH acknowledges support from STScI theory grant HST-AR-12640.01.

\bibliographystyle{aa}
\bibliography{var_narval}

\newpage
\begin{appendix}

\section{Journal of observations}

In Table \ref{tab_obs} we present the journal of observations for the target stars.

\begin{table*}
\begin{center}
\caption{Journal of observations. The signal to noise ratio (SNR) is given in the wavelength range 3800--6700 \AA\ and depends on the exact wavelength.} \label{tab_obs}
\begin{tabular}{llccc}
\hline
Star & Date  & HJD & exposure time & SNR \\
     &       & [d]-2450000 & [s]           &     \\
\hline
$\epsilon$ Ori  &  16 oct 2007      & 4389.70536 & 40                &   150--600 \\
(HD 37128)      &  18 oct 2007      & 4391.70278 & 60                &   150--650  \\
                &  18 oct 2007      & 4391.70587 & 60                &   150--650  \\
                &  18 oct 2007      & 4391.70896 & 60                &   150--650  \\
                &  18 oct 2007      & 4391.71205 & 60                &   100--400  \\
                &  18 oct 2007      & 4391.71516 & 60                &   100--350  \\
                &  18 oct 2007      & 4391.71826 & 60                &   100--500  \\
                &  19 oct 2007      & 4392.56330 & 180               &   150--900  \\
                &  19 oct 2007      & 4392.56889 & 180               &   150--900  \\
                &  19 oct 2007      & 4392.57334 & 180               &   150--900  \\
                &  19 oct 2007      & 4392.57781 & 180               &   150--800  \\
                &  19 oct 2007      & 4392.58228 & 180               &   200--950  \\
                &  19 oct 2007      & 4392.58675 & 180               &   200--950  \\
                &  19 oct 2007      & 4392.59121 & 180               &   200--1000  \\
                &  19 oct 2007      & 4392.59574 & 180               &   200--950  \\
                &  19 oct 2007      & 4392.60021 & 180               &   200--1000  \\
                &  19 oct 2007      & 4392.60467 & 180               &   200--1050  \\
                &  19 oct 2007      & 4392.60915 & 180               &   200--1100  \\
                &  19 oct 2007      & 4392.61361 & 180               &   200--1100  \\
                &  19 oct 2007      & 4392.61808 & 180               &   200--1100  \\
                &  19 oct 2007      & 4392.62254 & 180               &   200--1000  \\
                &  19 oct 2007      & 4392.62700 & 180               &   200--1050  \\
                &  19 oct 2007      & 4392.63147 & 180               &   200--1100  \\
                &  19 oct 2007      & 4392.63593 & 180               &   250--1200  \\
                &  19 oct 2007      & 4392.64039 & 180               &   300--1300  \\
                &  19 oct 2007      & 4392.64487 & 180               &   300--1300  \\
                &  19 oct 2007      & 4392.64936 & 180               &   300--1300  \\
                &  19 oct 2007      & 4392.65385 & 180               &   300--1300  \\
                &  19 oct 2007      & 4392.65834 & 180               &   250--1200  \\
                &  19 oct 2007      & 4392.66283 & 180               &   250--1200  \\
                &  19 oct 2007      & 4392.66733 & 180               &   250--1150  \\
                &  19 oct 2007      & 4392.67189 & 180               &   200--1100  \\
                &  19 oct 2007      & 4392.67640 & 180               &   200--1000  \\
                &  19 oct 2007      & 4392.68089 & 180               &   200--1000  \\
                &  19 oct 2007      & 4392.68538 & 180               &   200--1000  \\
                &  22 oct 2007      & 4395.71151 & 120               &   250--1300  \\
                &  22 oct 2007      & 4395.71529 & 120               &   250--1250  \\
                &  22 oct 2007      & 4395.71906 & 120               &   250--1200  \\
                &  22 oct 2007      & 4395.72285 & 120               &   250--1250  \\
                &  22 oct 2007      & 4395.72664 & 120               &   250--1200  \\
                &  22 oct 2007      & 4395.73044 & 120               &   250--1200  \\
                &  22 oct 2007      & 4395.73422 & 120               &   250--1200  \\
                &  22 oct 2007      & 4395.73800 & 120               &   250--1200  \\
                &  25 oct 2007      & 4398.71207 & 120               &   200--1100  \\
                &  25 oct 2007      & 4398.71586 & 120               &   200--1100  \\
                &  25 oct 2007      & 4398.71965 & 120               &   200--1000  \\
                &  25 oct 2007      & 4398.72344 & 120               &   200--900  \\
                &  25 oct 2007      & 4398.72723 & 120               &   200--900  \\
                &  25 oct 2007      & 4398.73102 & 120               &   200--1000  \\
\end{tabular}
\end{center}
\end{table*}

\setcounter{table}{0}

\begin{table*}
\begin{center}
\caption{Continued} \label{tab_obs}
\begin{tabular}{llccc}
\hline
Star & Date  & HJD & exposure time & SNR \\
     &       & [d]-2450000 & [s]           &     \\
\hline
HD 188209       &  21 jun 2008      & 4639.45048 & 2700              &   100--600  \\
                &  21 jun 2008      & 4639.48411 & 2700              &   100--500  \\
                &  21 jun 2008      & 4639.51776 & 2700              &   100--600  \\
                &  22 jun 2008      & 4640.45345 & 2700              &   200--1000  \\
                &  22 jun 2008      & 4640.48707 & 2700              &   200--1000  \\
                &  22 jun 2008      & 4640.52070 & 2700              &   200--1000  \\
                &  25 jun 2008      & 4643.41578 & 2700              &   200--1000  \\
                &  25 jun 2008      & 4643.44941 & 2700              &   200--950  \\
                &  25 jun 2008      & 4643.48304 & 2700              &   200--900  \\
                &  25 jun 2008      & 4643.51716 & 2700              &   200--900  \\
                &  25 jun 2008      & 4643.55079 & 2700              &   200--900  \\
                &  27 jun 2008      & 4645.39522 & 2700              &   200--1000  \\
                &  27 jun 2008      & 4645.42886 & 2700              &   200--1100  \\
                &  27 jun 2008      & 4645.46249 & 2700              &   200--1000  \\
                &  27 jun 2008      & 4645.49613 & 2700              &   200--1100  \\
                &  27 jun 2008      & 4645.52975 & 2700              &   200--1150  \\
                &  28 jun 2008      & 4646.39664 & 2700              &   200--1100  \\
                &  28 jun 2008      & 4646.43027 & 2700              &   200--1100  \\
                &  28 jun 2008      & 4646.46391 & 2700              &   200--1100  \\
                &  28 jun 2008      & 4646.49755 & 2700              &   200--1100  \\
                &  28 jun 2008      & 4646.53120 & 2700              &   200--1050  \\
                &  29 jun 2008      & 4647.59313 & 2700              &   200--1000  \\
                &  29 jun 2008      & 4647.62682 & 2700              &   200--1050  \\
                &  29 jun 2008      & 4647.66051 & 2700              &   200--1000  \\
                &  30 jun 2008      & 4648.41724 & 2700              &   150--600  \\
                &  30 jun 2008      & 4648.45106 & 2700              &   200--900  \\
                &  30 jun 2008      & 4648.48536 & 2700              &   200--950  \\
\hline
15 Sgr          &  25 jul 2009      & 5038.49492 & 2700              &   200--1000 \\
(HD 167264)     &  26 jul 2009      & 5039.39600 & 2400              &   100--600 \\
                &  27 jul 2009      & 5040.37934 & 2400              &   150--850 \\
                &  28 jul 2009      & 5041.37153 & 2400              &   200--1000 \\
                &  29 jul 2009      & 5042.37736 & 2400              &   200--1150 \\
                &  31 jul 2009      & 5044.38242 & 2300              &   200--1100 \\
                &  04 aug 2009      & 5048.43269 & 2300              &   200--900 \\
\hline
HD 207198       &  24 jul 2009      & 5037.44575 & 2700              &   150--800  \\
                &  25 jul 2009      & 5038.42181 & 2700              &   150--900  \\
                &  25 jul 2009      & 5038.45546 & 2700              &   150--850  \\
                &  27 jul 2009      & 5040.42000 & 2700              &   150--800  \\
                &  27 jul 2009      & 5040.45362 & 2700              &   100--550  \\
                &  28 jul 2009      & 5041.41078 & 2700              &   200--1000  \\
                &  28 jul 2009      & 5041.44442 & 2700              &   200--1000  \\
                &  29 jul 2009      & 5042.42667 & 2700              &   200--1050  \\
                &  29 jul 2009      & 5042.46030 & 2700              &   200--1050  \\
                &  30 jul 2009      & 5043.40589 & 2700              &   200--1000  \\
                &  30 jul 2009      & 5043.43952 & 2700              &   200--1000  \\
                &  31 jul 2009      & 5044.42425 & 2700              &   200--1100  \\
                &  31 jul 2009      & 5044.45789 & 2700              &   200--1050  \\
                &  01 aug 2009      & 5045.44377 & 2700              &   100--700 \\
                &  04 aug 2009      & 5048.47207 & 2700              &   200--1000 \\
\hline
HD 34078        &  15 oct 2007      & 4389.52124 & 2000              &   210--600 \\
(AE Aur)        &  15 oct 2007      & 4389.59673 & 2000              &   300--750 \\
                &  15 oct 2007      & 4389.62227 & 2000              &   300--800 \\
                &  15 oct 2007      & 4389.64780 & 2000              &   300--800 \\
                &  15 oct 2007      & 4389.67333 & 2000              &   300--800 \\
                &  18 oct 2007      & 4392.46172 & 2000              &   200--600 \\
                &  18 oct 2007      & 4392.48726 & 2000              &   200--550 \\
                &  18 oct 2007      & 4392.51279 & 2000              &   200--560 \\
                &  18 oct 2007      & 4392.53833 & 2000              &   130--370 \\
                &  19 oct 2007      & 4393.49062 & 2000              &   320--830 \\
                &  19 oct 2007      & 4393.51616 & 2000              &   310--810 \\

\end{tabular}
\end{center}
\end{table*}

\setcounter{table}{0}

\begin{table*}
\begin{center}
\caption{Continued} \label{tab_obs}
\begin{tabular}{llccc}
\hline
Star & Date  & HJD & exposure time & SNR \\
     &       & [d]-2450000 & [s]           &     \\
\hline
HD 209975       &  21 jun 2008      & 4639.54794 & 1800              &   300--800 \\
                &  21 jun 2008      & 4639.57115 & 1800              &   280--700 \\
                &  21 jun 2008      & 4639.59438 & 1800              &   280--680 \\
                &  21 jun 2008      & 4639.61760 & 1800              &   380--900 \\
                &  21 jun 2008      & 4639.64083 & 1800              &   350--850 \\
                &  22 jun 2008      & 4640.55014 & 1800              &   500--1200 \\
                &  22 jun 2008      & 4640.57336 & 1800              &   500--1200 \\
                &  22 jun 2008      & 4640.59659 & 1800              &   550--1250 \\
                &  22 jun 2008      & 4640.61983 & 1800              &   450--1150 \\
                &  22 jun 2008      & 4640.64305 & 1800              &   400--1000 \\
                &  25 jun 2008      & 4643.59284 & 1800              &   550--1250 \\
                &  25 jun 2008      & 4643.61605 & 1800              &   550--1250 \\
                &  25 jun 2008      & 4643.64215 & 1800              &   550--1350 \\
                &  27 jun 2008      & 4645.56781 & 1800              &   650--1450 \\
                &  27 jun 2008      & 4645.59102 & 1800              &   650--1450 \\
                &  27 jun 2008      & 4645.61424 & 1800              &   650--1450 \\
                &  27 jun 2008      & 4645.63782 & 1800              &   650--1450 \\
                &  27 jun 2008      & 4645.66102 & 1800              &   650--1450 \\
                &  28 jun 2008      & 4646.56284 & 1800              &   520--1250 \\
                &  28 jun 2008      & 4646.58606 & 1800              &   470--1150 \\
                &  28 jun 2008      & 4646.60928 & 1800              &   470--1180 \\
                &  28 jun 2008      & 4646.63249 & 1800              &   570--1370 \\
                &  28 jun 2008      & 4646.65571 & 1800              &   580--1400 \\
\hline
HD 214680       &  15 oct 2007      & 4389.31523 & 1200              &   500--960 \\
(10 Lac)        &  15 oct 2007      & 4389.33151 & 1200              &   550--1050 \\
                &  15 oct 2007      & 4389.34780 & 1200              &   550--1050 \\
                &  16 oct 2007      & 4390.41320 & 1200              &   520--1000 \\
                &  16 oct 2007      & 4390.42949 & 1200              &   410--800 \\
                &  16 oct 2007      & 4390.44577 & 1200              &   450--880 \\
                &  17 oct 2007      & 4391.46364 & 2000              &   710--1410 \\
                &  17 oct 2007      & 4391.48917 & 2000              &   700--1400 \\
                &  17 oct 2007      & 4391.51470 & 2000              &   680--1380 \\
                &  18 oct 2007      & 4392.25412 & 1200              &   400--800 \\
                &  18 oct 2007      & 4392.27040 & 1200              &   360--720 \\
                &  18 oct 2007      & 4392.28669 & 1200              &   340--680 \\
                &  19 oct 2007      & 4393.26458 & 2000              &   620--1250 \\
                &  19 oct 2007      & 4393.29011 & 2000              &   640--1280 \\
                &  19 oct 2007      & 4393.31563 & 2000              &   660--1300 \\
                &  20 oct 2007      & 4394.24877 & 2000              &   480--1200 \\
                &  20 oct 2007      & 4394.27430 & 2000              &   680--1330 \\
                &  20 oct 2007      & 4394.29984 & 2000              &   700--1360 \\
                &  21 oct 2007      & 4395.26345 & 2000              &   500--1000 \\
                &  21 oct 2007      & 4395.28899 & 2000              &   600--1200 \\
                &  21 oct 2007      & 4395.31454 & 2000              &   640--1250 \\
                &  23 oct 2007      & 4397.25262 & 2000              &   580--1170 \\
                &  23 oct 2007      & 4397.27815 & 2000              &   580--1150 \\
                &  23 oct 2007      & 4397.30369 & 2000              &   560--1340 \\
                &  24 oct 2007      & 4398.25670 & 2000              &   640--1300 \\
                &  24 oct 2007      & 4398.28224 & 2000              &   650--1300 \\
                &  24 oct 2007      & 4398.30777 & 2000              &   750--1450 \\

\end{tabular}
\end{center}
\end{table*}

\clearpage

\section{Temporal variance spectra of sample stars}
\label{ap_var}

\begin{figure*}
     \centering
     \subfigure[\ha]{
          \includegraphics[width=.28\textwidth]{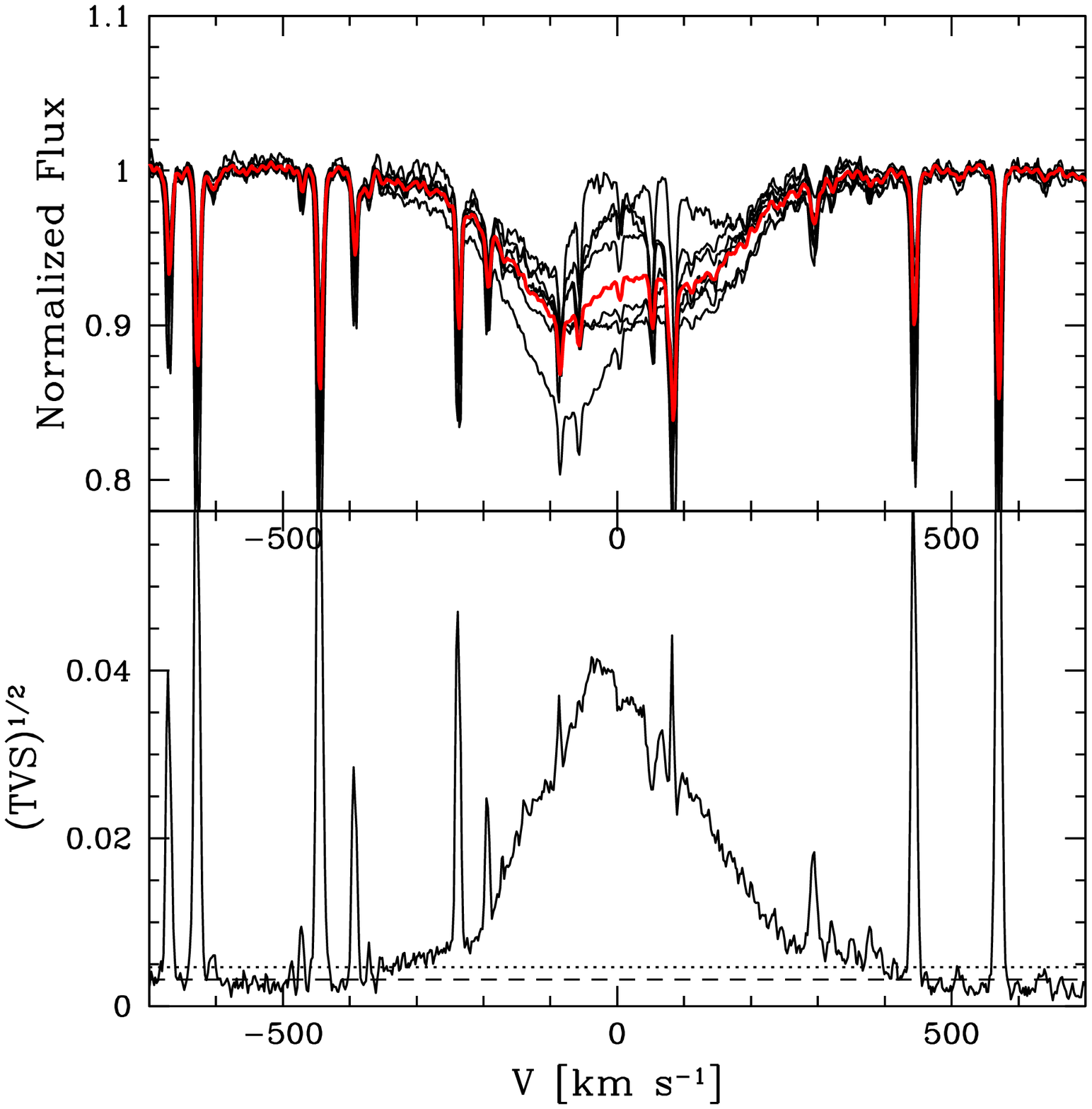}}
     \hspace{0.2cm}
     \subfigure[H$_{\beta}$]{
          \includegraphics[width=.28\textwidth]{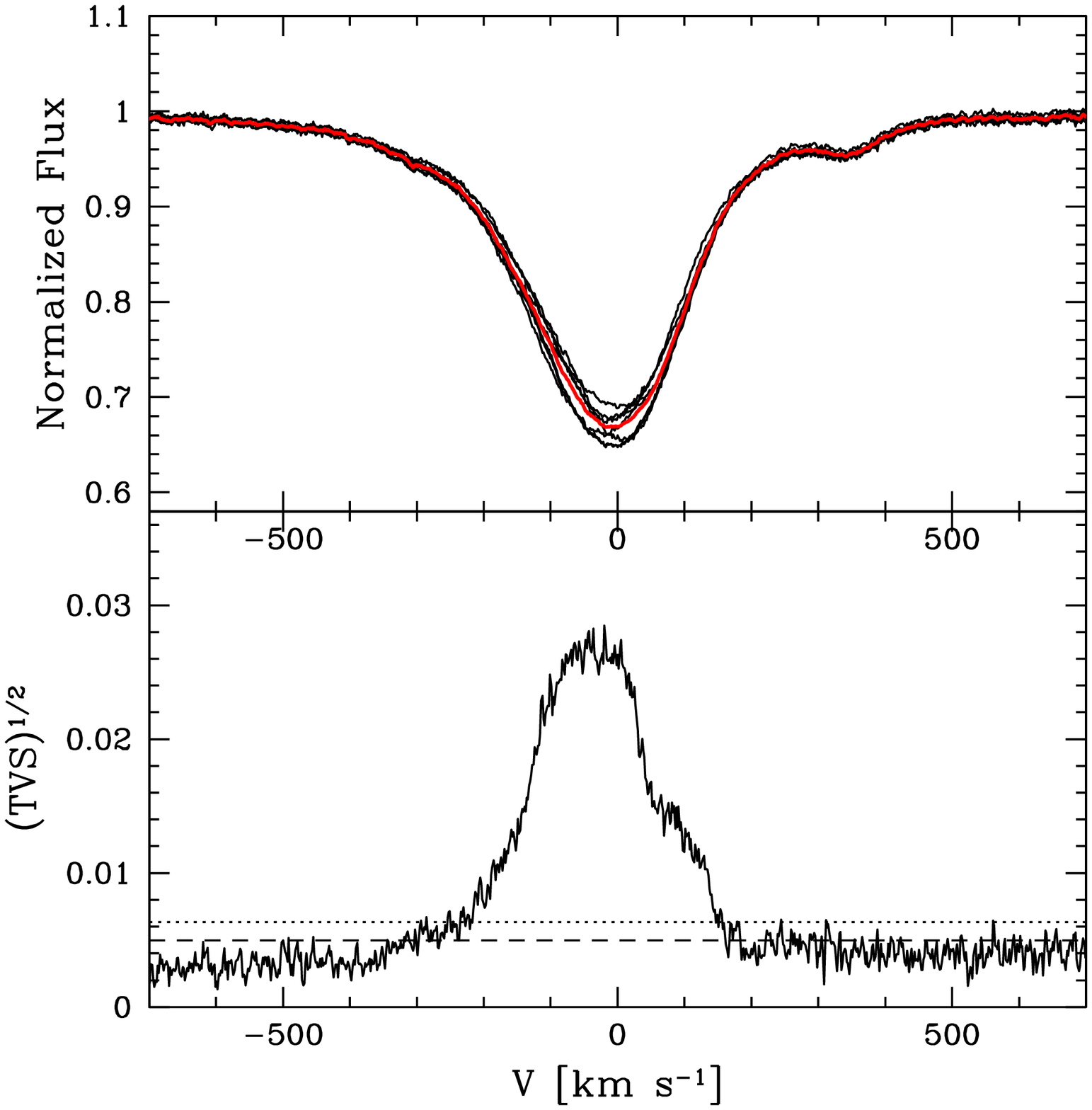}}
     \hspace{0.2cm}
     \subfigure[H$_{\gamma}$]{
          \includegraphics[width=.28\textwidth]{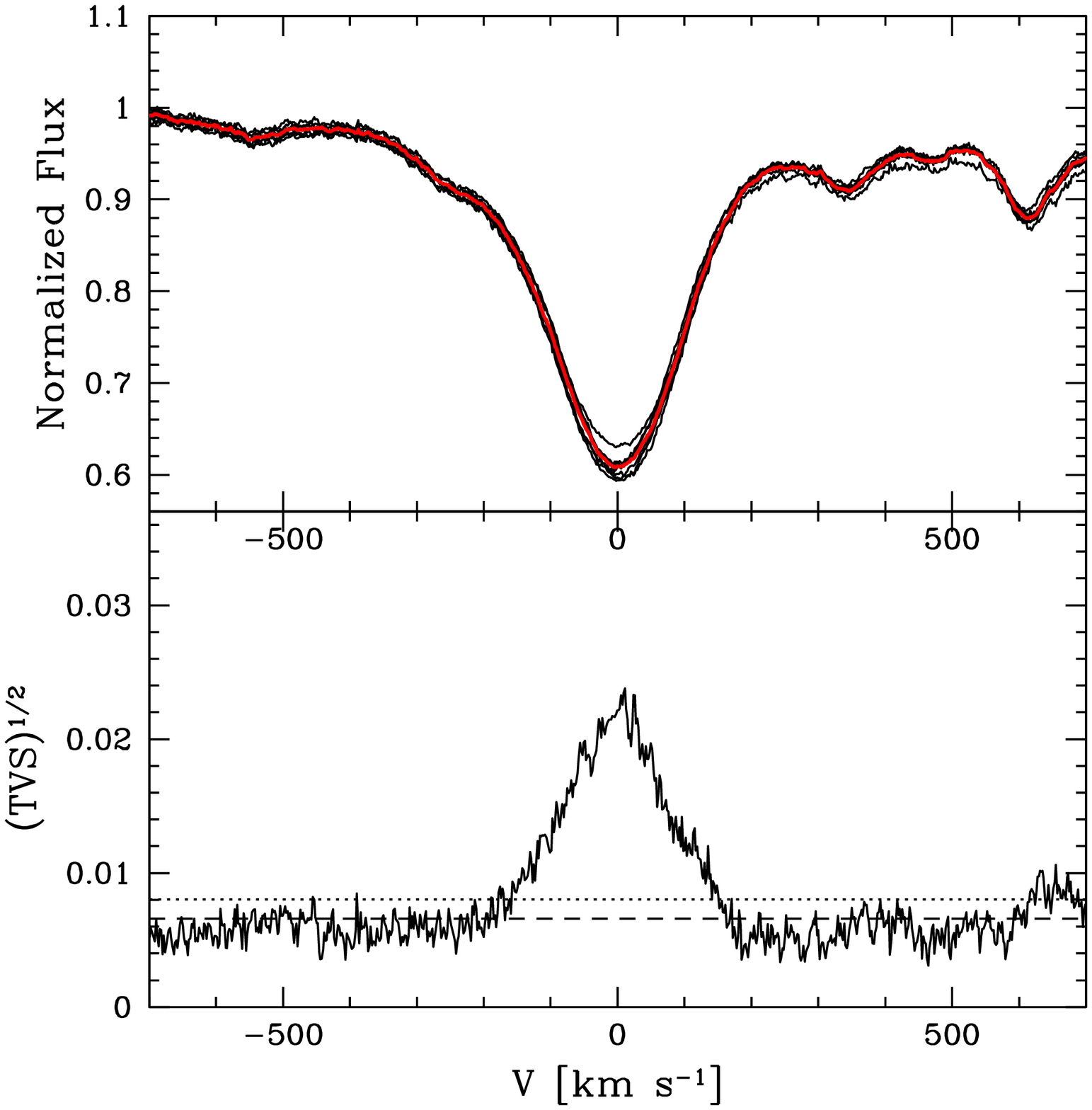}}\\
     \subfigure[\ion{He}{I} 4026]{
          \includegraphics[width=.28\textwidth]{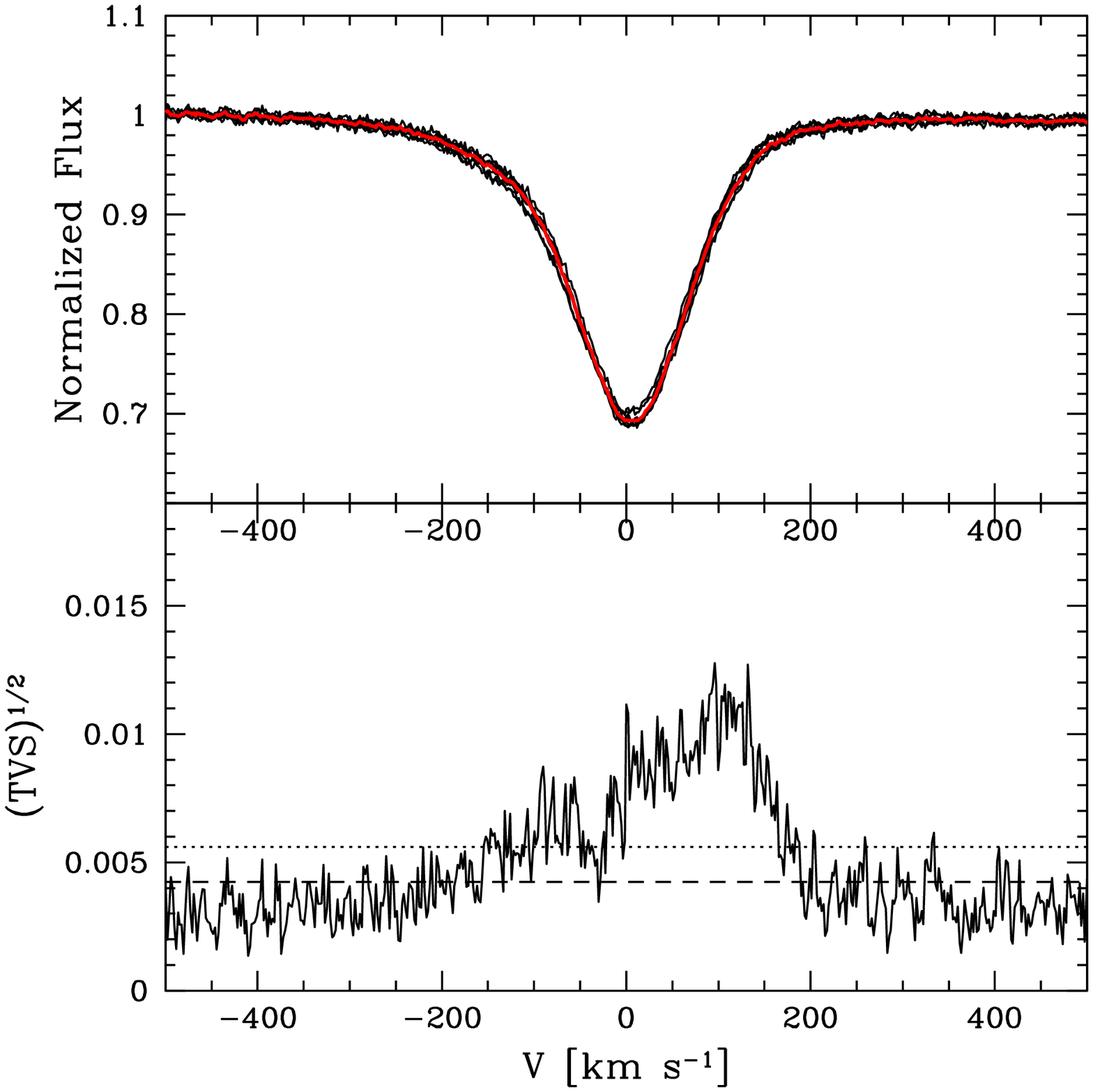}}
     \hspace{0.2cm}
     \subfigure[\ion{He}{I} 4471]{
          \includegraphics[width=.28\textwidth]{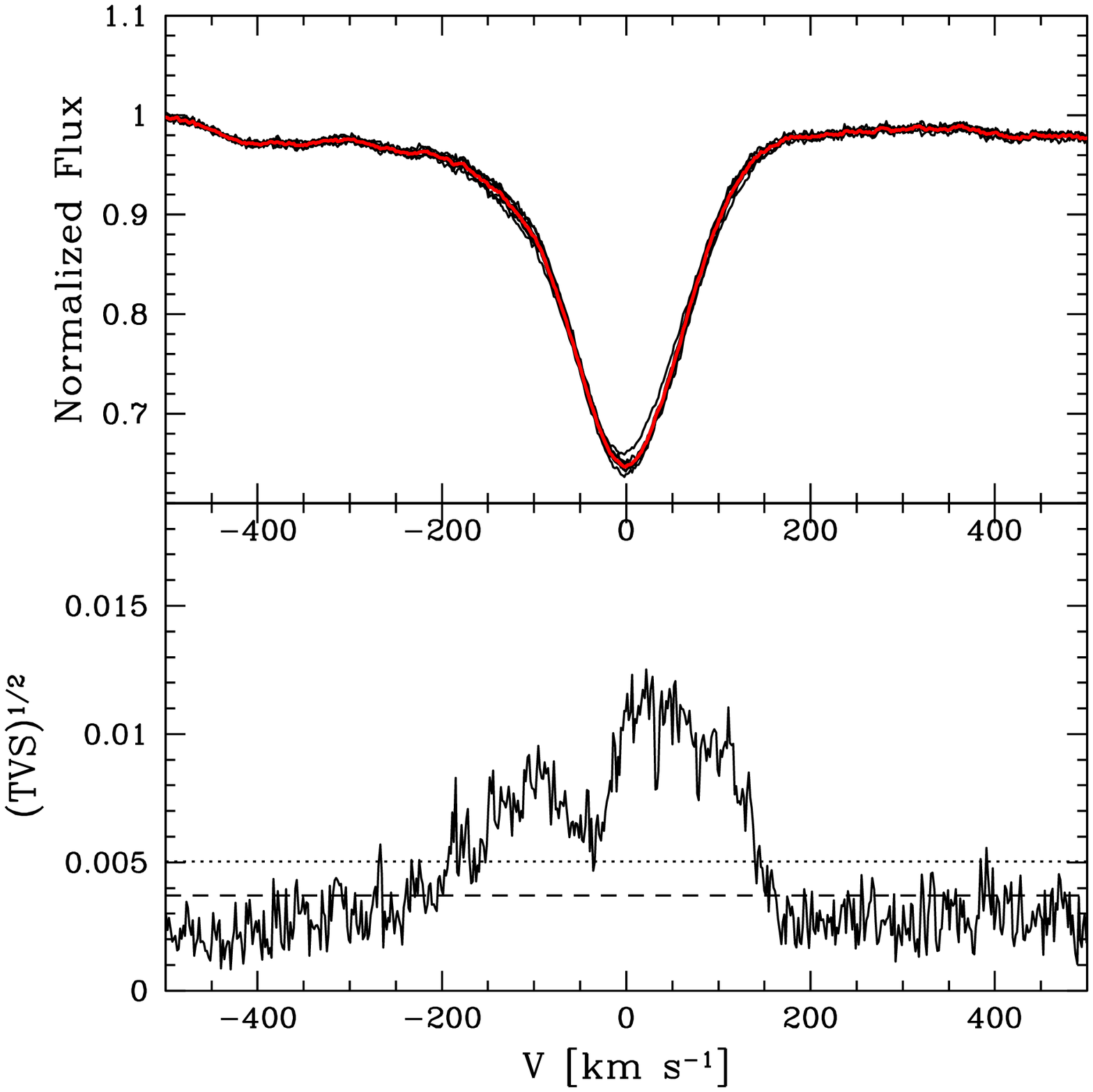}}
     \hspace{0.2cm}
     \subfigure[\ion{He}{I} 4712]{
          \includegraphics[width=.28\textwidth]{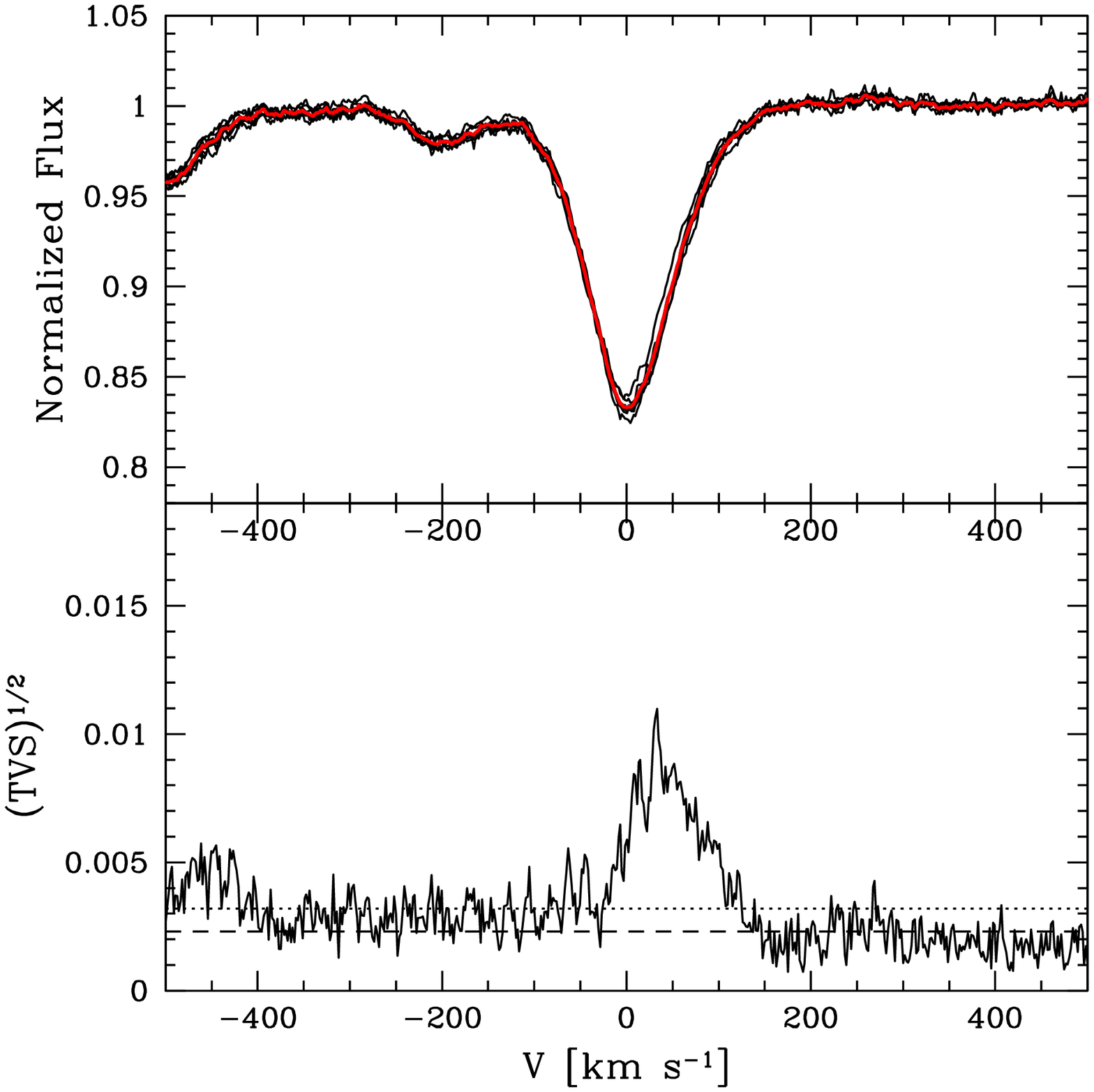}}\\
     \subfigure[\ion{He}{I} 5876]{
          \includegraphics[width=.28\textwidth]{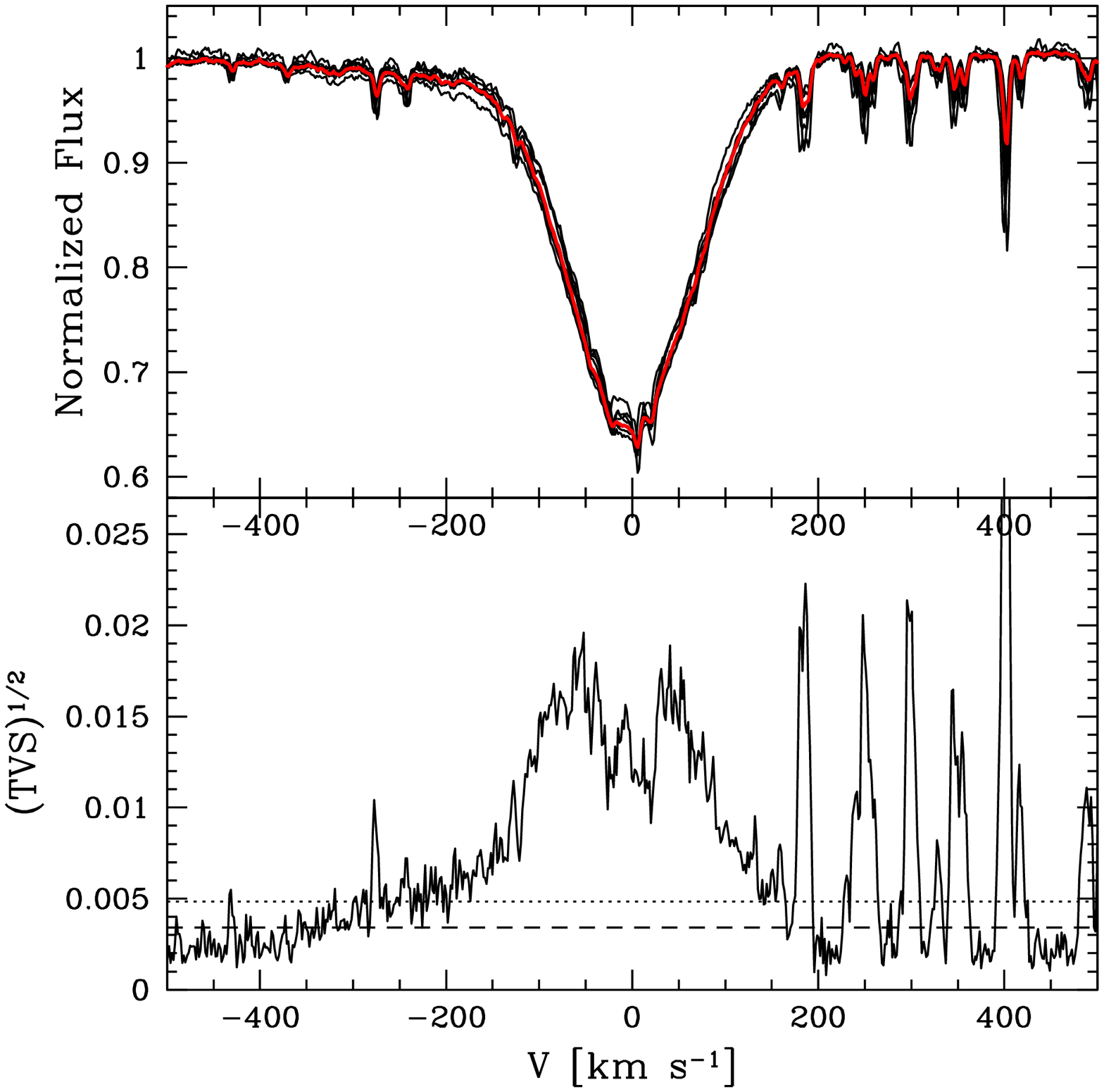}}
     \hspace{0.2cm}
     \subfigure[\ion{He}{II} 4542]{
          \includegraphics[width=.28\textwidth]{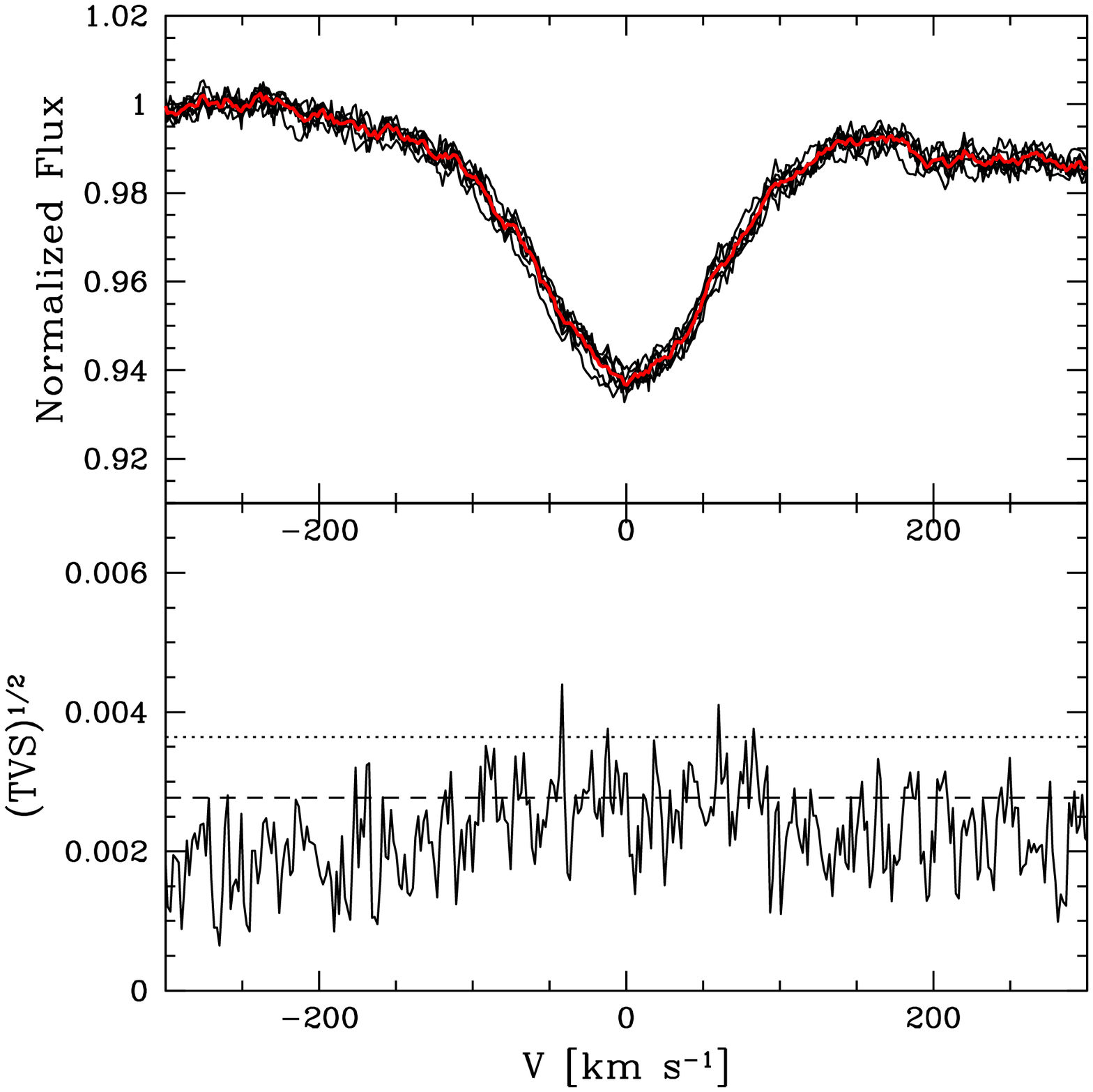}}
     \hspace{0.2cm}
     \subfigure[\ion{He}{II} 4686]{
          \includegraphics[width=.28\textwidth]{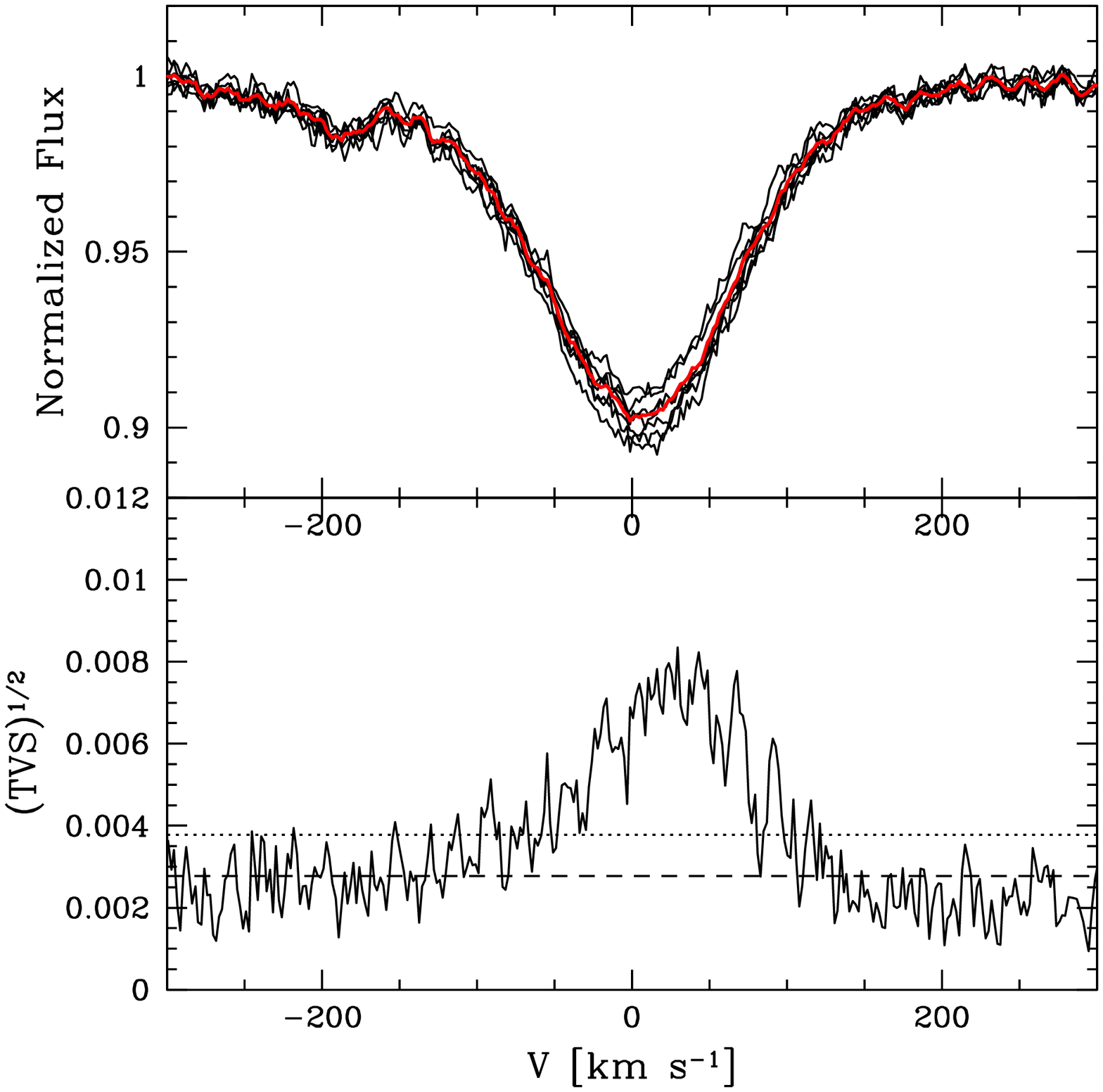}}\\
     \subfigure[\ion{He}{II} 5412]{
          \includegraphics[width=.28\textwidth]{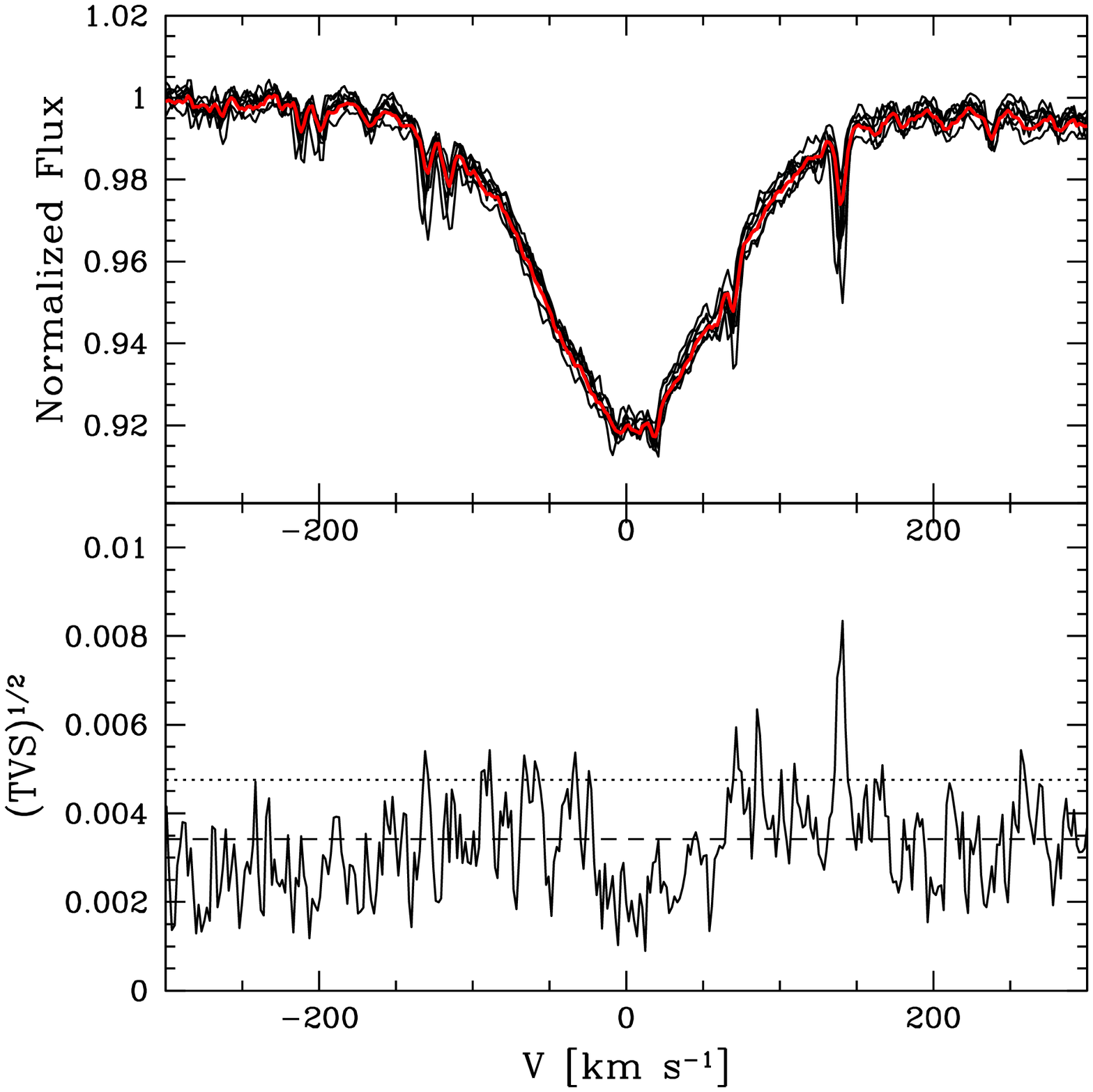}}
     \hspace{0.2cm}
     \subfigure[\ion{O}{III} 5592]{
          \includegraphics[width=.28\textwidth]{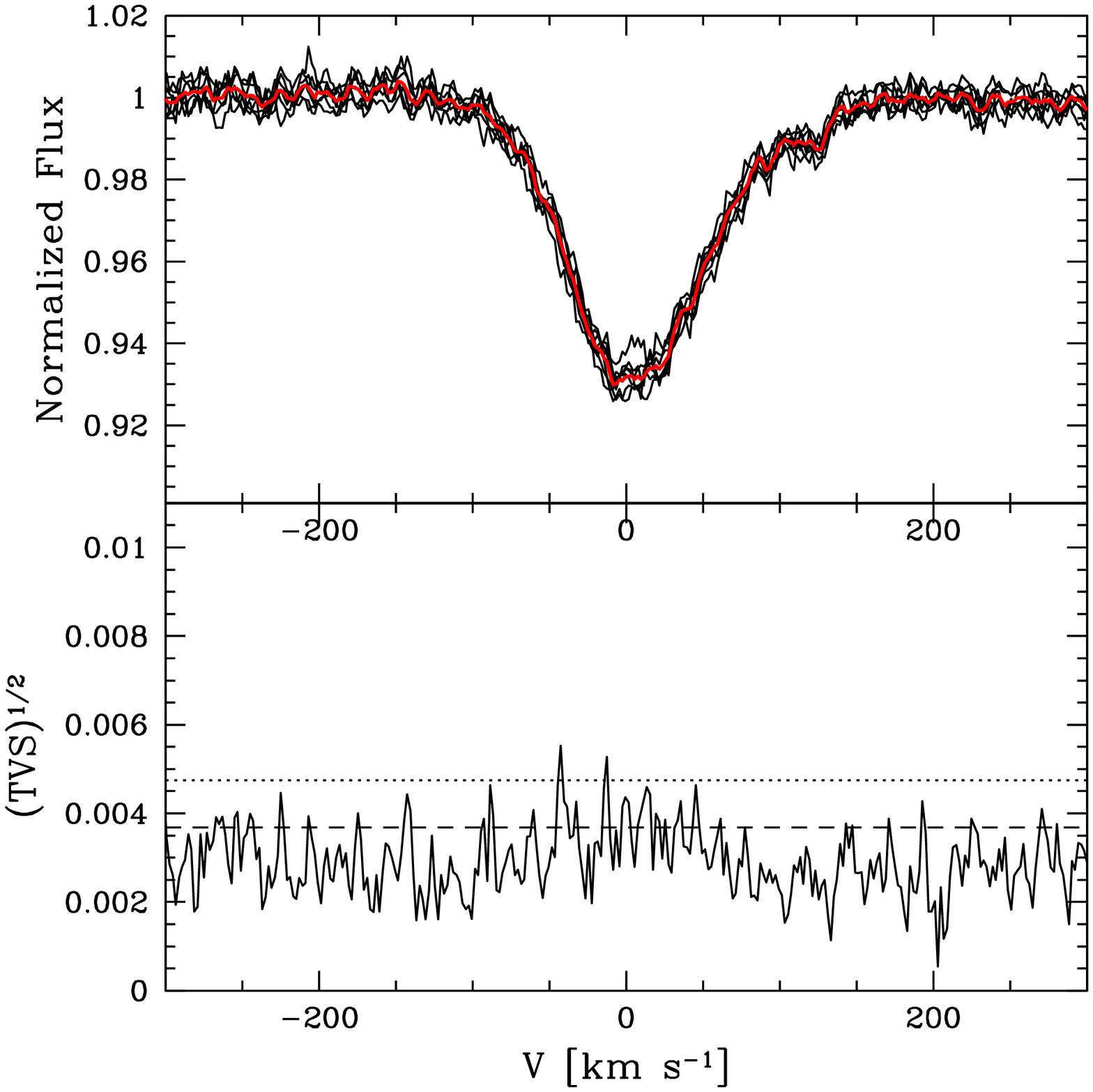}}
     \hspace{0.2cm}
     \subfigure[\ion{C}{IV} 5802]{
          \includegraphics[width=.28\textwidth]{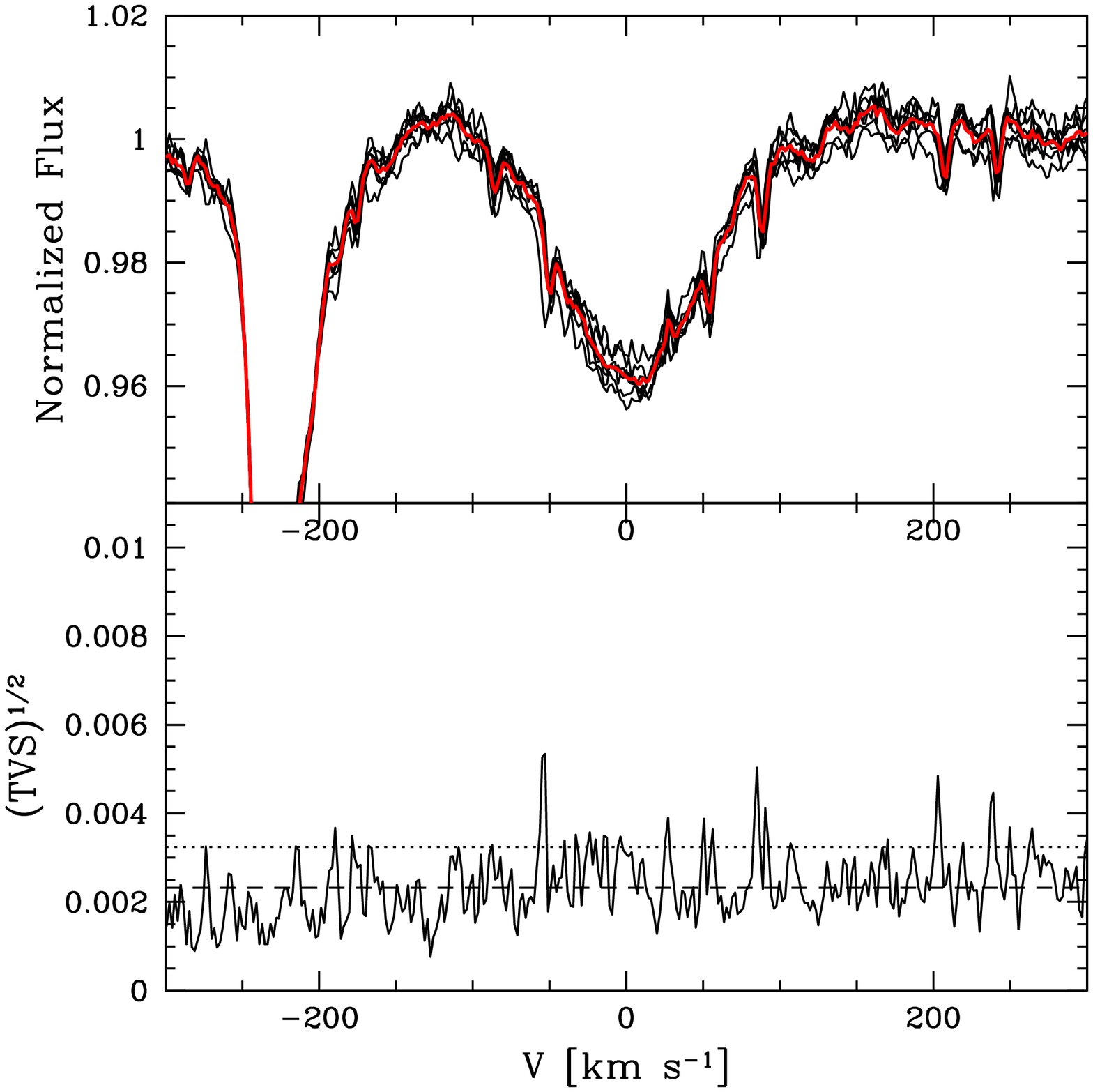}}\\
     \caption{Variability of HD~167264 between July 25$^{th}$ and August 4$^{th}$ 2009.}
     \label{fig_var_15sgr}
\end{figure*}

\newpage

\begin{figure*}
     \centering
     \subfigure[\ha]{
          \includegraphics[width=.28\textwidth]{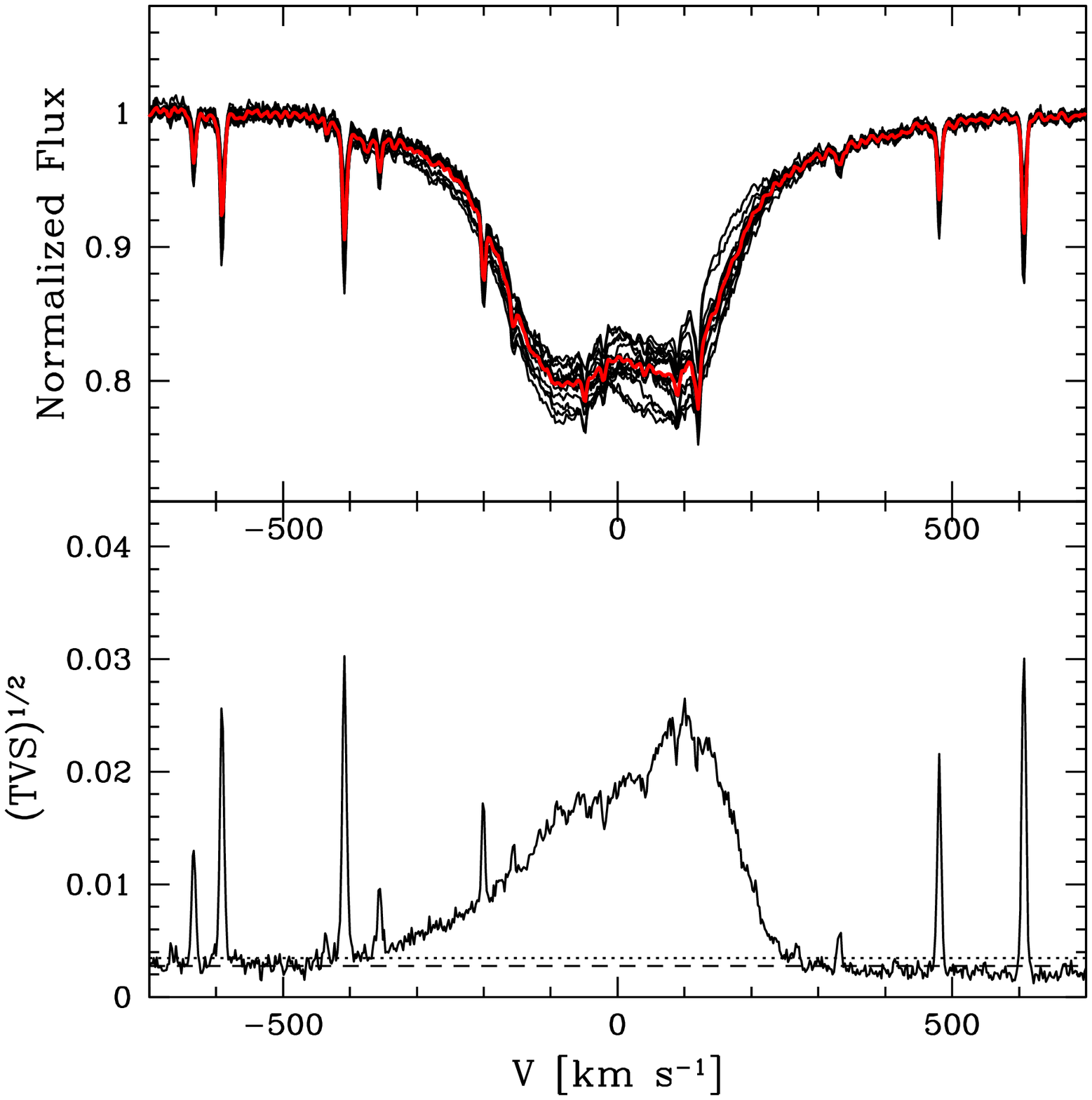}}
     \hspace{0.2cm}
     \subfigure[H$_{\beta}$]{
          \includegraphics[width=.28\textwidth]{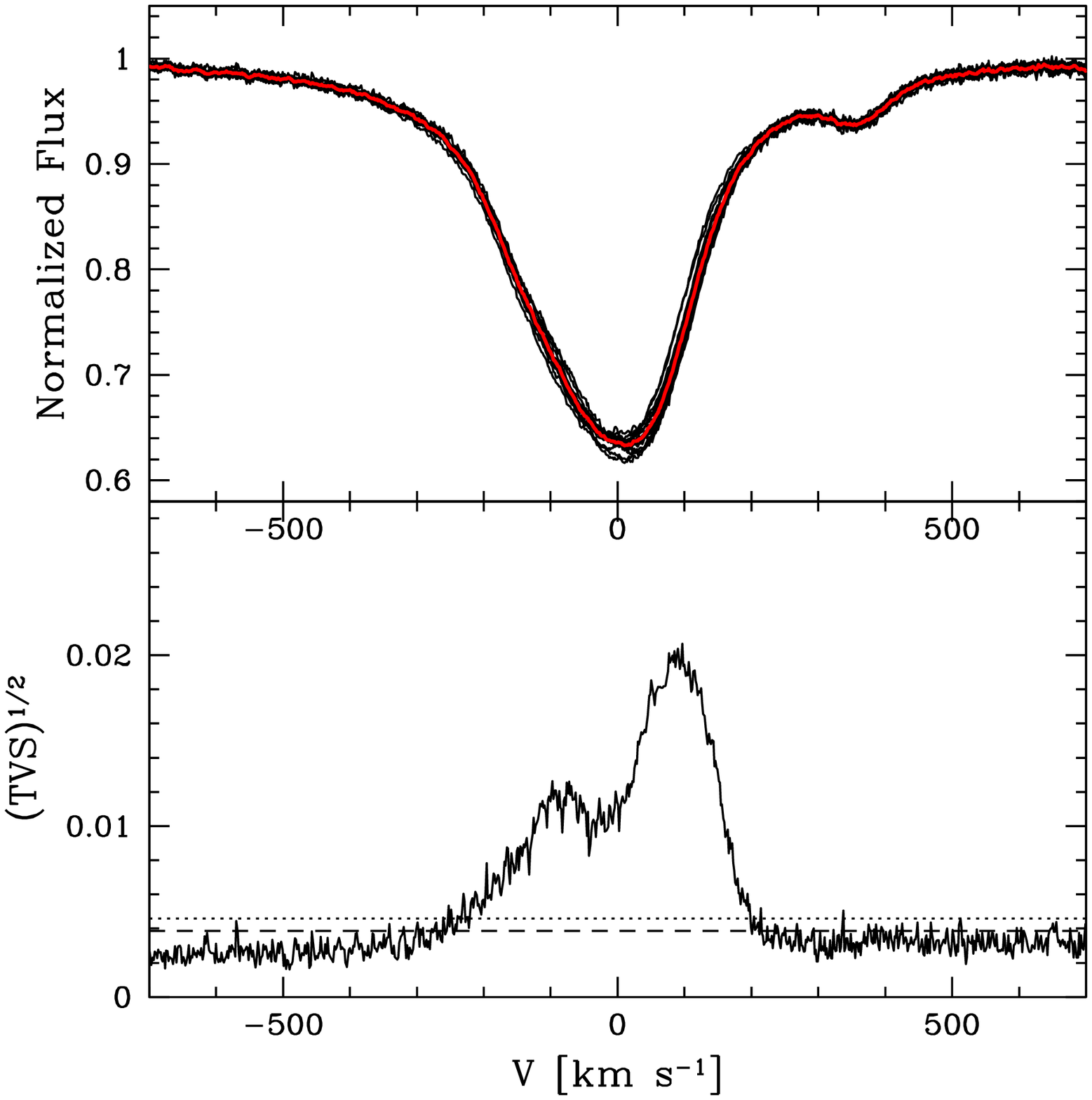}}
     \hspace{0.2cm}
     \subfigure[H$_{\gamma}$]{
          \includegraphics[width=.28\textwidth]{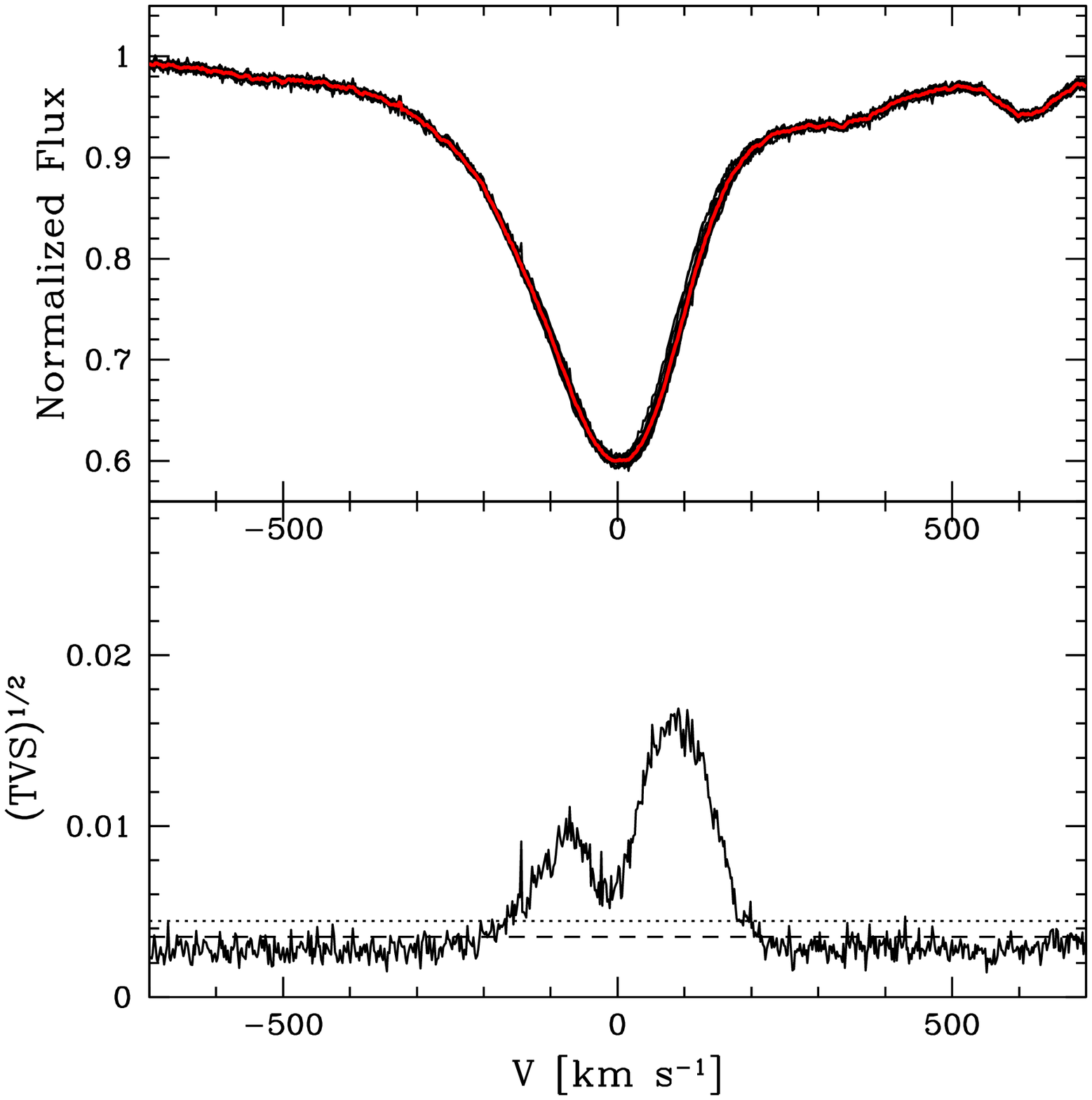}}\\
     \subfigure[\ion{He}{I} 4026]{
          \includegraphics[width=.28\textwidth]{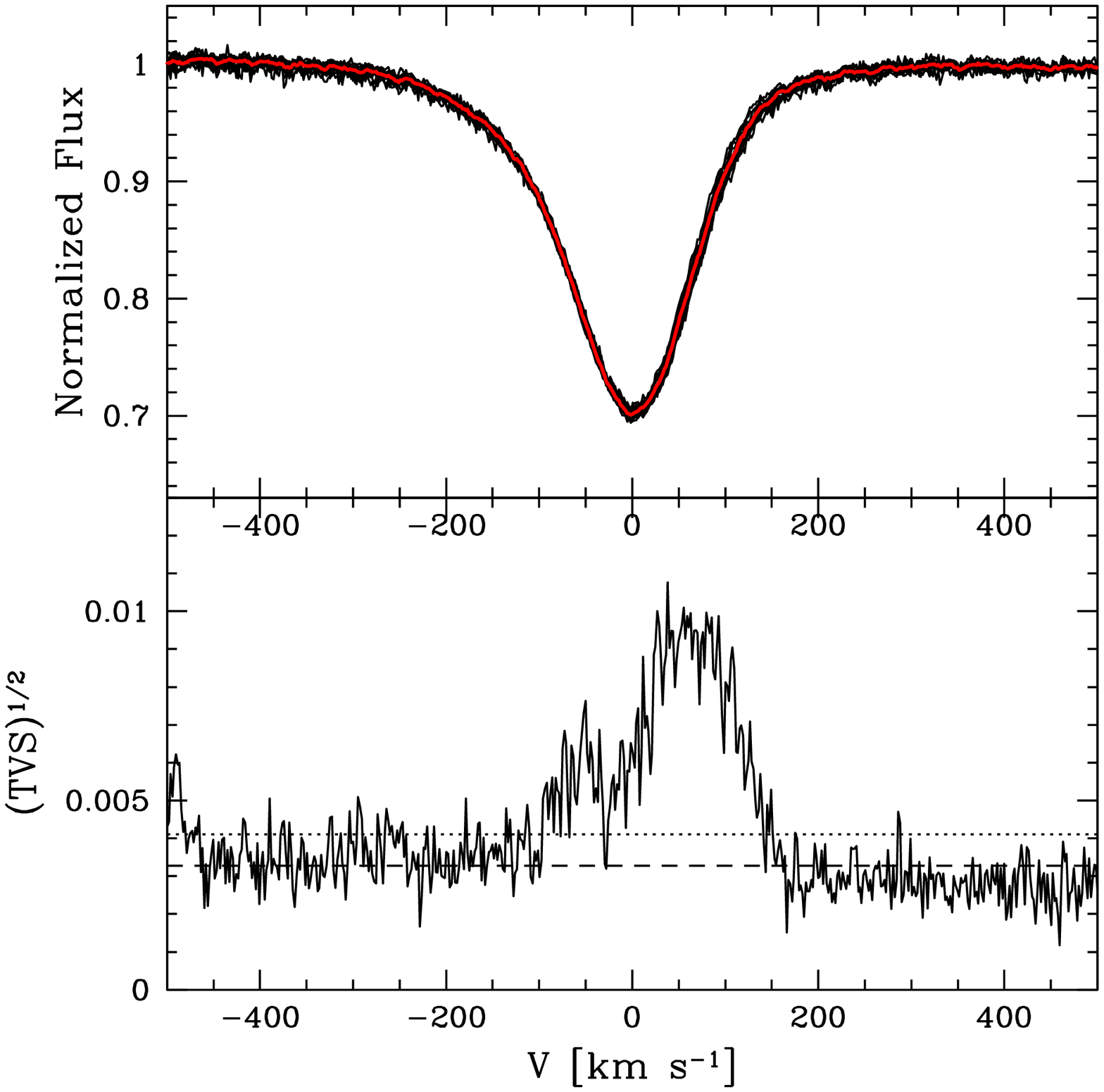}}
     \hspace{0.2cm}
     \subfigure[\ion{He}{I} 4471]{
          \includegraphics[width=.28\textwidth]{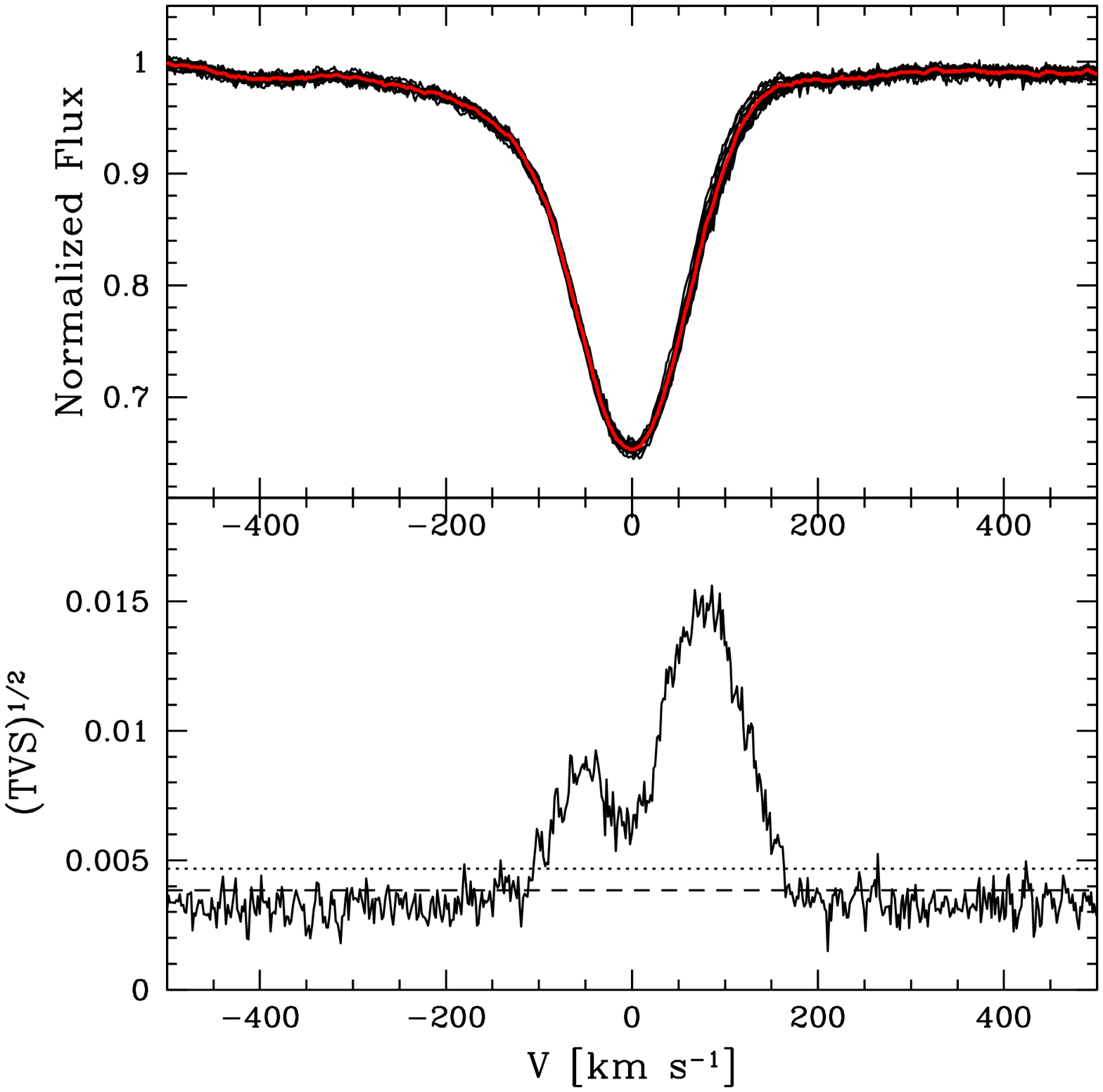}}
     \hspace{0.2cm}
     \subfigure[\ion{He}{I} 4712]{
          \includegraphics[width=.28\textwidth]{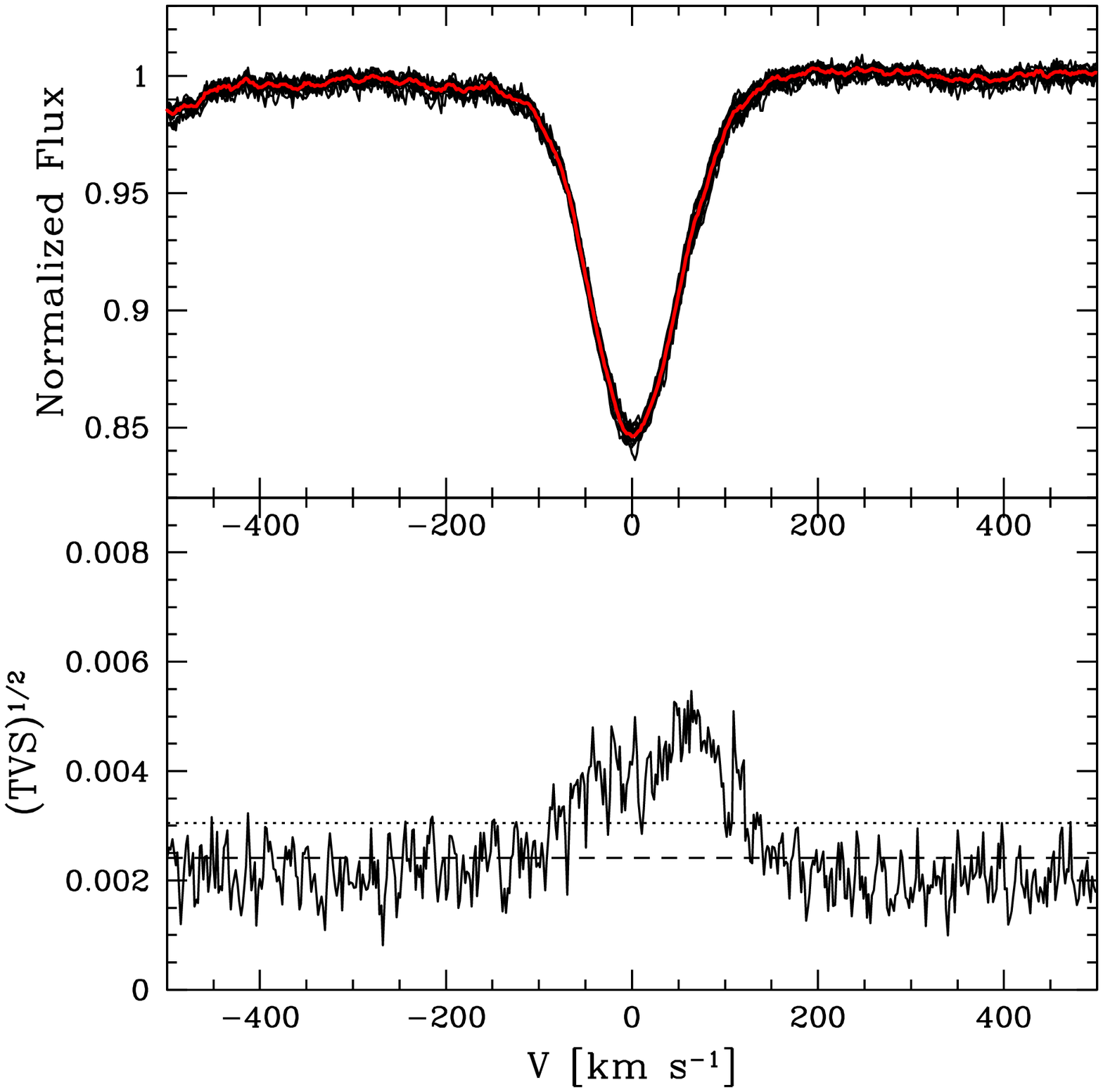}}\\
     \subfigure[\ion{He}{I} 5876]{
          \includegraphics[width=.28\textwidth]{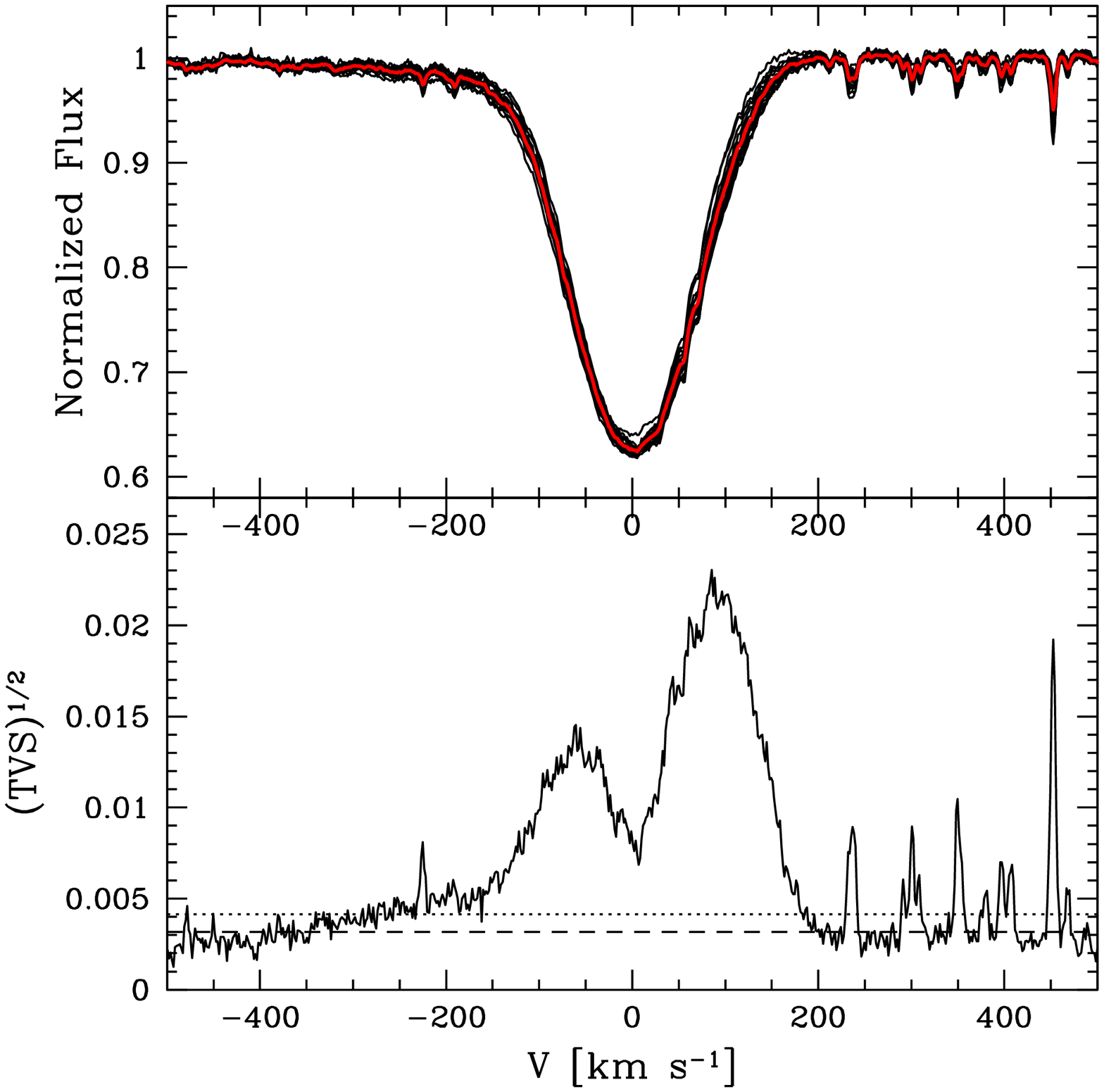}}
     \hspace{0.2cm}
     \subfigure[\ion{He}{II} 4542]{
          \includegraphics[width=.28\textwidth]{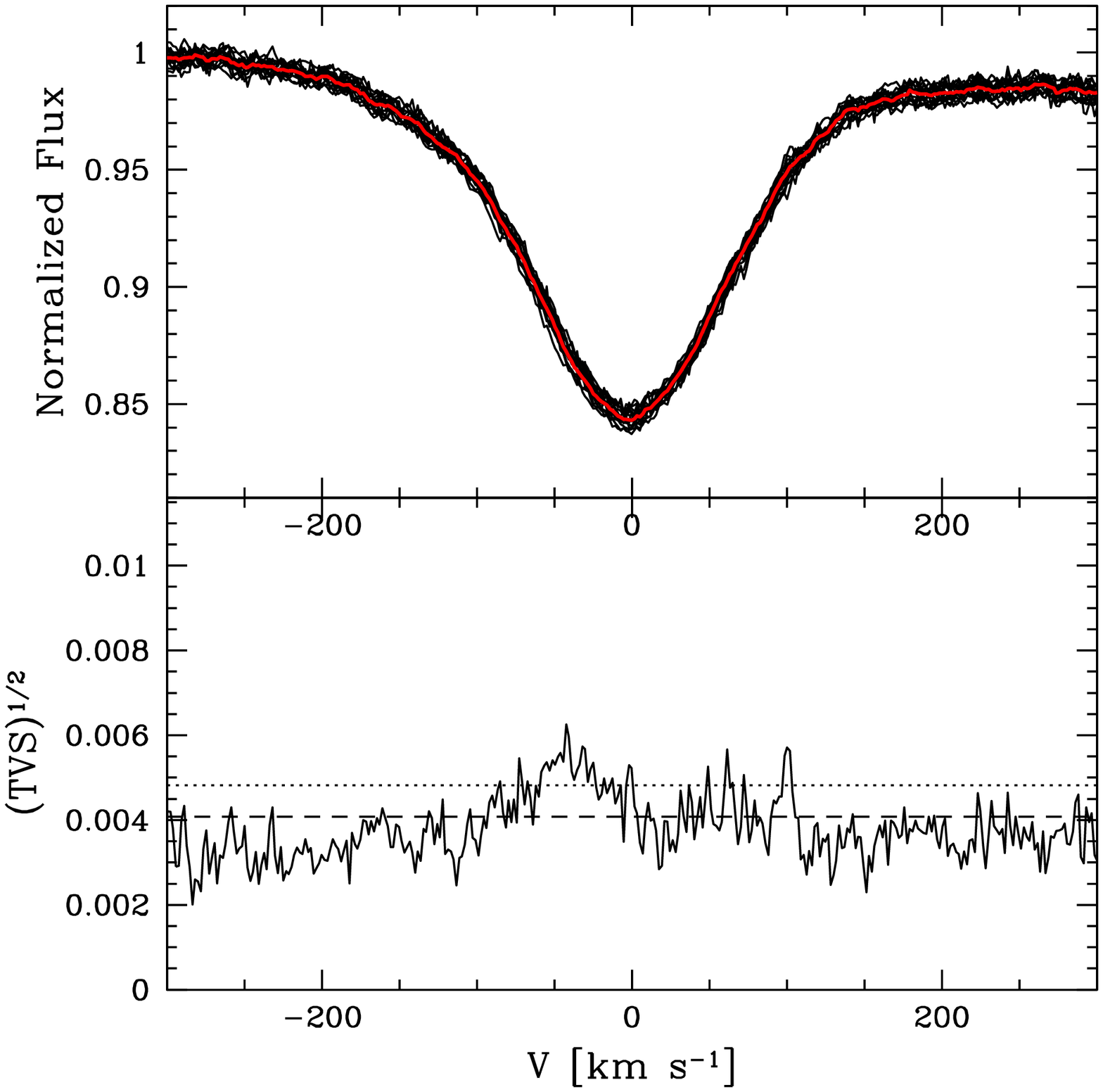}}
     \hspace{0.2cm}
     \subfigure[\ion{He}{II} 4686]{
          \includegraphics[width=.28\textwidth]{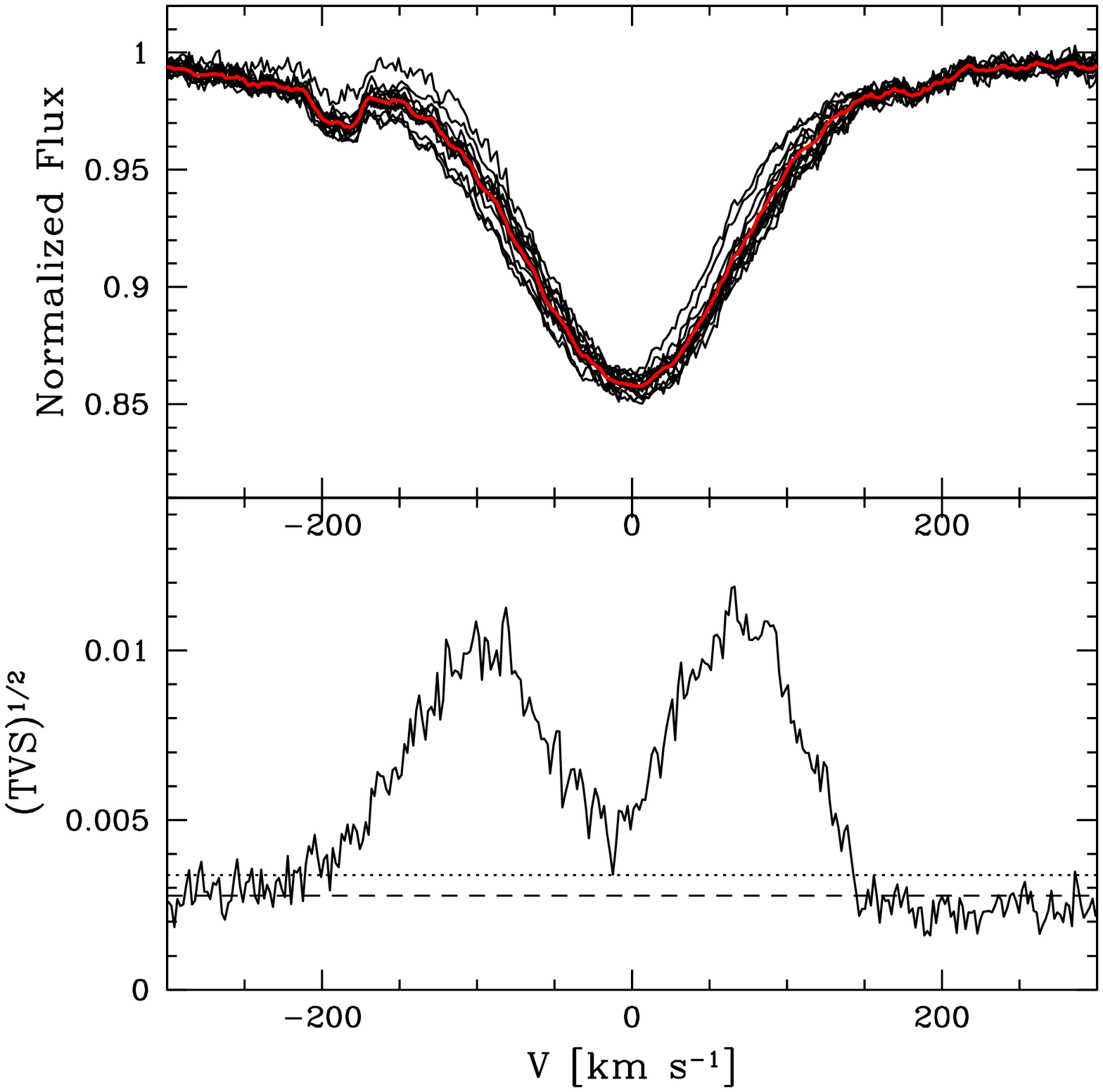}}\\
     \subfigure[\ion{He}{II} 5412]{
          \includegraphics[width=.28\textwidth]{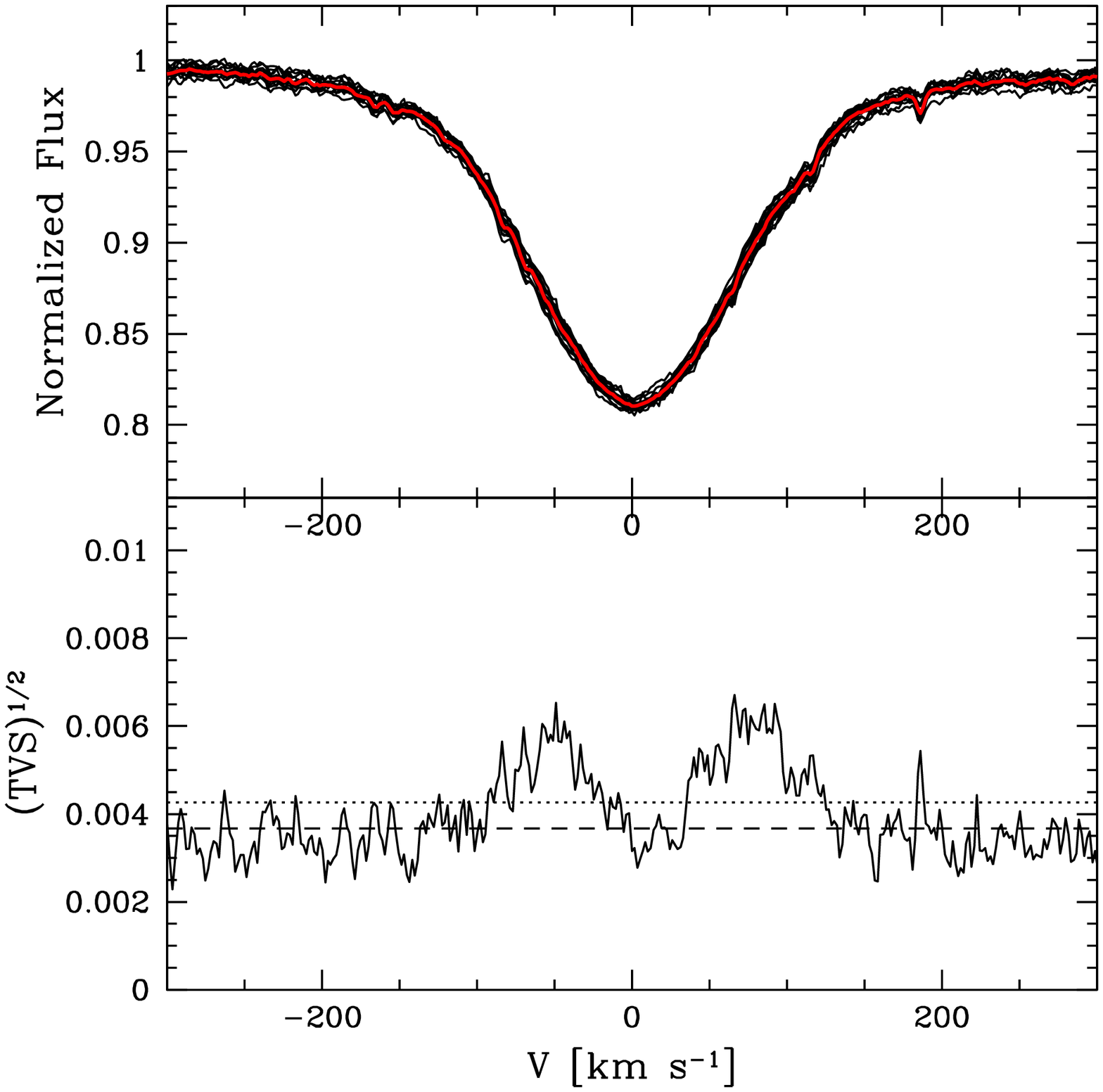}}
     \hspace{0.2cm}
     \subfigure[\ion{O}{III} 5592]{
          \includegraphics[width=.28\textwidth]{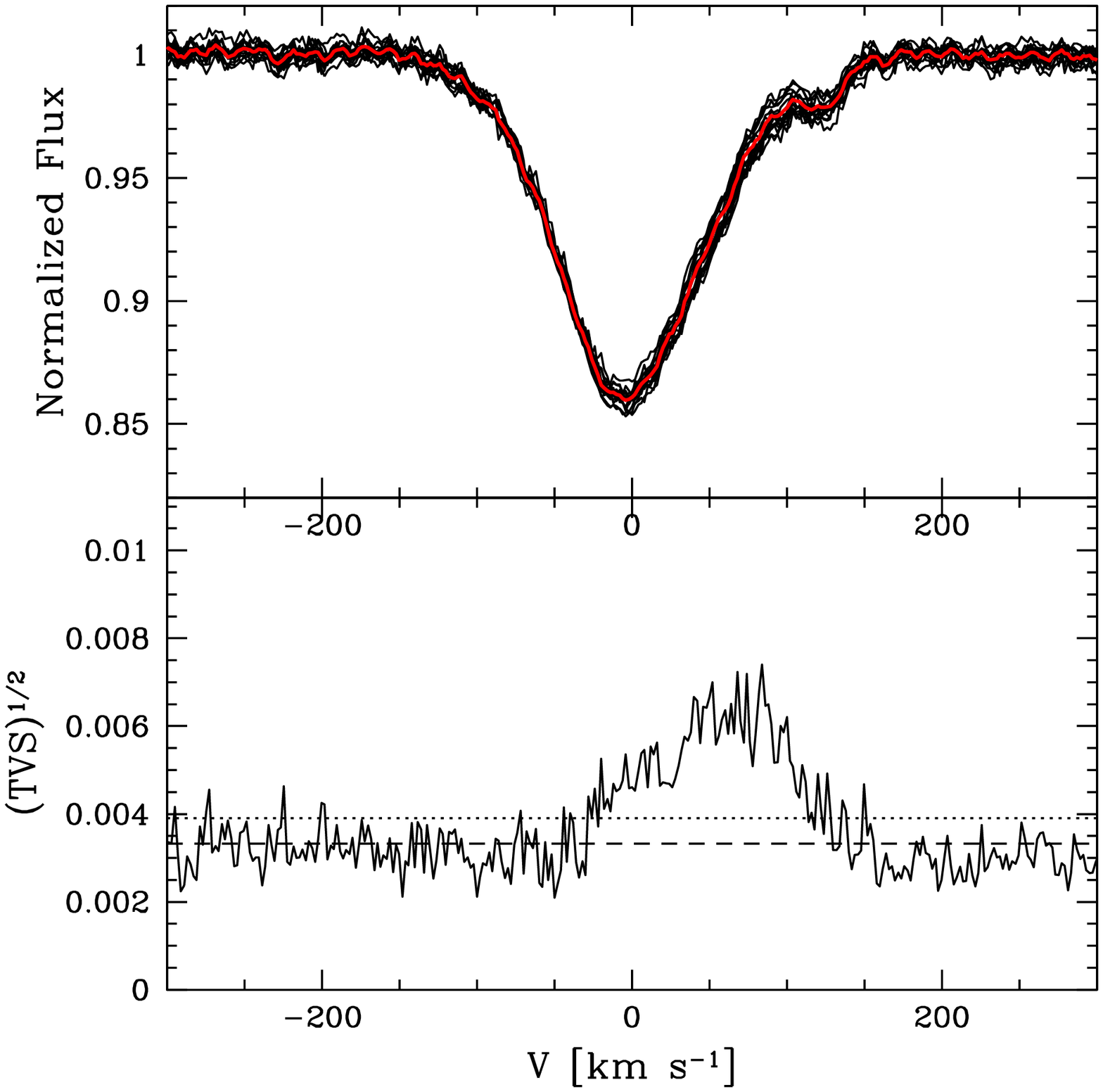}}
     \hspace{0.2cm}
     \subfigure[\ion{C}{IV} 5802]{
          \includegraphics[width=.28\textwidth]{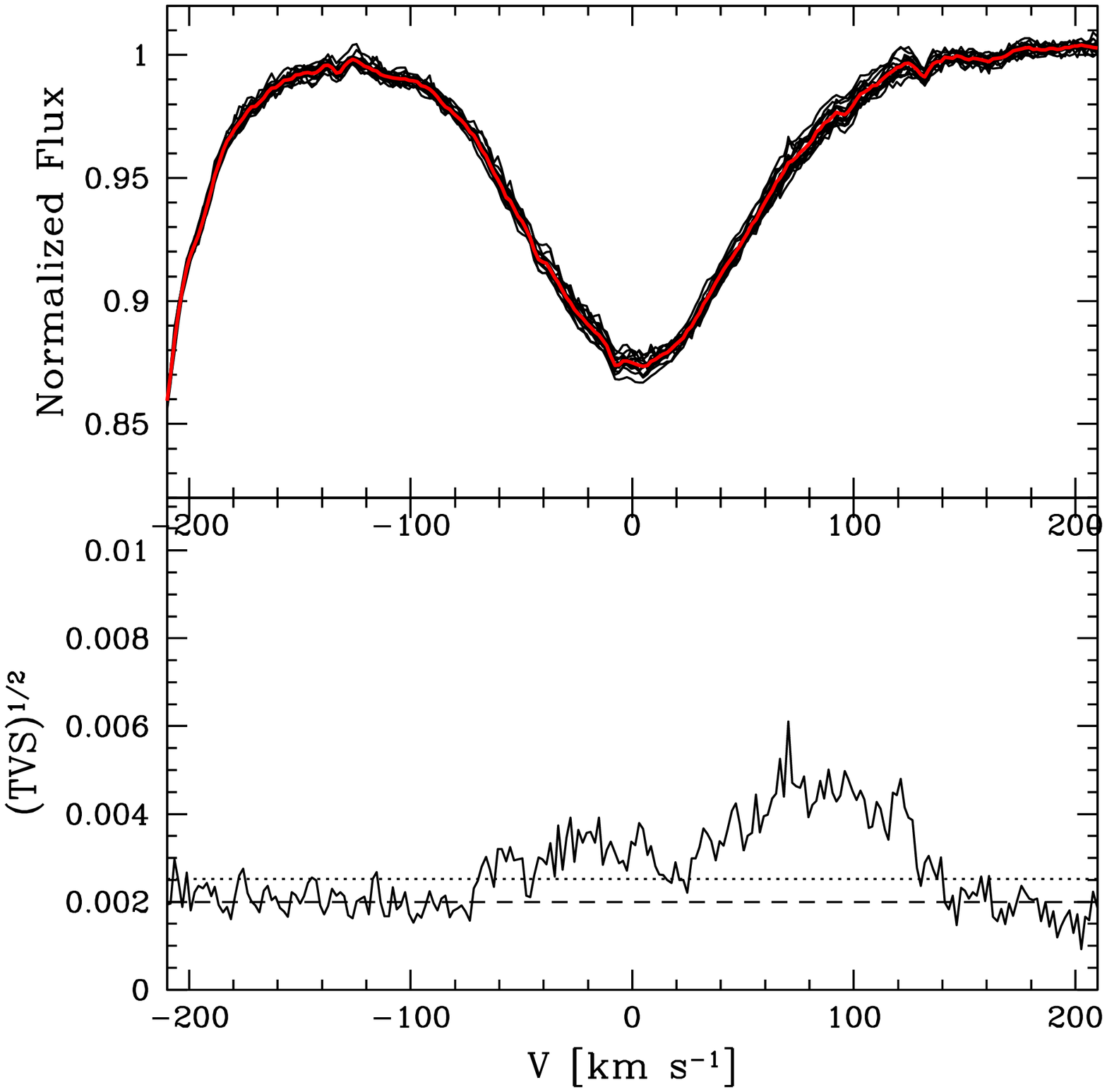}}\\
     \caption{Variability of HD207198 between July 25$^{th}$ and August 4$^{th}$ 2009.}
     \label{fig_var_207198}
\end{figure*}

\newpage

\begin{figure*}
     \centering
     \subfigure[\ha]{
          \includegraphics[width=.28\textwidth]{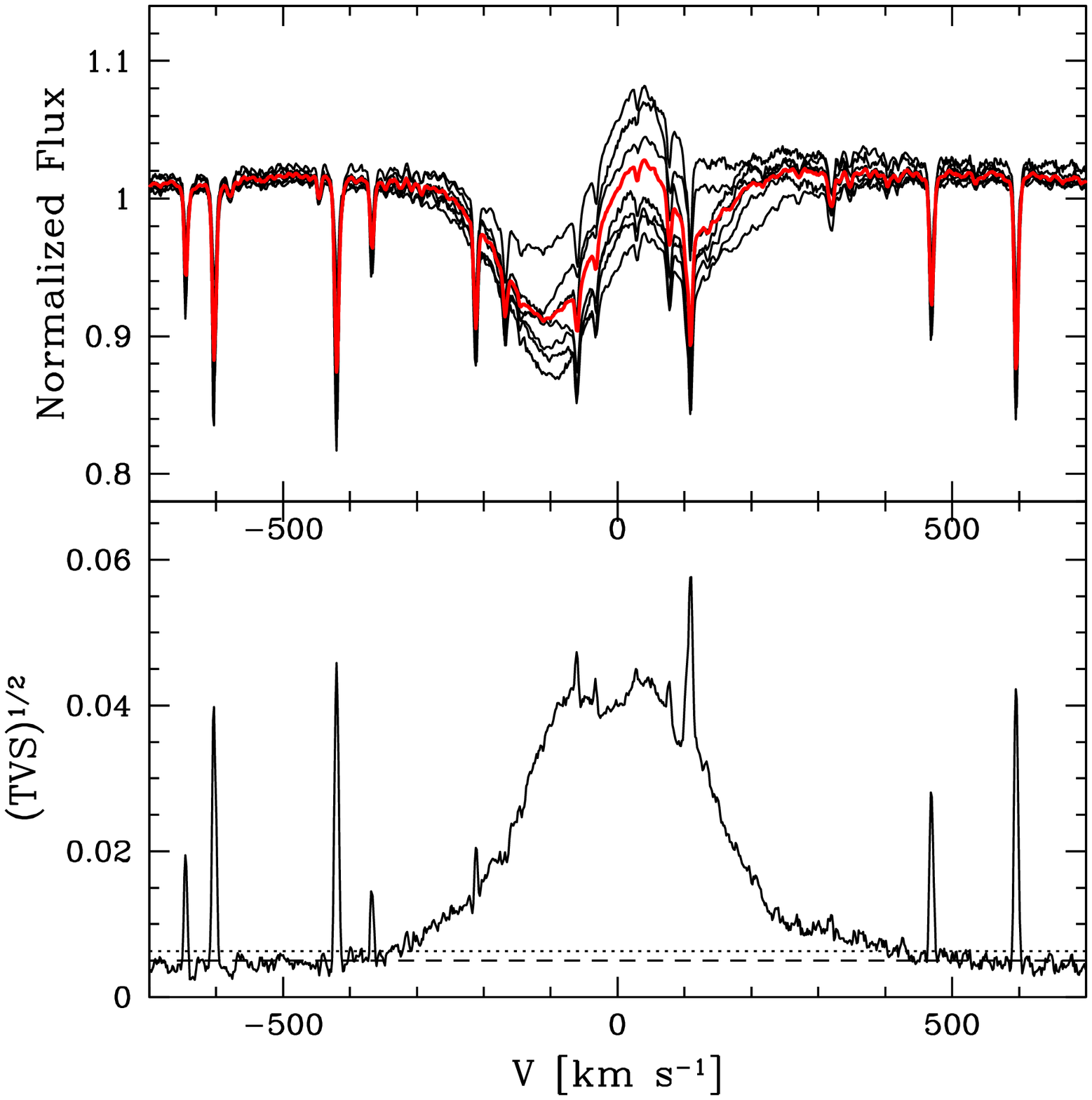}}
     \hspace{0.2cm}
     \subfigure[H$_{\beta}$]{
          \includegraphics[width=.28\textwidth]{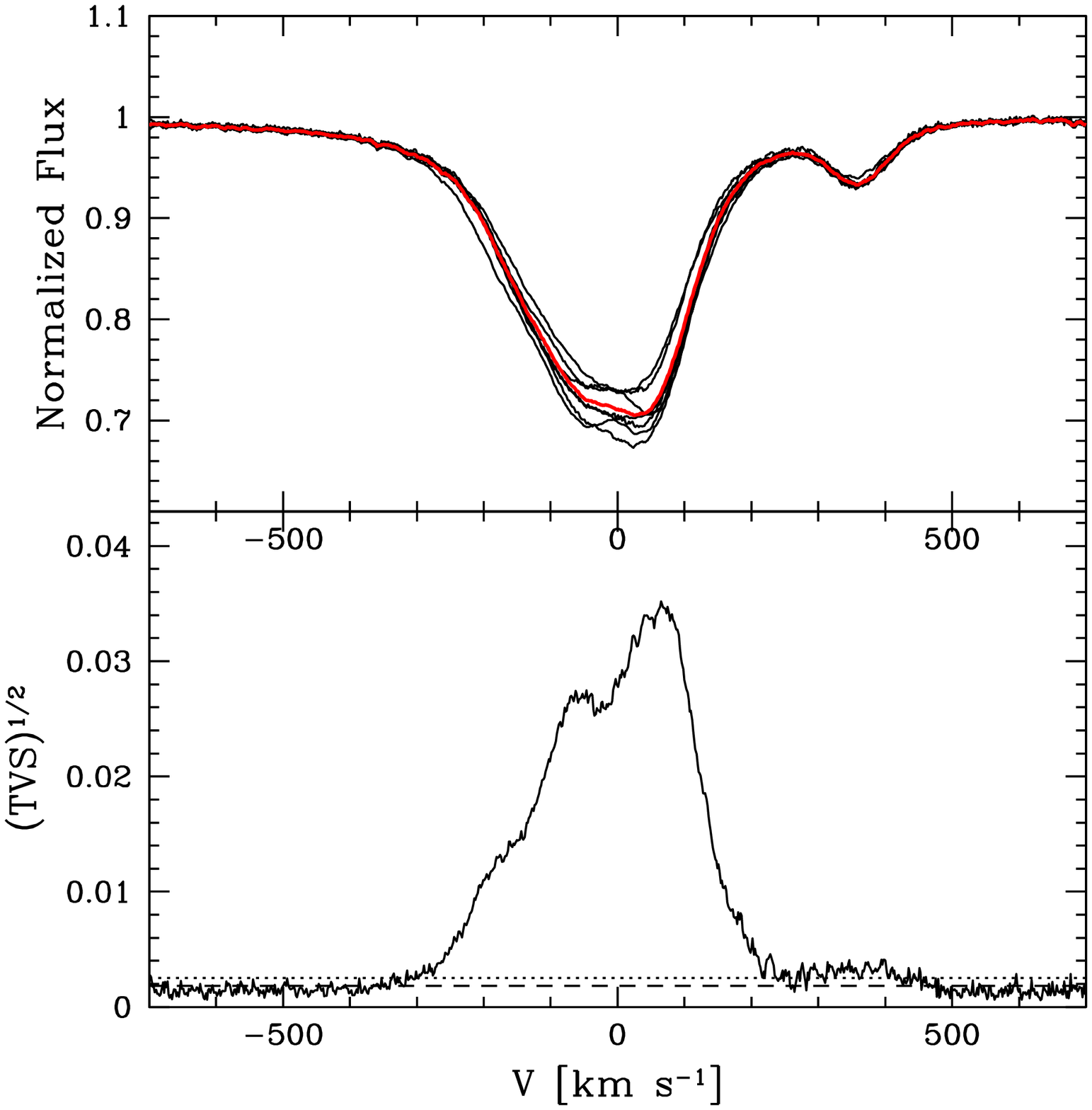}}
     \hspace{0.2cm}
     \subfigure[H$_{\gamma}$]{
          \includegraphics[width=.28\textwidth]{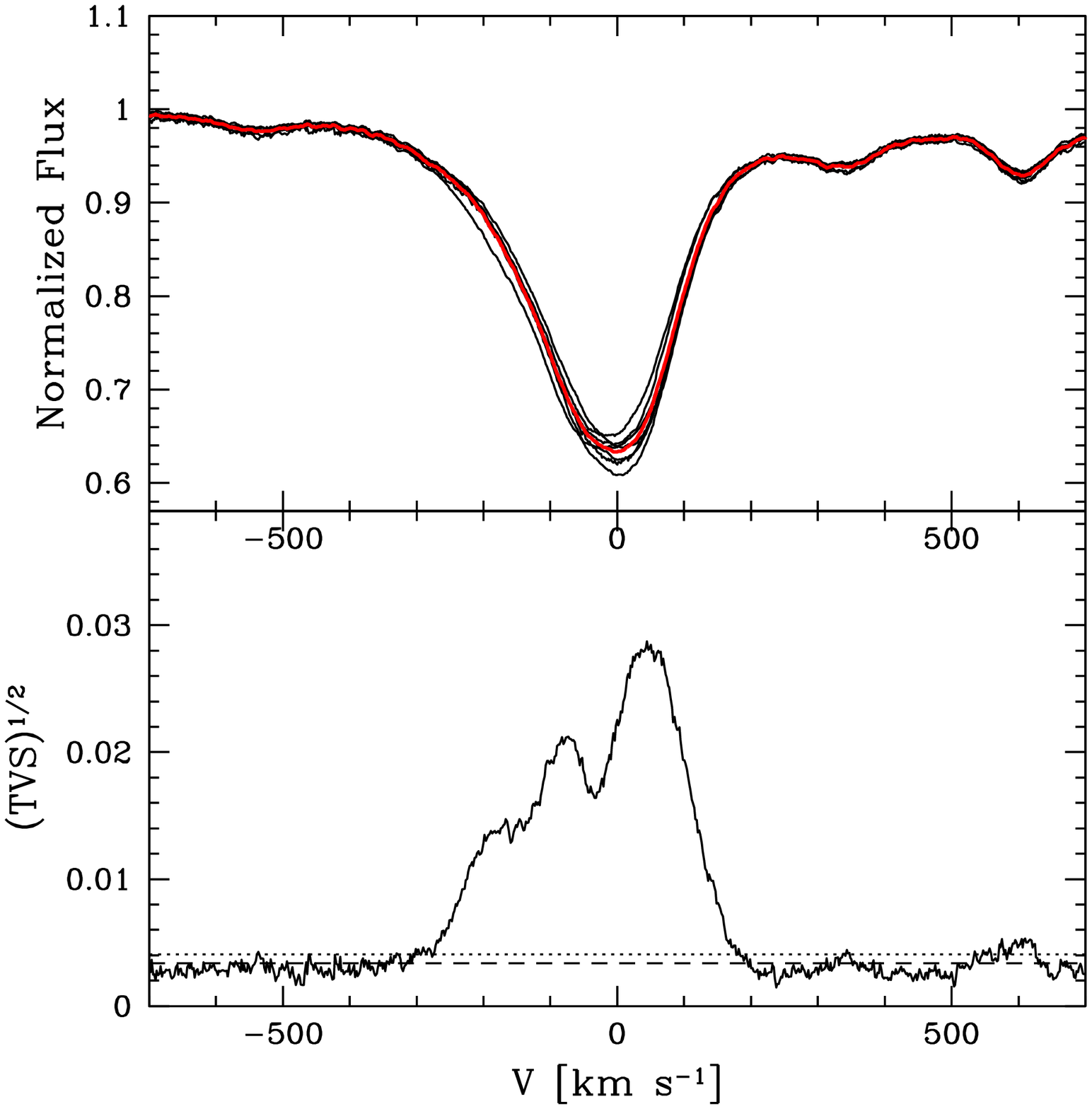}}\\
     \subfigure[\ion{He}{I} 4026]{
          \includegraphics[width=.28\textwidth]{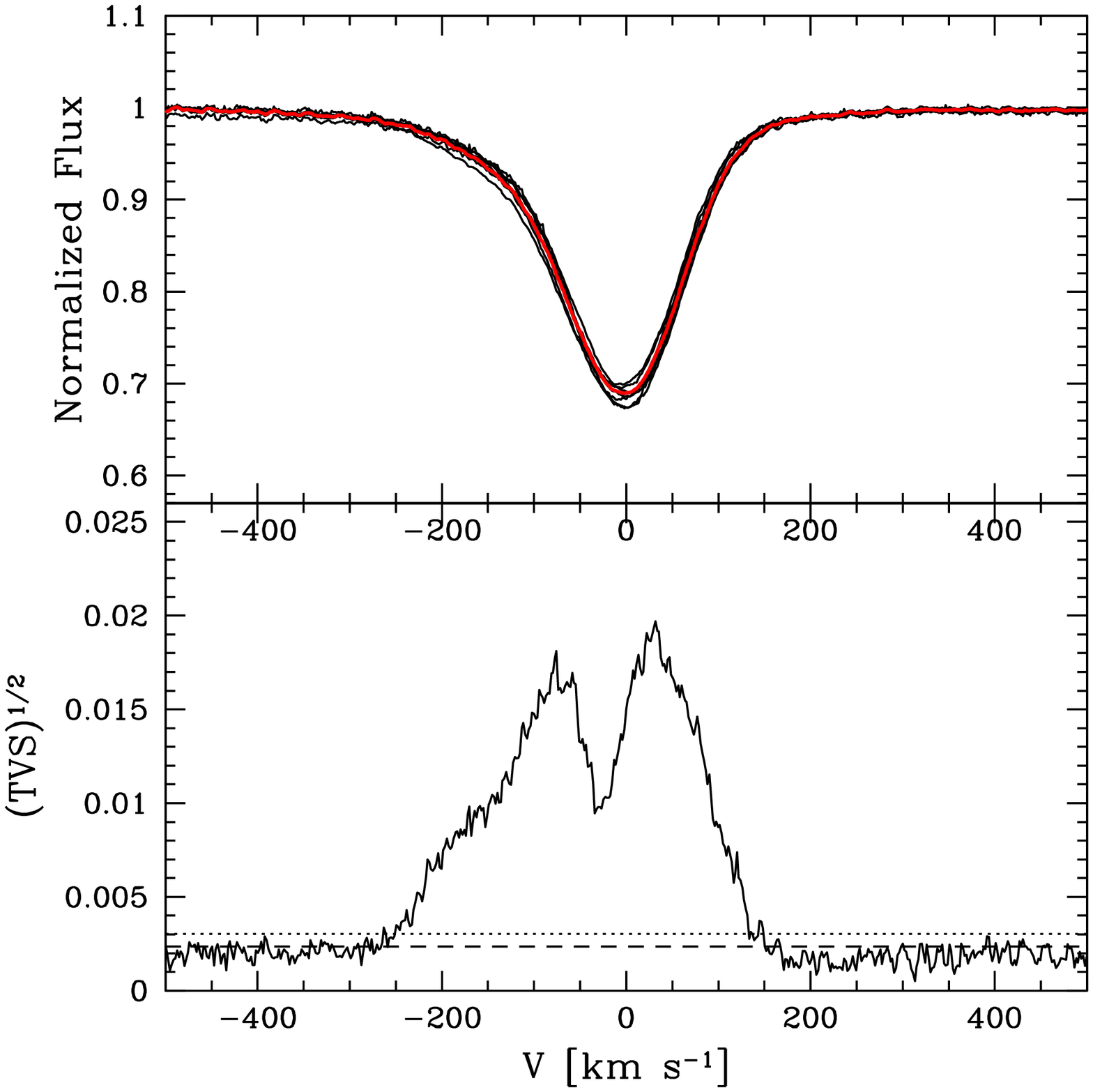}}
     \hspace{0.2cm}
     \subfigure[\ion{He}{I} 4471]{
          \includegraphics[width=.28\textwidth]{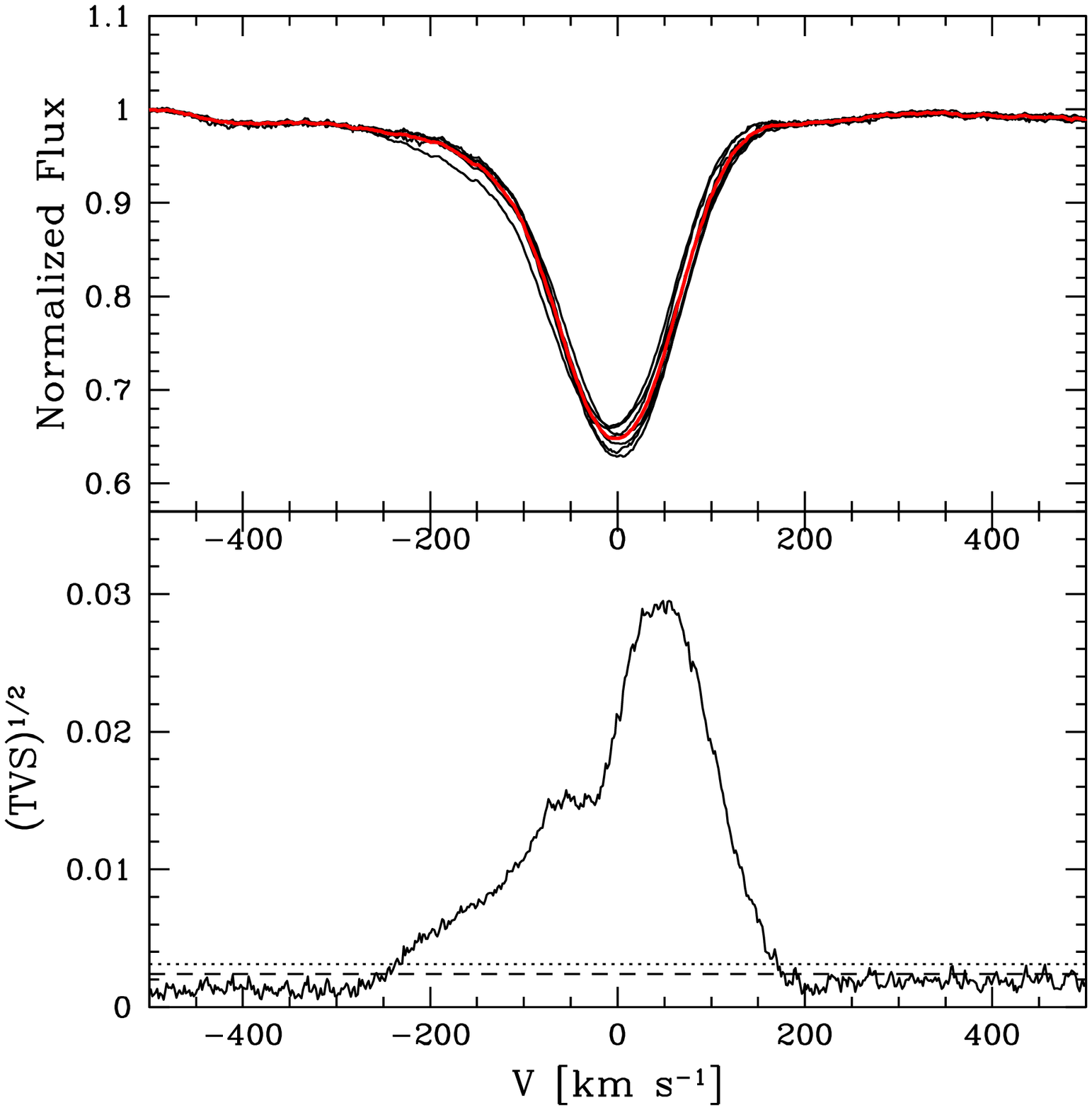}}
     \hspace{0.2cm}
     \subfigure[\ion{He}{I} 4712]{
          \includegraphics[width=.28\textwidth]{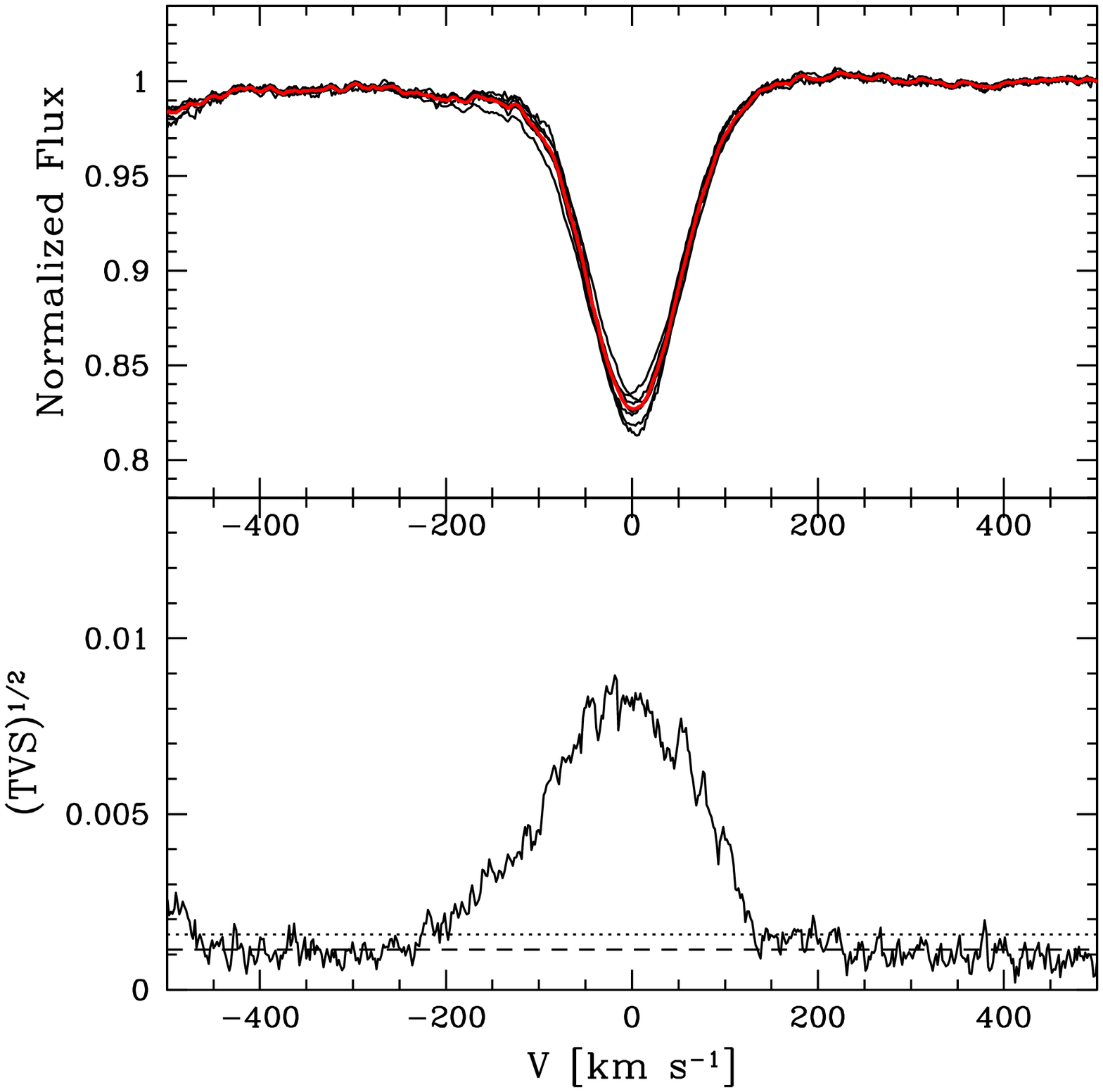}}\\
     \subfigure[\ion{He}{I} 5876]{
          \includegraphics[width=.28\textwidth]{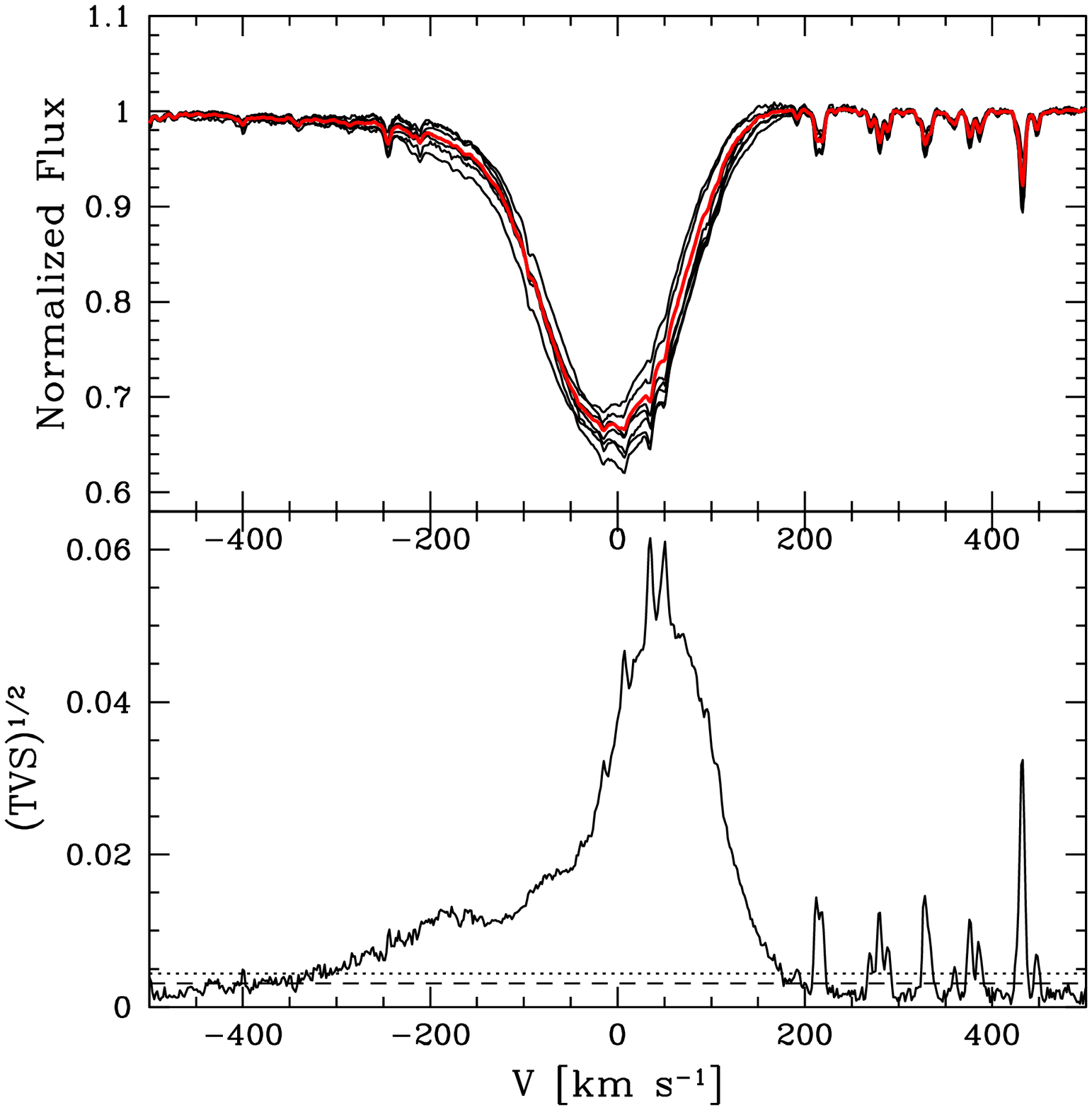}}
     \hspace{0.2cm}
     \subfigure[\ion{He}{II} 4542]{
          \includegraphics[width=.28\textwidth]{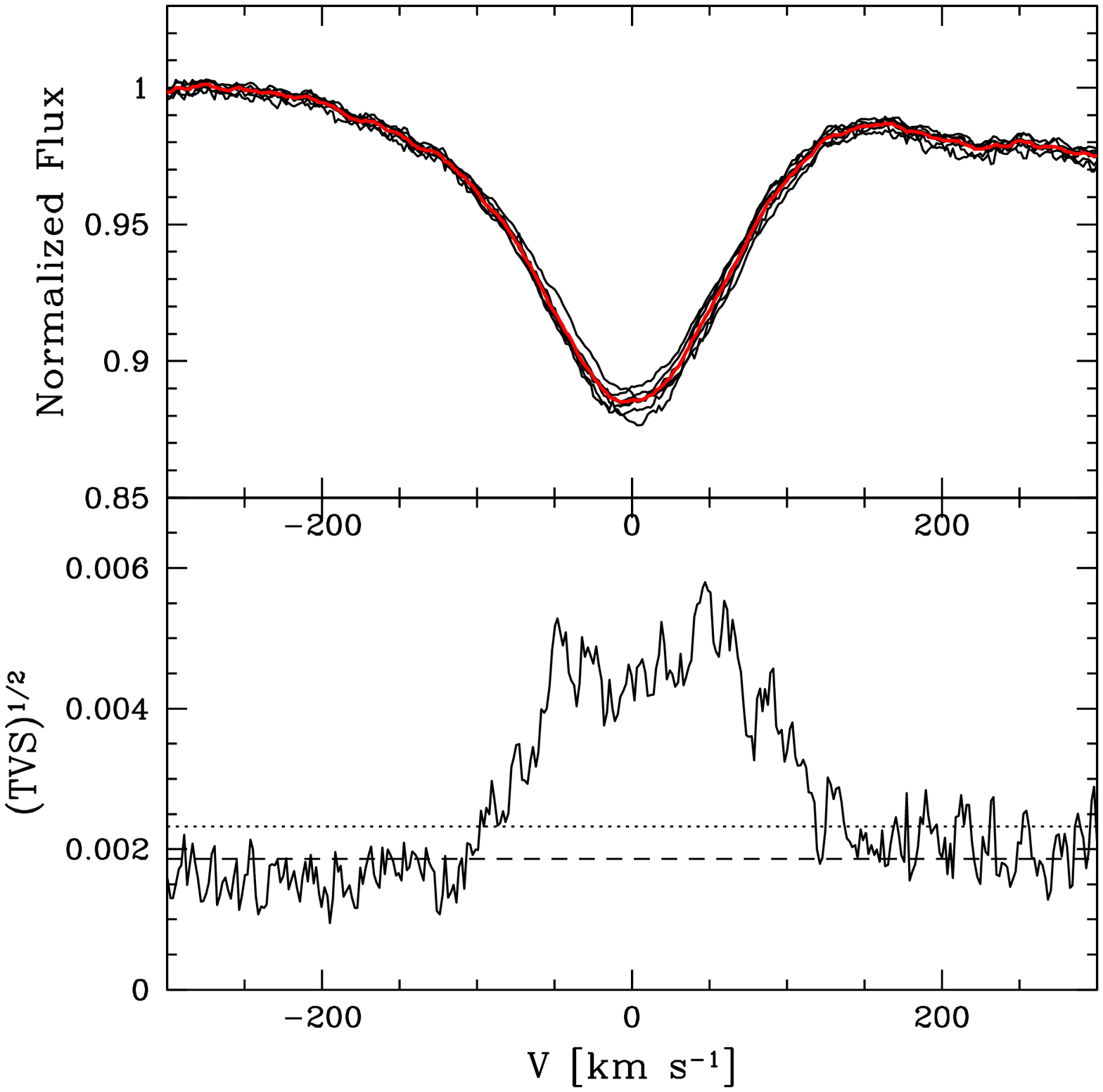}}
     \hspace{0.2cm}
     \subfigure[\ion{He}{II} 4686]{
          \includegraphics[width=.28\textwidth]{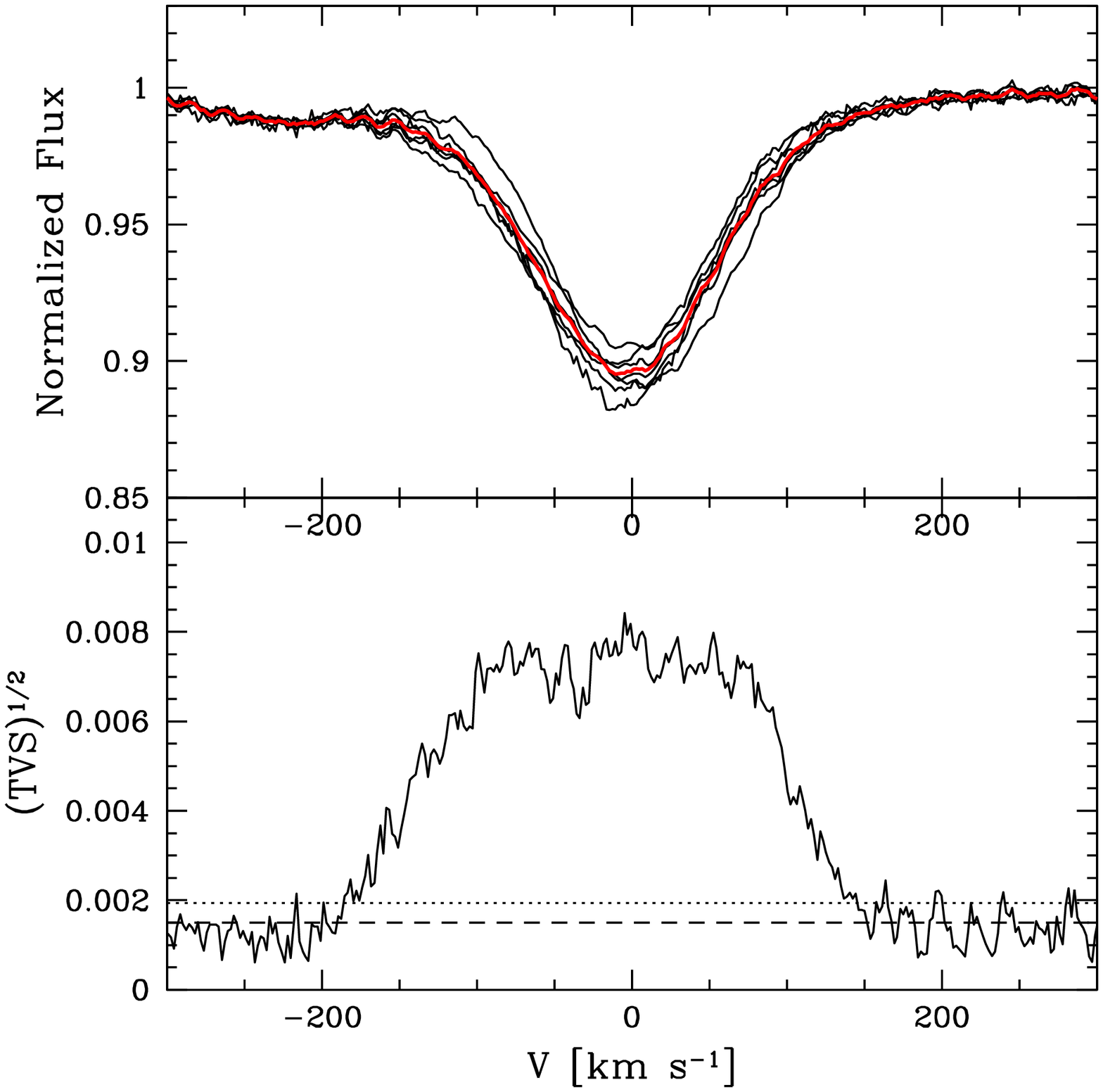}}\\
     \subfigure[\ion{He}{II} 5412]{
          \includegraphics[width=.28\textwidth]{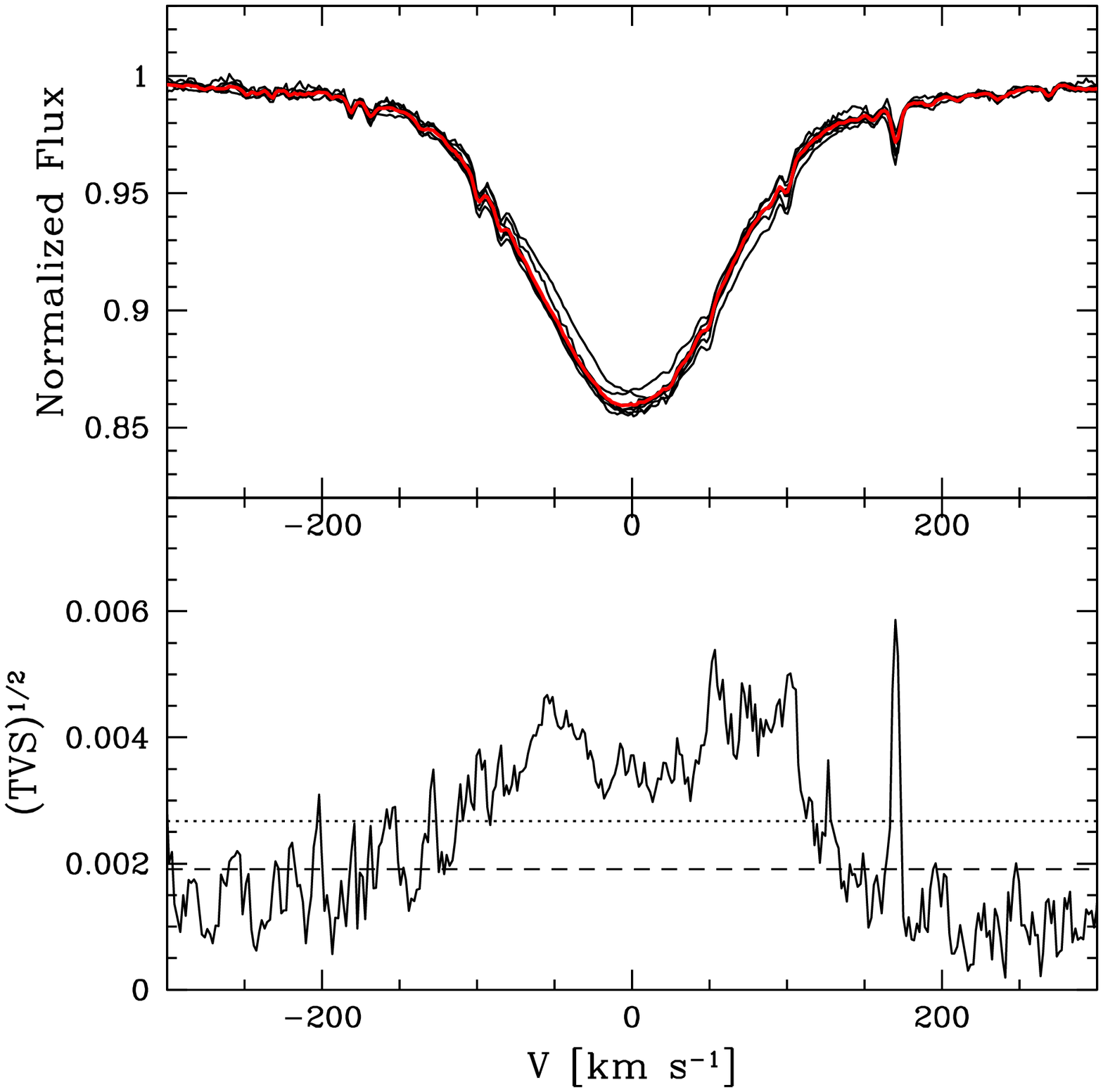}}
     \hspace{0.2cm}
     \subfigure[\ion{O}{III} 5592]{
          \includegraphics[width=.28\textwidth]{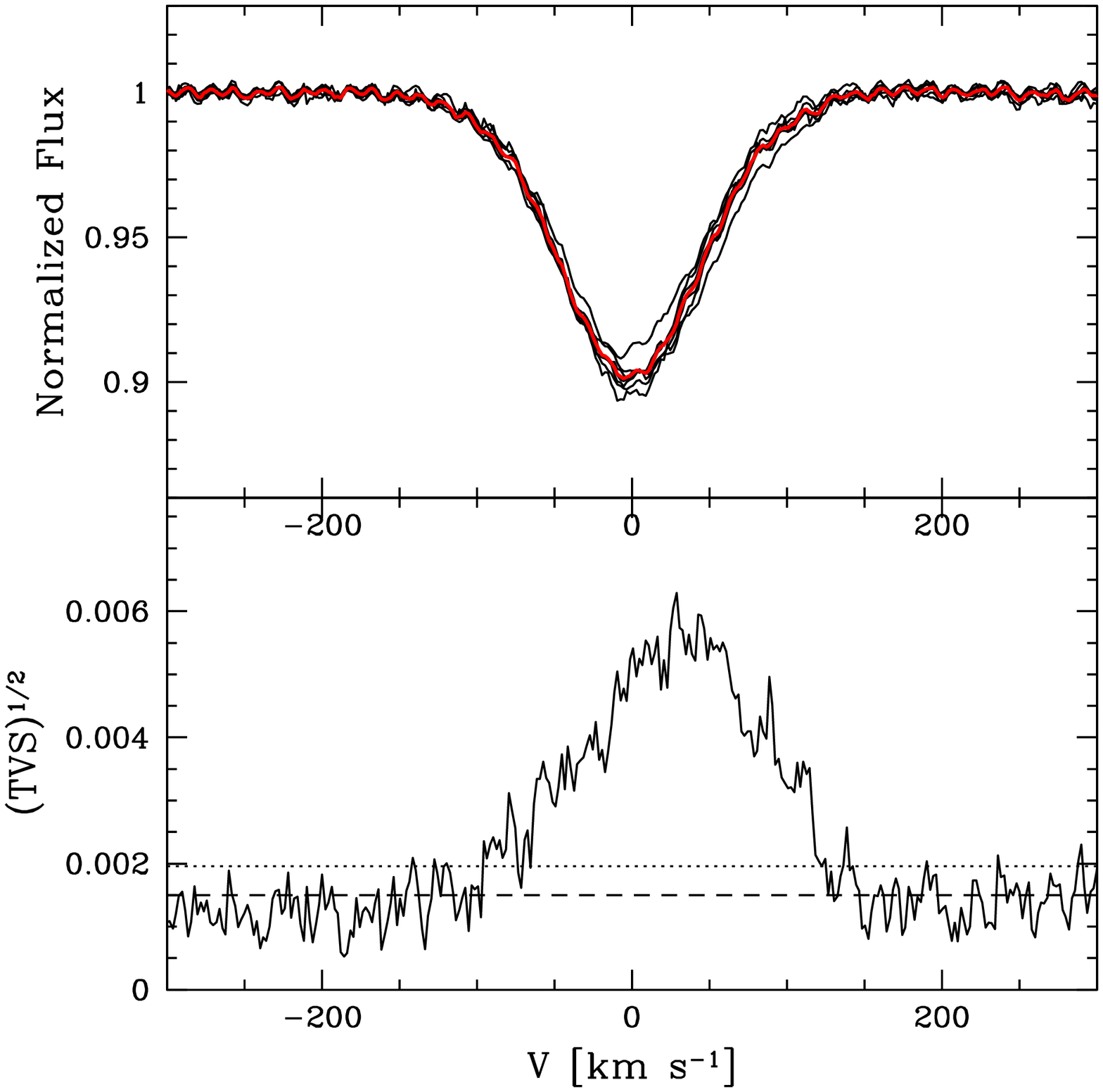}}
     \hspace{0.2cm}
     \subfigure[\ion{C}{IV} 5802]{
          \includegraphics[width=.28\textwidth]{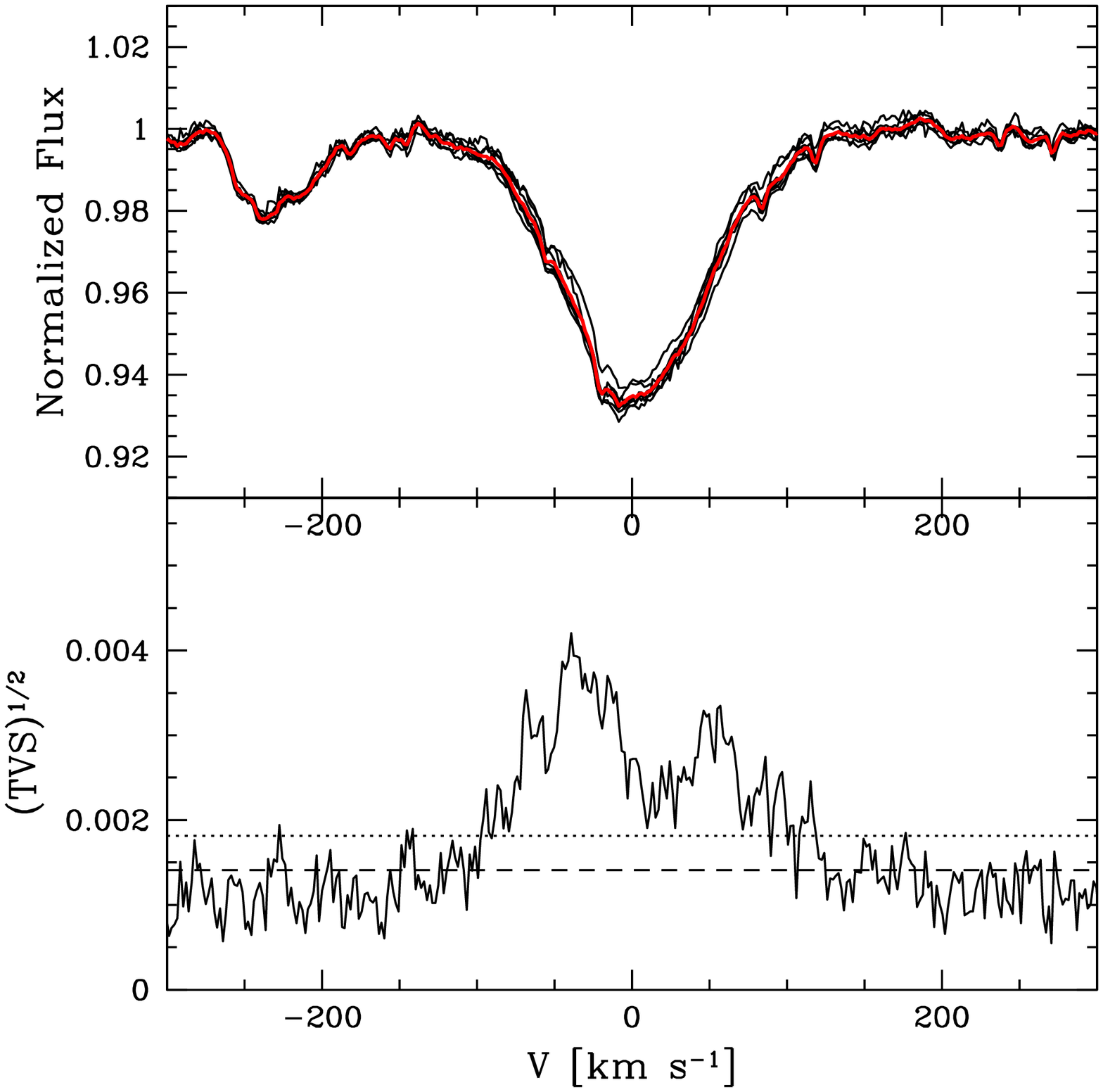}}\\
     \caption{Variability of HD188209 between June 21$^{st}$ and June 30$^{th}$ 2008.}
     \label{fig_var_188209_month}
\end{figure*}

\newpage

\begin{figure*}
     \centering
     \subfigure[\ha]{
          \includegraphics[width=.28\textwidth]{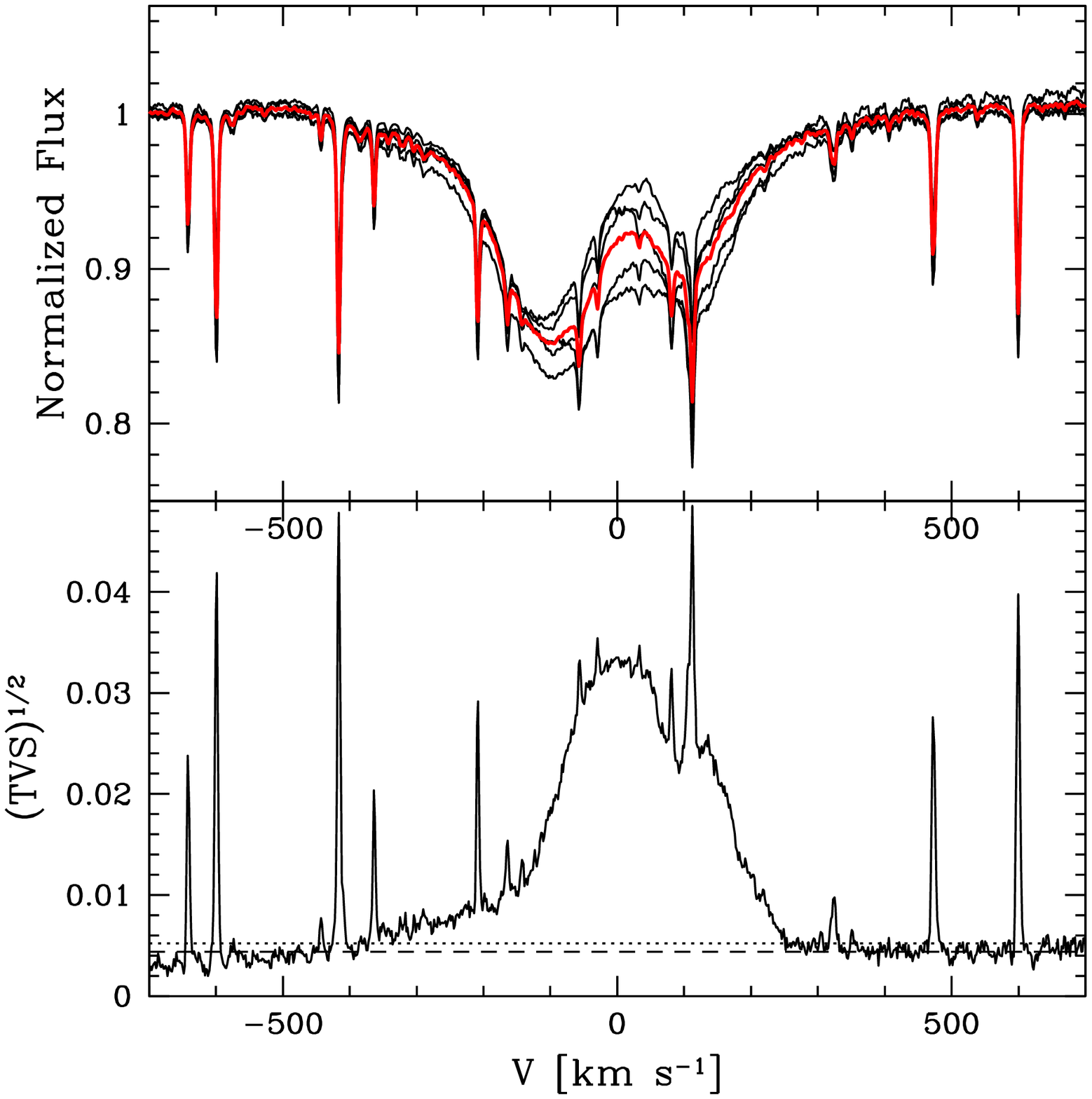}}
     \hspace{0.2cm}
     \subfigure[H$_{\beta}$]{
          \includegraphics[width=.28\textwidth]{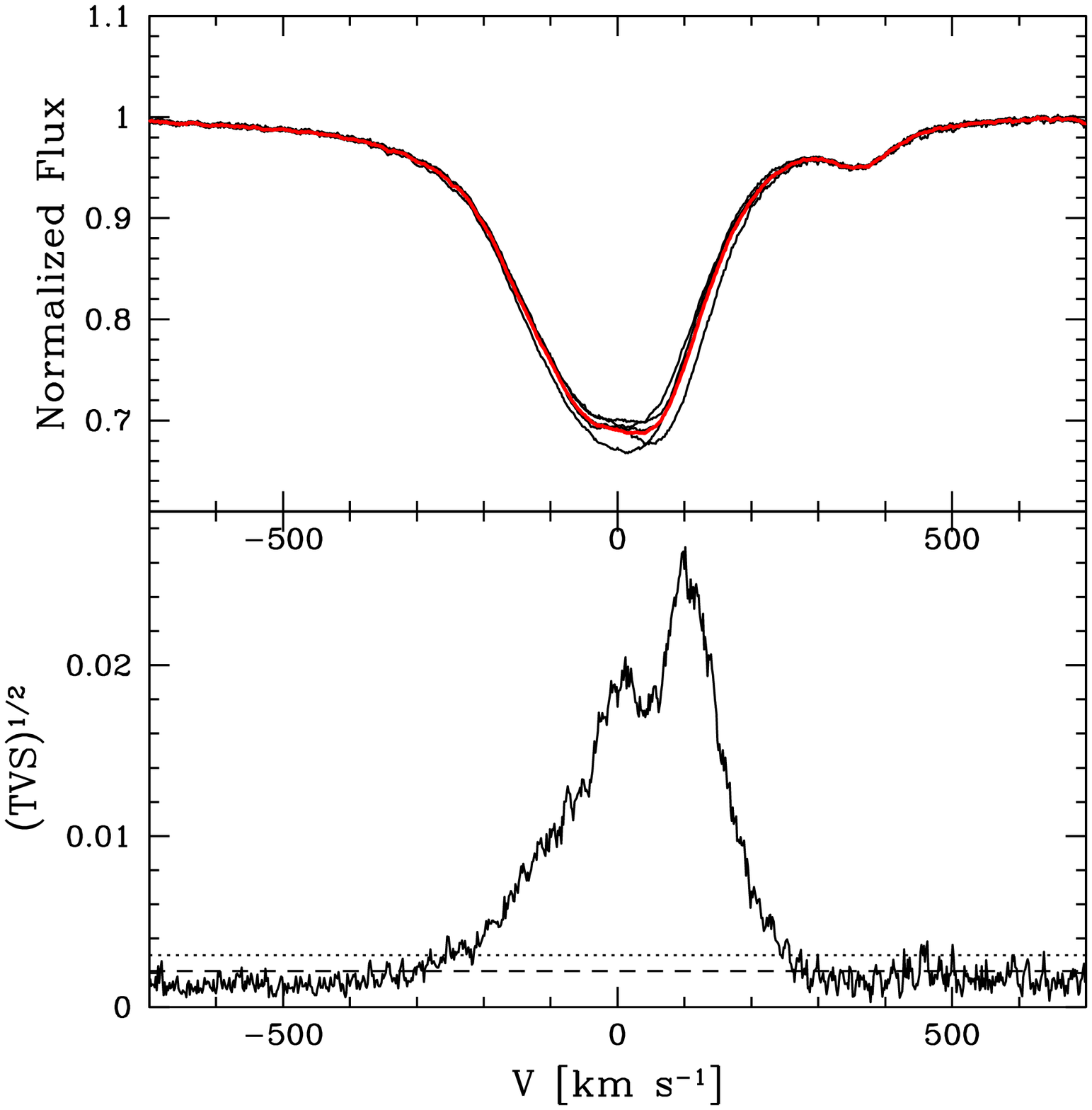}}
     \hspace{0.2cm}
     \subfigure[H$_{\gamma}$]{
          \includegraphics[width=.28\textwidth]{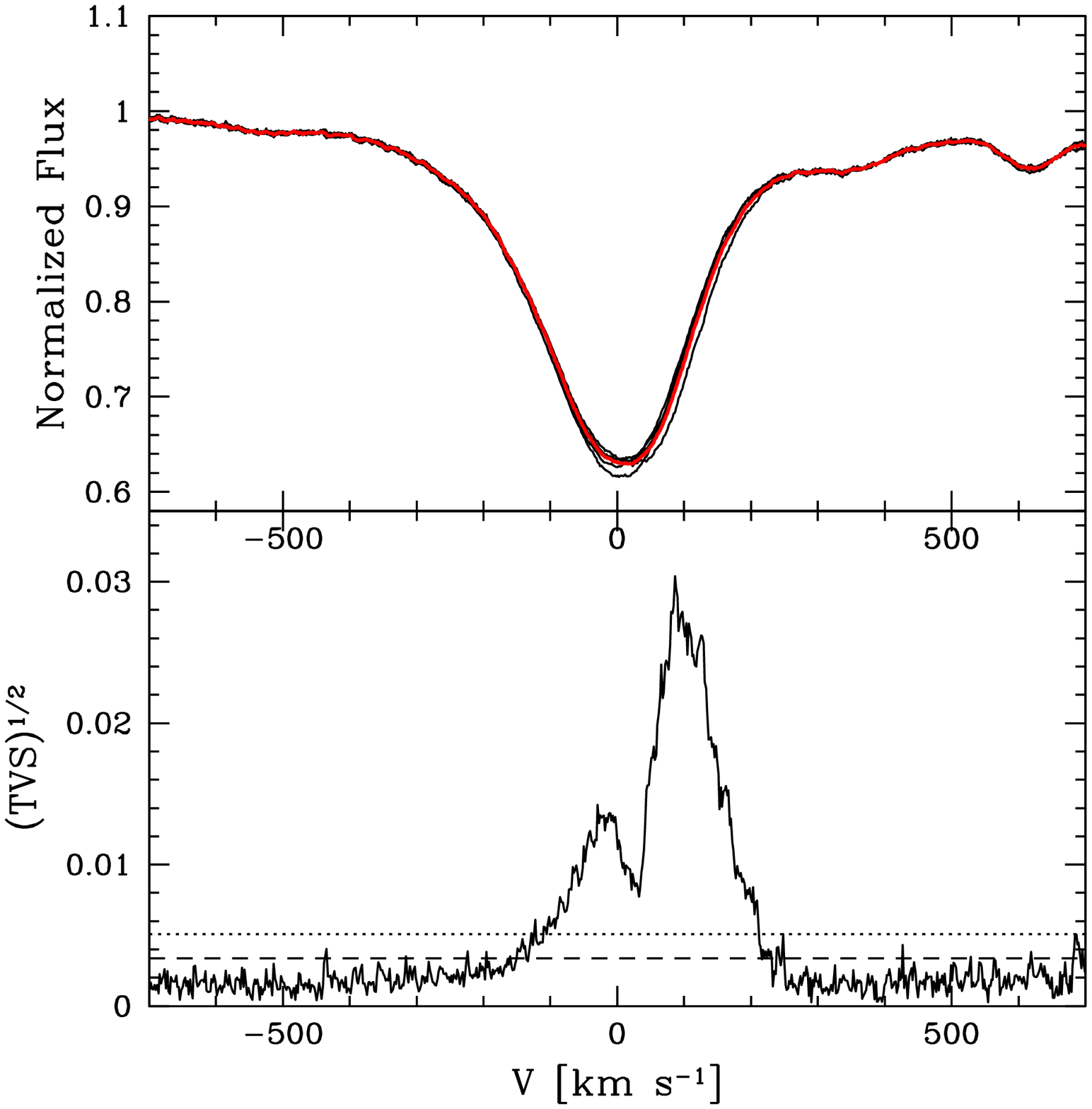}}\\
     \subfigure[\ion{He}{I} 4026]{
          \includegraphics[width=.28\textwidth]{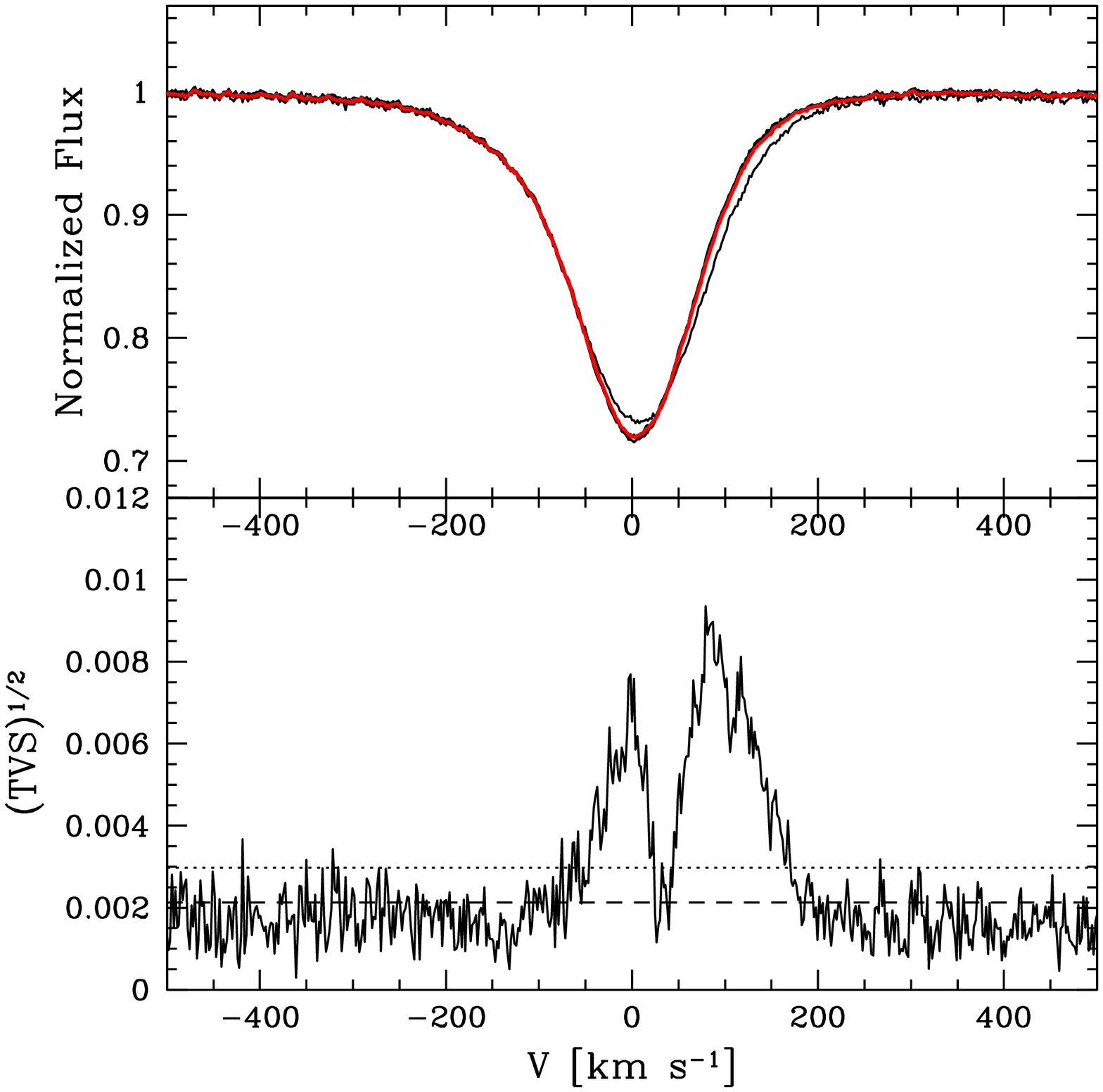}}
     \hspace{0.2cm}
     \subfigure[\ion{He}{I} 4471]{
          \includegraphics[width=.28\textwidth]{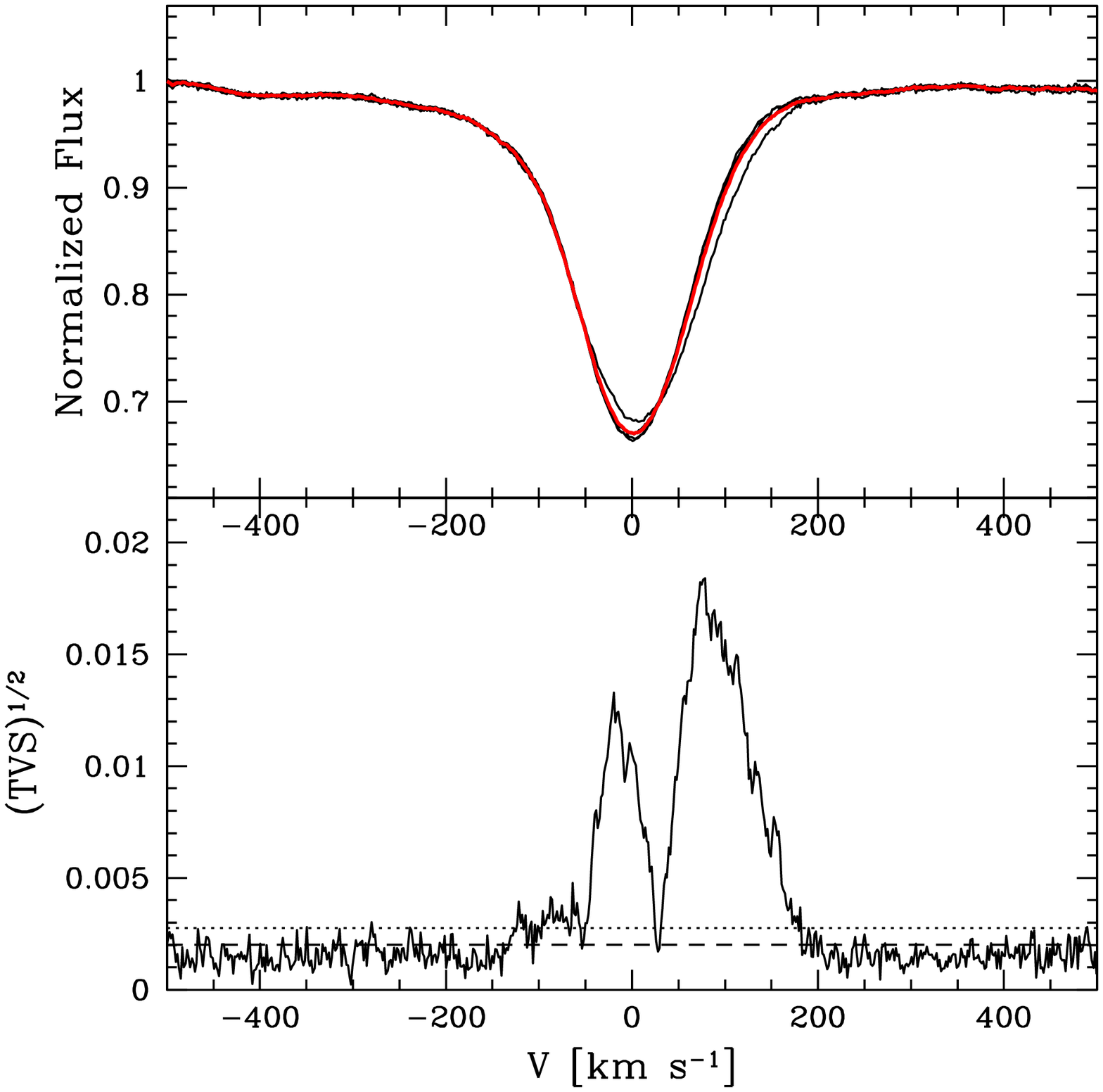}}
     \hspace{0.2cm}
     \subfigure[\ion{He}{I} 4712]{
          \includegraphics[width=.28\textwidth]{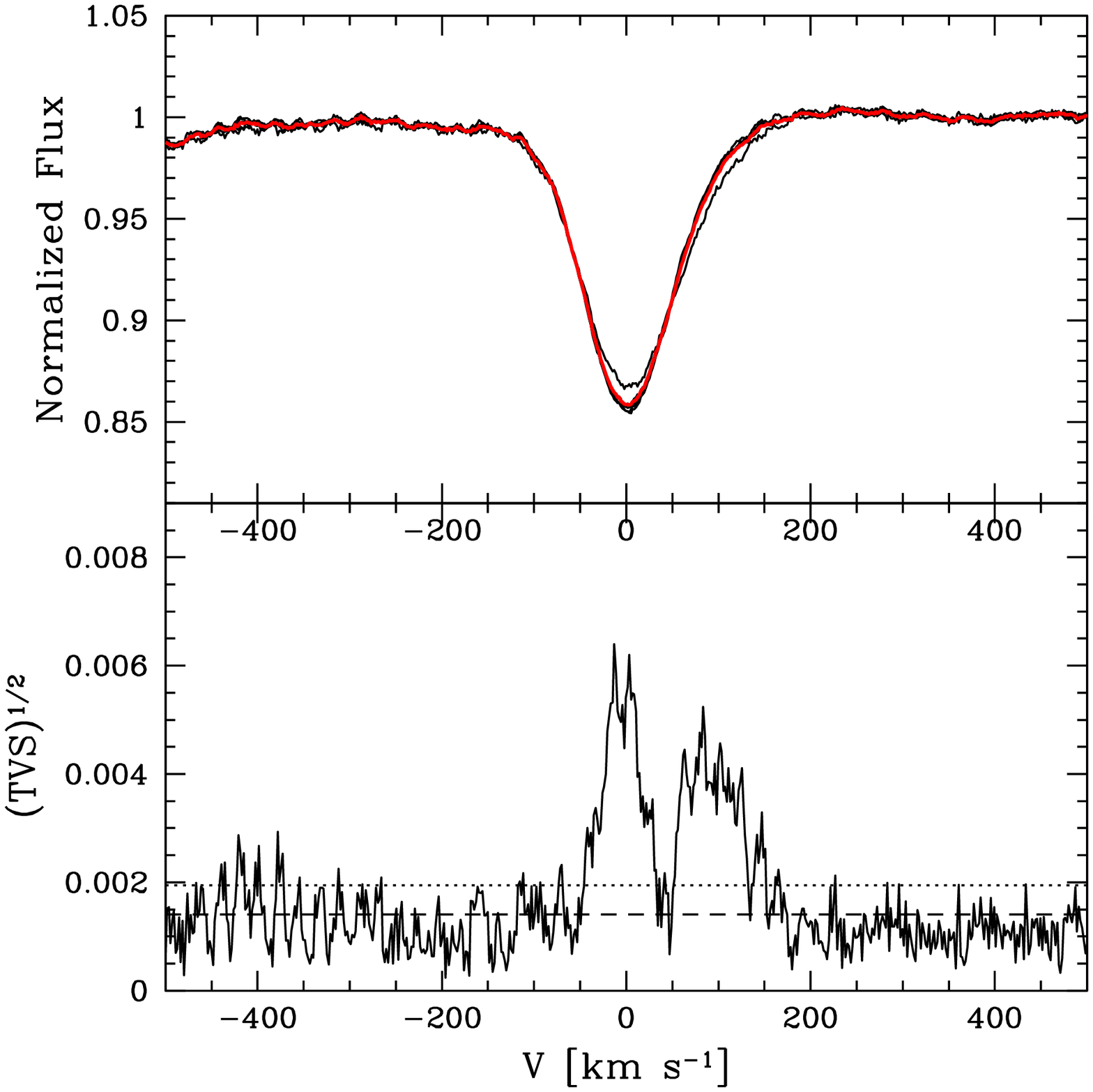}}\\
     \subfigure[\ion{He}{I} 5876]{
          \includegraphics[width=.28\textwidth]{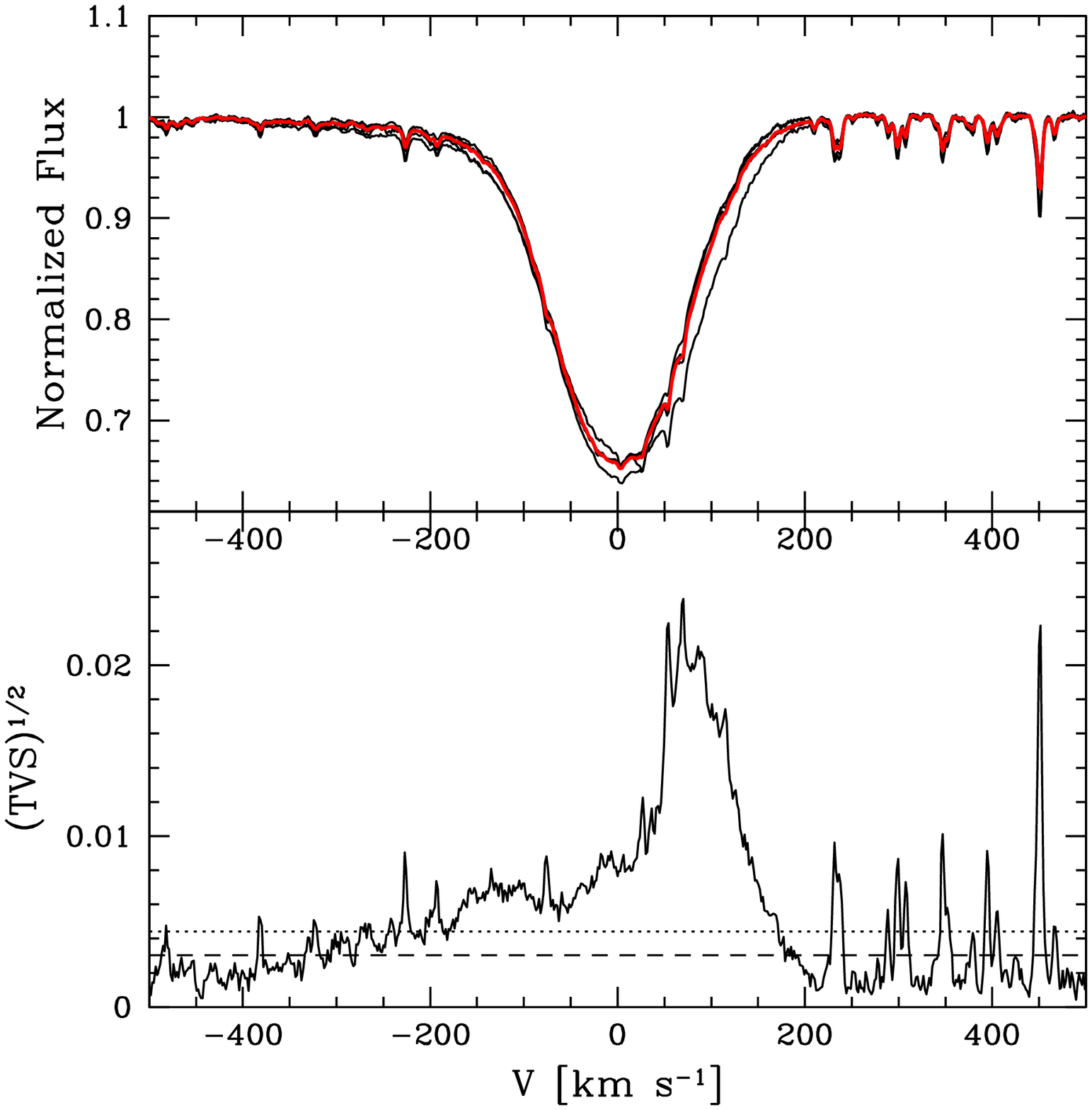}}
     \hspace{0.2cm}
     \subfigure[\ion{He}{II} 4542]{
          \includegraphics[width=.28\textwidth]{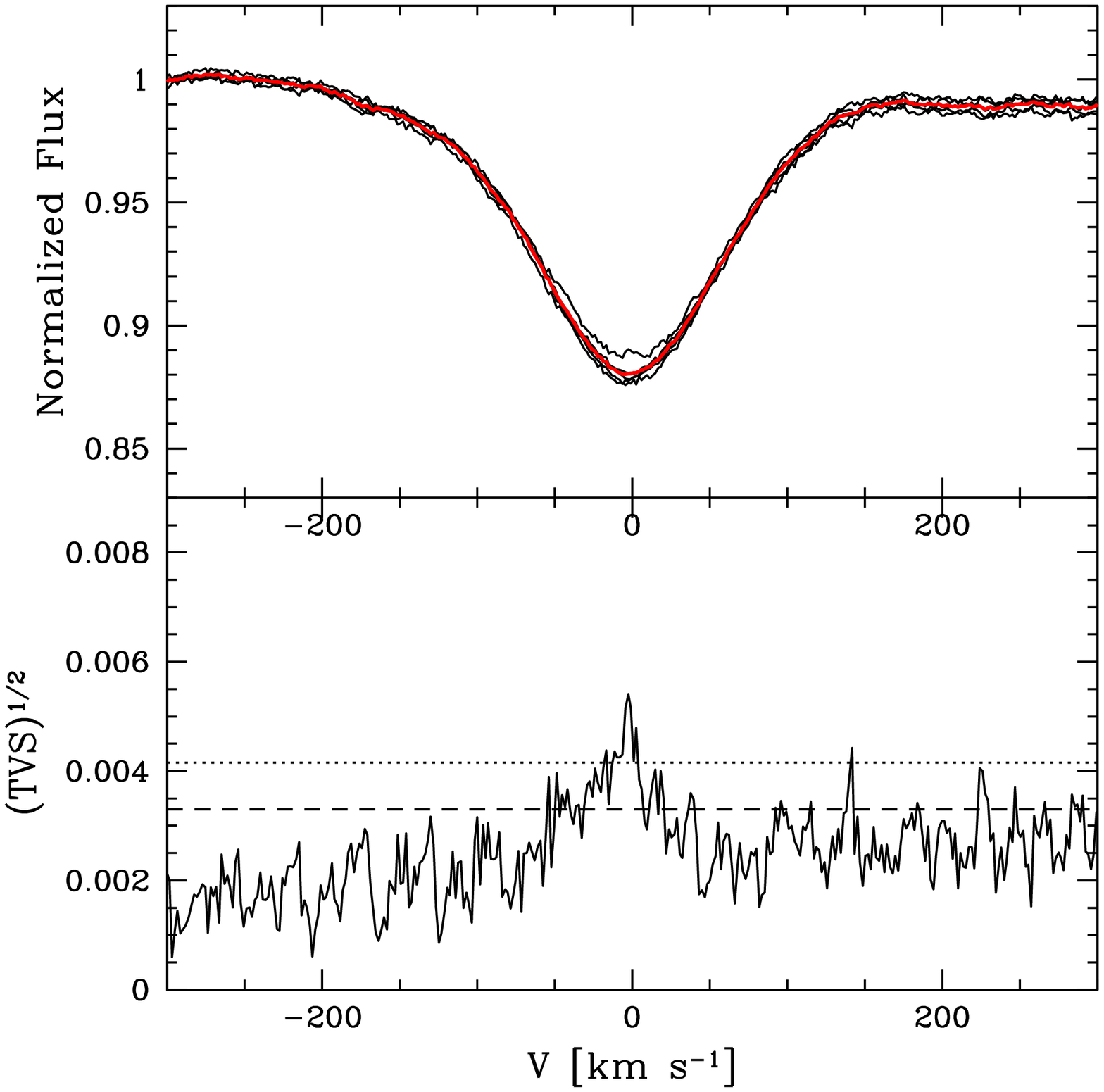}}
     \hspace{0.2cm}
     \subfigure[\ion{He}{II} 4686]{
          \includegraphics[width=.28\textwidth]{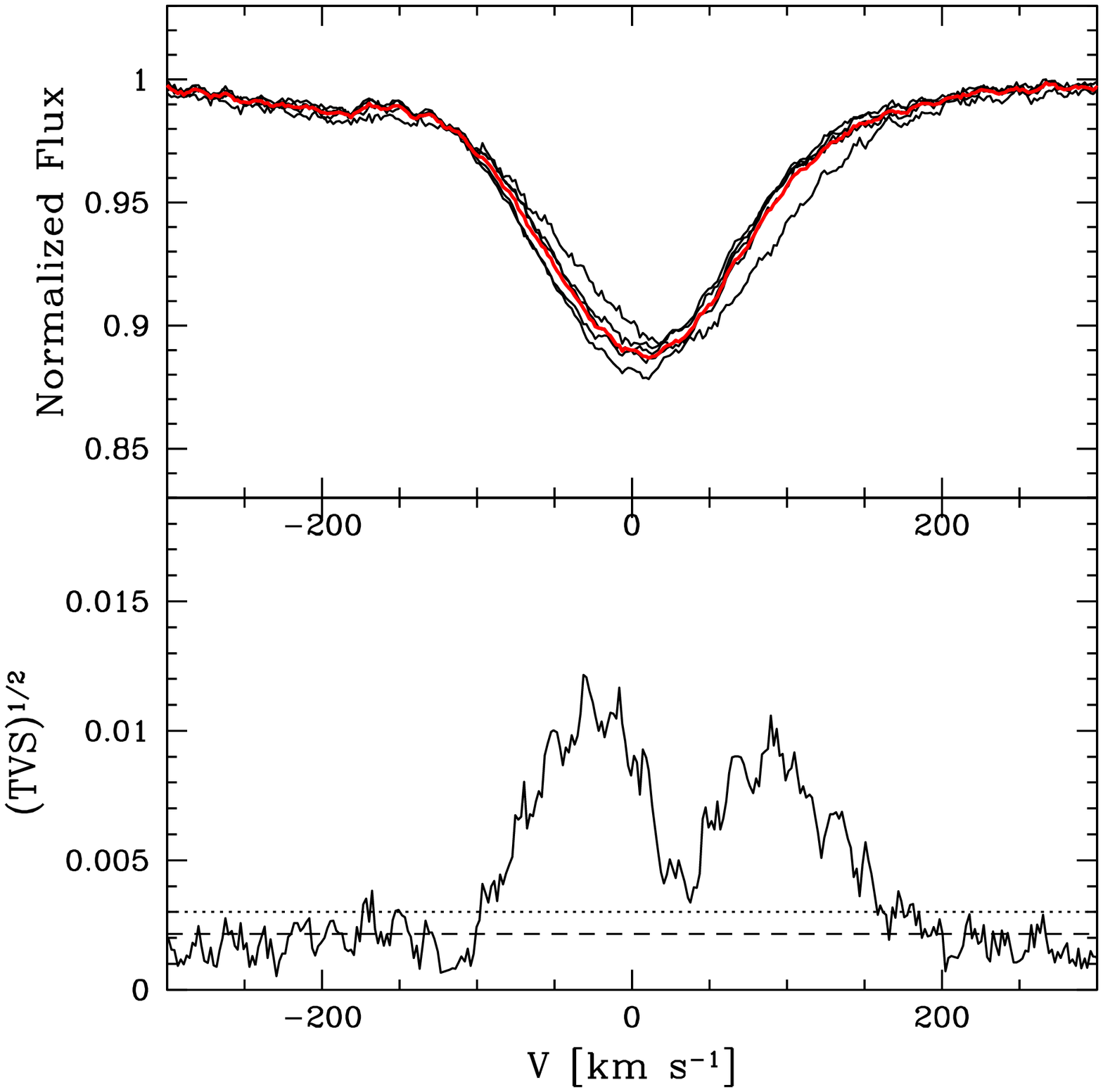}}\\
     \subfigure[\ion{He}{II} 5412]{
          \includegraphics[width=.28\textwidth]{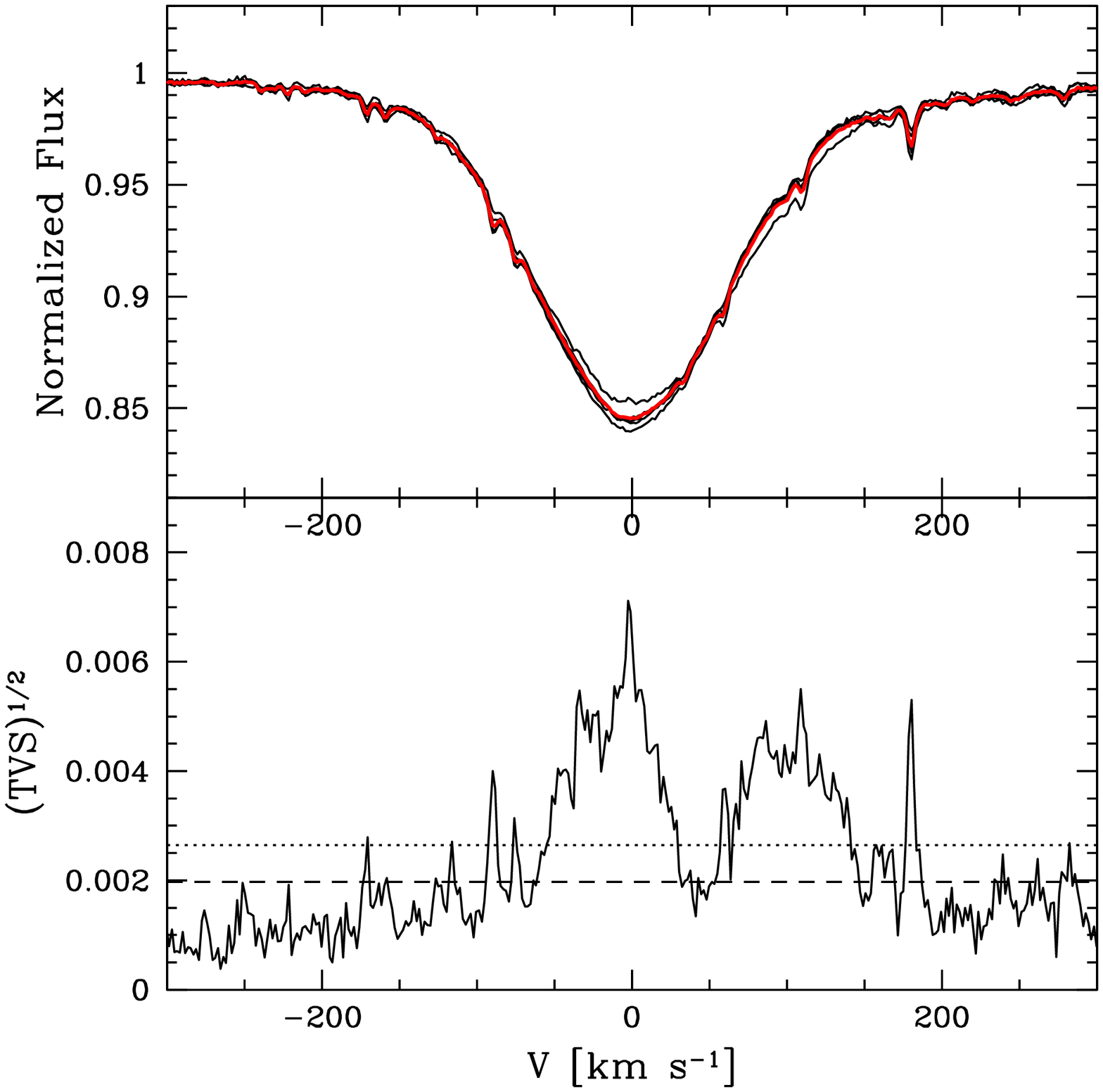}}
     \hspace{0.2cm}
     \subfigure[\ion{O}{III} 5592]{
          \includegraphics[width=.28\textwidth]{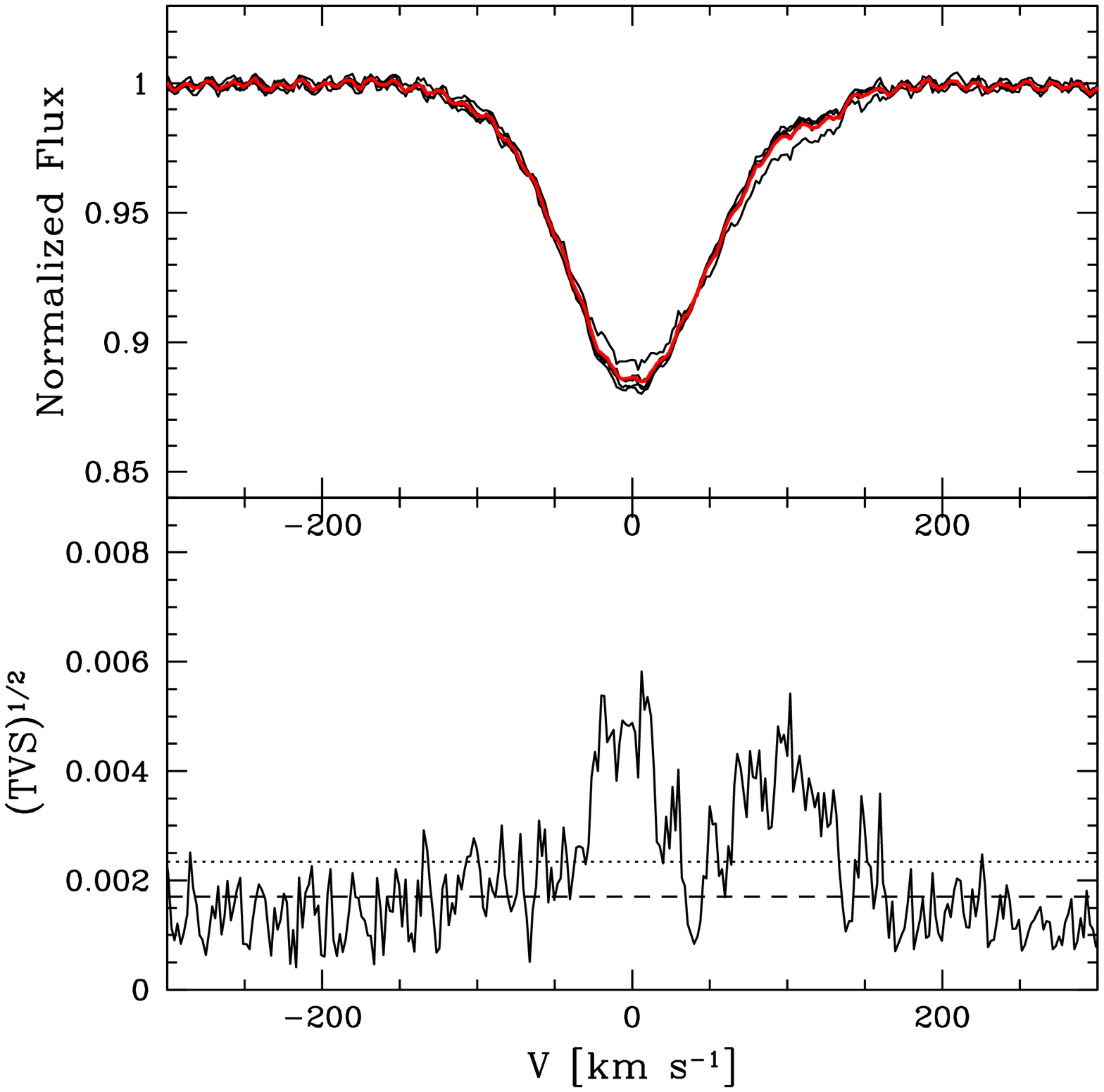}}
     \hspace{0.2cm}
     \subfigure[\ion{C}{IV} 5802]{
          \includegraphics[width=.28\textwidth]{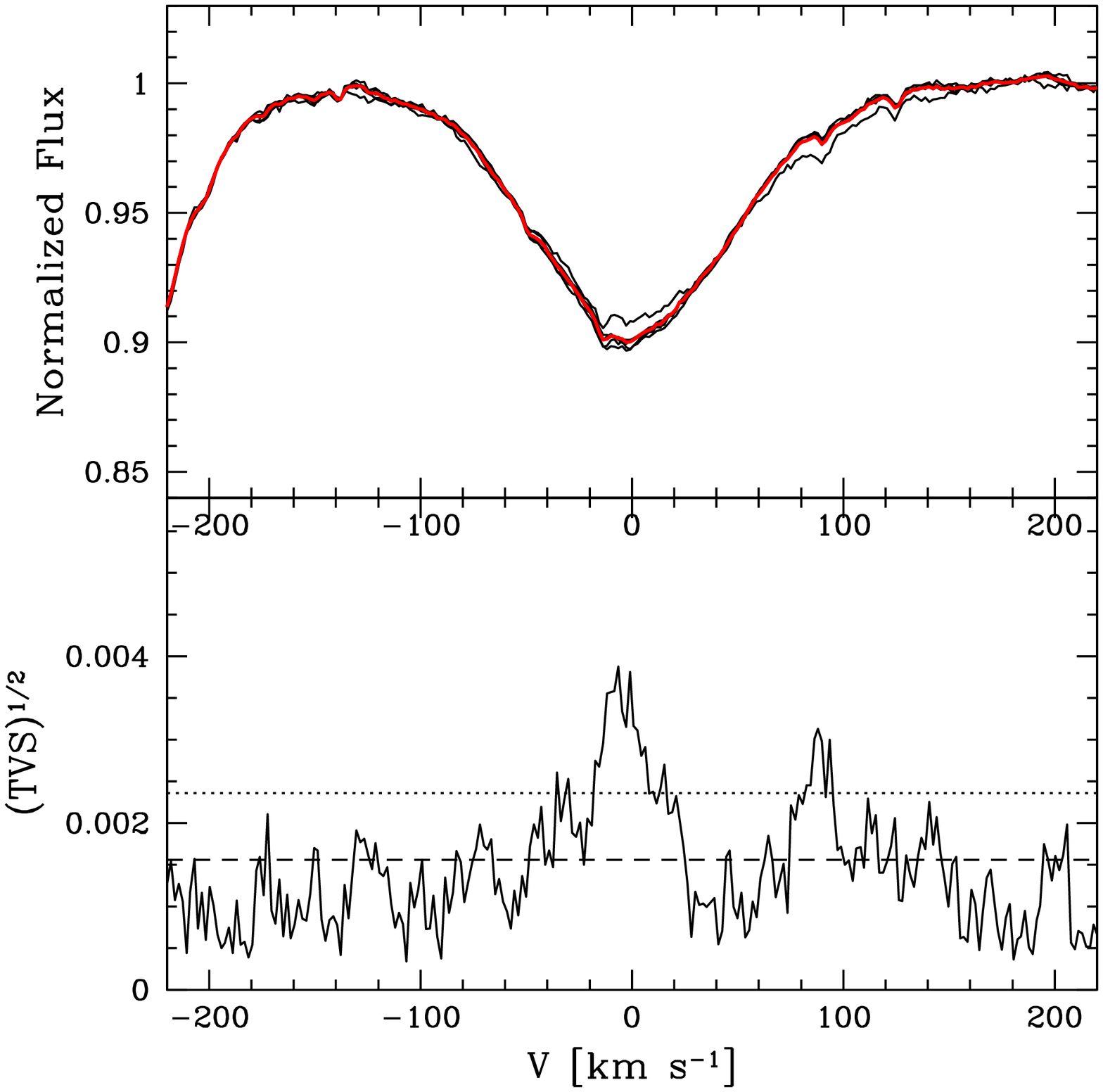}}\\
     \caption{Variability of HD209975 between June 21$^{st}$ and June 27$^{th}$ 2008.}
     \label{fig_var_209975}
\end{figure*}

\newpage

\begin{figure*}
     \centering
     \subfigure[\ha]{
          \includegraphics[width=.28\textwidth]{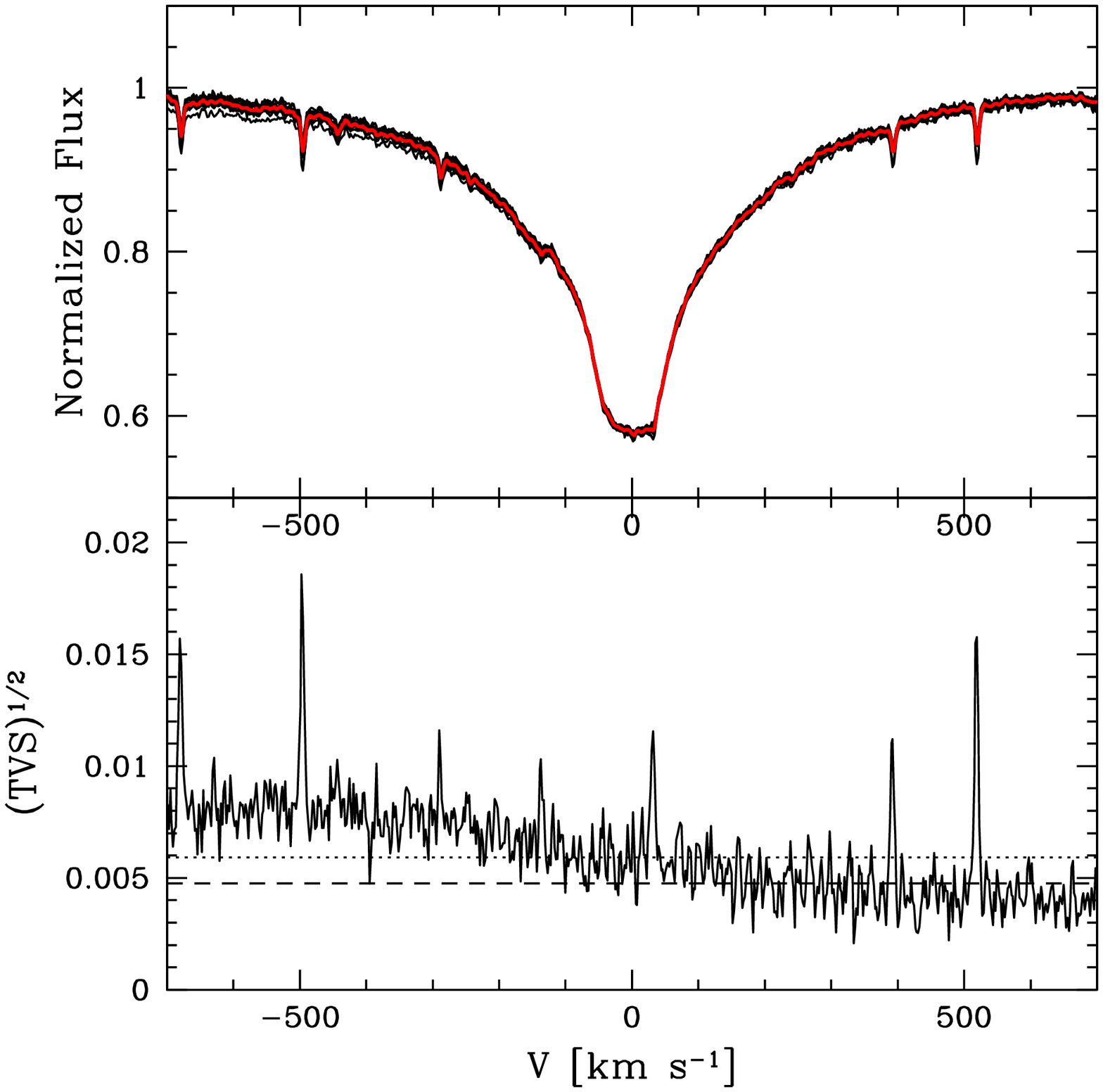}}
     \hspace{0.2cm}
     \subfigure[H$_{\beta}$]{
          \includegraphics[width=.28\textwidth]{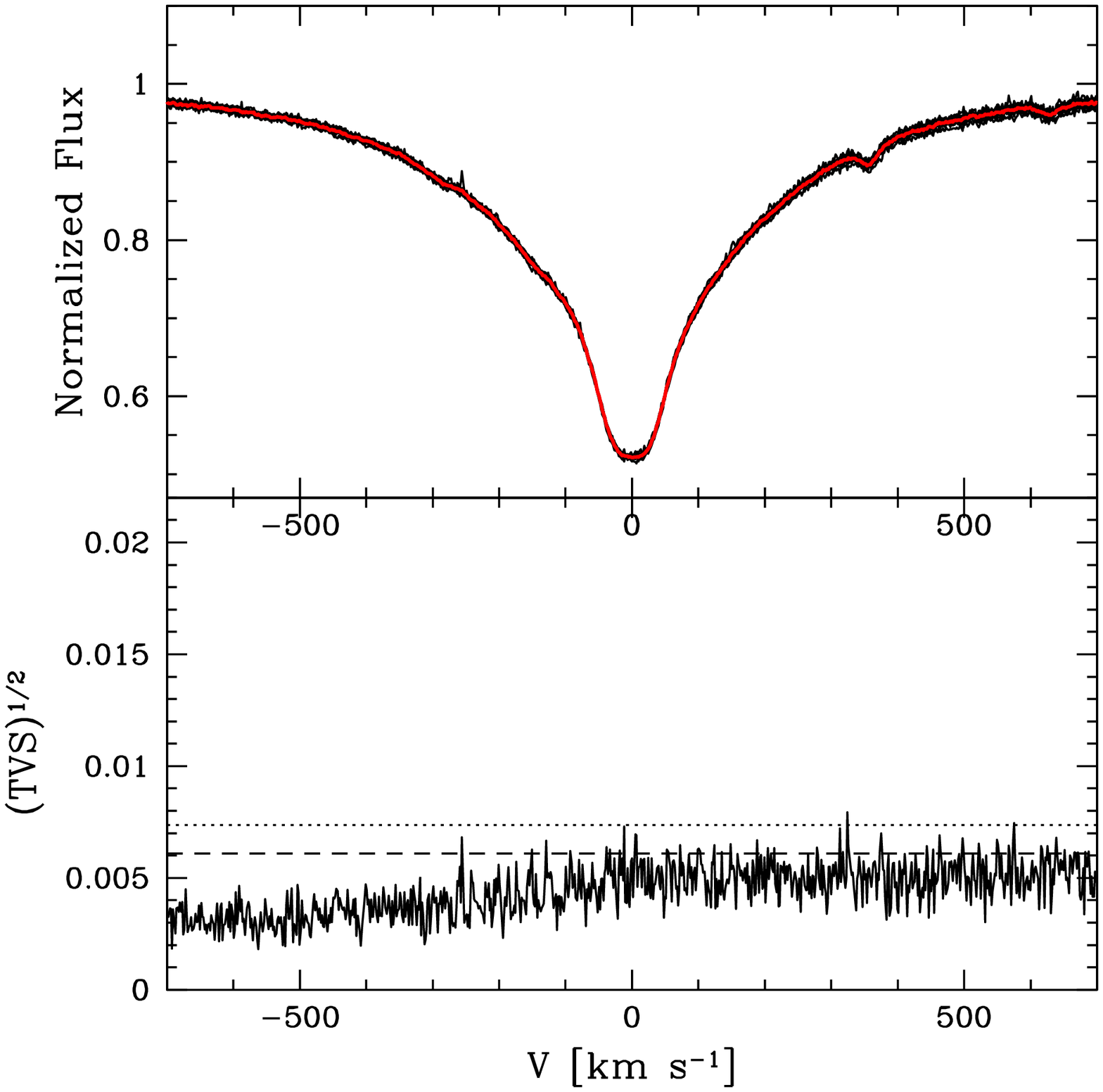}}
     \hspace{0.2cm}
     \subfigure[H$_{\gamma}$]{
          \includegraphics[width=.28\textwidth]{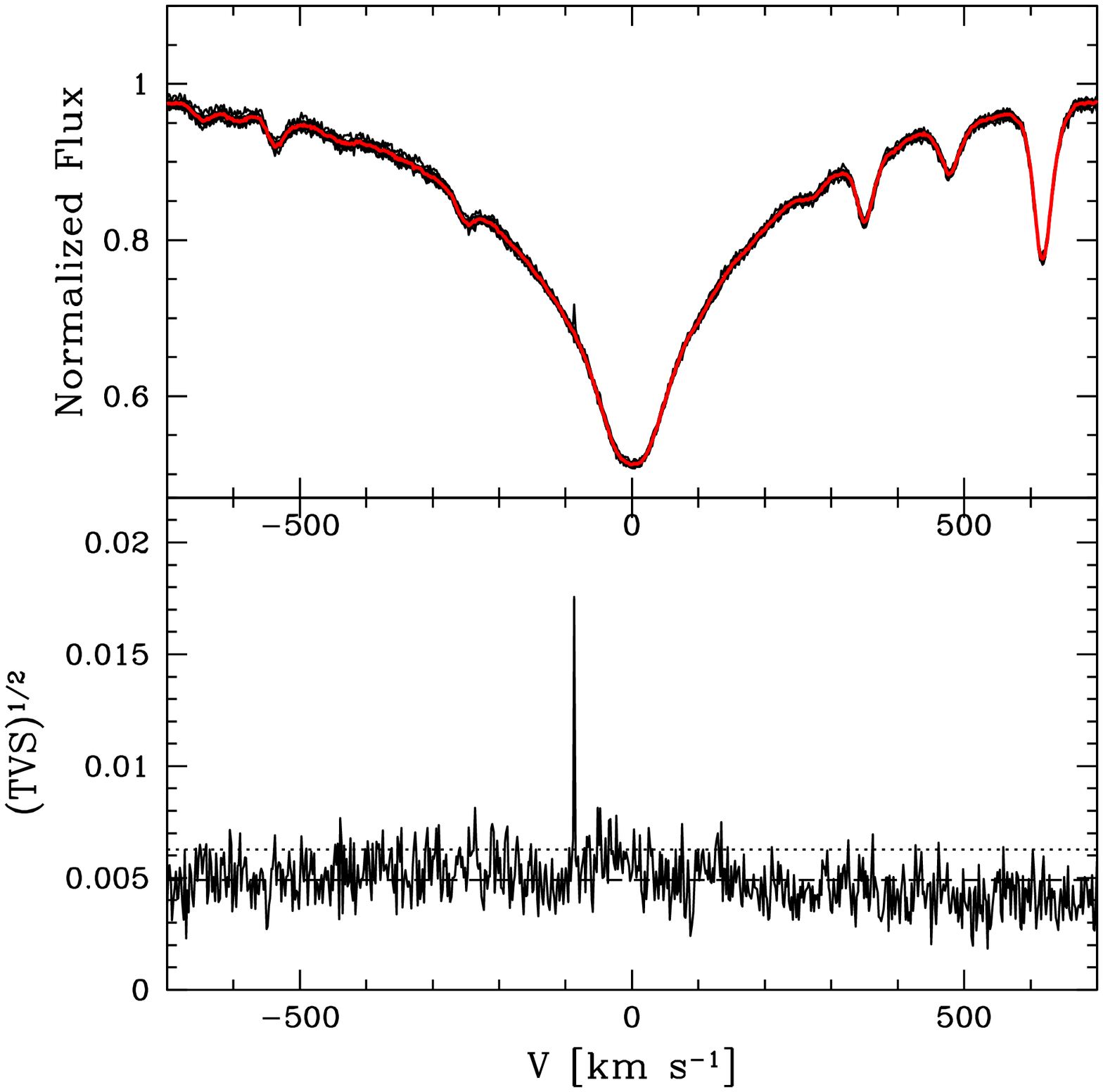}}\\
     \subfigure[\ion{He}{I} 4026]{
          \includegraphics[width=.28\textwidth]{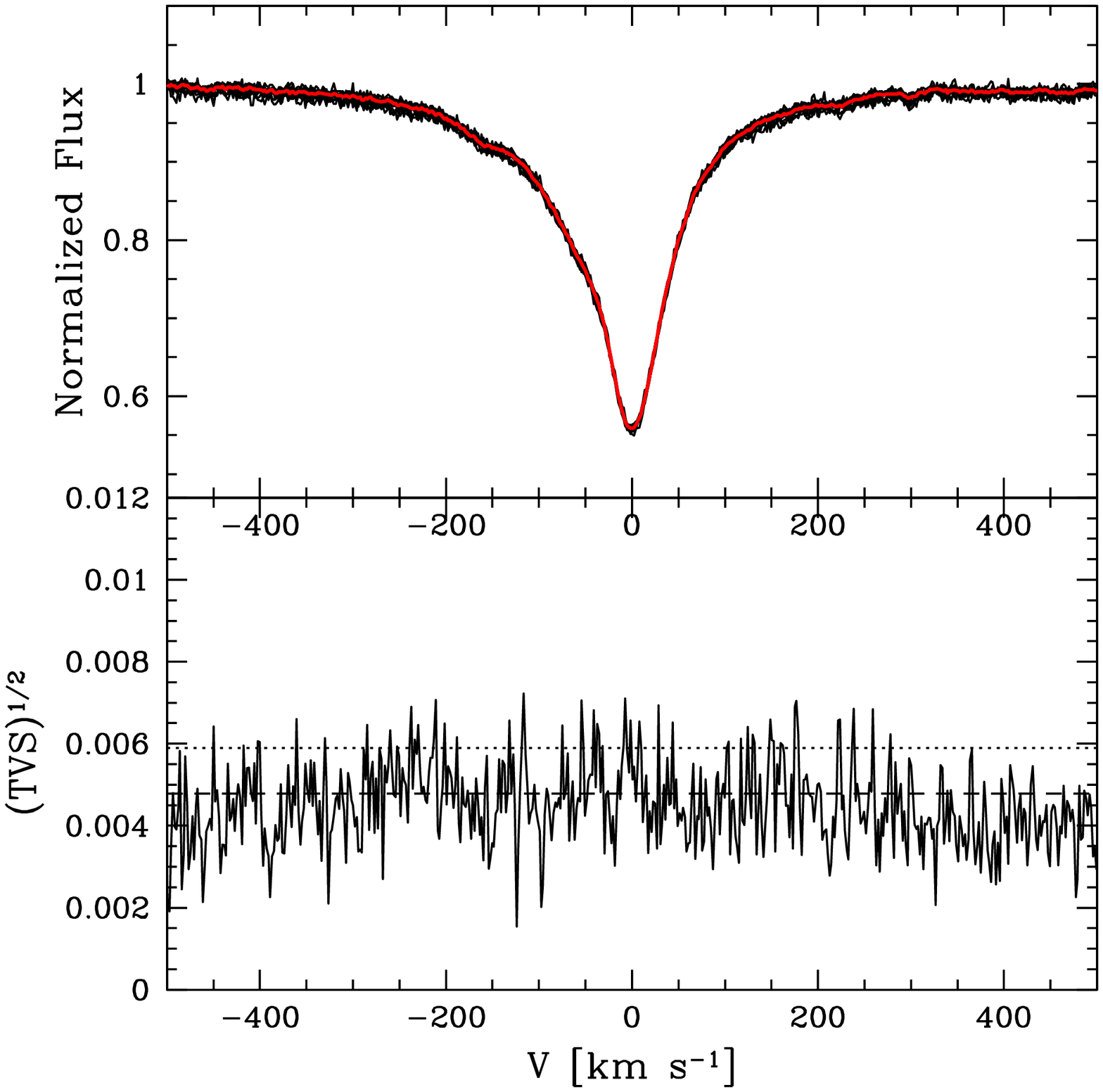}}
     \hspace{0.2cm}
     \subfigure[\ion{He}{I} 4471]{
          \includegraphics[width=.28\textwidth]{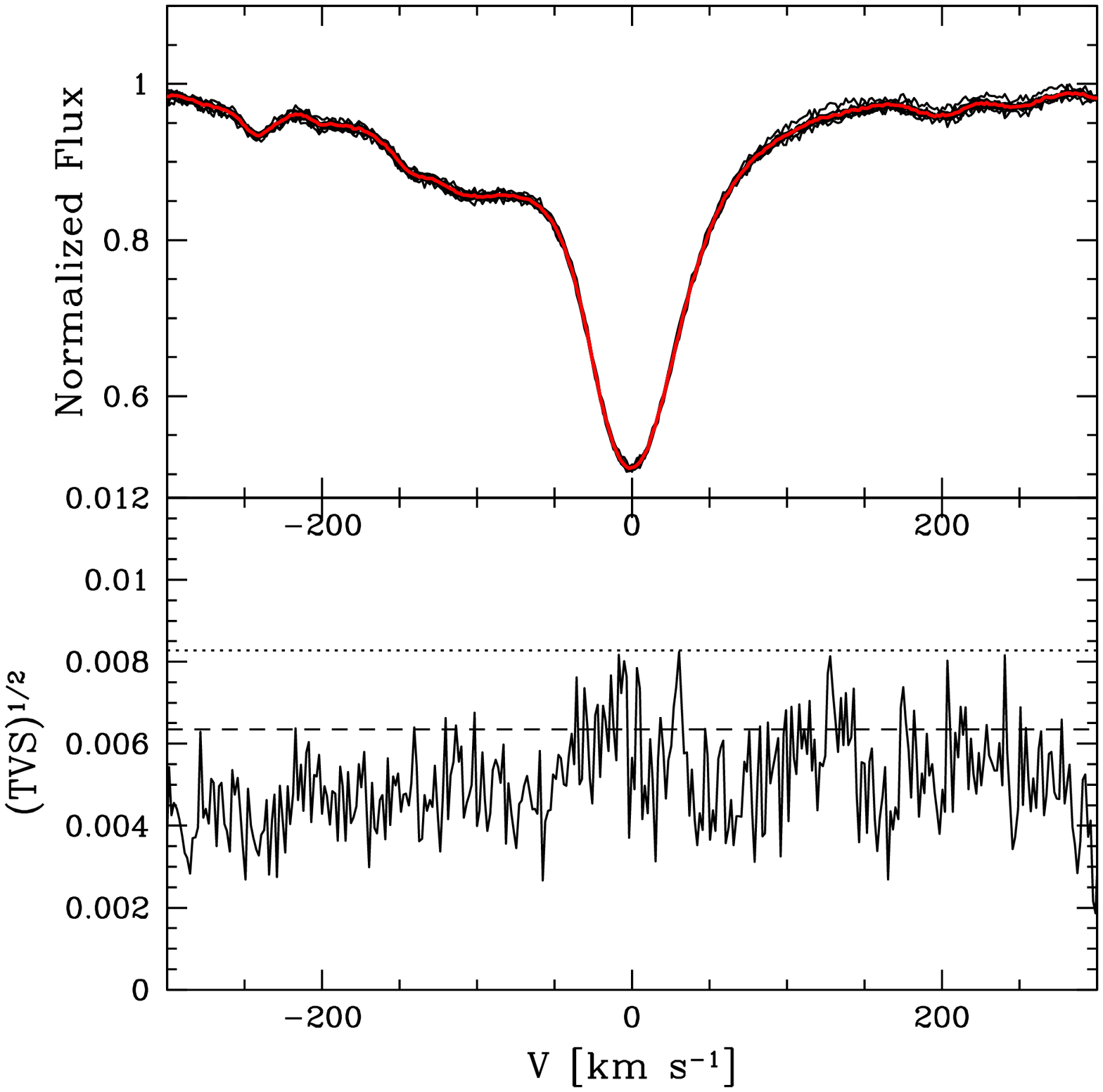}}
     \hspace{0.2cm}
     \subfigure[\ion{He}{I} 4712]{
          \includegraphics[width=.28\textwidth]{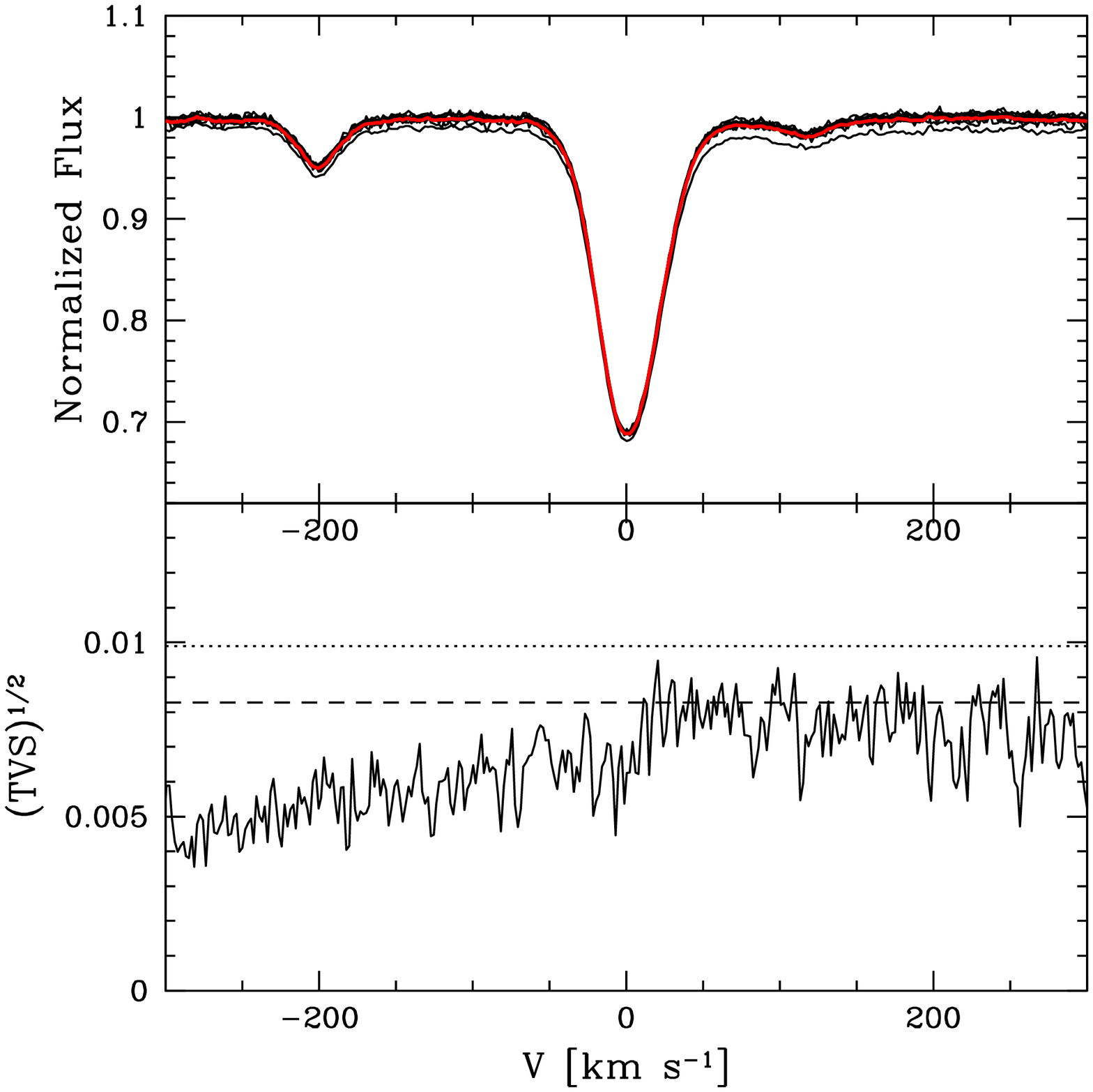}}\\
     \subfigure[\ion{He}{I} 5876]{
          \includegraphics[width=.28\textwidth]{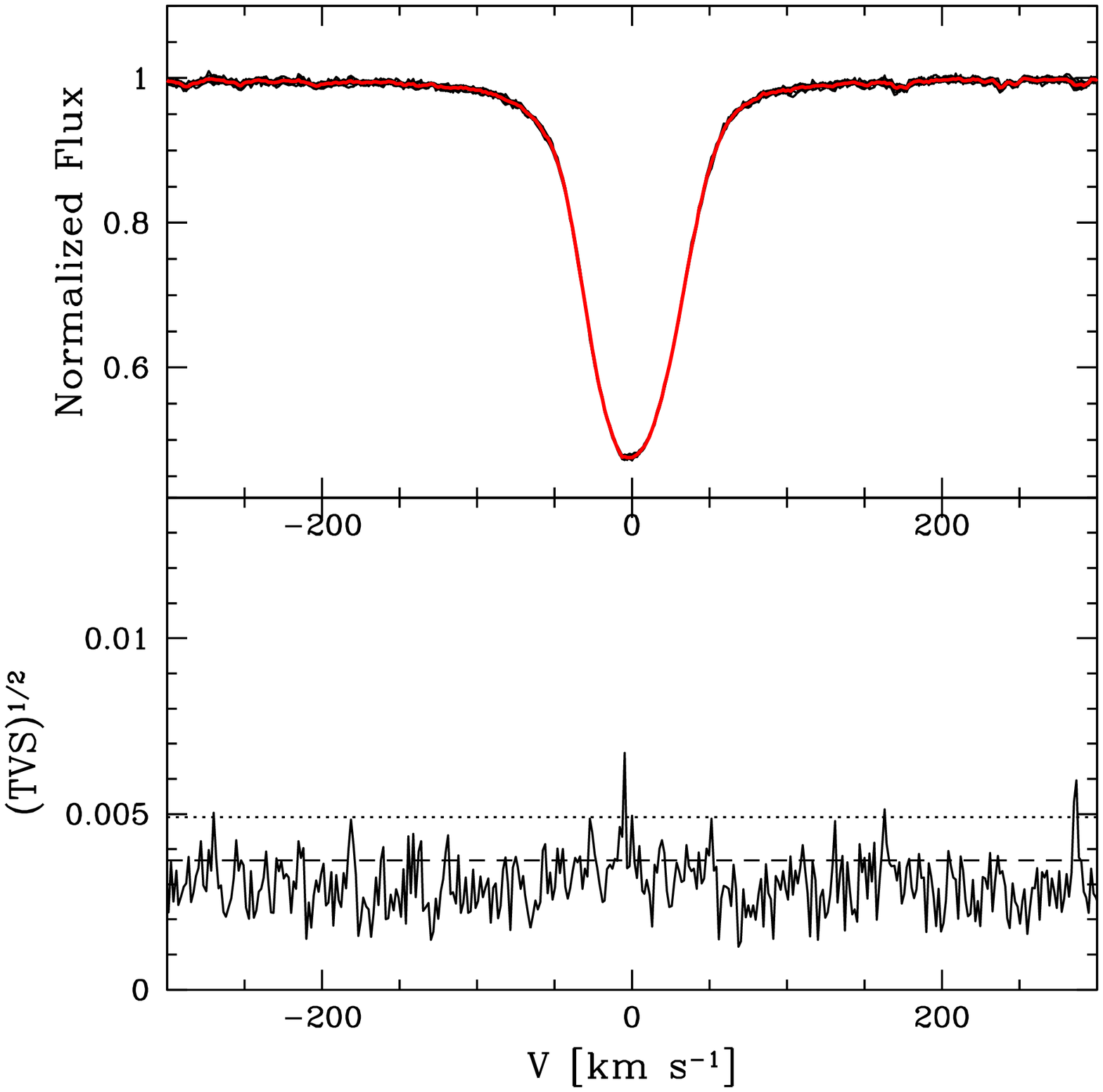}}
     \hspace{0.2cm}
     \subfigure[\ion{He}{II} 4542]{
          \includegraphics[width=.28\textwidth]{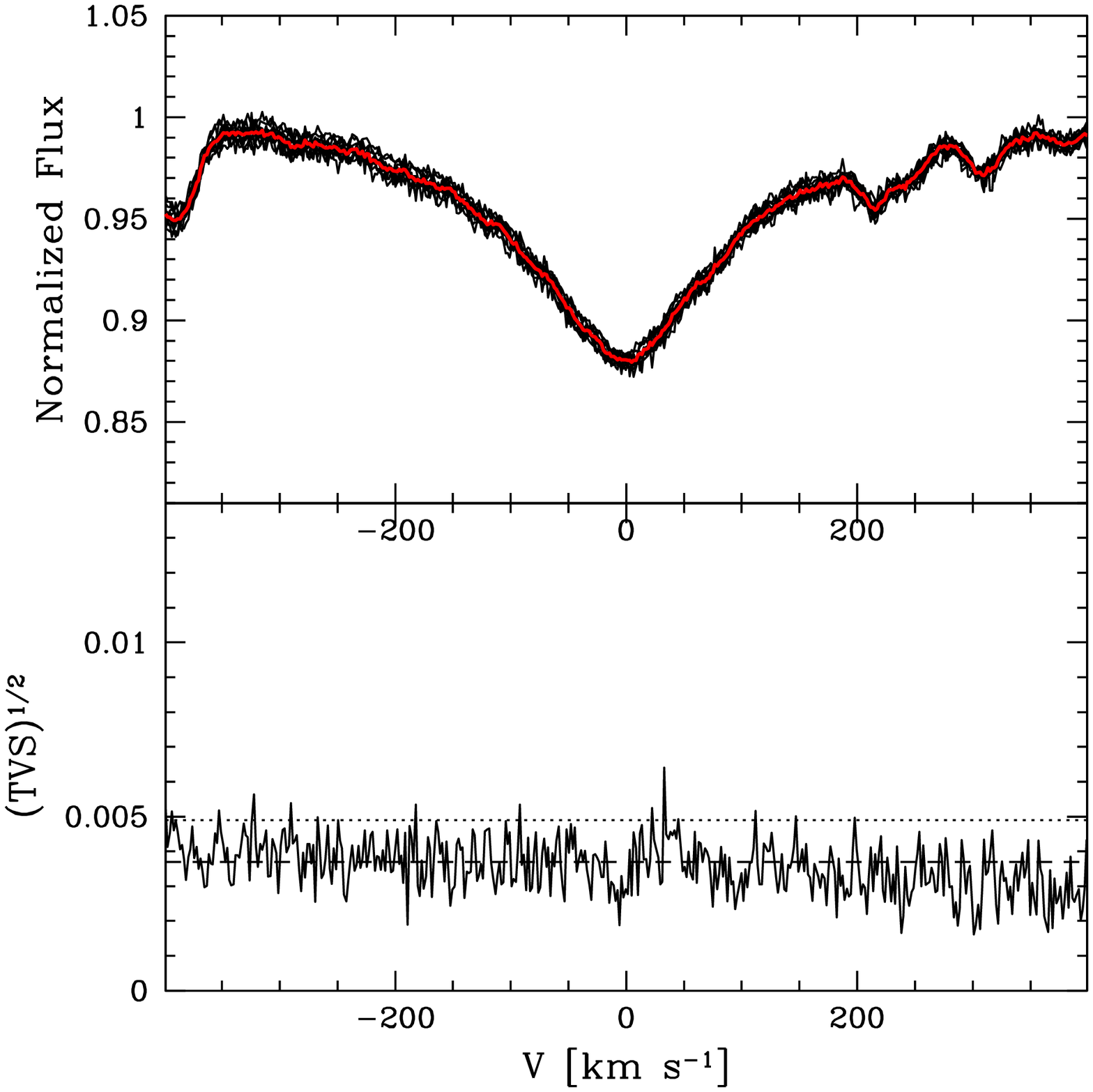}}
     \hspace{0.2cm}
     \subfigure[\ion{He}{II} 4686]{
          \includegraphics[width=.28\textwidth]{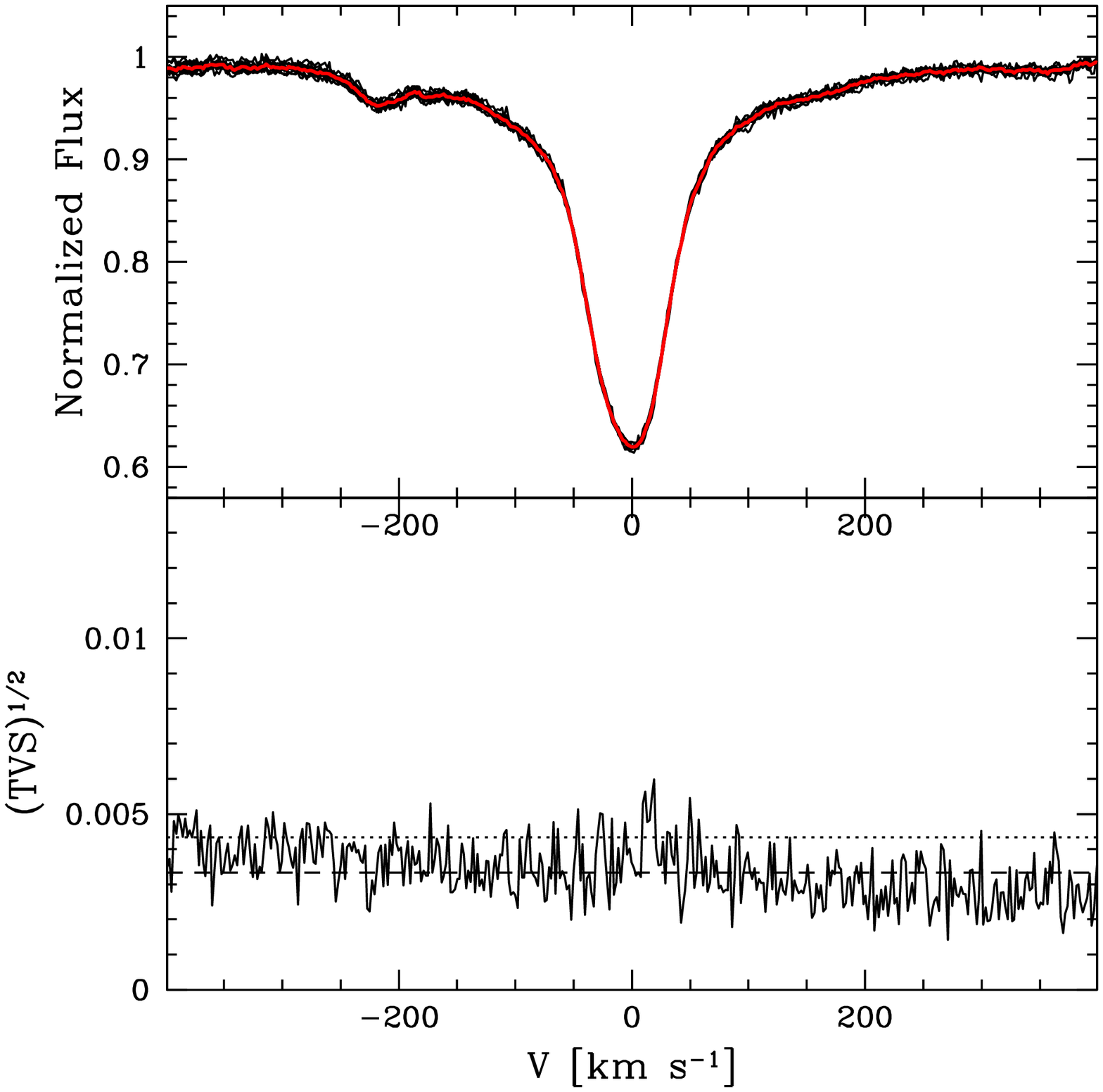}}\\
     \subfigure[\ion{He}{II} 5412]{
          \includegraphics[width=.28\textwidth]{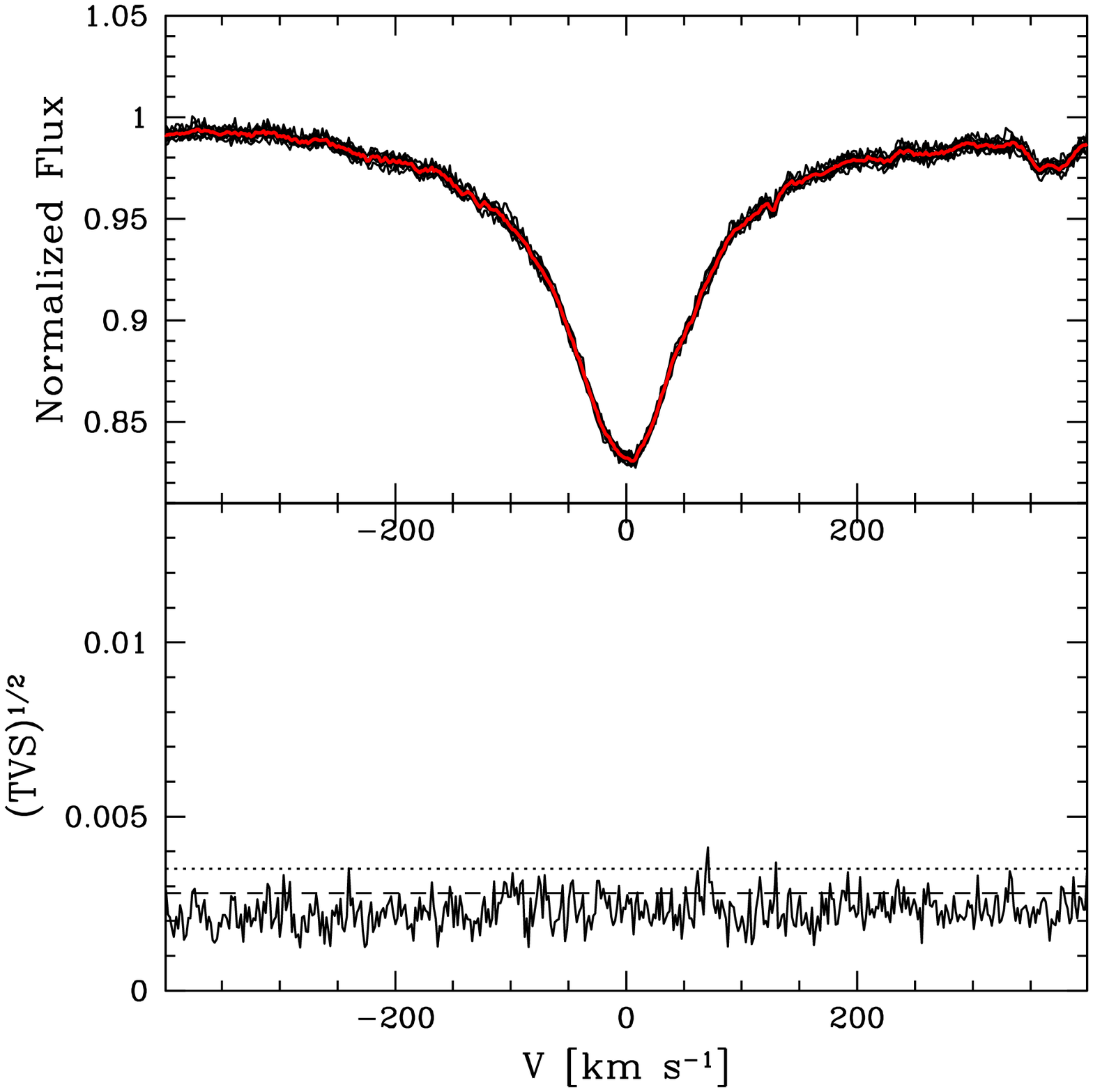}}
     \hspace{0.2cm}
     \subfigure[\ion{O}{III} 5592]{
          \includegraphics[width=.28\textwidth]{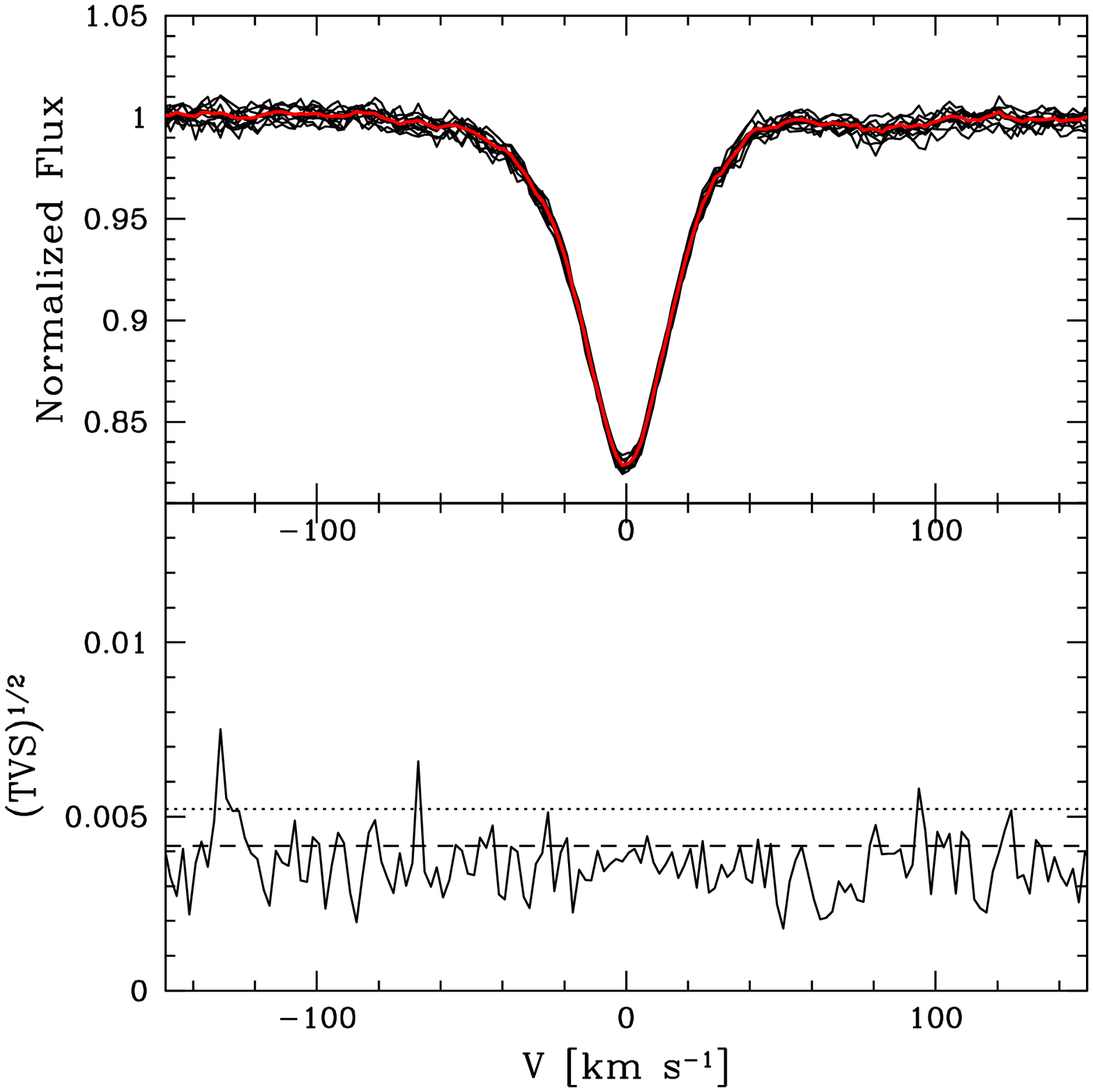}}
     \hspace{0.2cm}
     \subfigure[\ion{C}{IV} 5802]{
          \includegraphics[width=.28\textwidth]{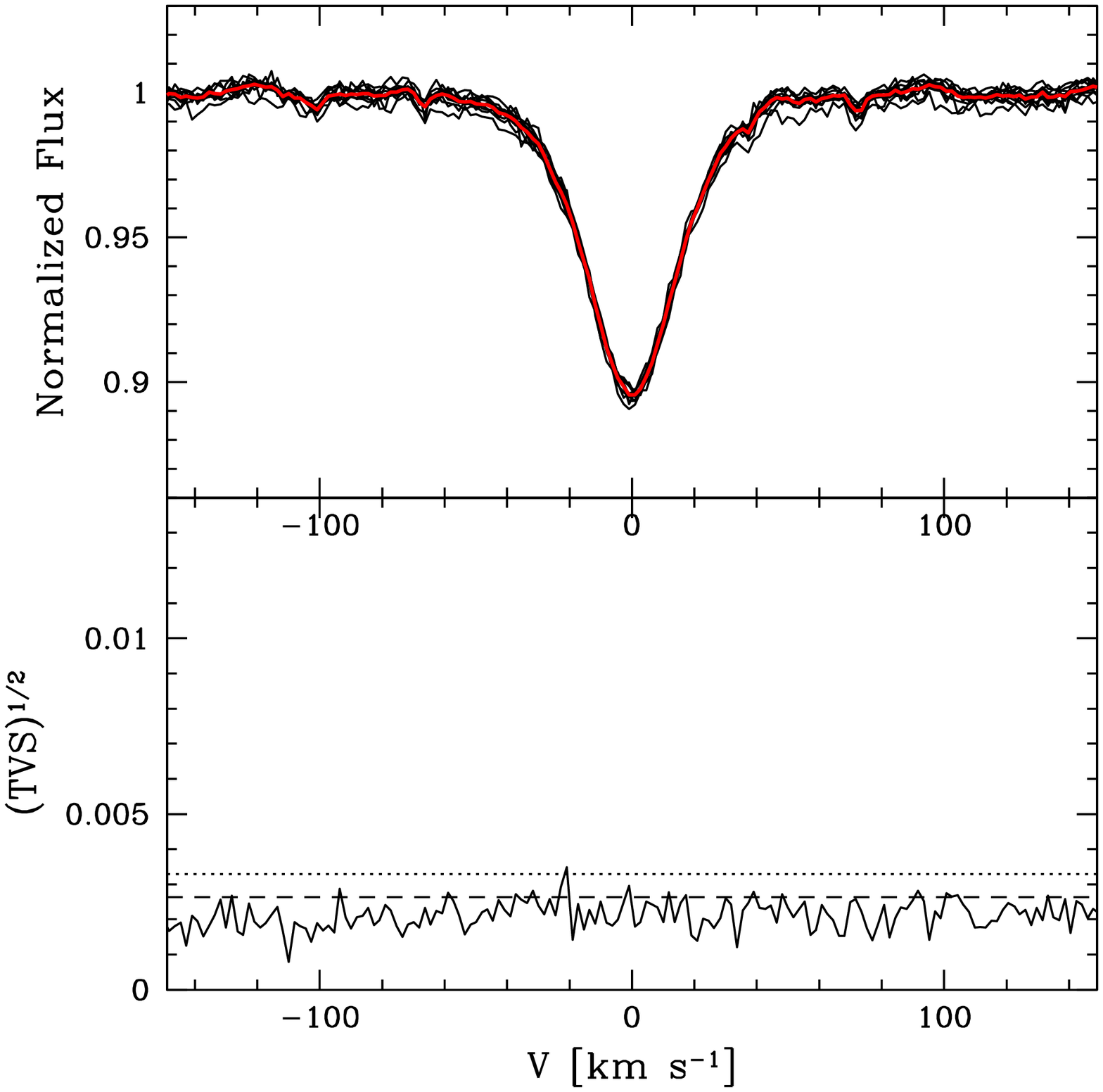}}\\
     \caption{Variability of AE~Aur between October 15$^{th}$ and October 19$^{th}$ 2007.}
     \label{fig_var_aeaur}
\end{figure*}

\newpage

\begin{figure*}
     \centering
     \subfigure[\ha]{
          \includegraphics[width=.28\textwidth]{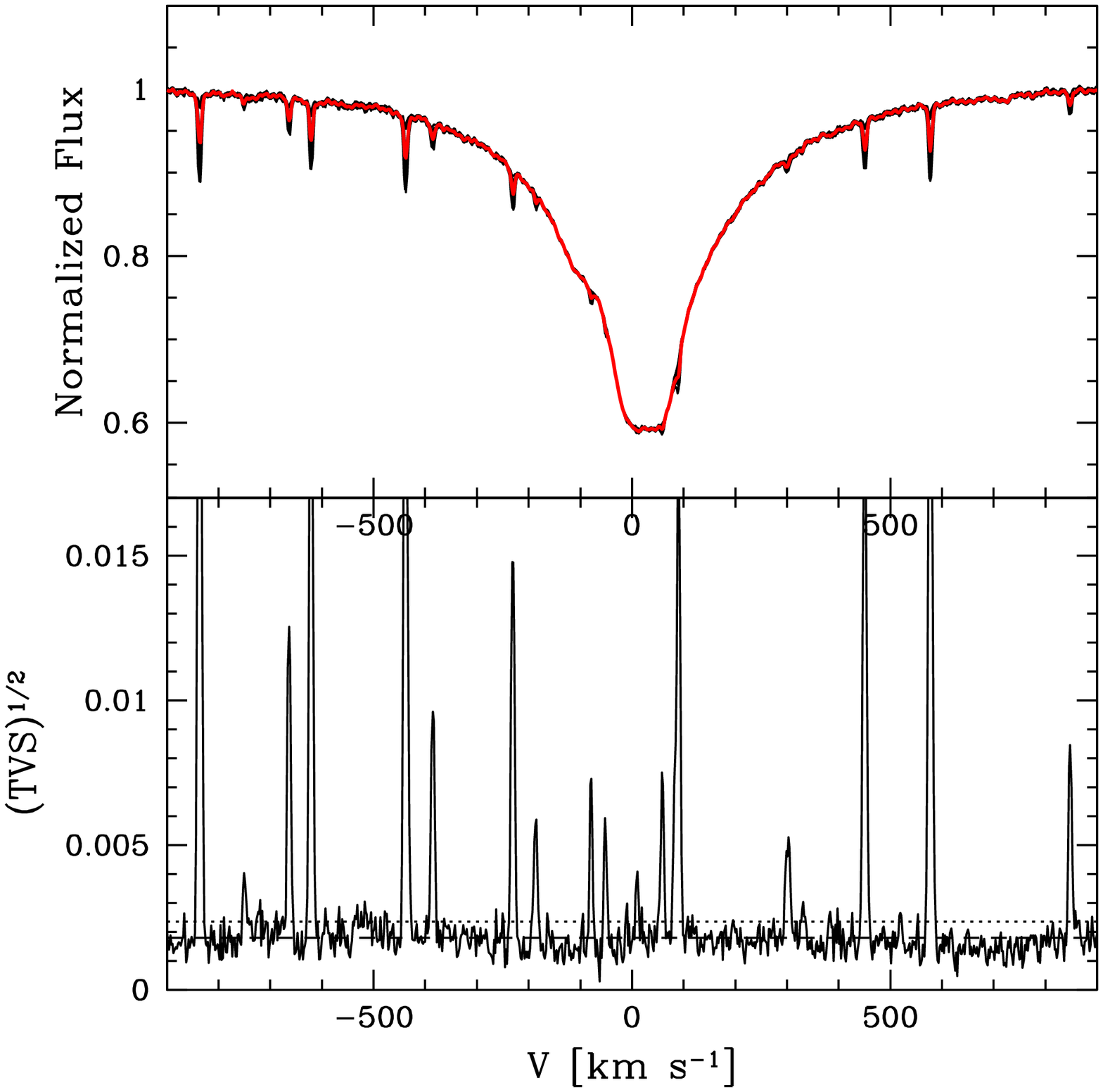}}
     \hspace{0.2cm}
     \subfigure[H$_{\beta}$]{
          \includegraphics[width=.28\textwidth]{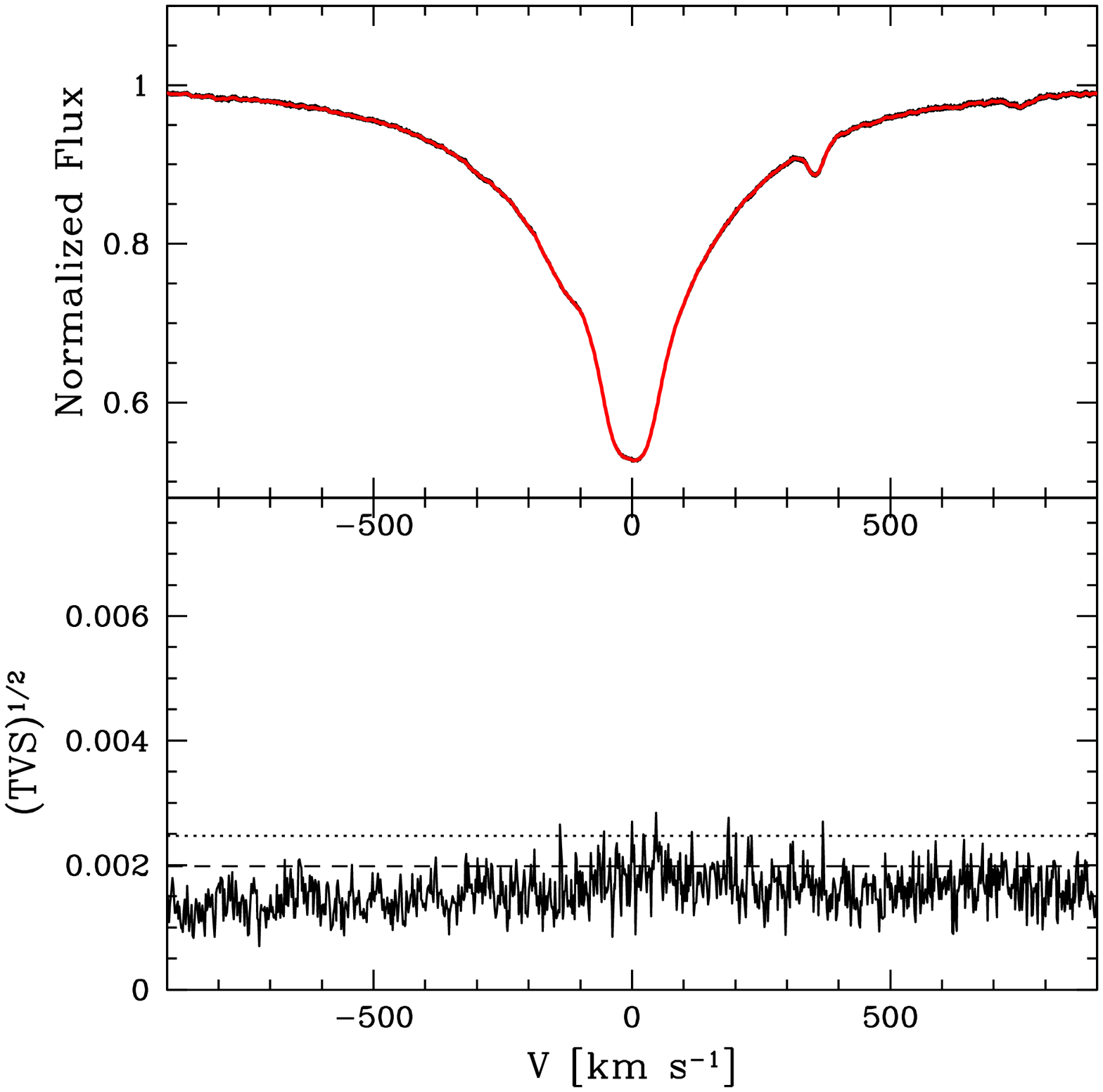}}
     \hspace{0.2cm}
     \subfigure[H$_{\gamma}$]{
          \includegraphics[width=.28\textwidth]{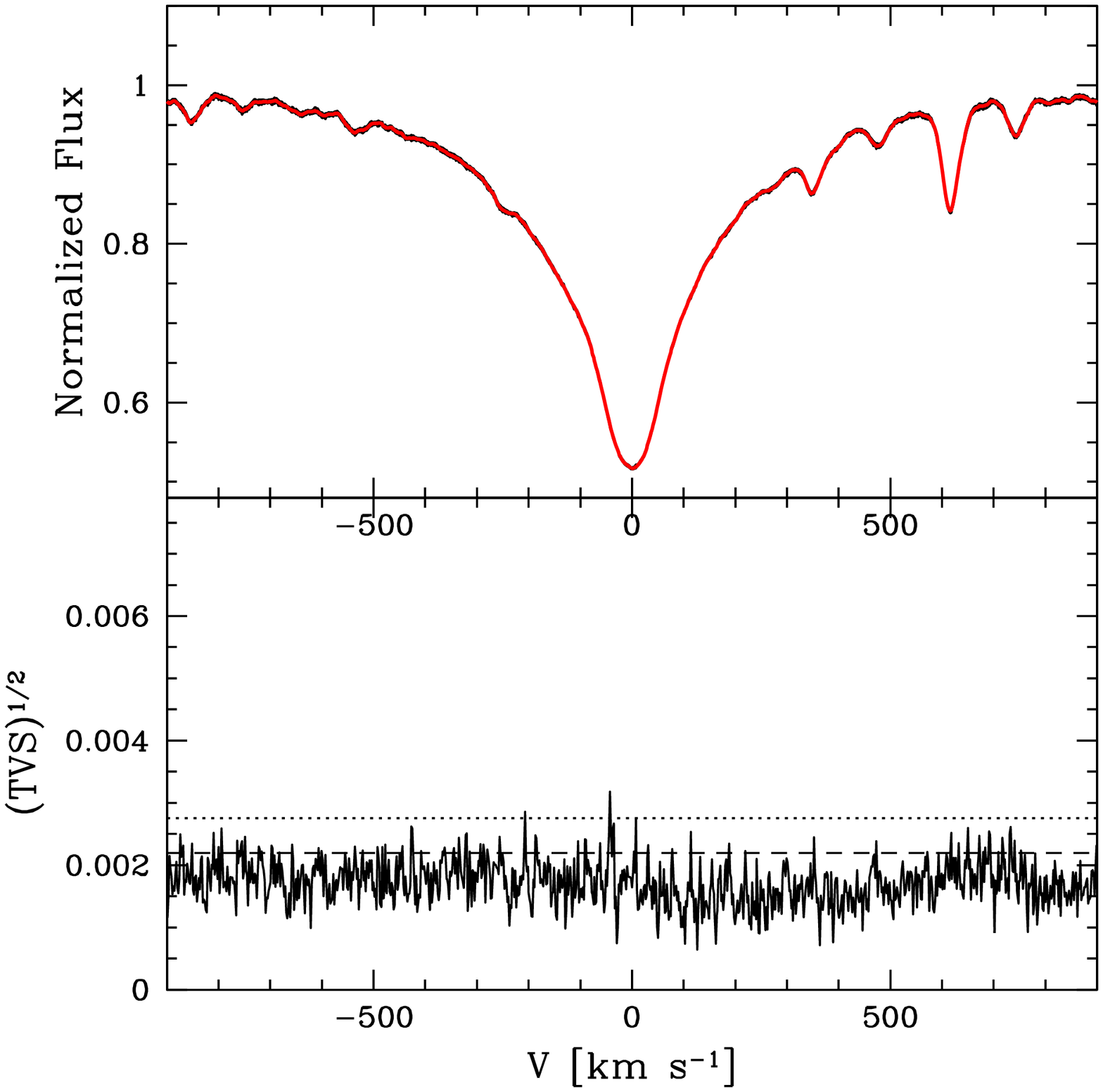}}\\
     \subfigure[\ion{He}{I} 4026]{
          \includegraphics[width=.28\textwidth]{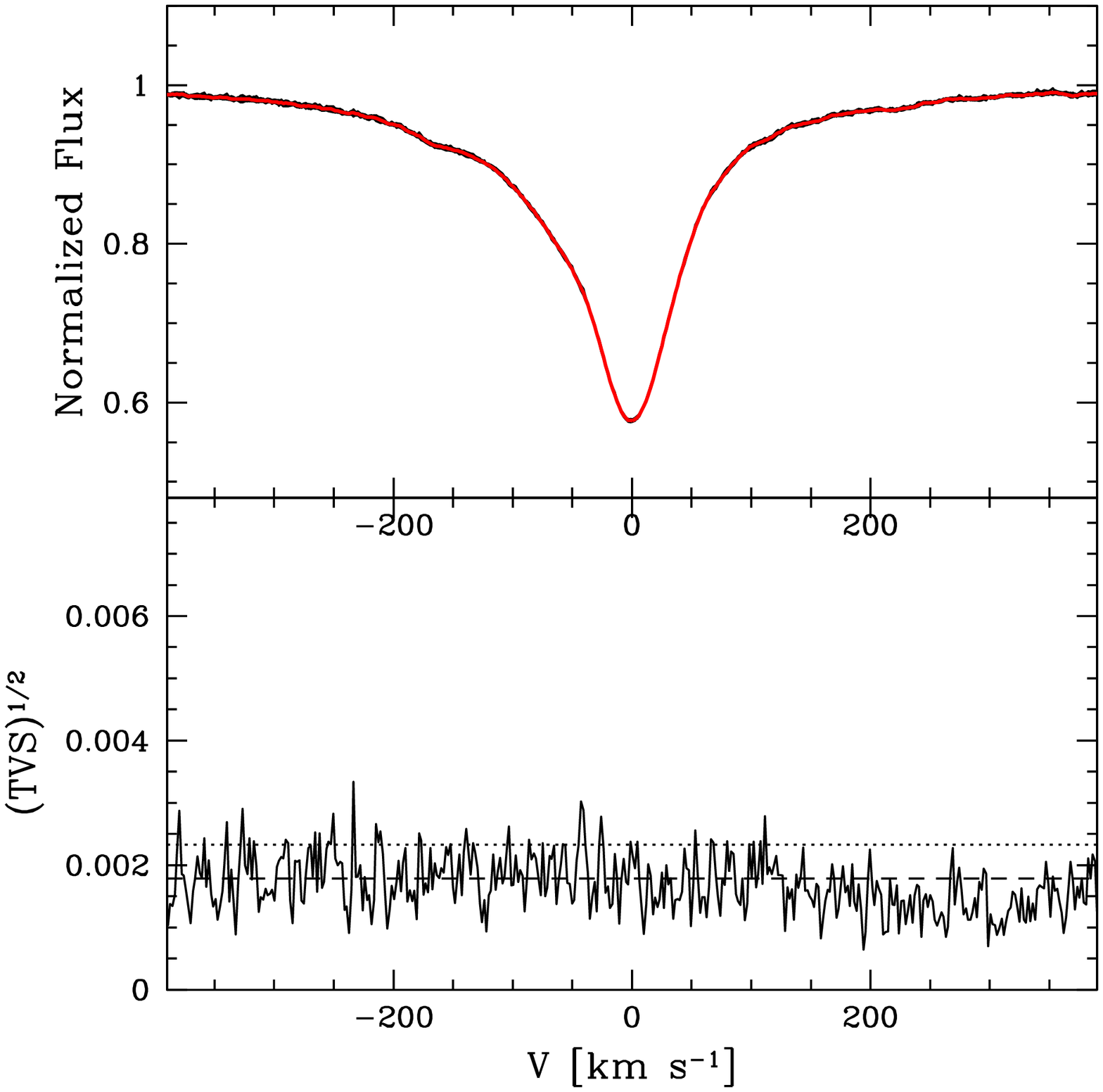}}
     \hspace{0.2cm}
     \subfigure[\ion{He}{I} 4471]{
          \includegraphics[width=.28\textwidth]{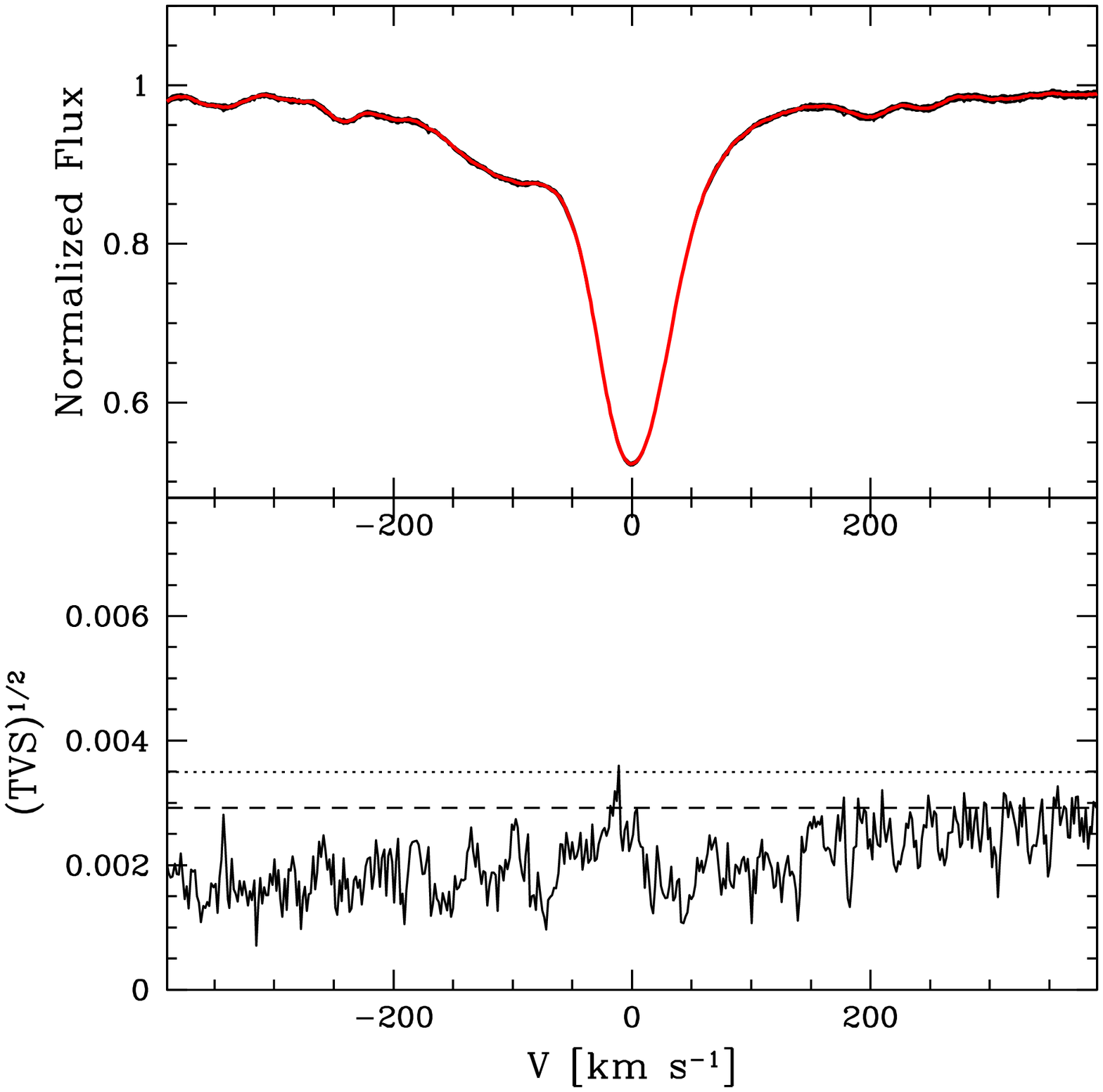}}
     \hspace{0.2cm}
     \subfigure[\ion{He}{I} 4712]{
          \includegraphics[width=.28\textwidth]{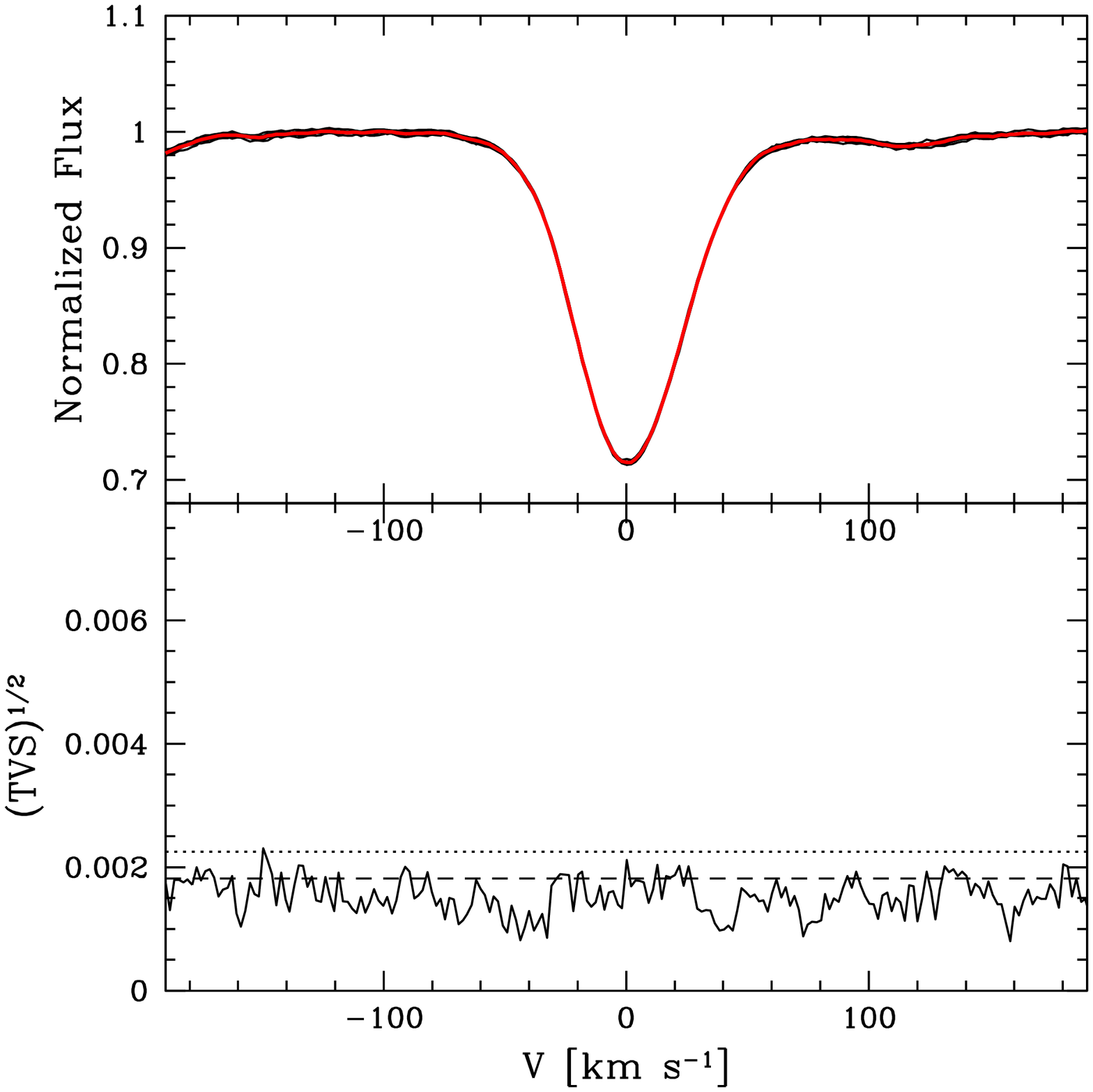}}\\
     \subfigure[\ion{He}{I} 5876]{
          \includegraphics[width=.28\textwidth]{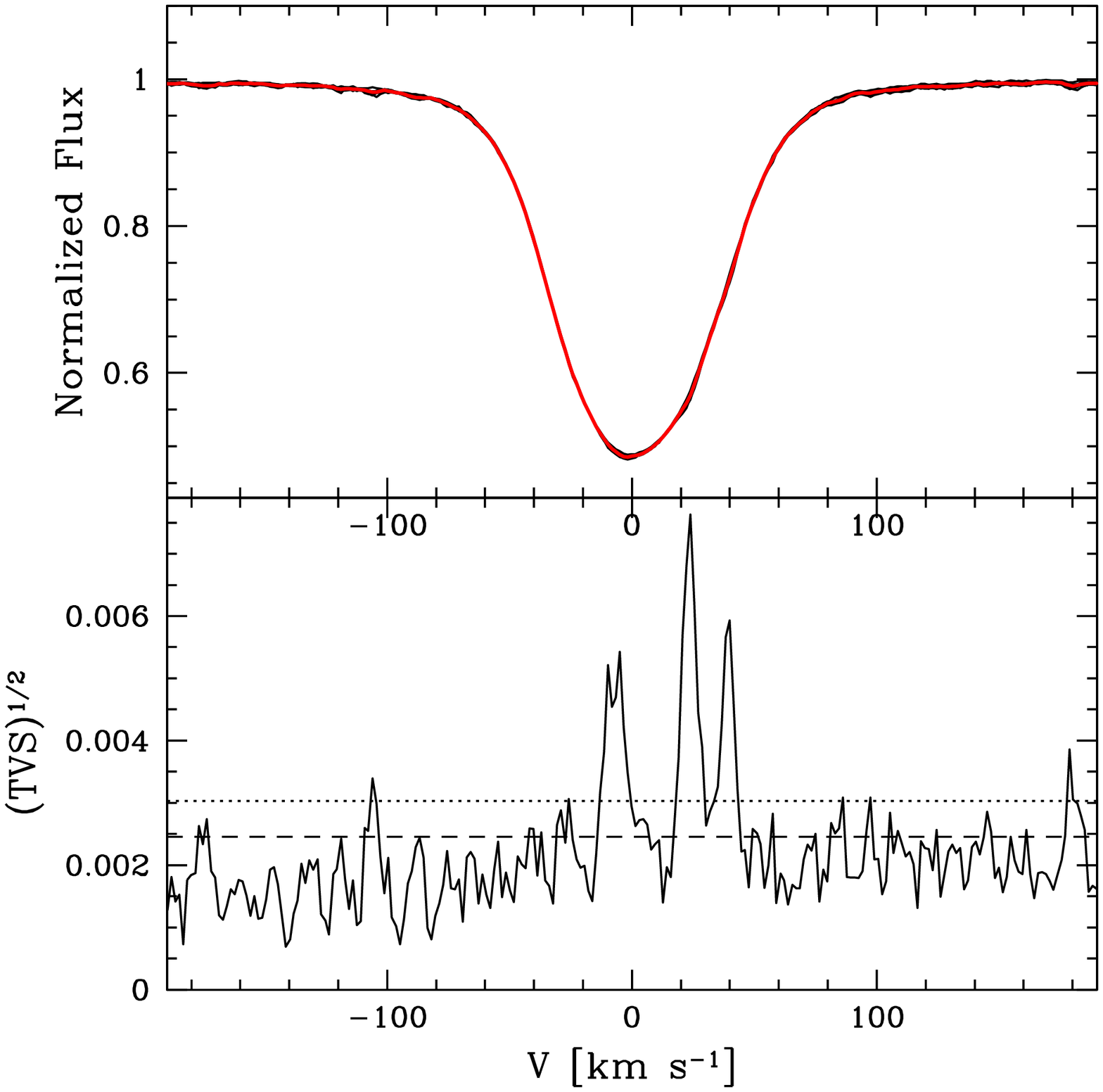}}
     \hspace{0.2cm}
     \subfigure[\ion{He}{II} 4542]{
          \includegraphics[width=.28\textwidth]{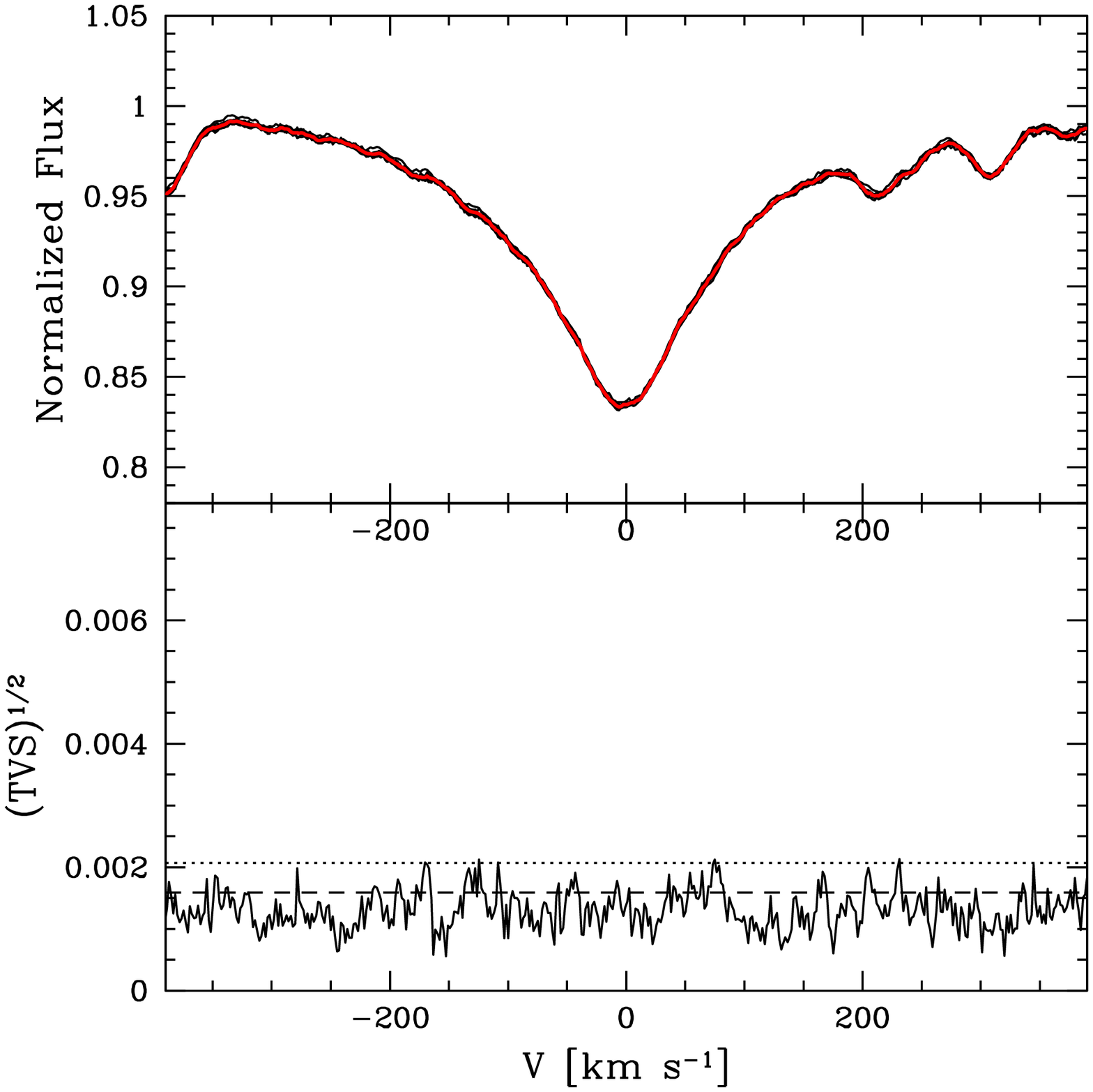}}
     \hspace{0.2cm}
     \subfigure[\ion{He}{II} 4686]{
          \includegraphics[width=.28\textwidth]{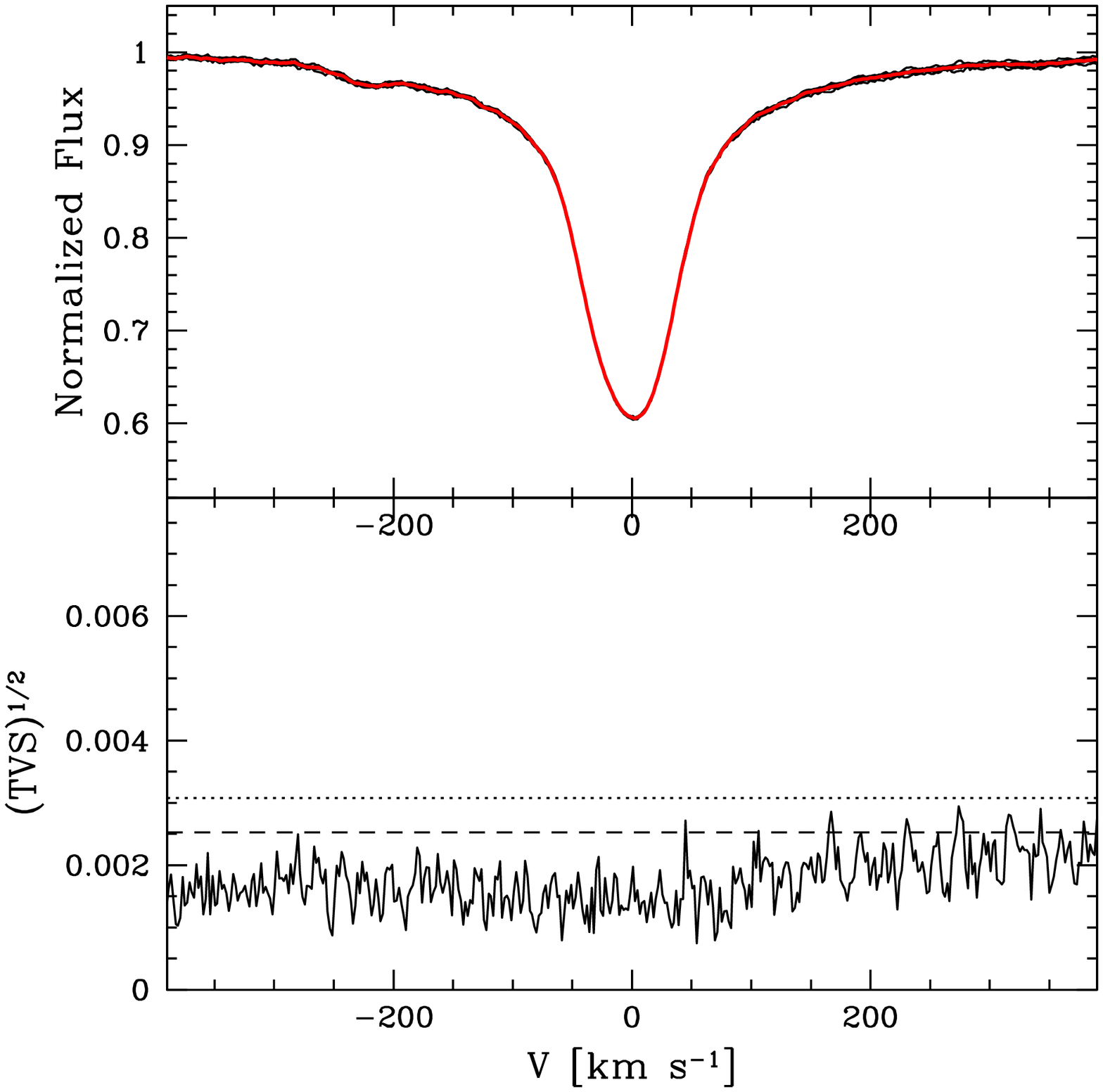}}\\
     \subfigure[\ion{He}{II} 5412]{
          \includegraphics[width=.28\textwidth]{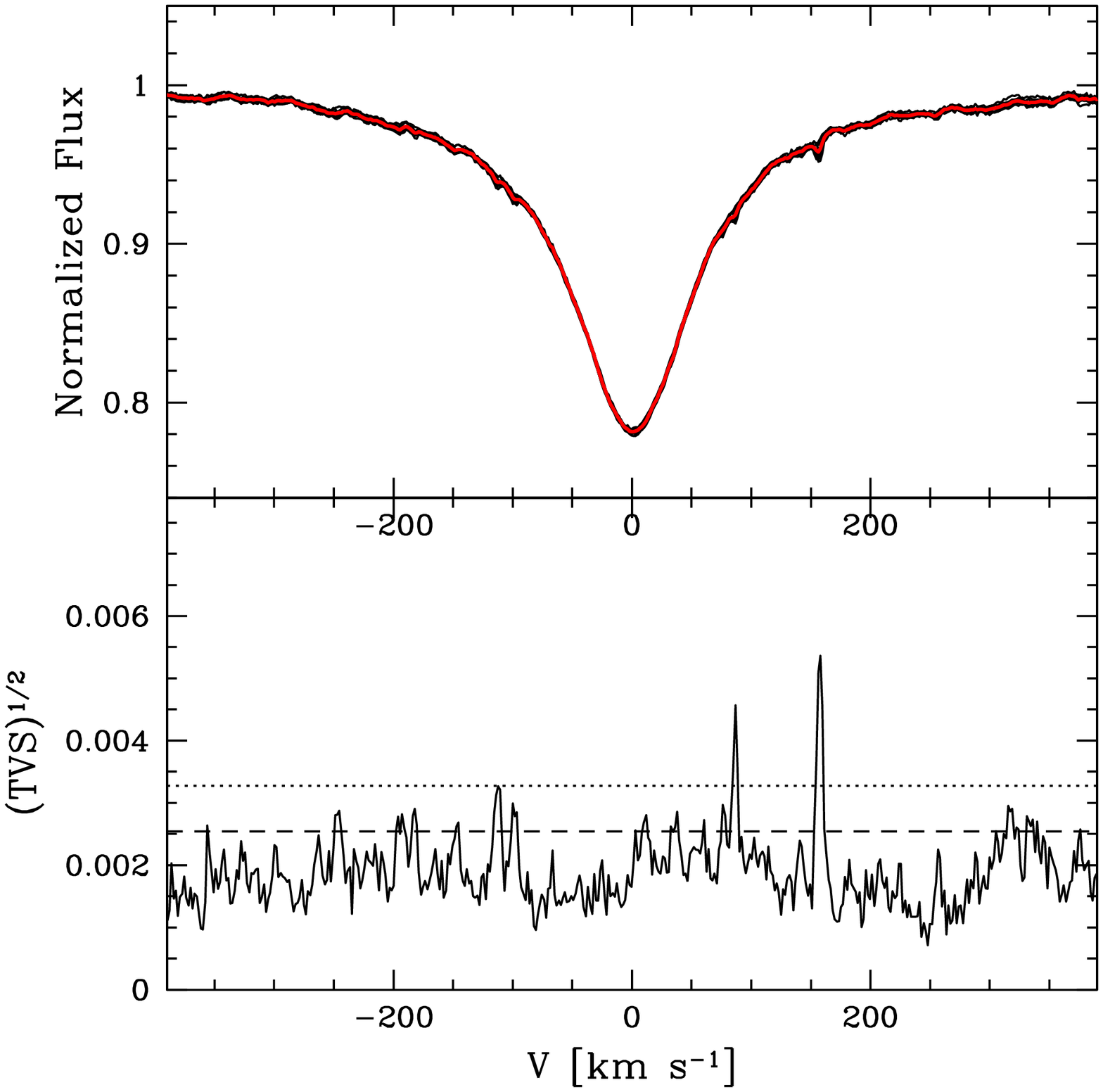}}
     \hspace{0.2cm}
     \subfigure[\ion{O}{III} 5592]{
          \includegraphics[width=.28\textwidth]{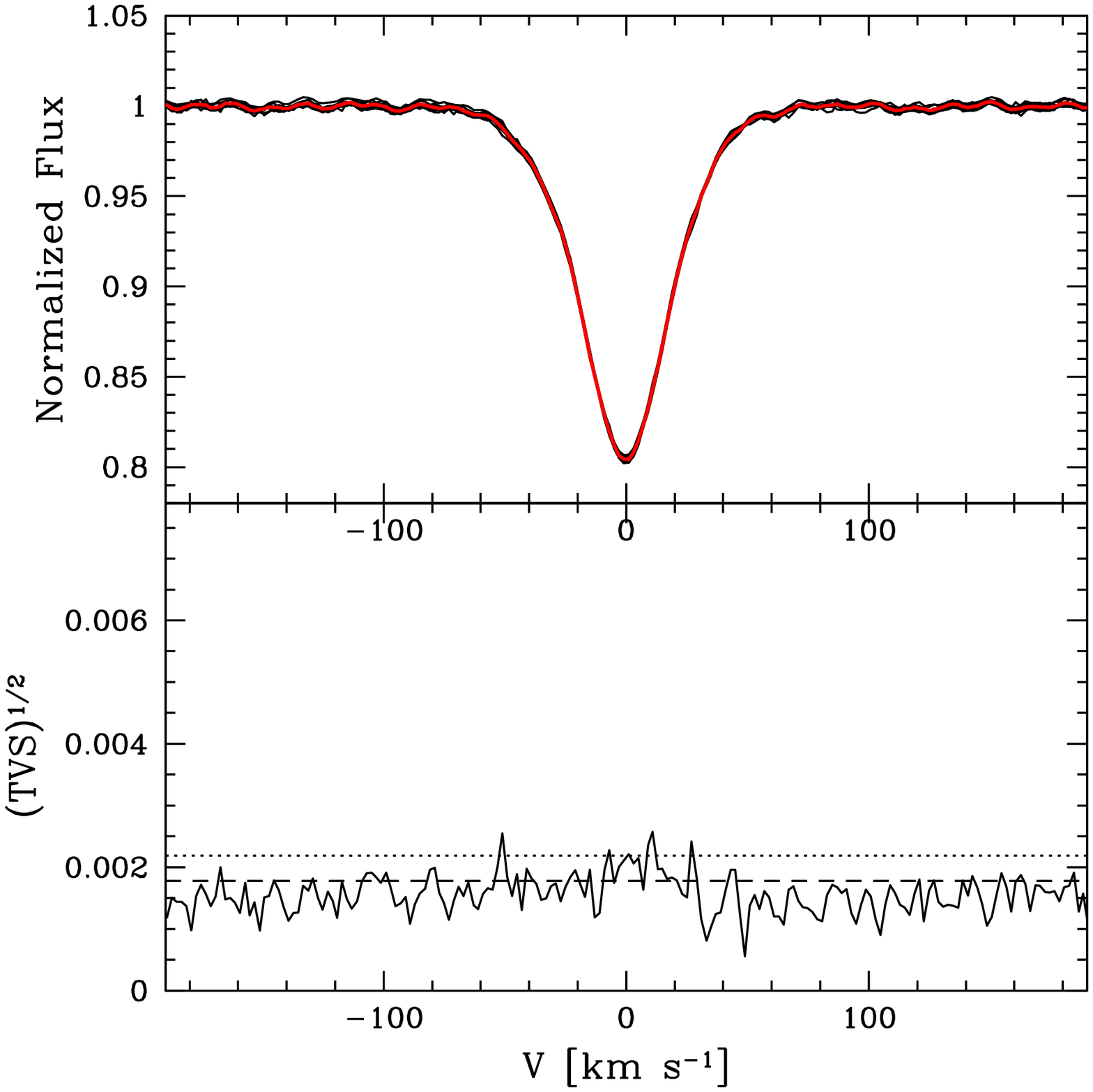}}
     \hspace{0.2cm}
     \subfigure[\ion{C}{IV} 5802]{
          \includegraphics[width=.28\textwidth]{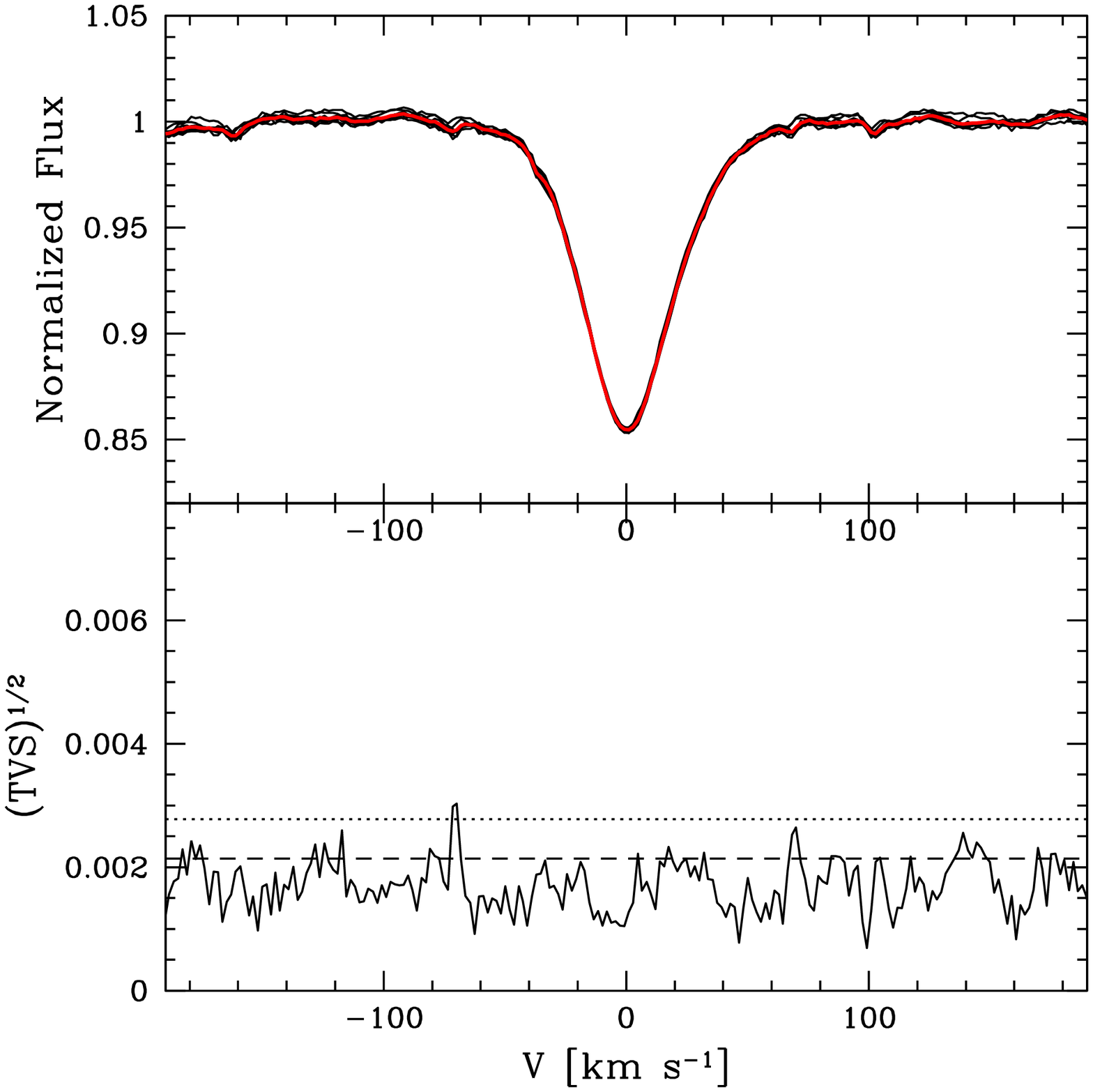}}\\
     \caption{Variability of 10~Lac between October 15$^{th}$ and October 24$^{th}$ 2007.}
     \label{fig_var_10lac}
\end{figure*}

\newpage

\begin{figure*}
     \centering
     \subfigure[\ha]{
          \includegraphics[width=.28\textwidth]{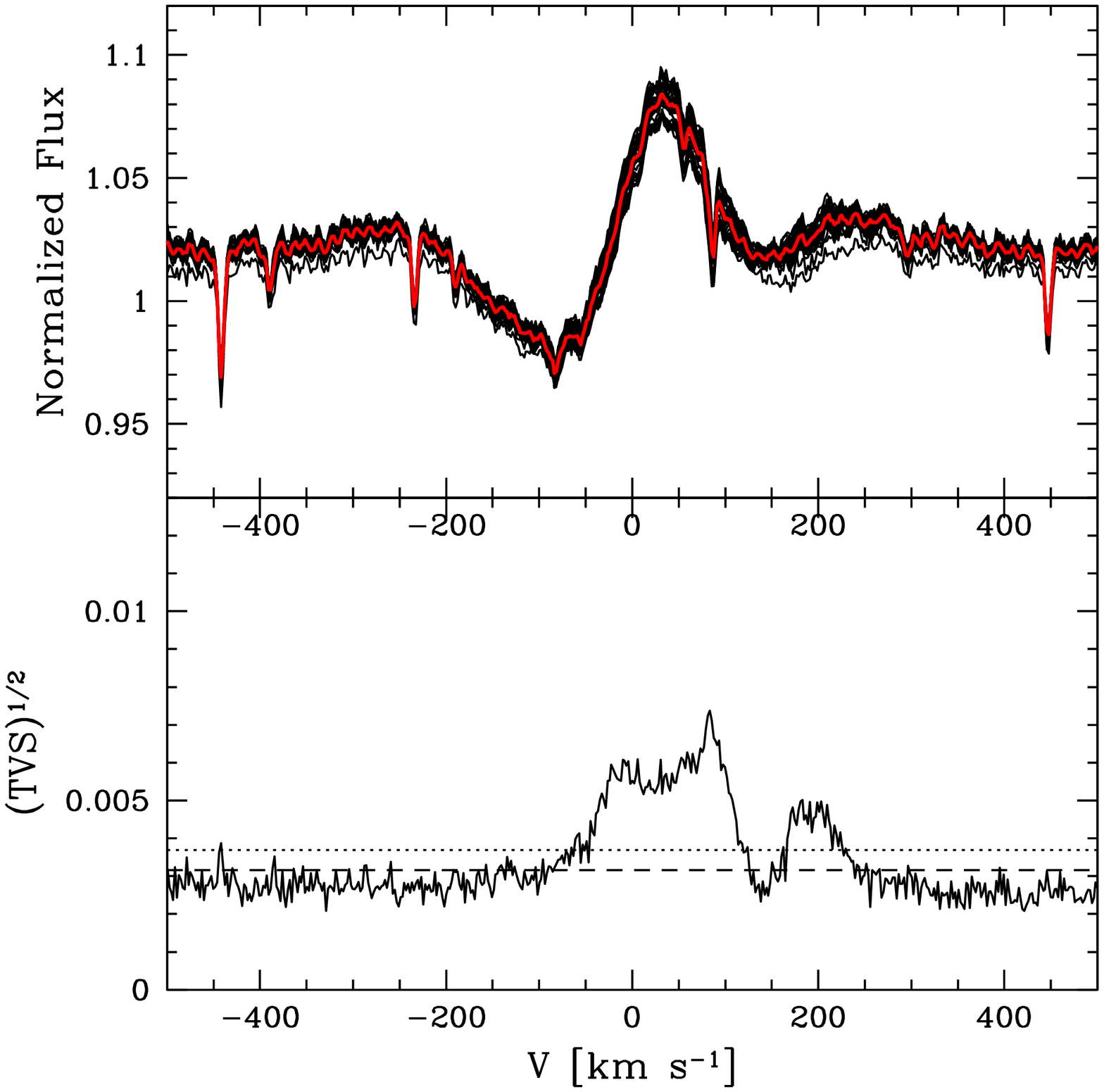}}
     \hspace{0.2cm}
     \subfigure[H$_{\beta}$]{
          \includegraphics[width=.28\textwidth]{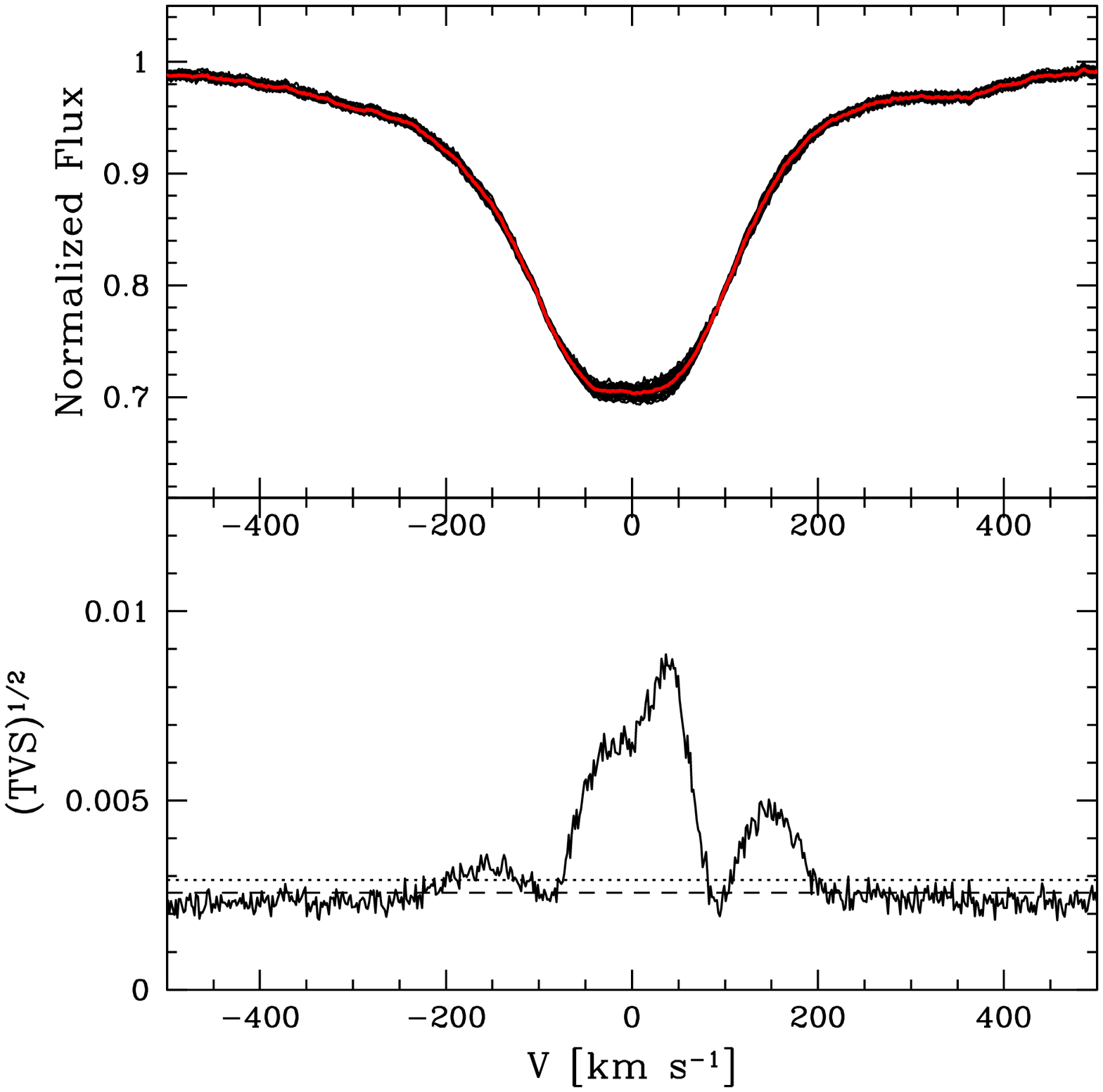}}
     \hspace{0.2cm}
     \subfigure[H$_{\gamma}$]{
          \includegraphics[width=.28\textwidth]{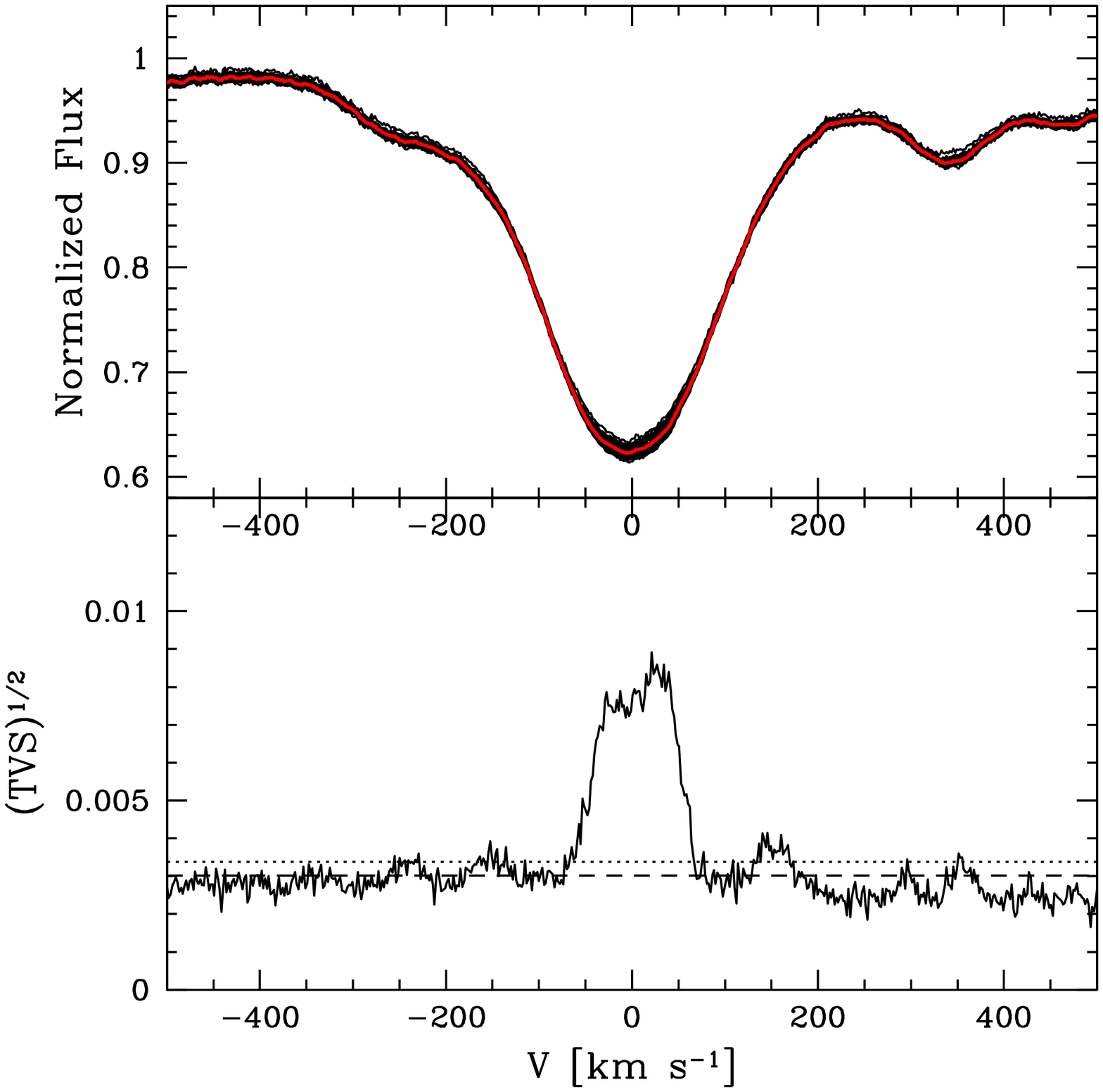}}\\
     \subfigure[\ion{He}{I} 4026]{
          \includegraphics[width=.28\textwidth]{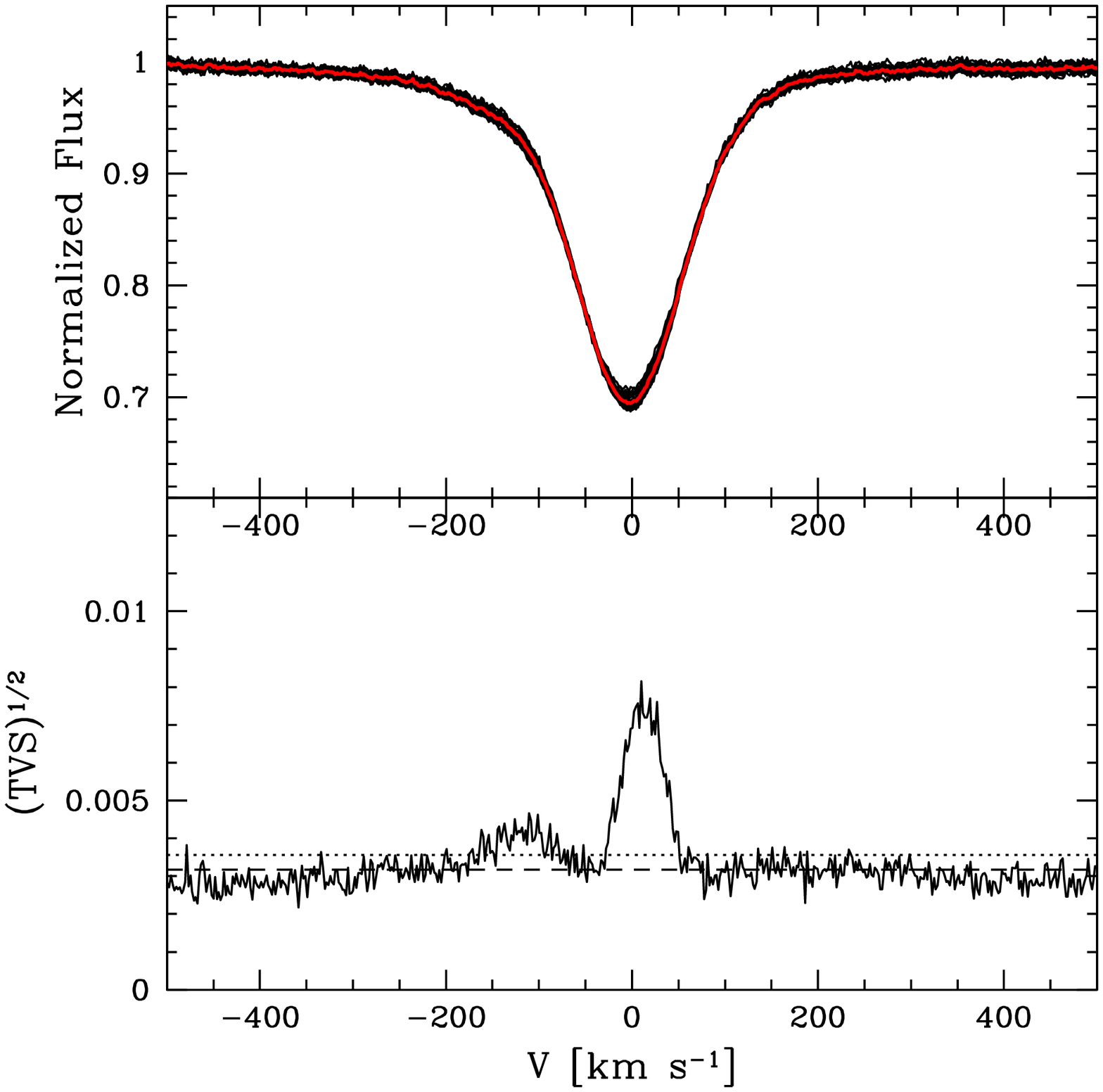}}
     \hspace{0.2cm}
     \subfigure[\ion{He}{I} 4471]{
          \includegraphics[width=.28\textwidth]{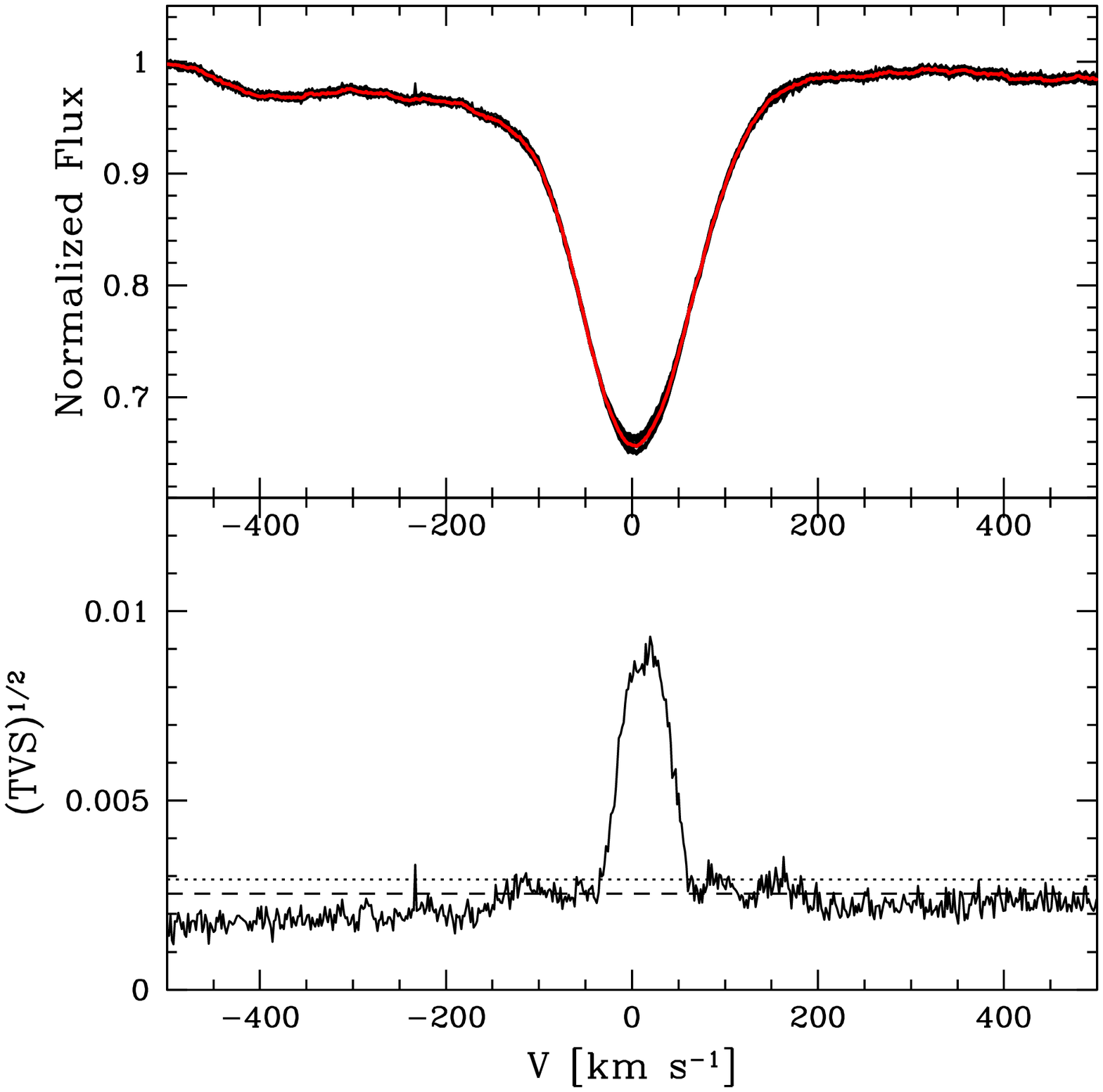}}
     \hspace{0.2cm}
     \subfigure[\ion{He}{I} 4712]{
          \includegraphics[width=.28\textwidth]{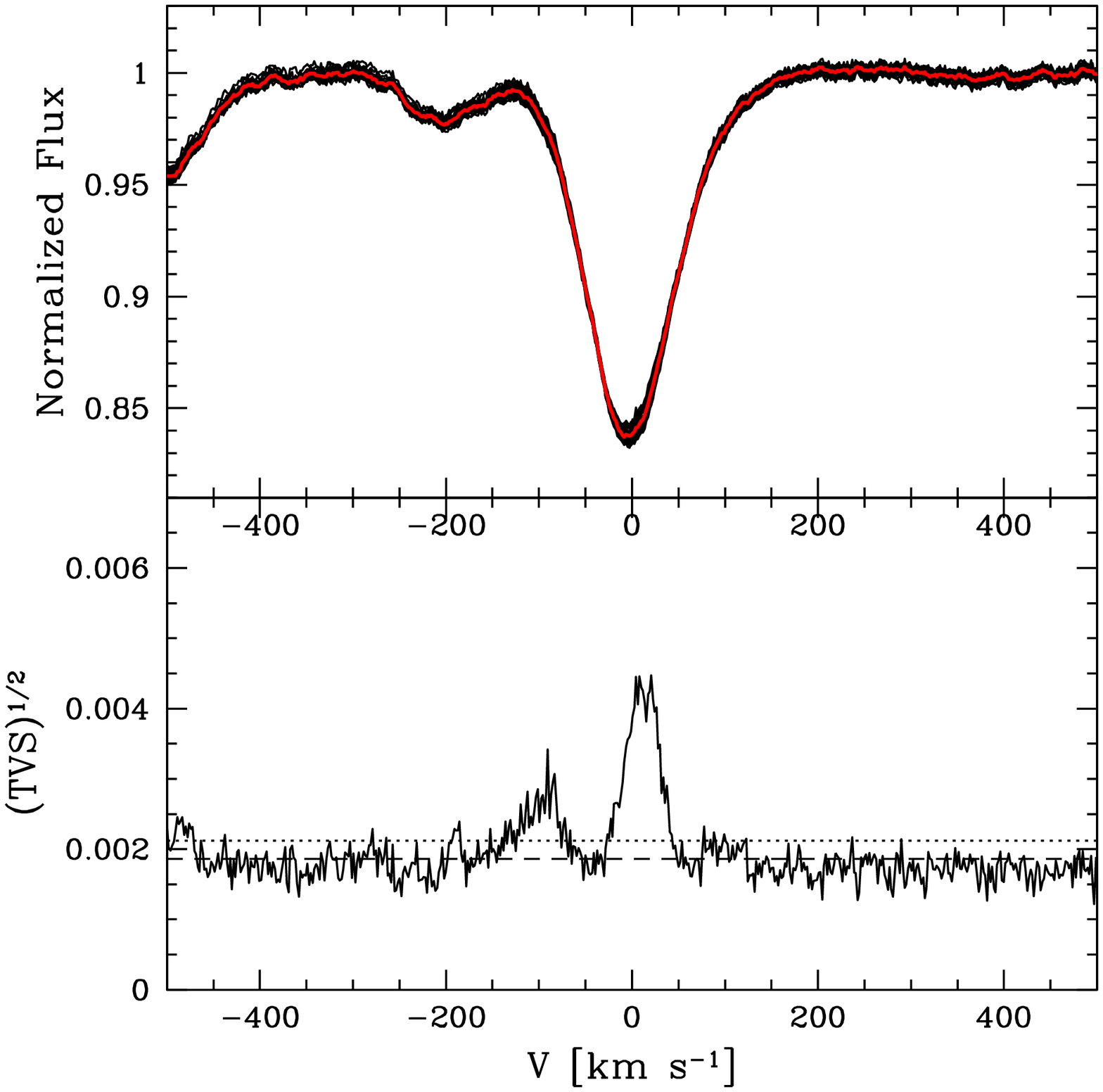}}\\
     \subfigure[\ion{He}{I} 5876]{
          \includegraphics[width=.28\textwidth]{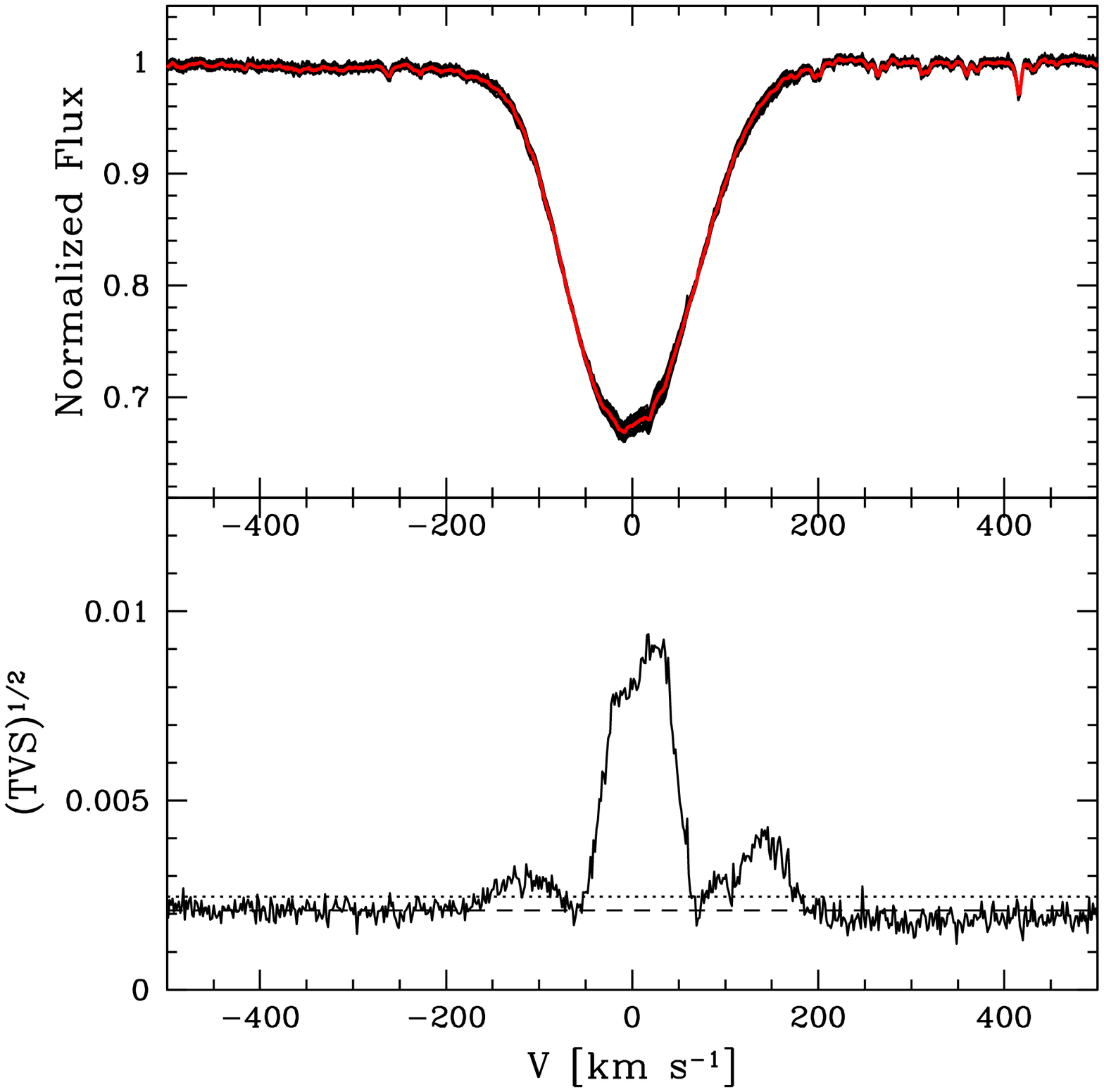}}
     \hspace{0.2cm}
     \subfigure[\ion{He}{II} 4542]{
          \includegraphics[width=.28\textwidth]{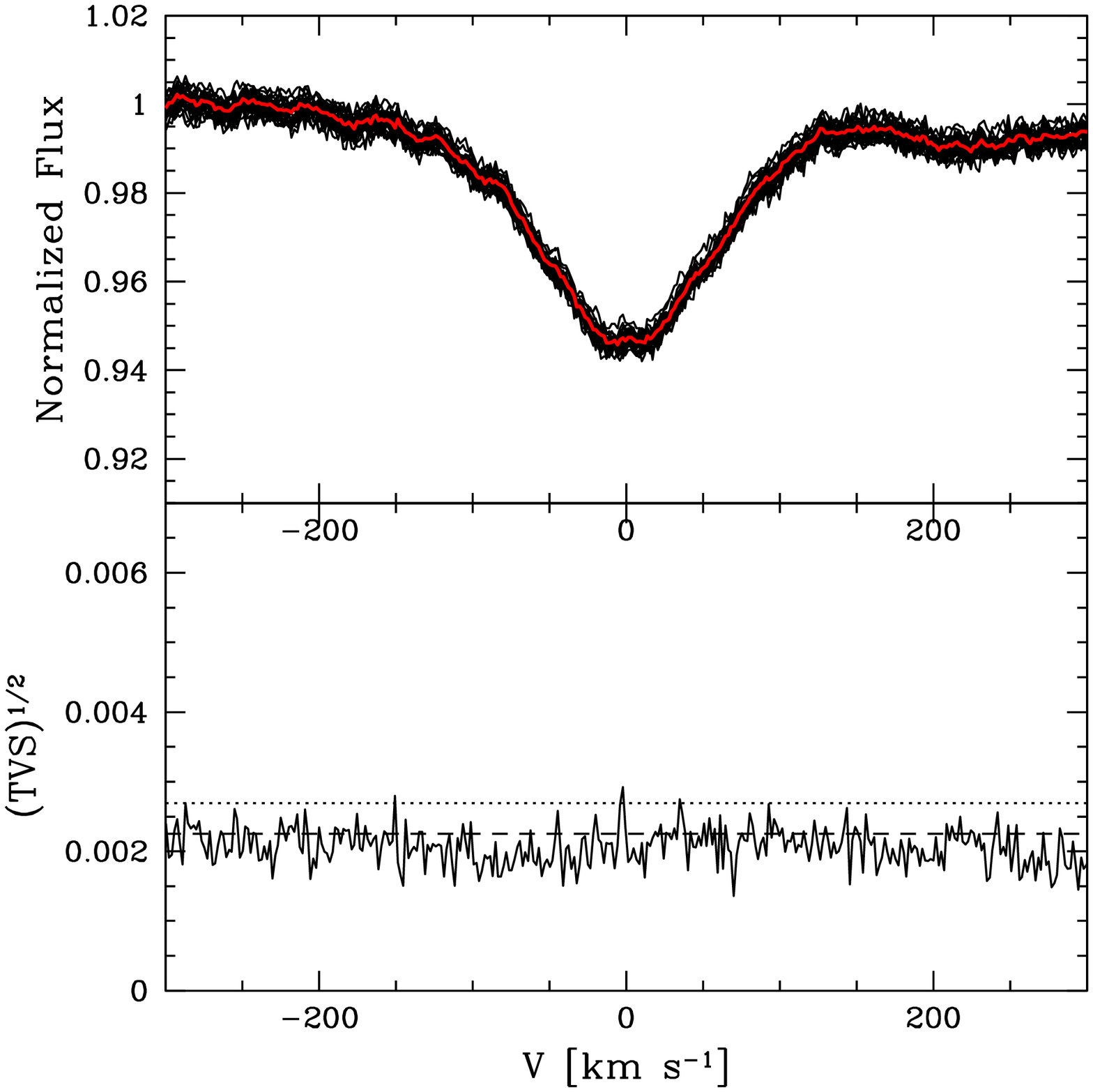}}
     \hspace{0.2cm}
     \subfigure[\ion{He}{II} 4686]{
          \includegraphics[width=.28\textwidth]{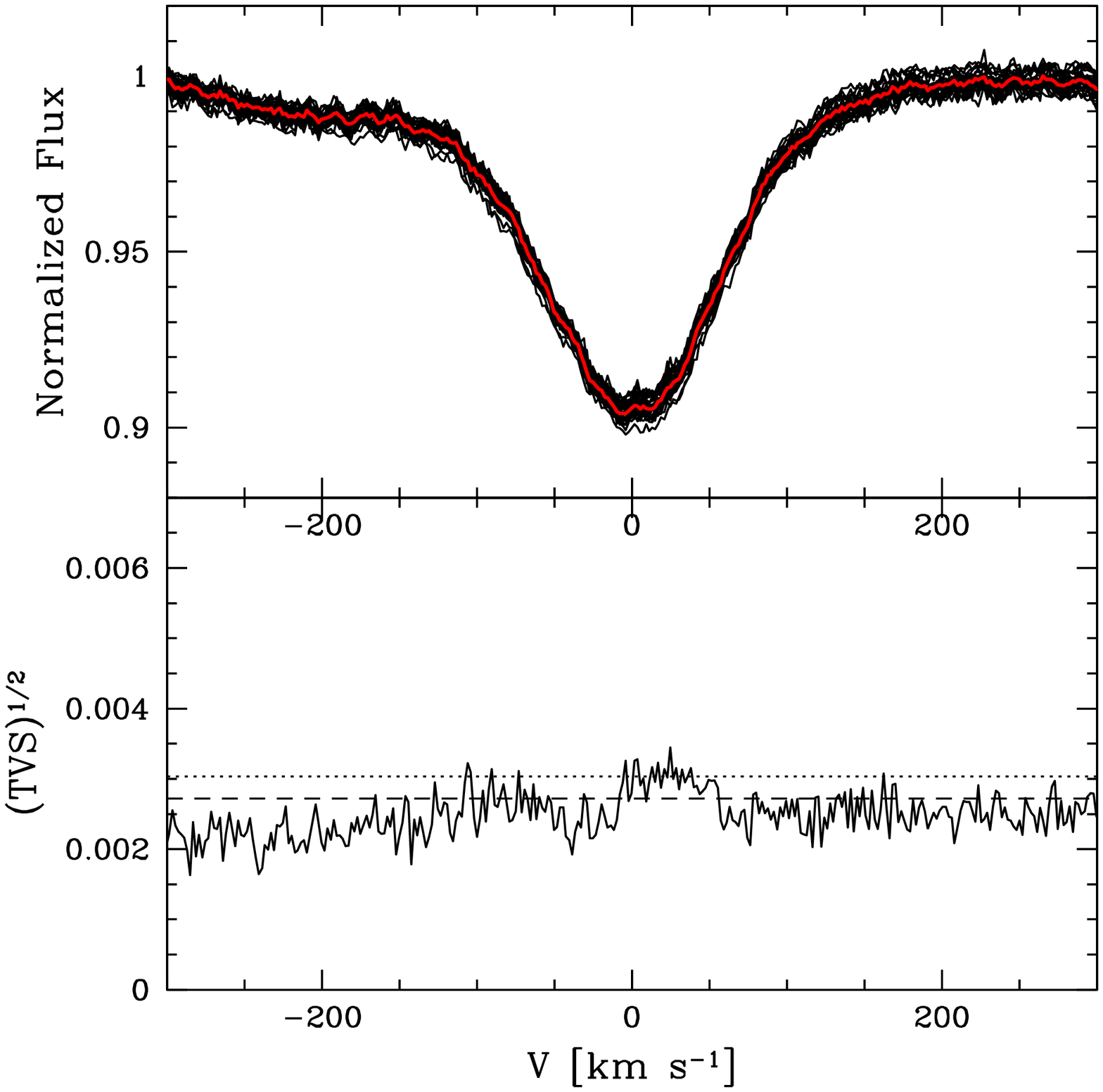}}\\
     \subfigure[\ion{He}{II} 5412]{
          \includegraphics[width=.28\textwidth]{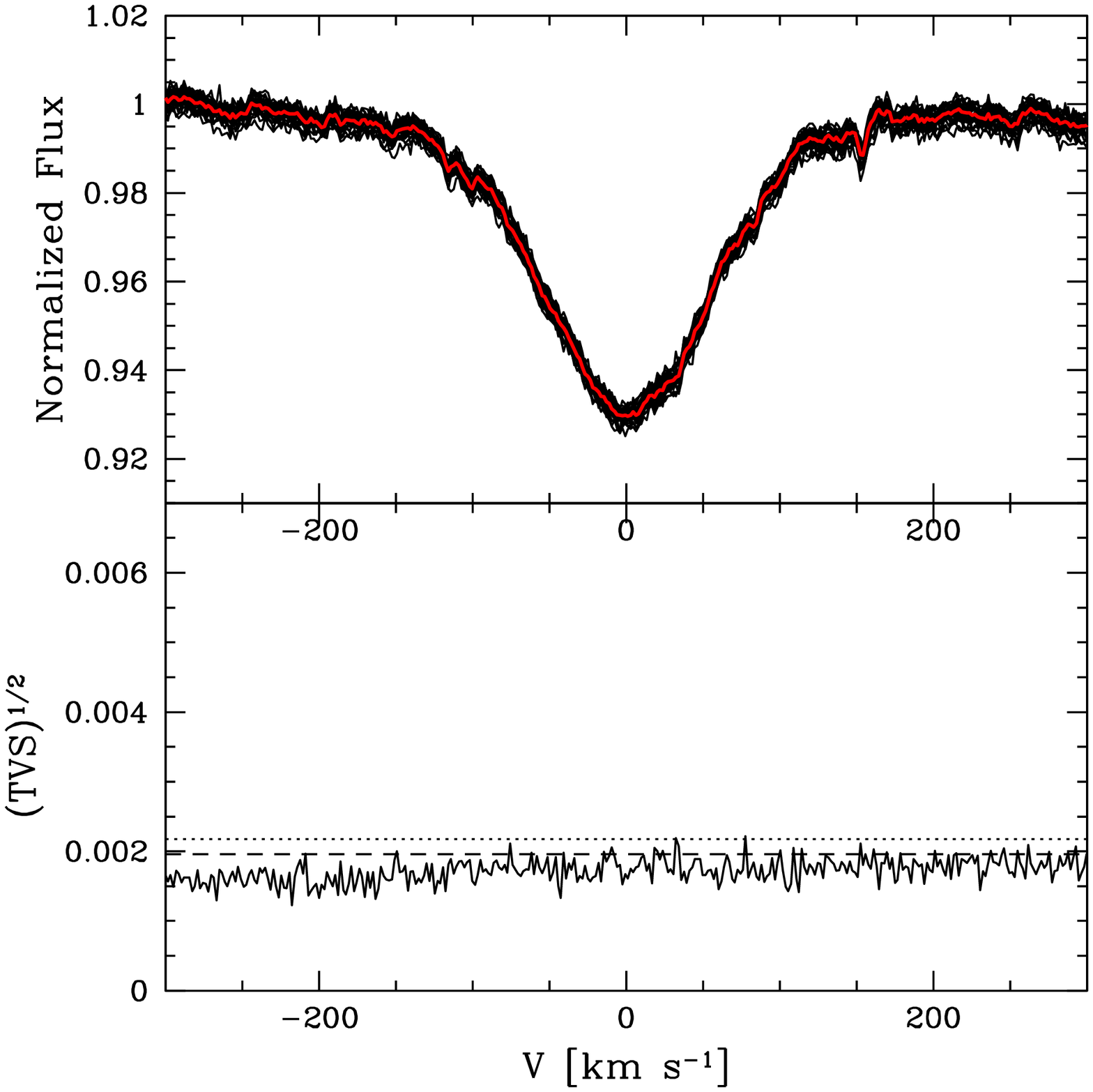}}
     \hspace{0.2cm}
     \subfigure[\ion{O}{III} 5592]{
          \includegraphics[width=.28\textwidth]{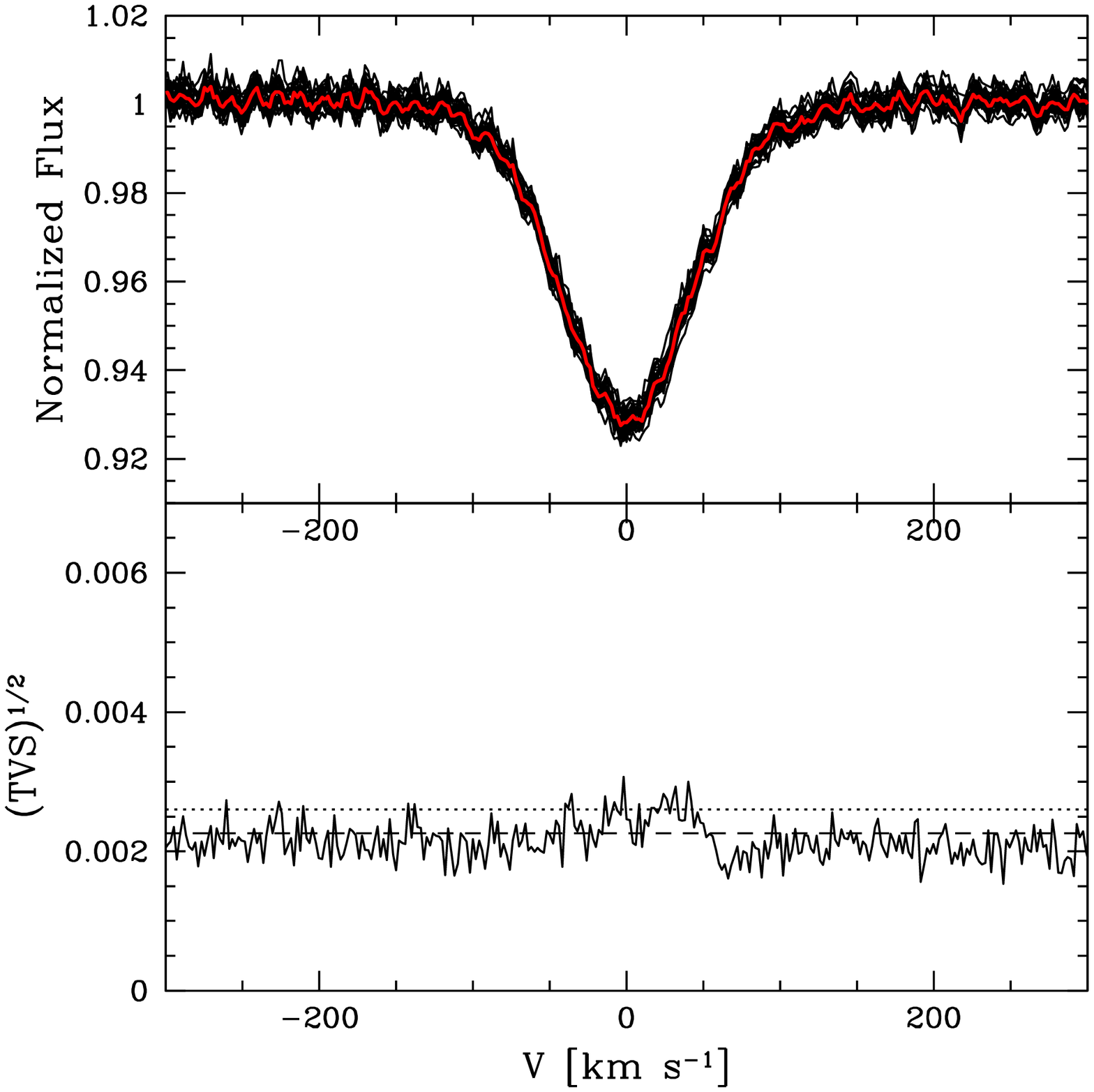}}
     \hspace{0.2cm}
     \subfigure[\ion{C}{IV} 5802]{
          \includegraphics[width=.28\textwidth]{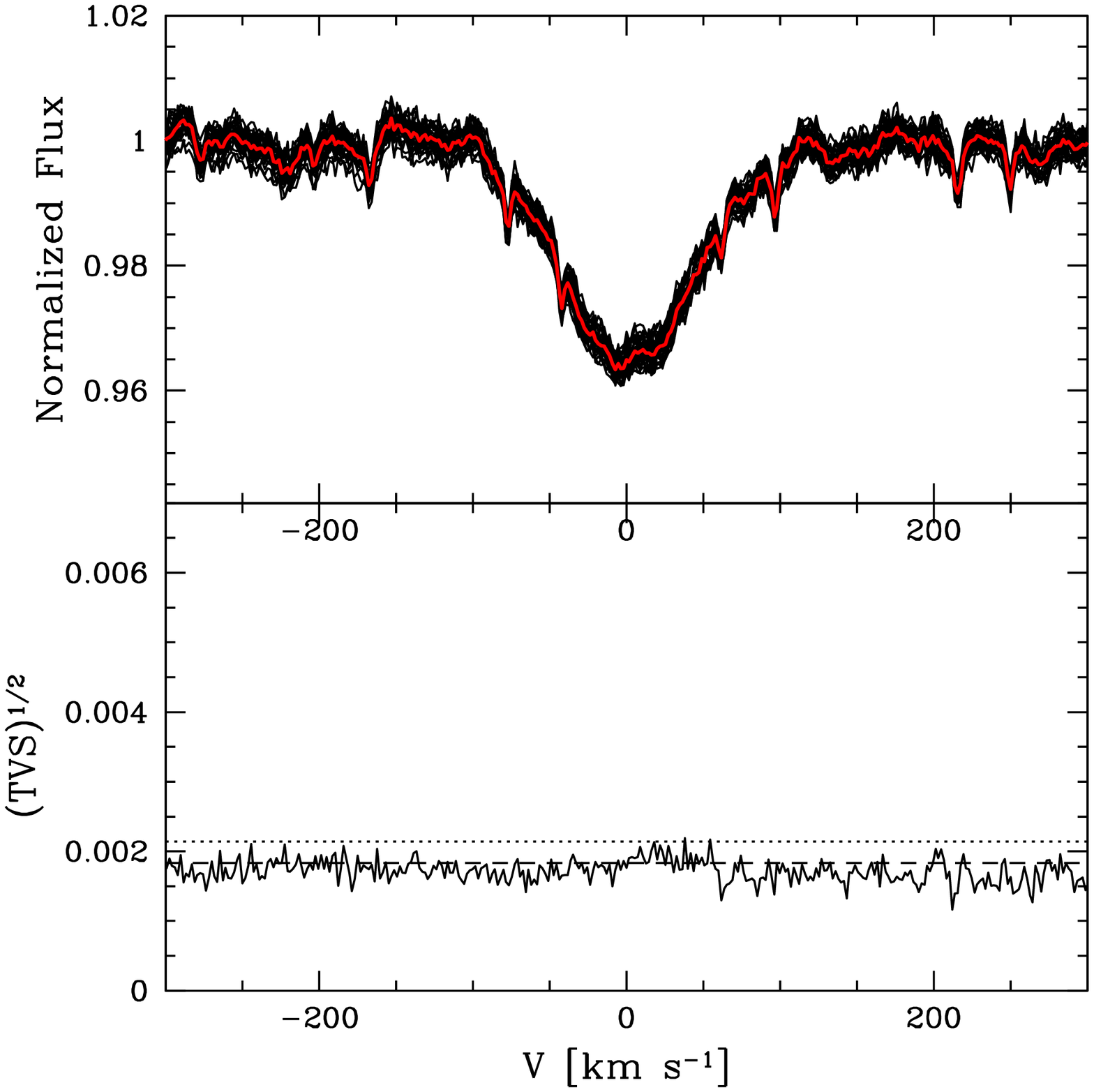}}\\
     \caption{Variability of $\epsilon$ Ori on the night of 19$^{th}$ to 20$^{th}$ October 2007.}
     \label{fig_var_epsori_day}
\end{figure*}

\newpage

\begin{figure*}
     \centering
     \subfigure[\ha]{
          \includegraphics[width=.28\textwidth]{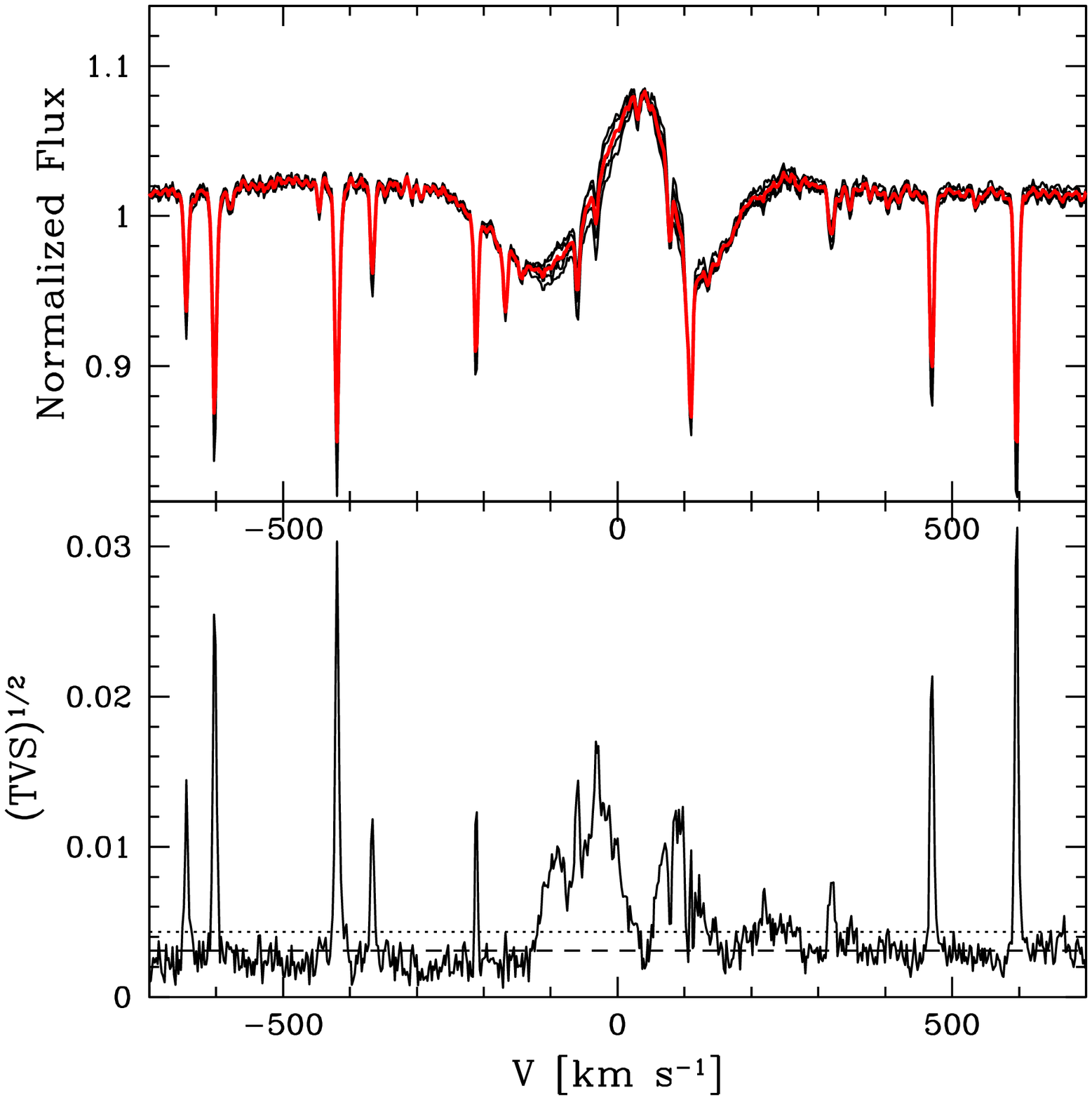}}
     \hspace{0.2cm}
     \subfigure[H$_{\beta}$]{
          \includegraphics[width=.28\textwidth]{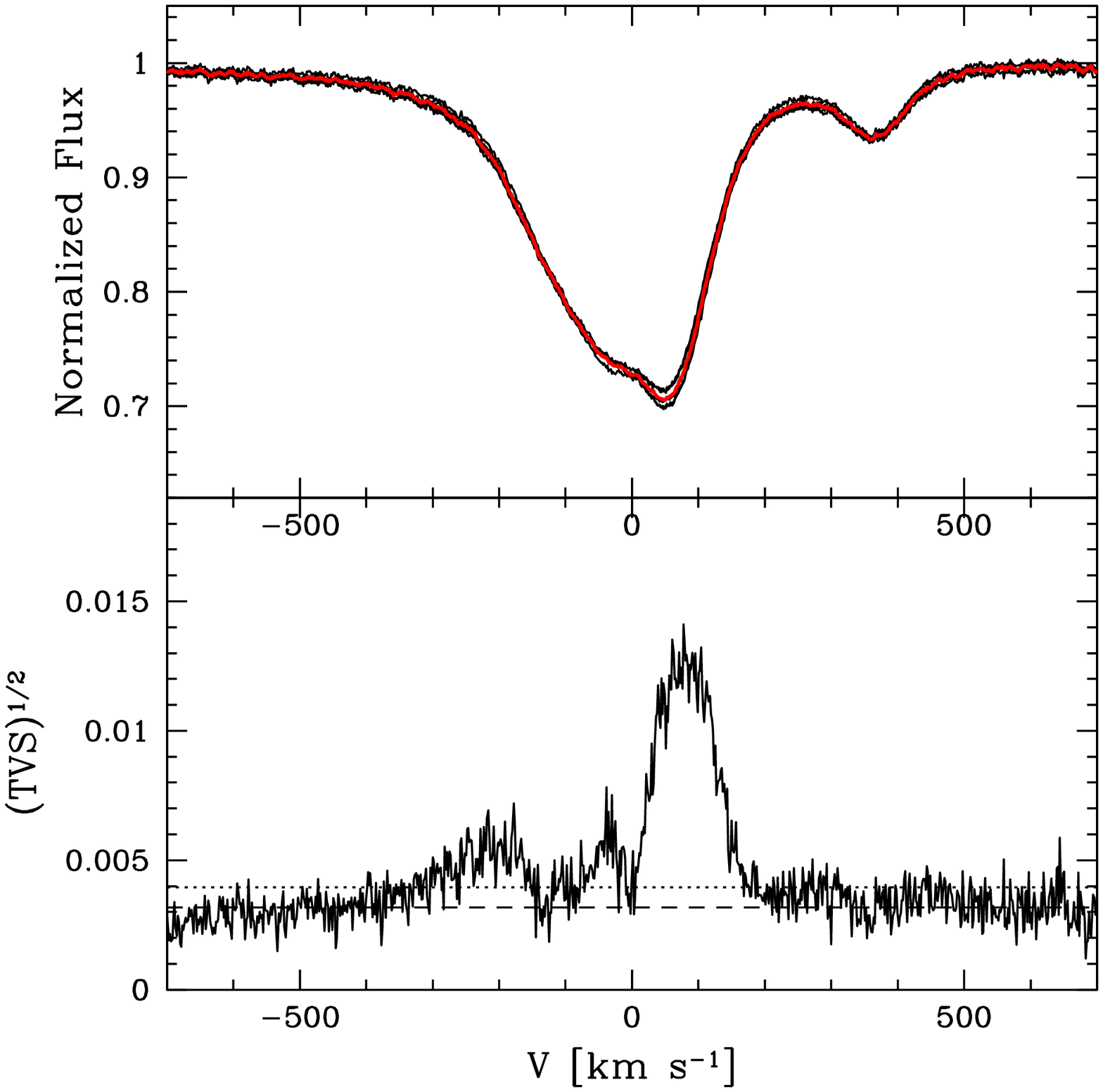}}
     \hspace{0.2cm}
     \subfigure[H$_{\gamma}$]{
          \includegraphics[width=.28\textwidth]{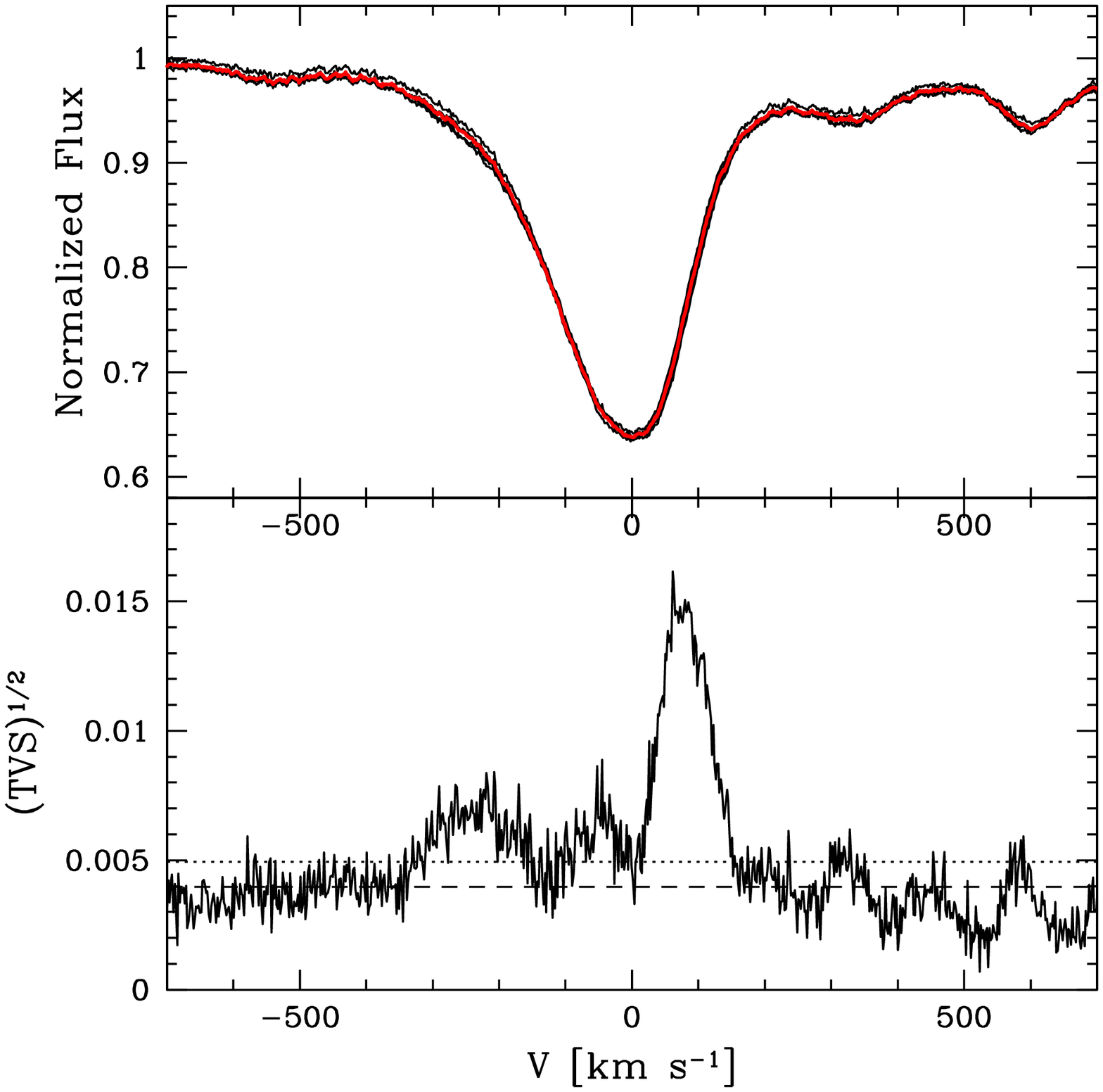}}\\
     \subfigure[\ion{He}{I} 4026]{
          \includegraphics[width=.28\textwidth]{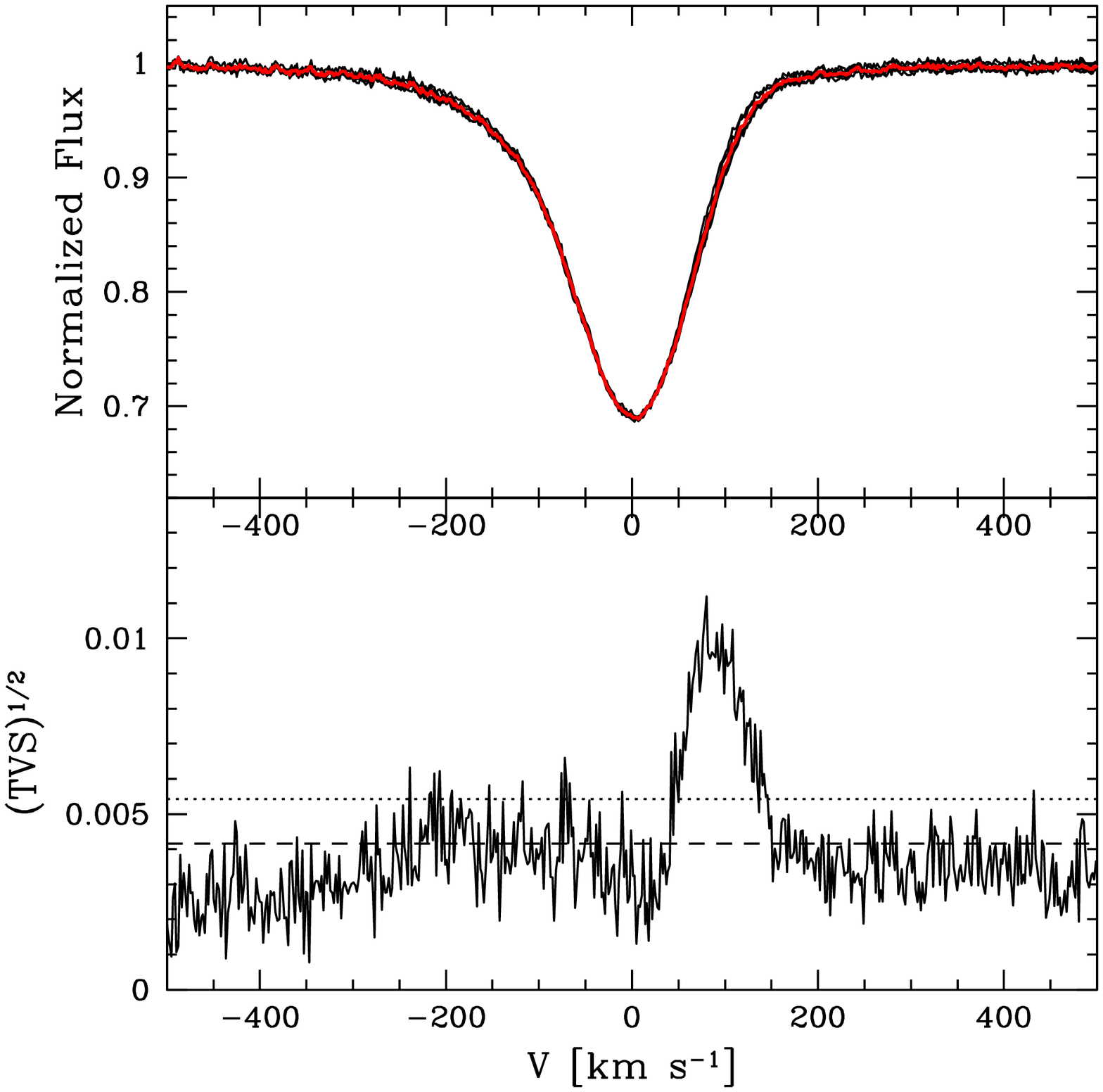}}
     \hspace{0.2cm}
     \subfigure[\ion{He}{I} 4471]{
          \includegraphics[width=.28\textwidth]{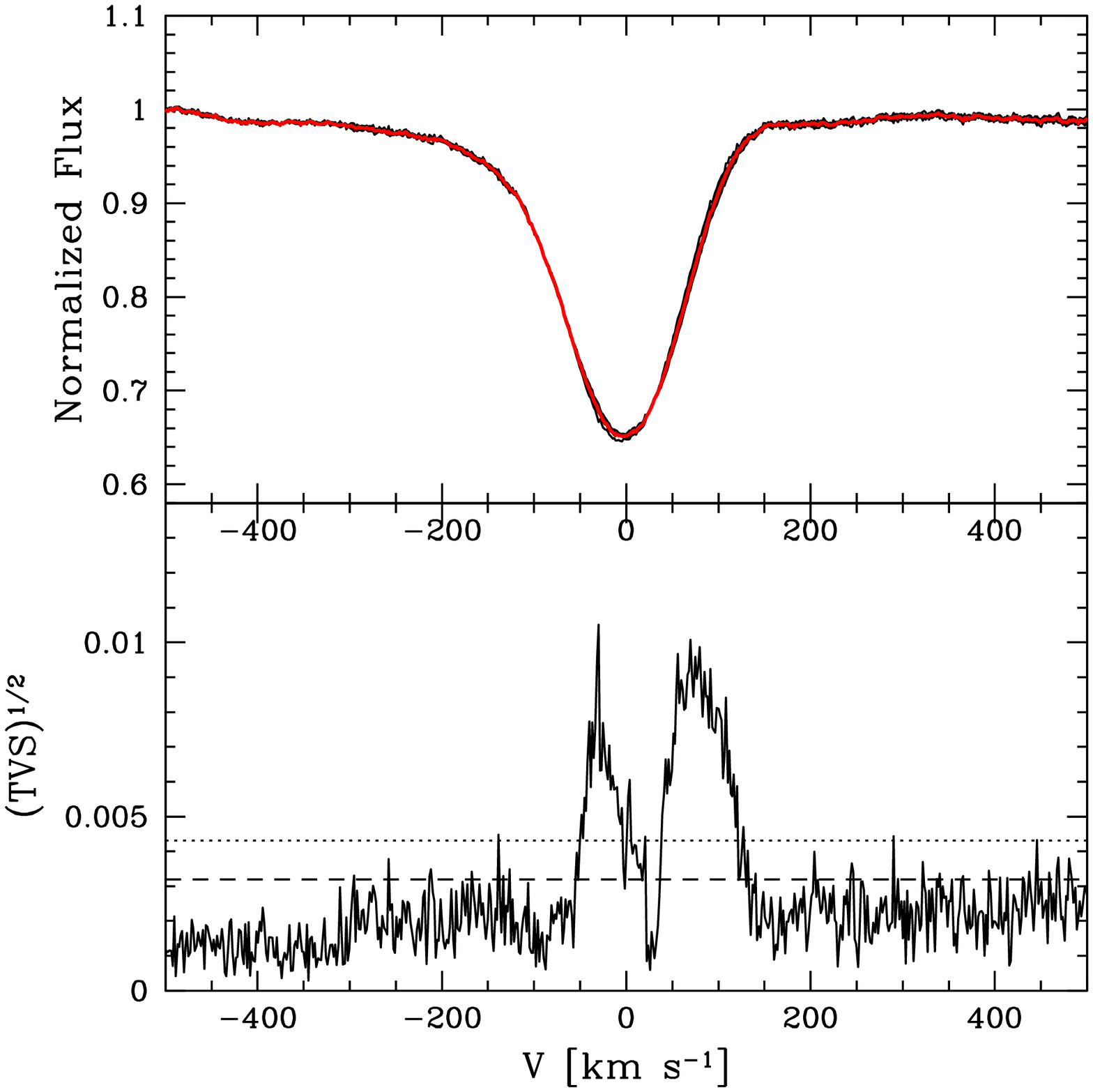}}
     \hspace{0.2cm}
     \subfigure[\ion{He}{I} 4712]{
          \includegraphics[width=.28\textwidth]{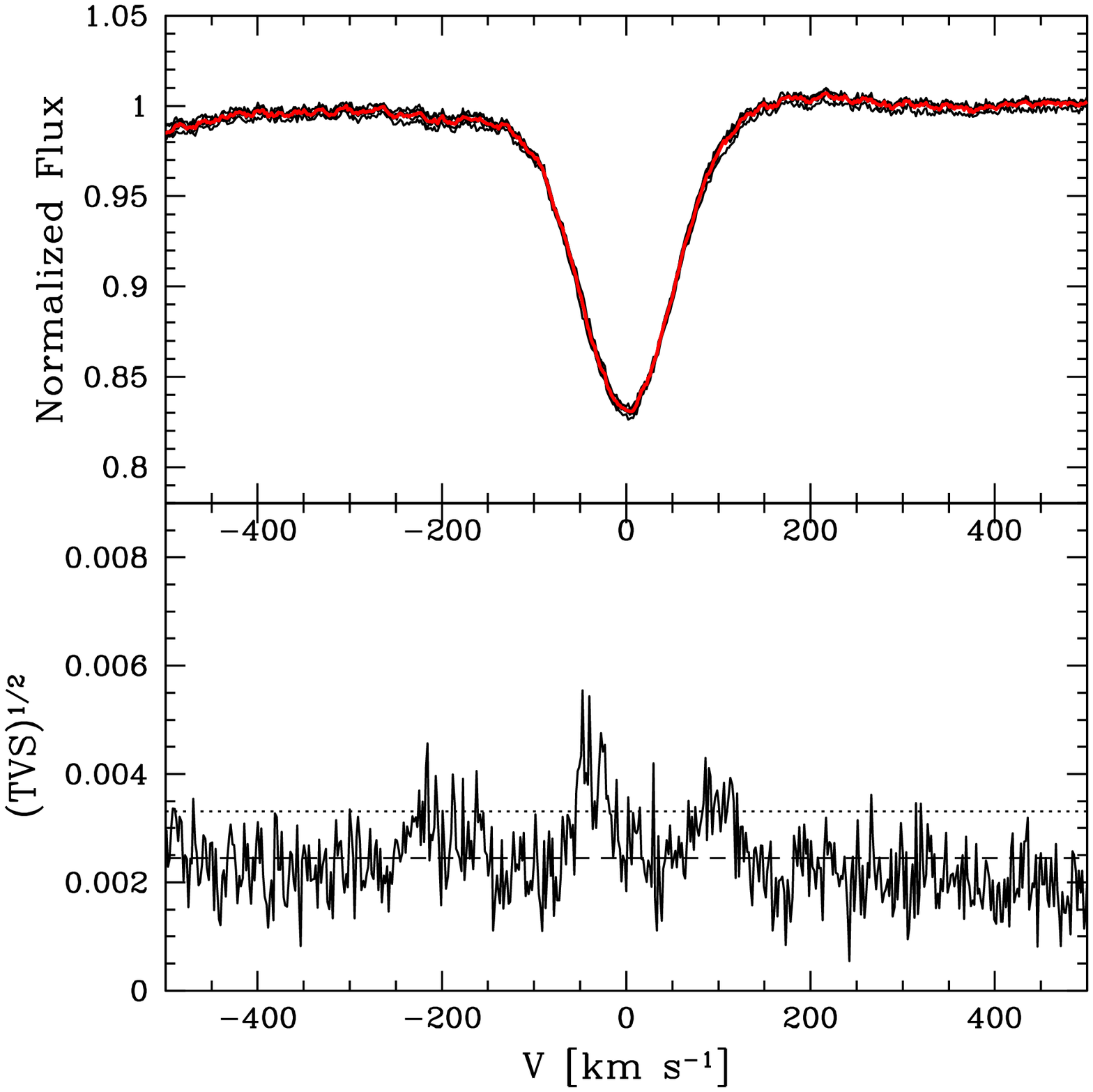}}\\
     \subfigure[\ion{He}{I} 5876]{
          \includegraphics[width=.28\textwidth]{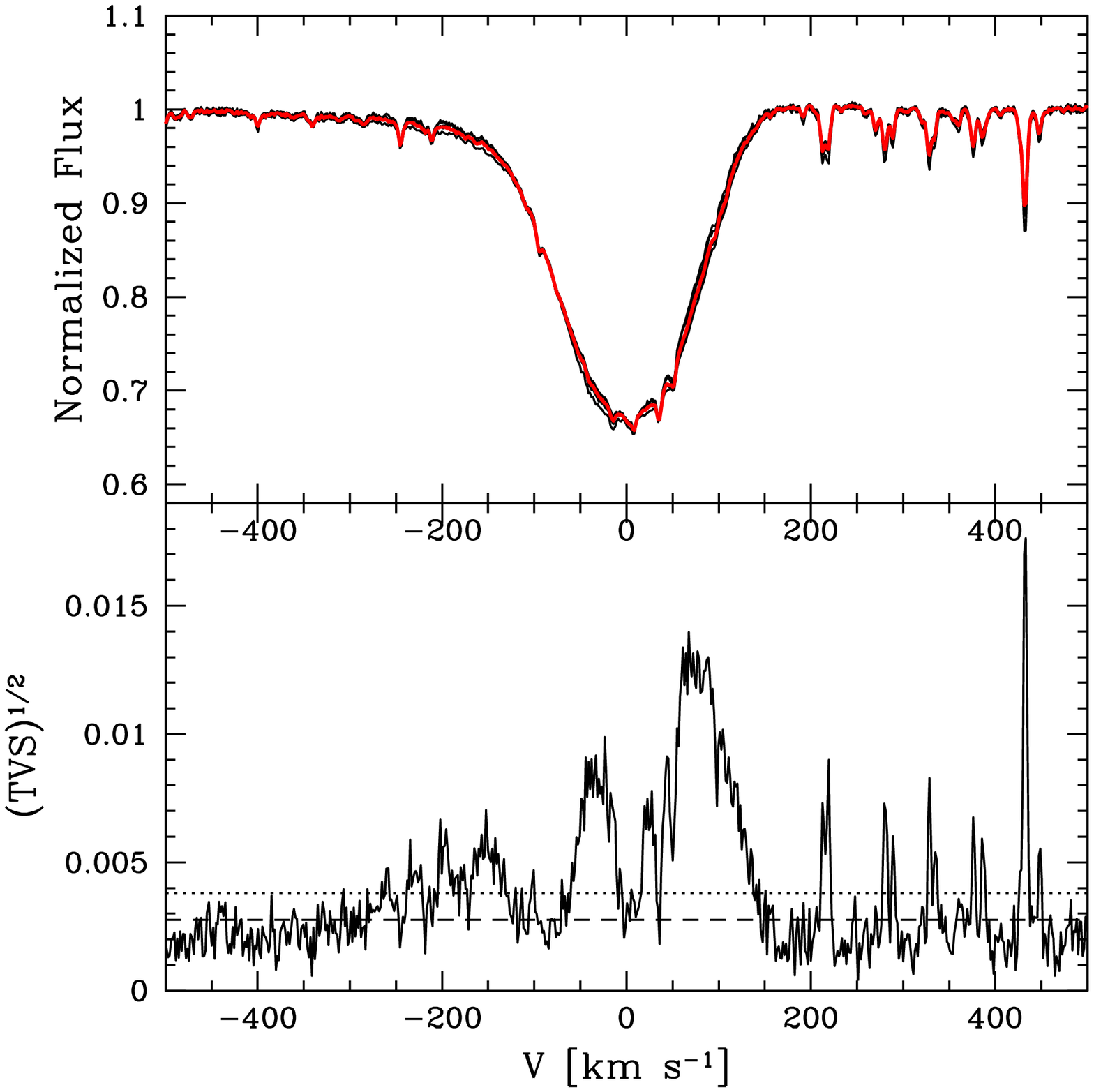}}
     \hspace{0.2cm}
     \subfigure[\ion{He}{II} 4542]{
          \includegraphics[width=.28\textwidth]{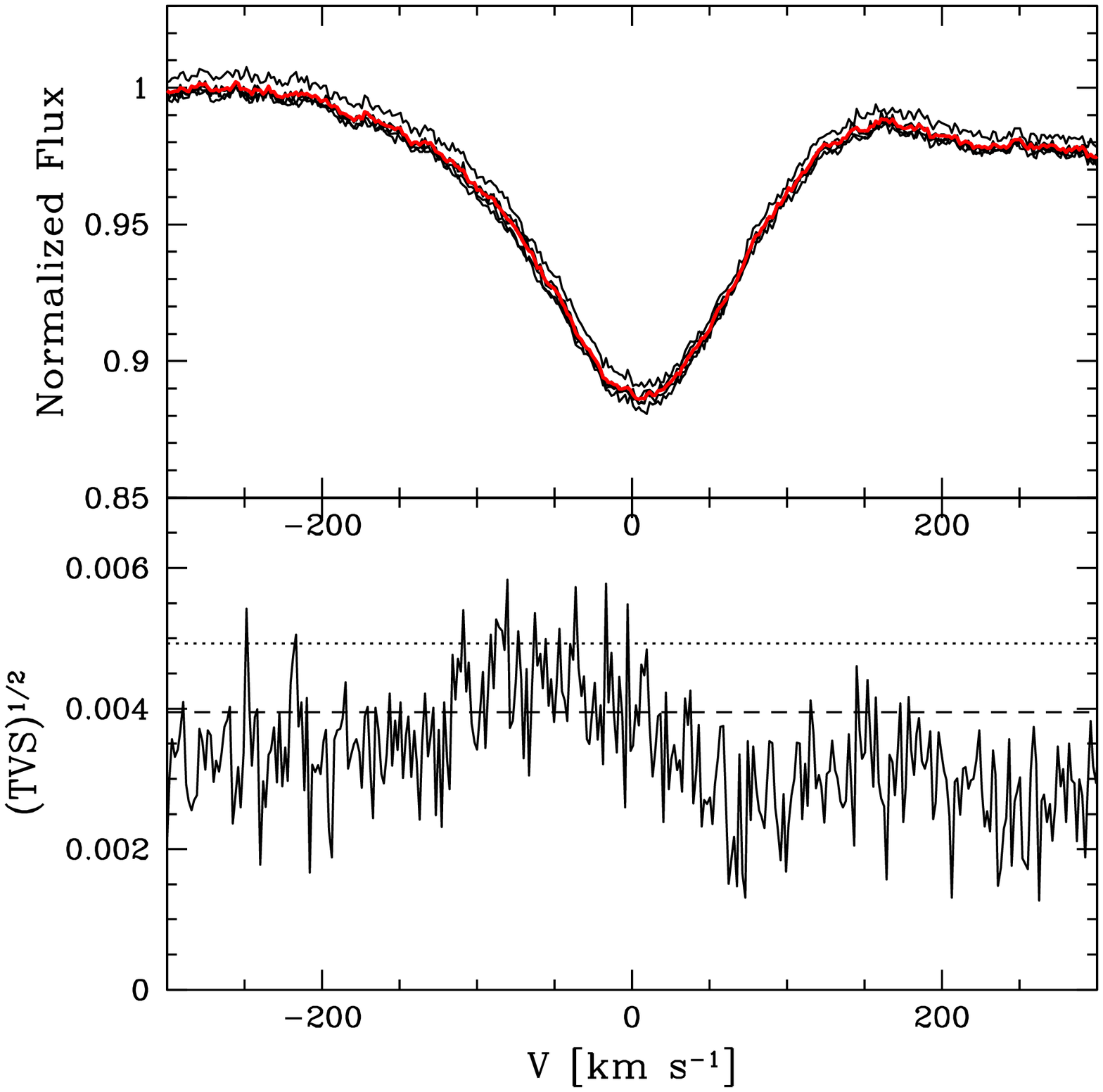}}
     \hspace{0.2cm}
     \subfigure[\ion{He}{II} 4686]{
          \includegraphics[width=.28\textwidth]{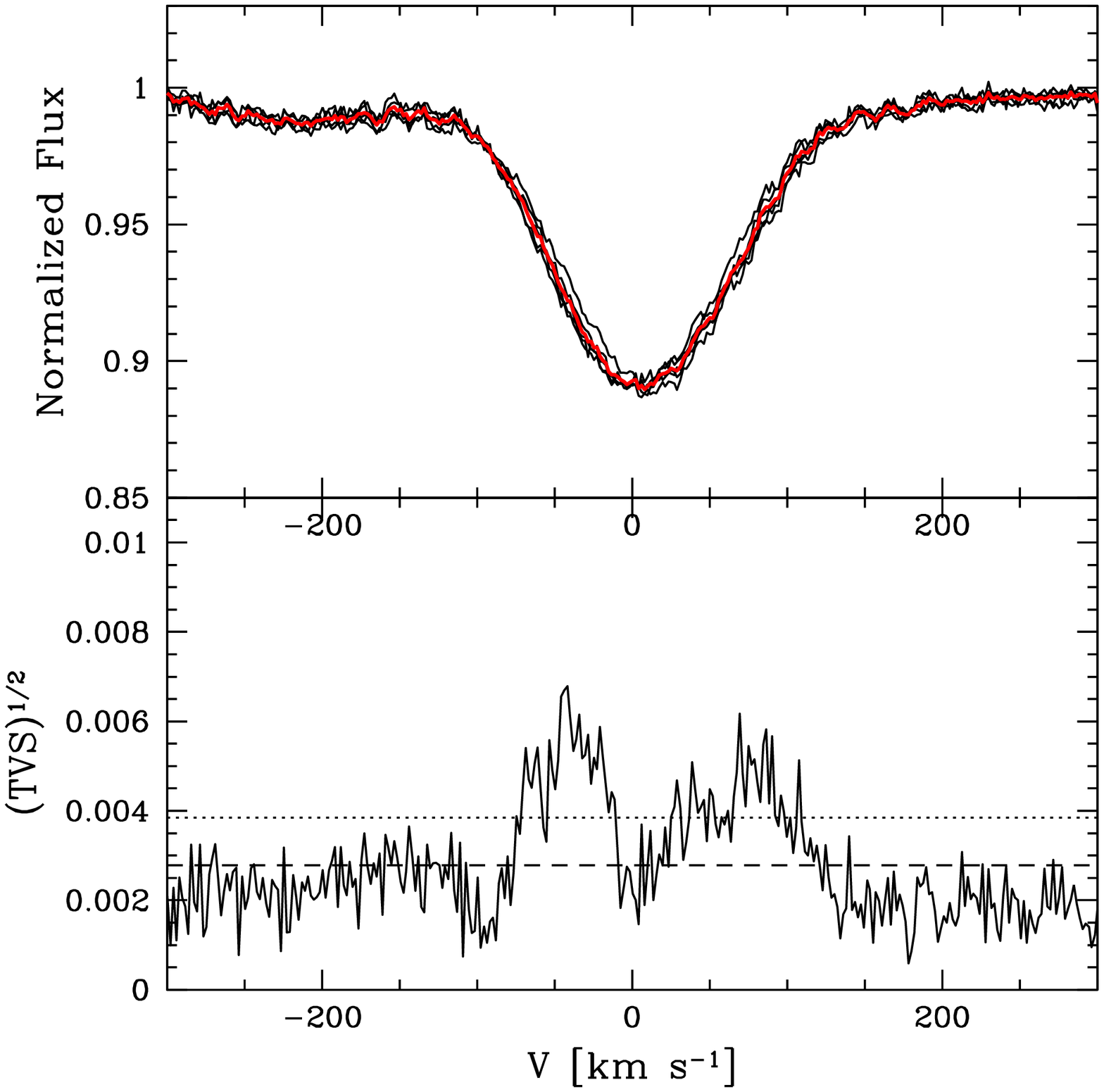}}\\
     \subfigure[\ion{He}{II} 5412]{
          \includegraphics[width=.28\textwidth]{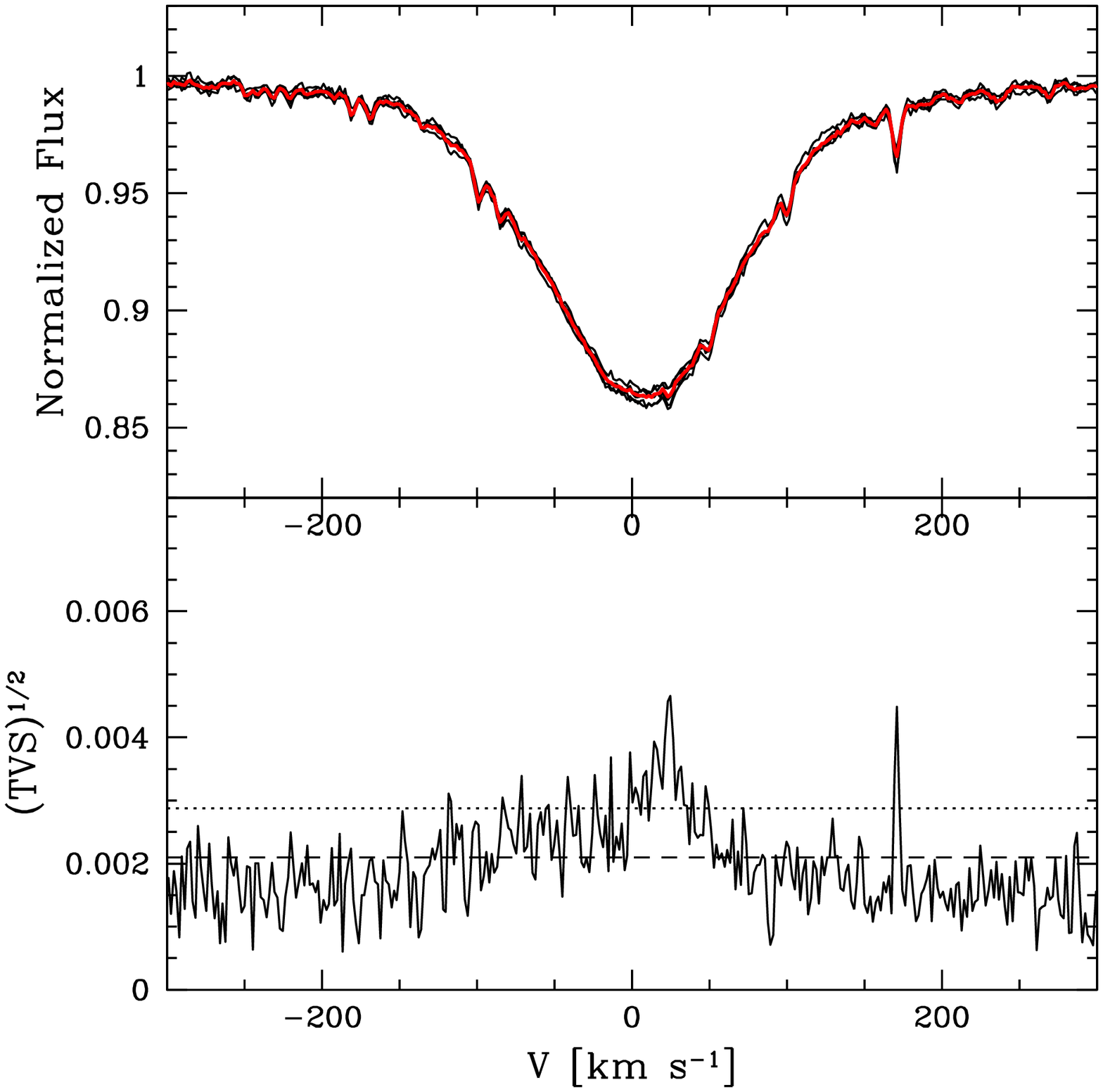}}
     \hspace{0.2cm}
     \subfigure[\ion{O}{III} 5592]{
          \includegraphics[width=.28\textwidth]{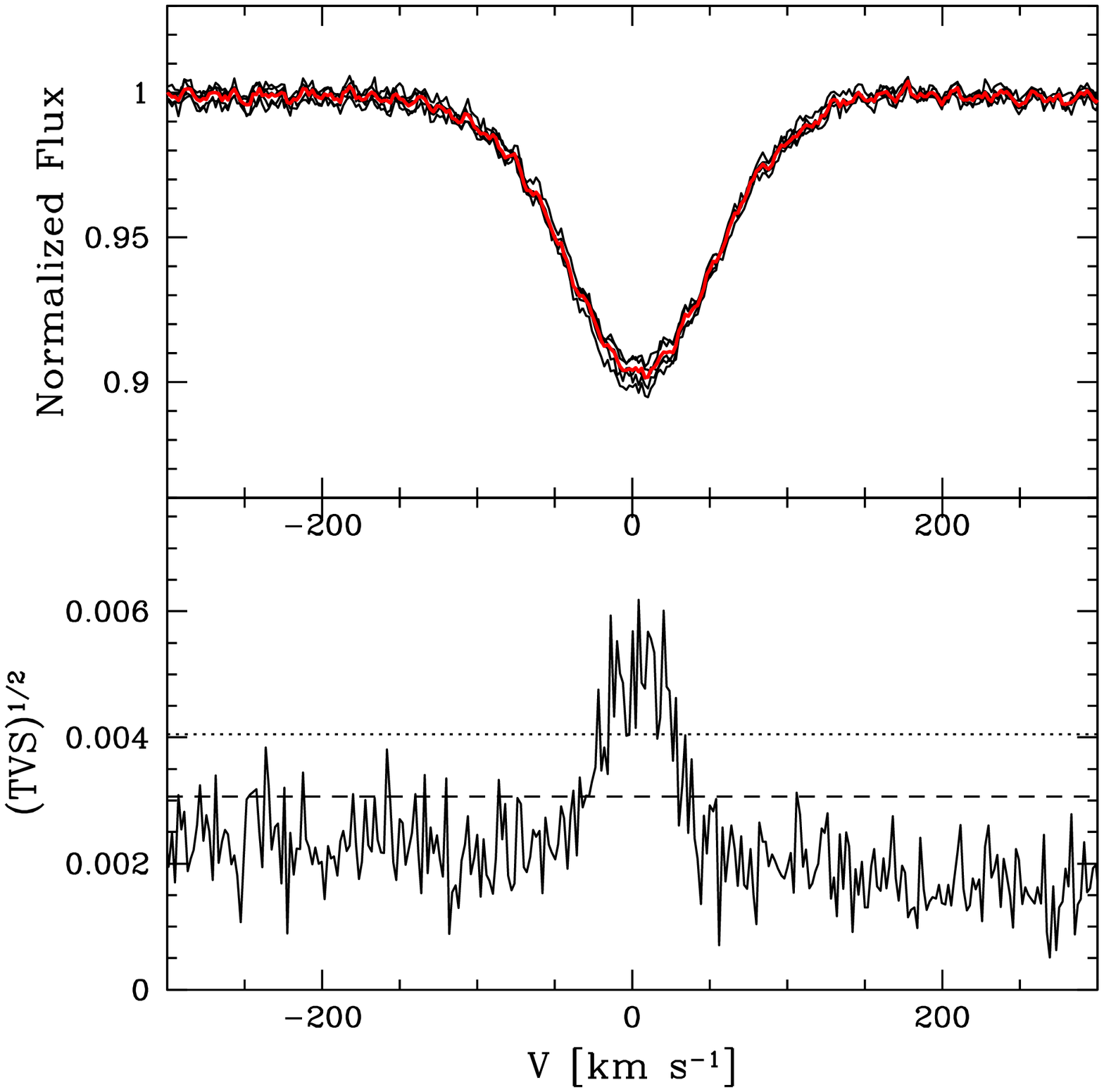}}
     \hspace{0.2cm}
     \subfigure[\ion{C}{IV} 5802]{
          \includegraphics[width=.28\textwidth]{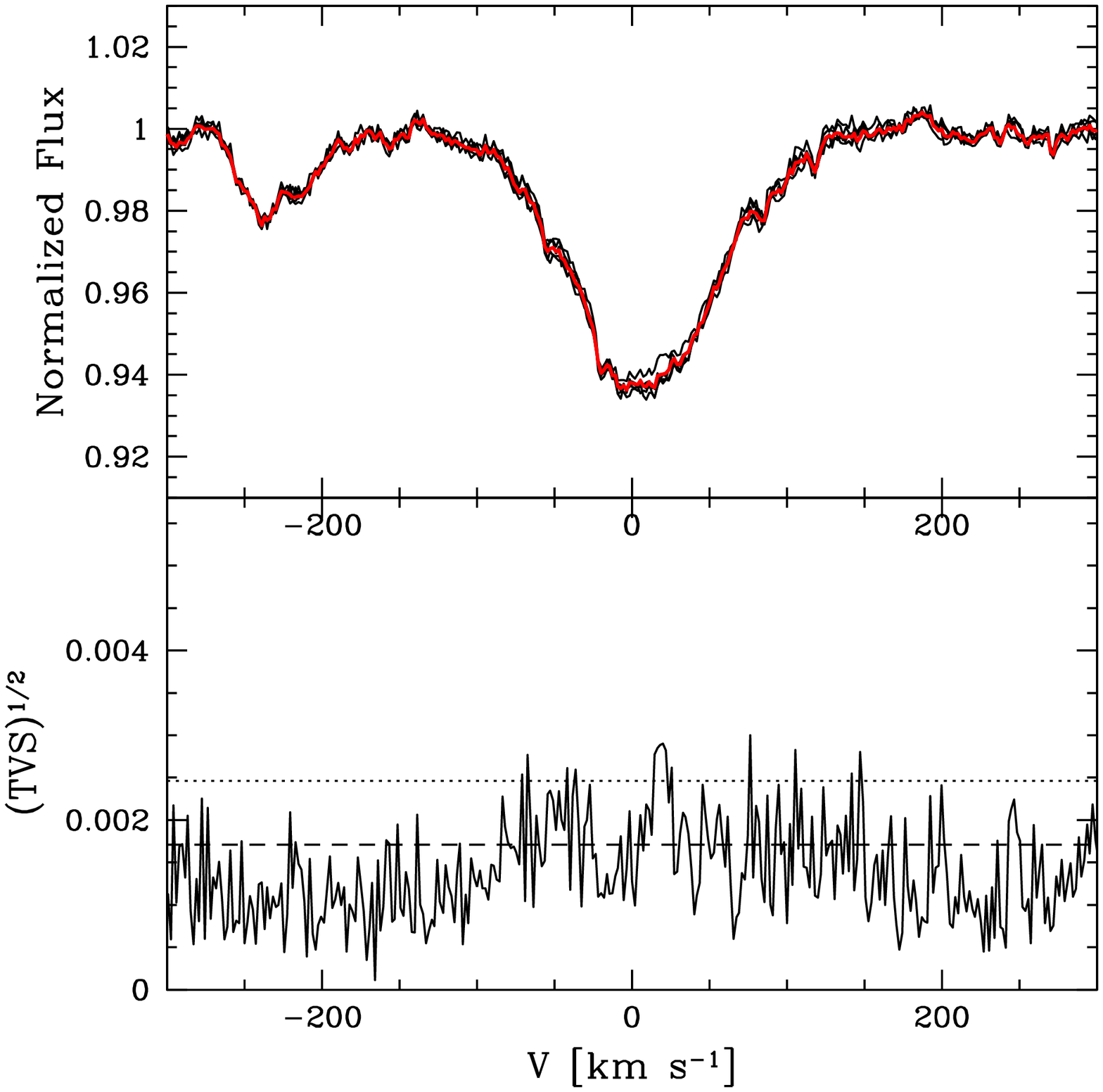}}\\
     \caption{Variability of HD188209 during the night of June 25--26$^{th}$ 2008.}
     \label{fig_var_188209_day}
\end{figure*}

\newpage

\begin{figure*}
     \centering
     \subfigure[\ha]{
          \includegraphics[width=.28\textwidth]{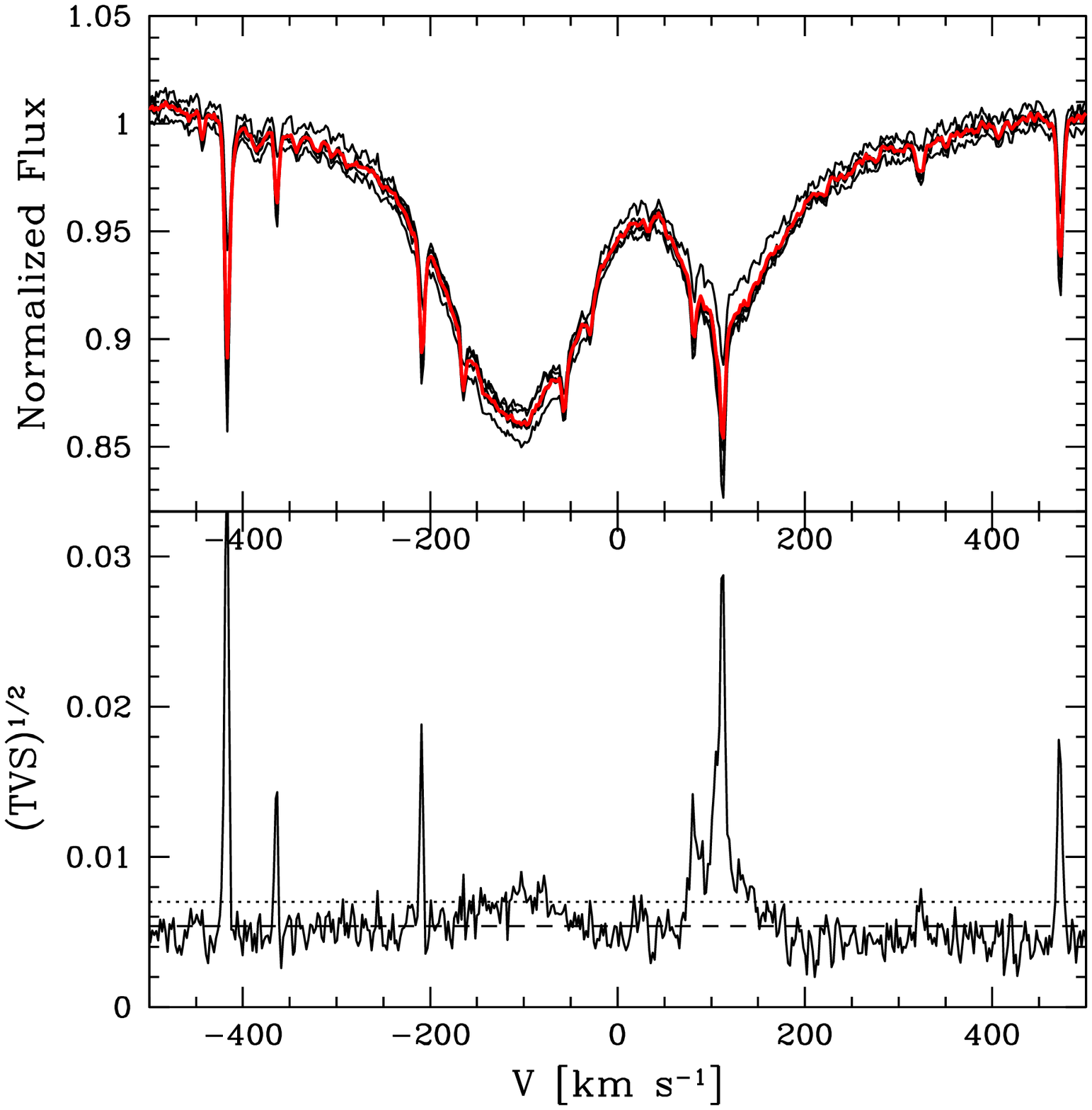}}
     \hspace{0.2cm}
     \subfigure[H$_{\beta}$]{
          \includegraphics[width=.28\textwidth]{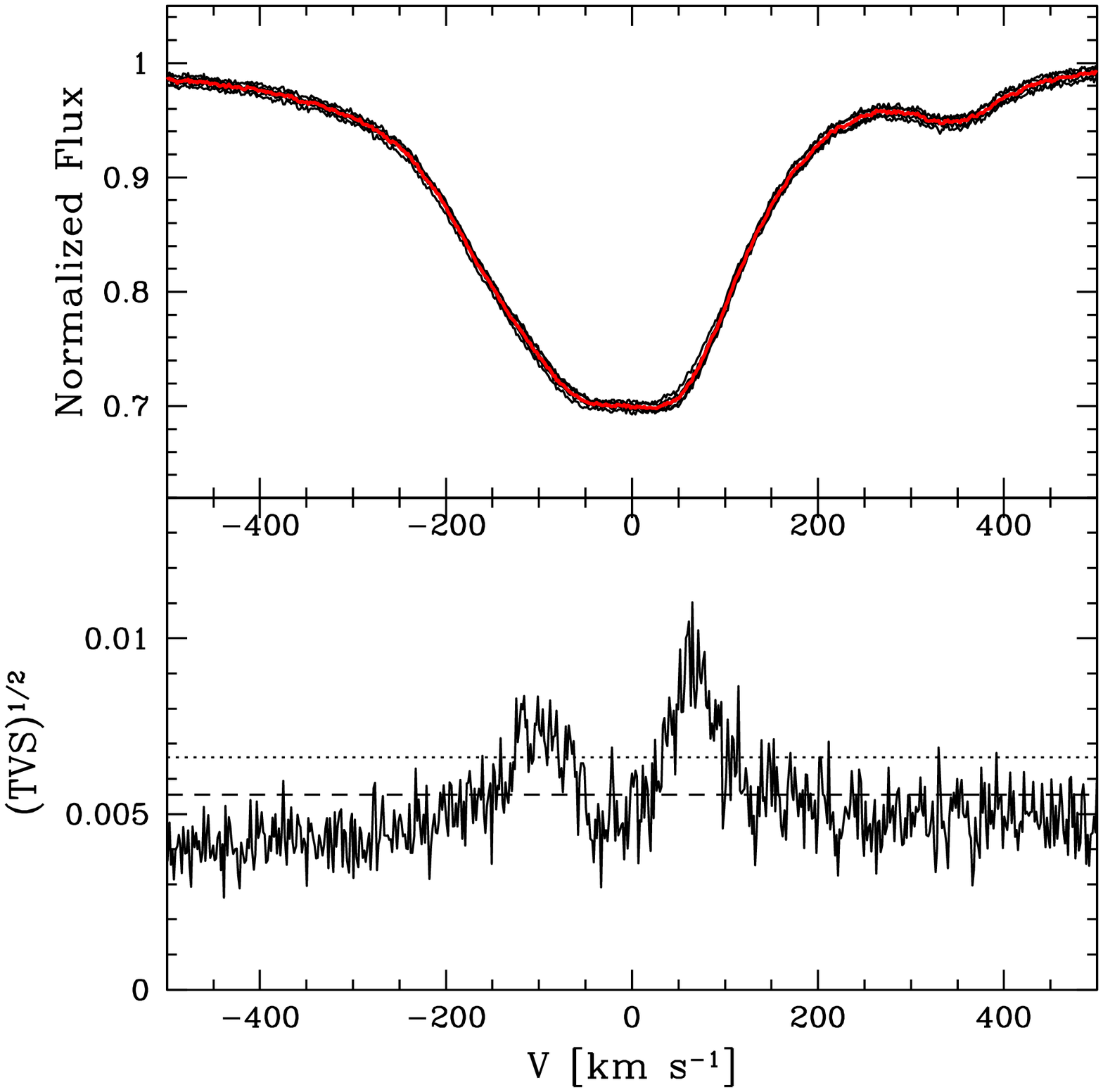}}
     \hspace{0.2cm}
     \subfigure[H$_{\gamma}$]{
          \includegraphics[width=.28\textwidth]{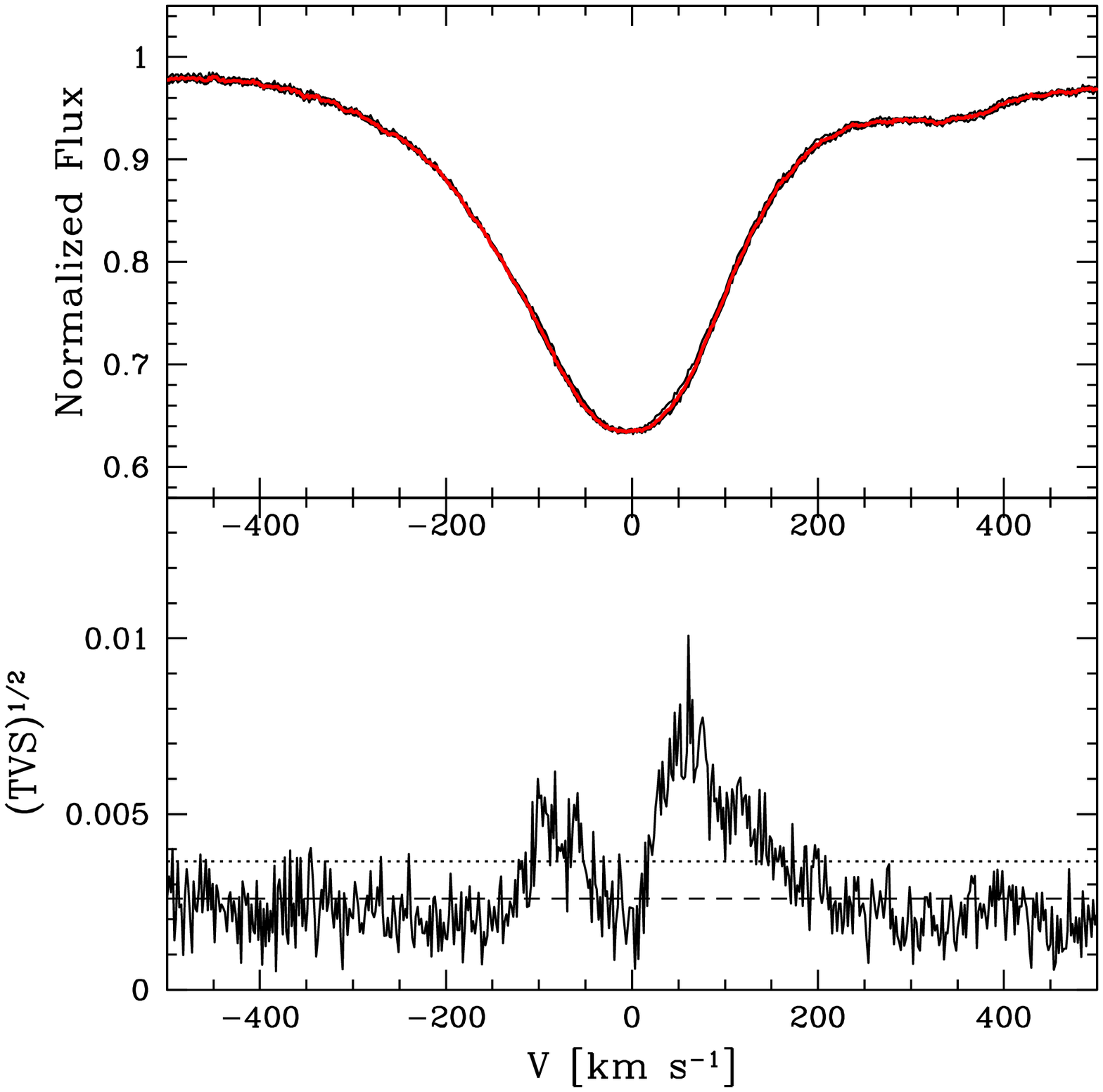}}\\
     \subfigure[\ion{He}{I} 4026]{
          \includegraphics[width=.28\textwidth]{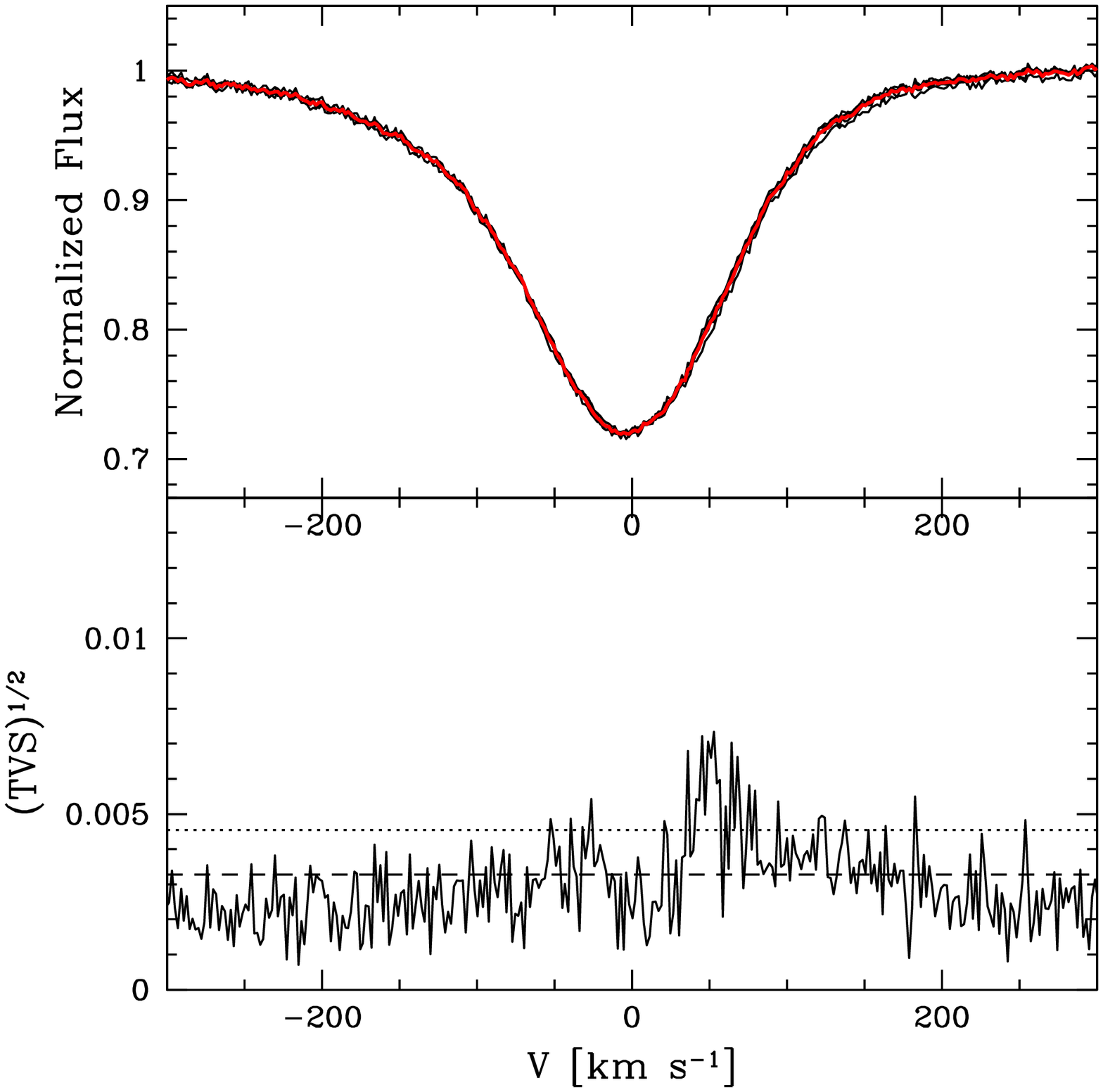}}
     \hspace{0.2cm}
     \subfigure[\ion{He}{I} 4471]{
          \includegraphics[width=.28\textwidth]{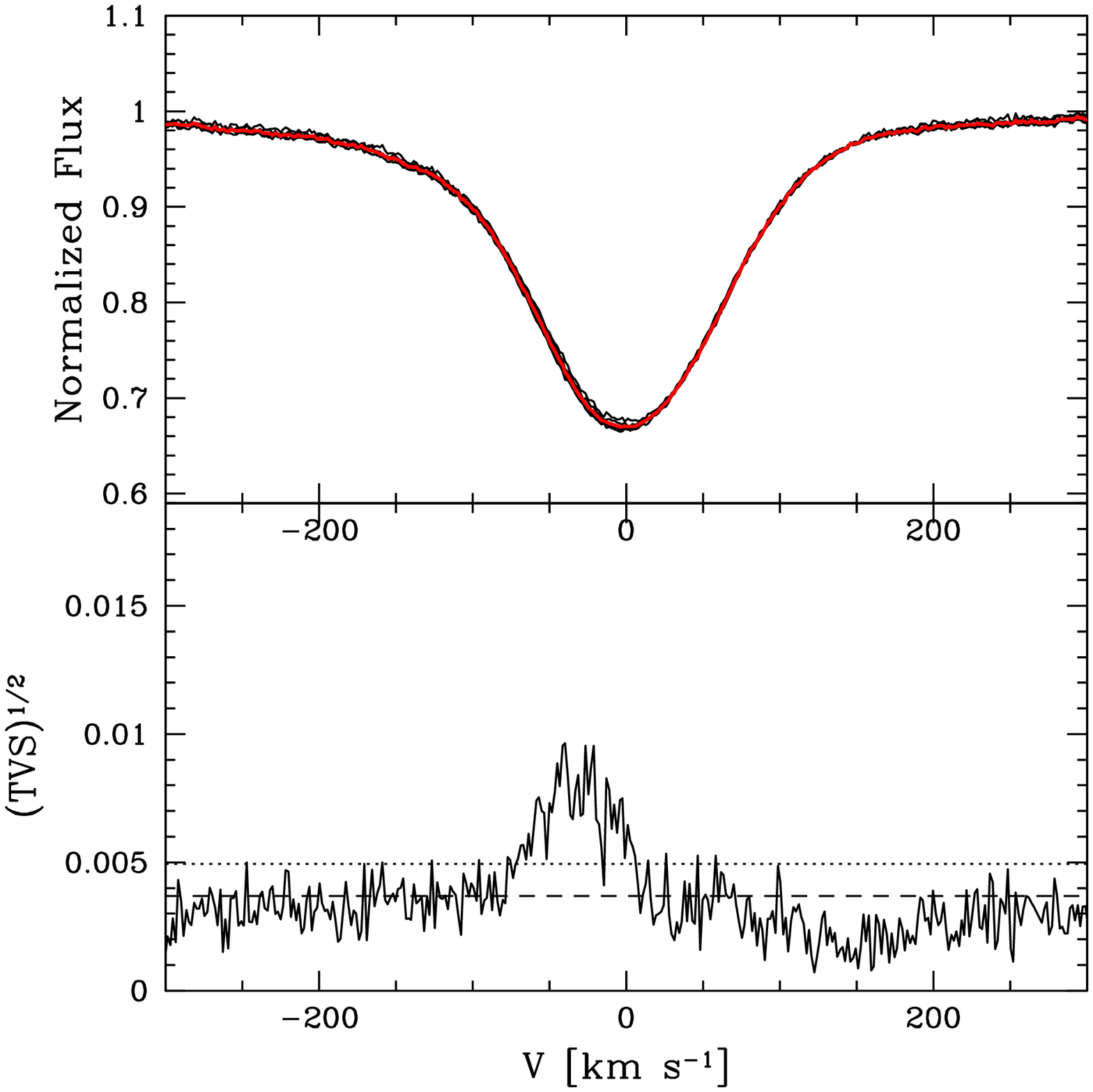}}
     \hspace{0.2cm}
     \subfigure[\ion{He}{I} 4712]{
          \includegraphics[width=.28\textwidth]{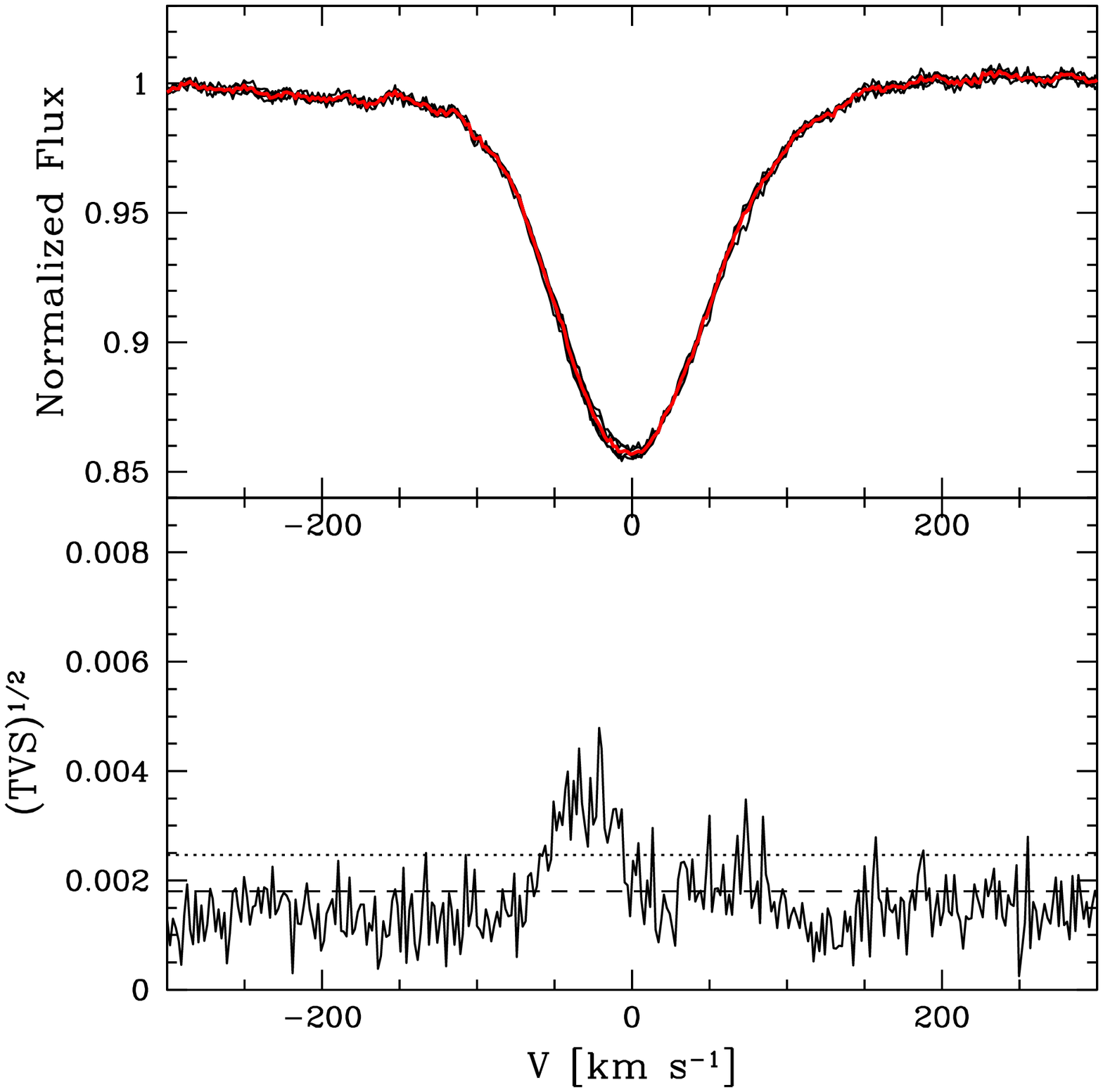}}\\
     \subfigure[\ion{He}{I} 5876]{
          \includegraphics[width=.28\textwidth]{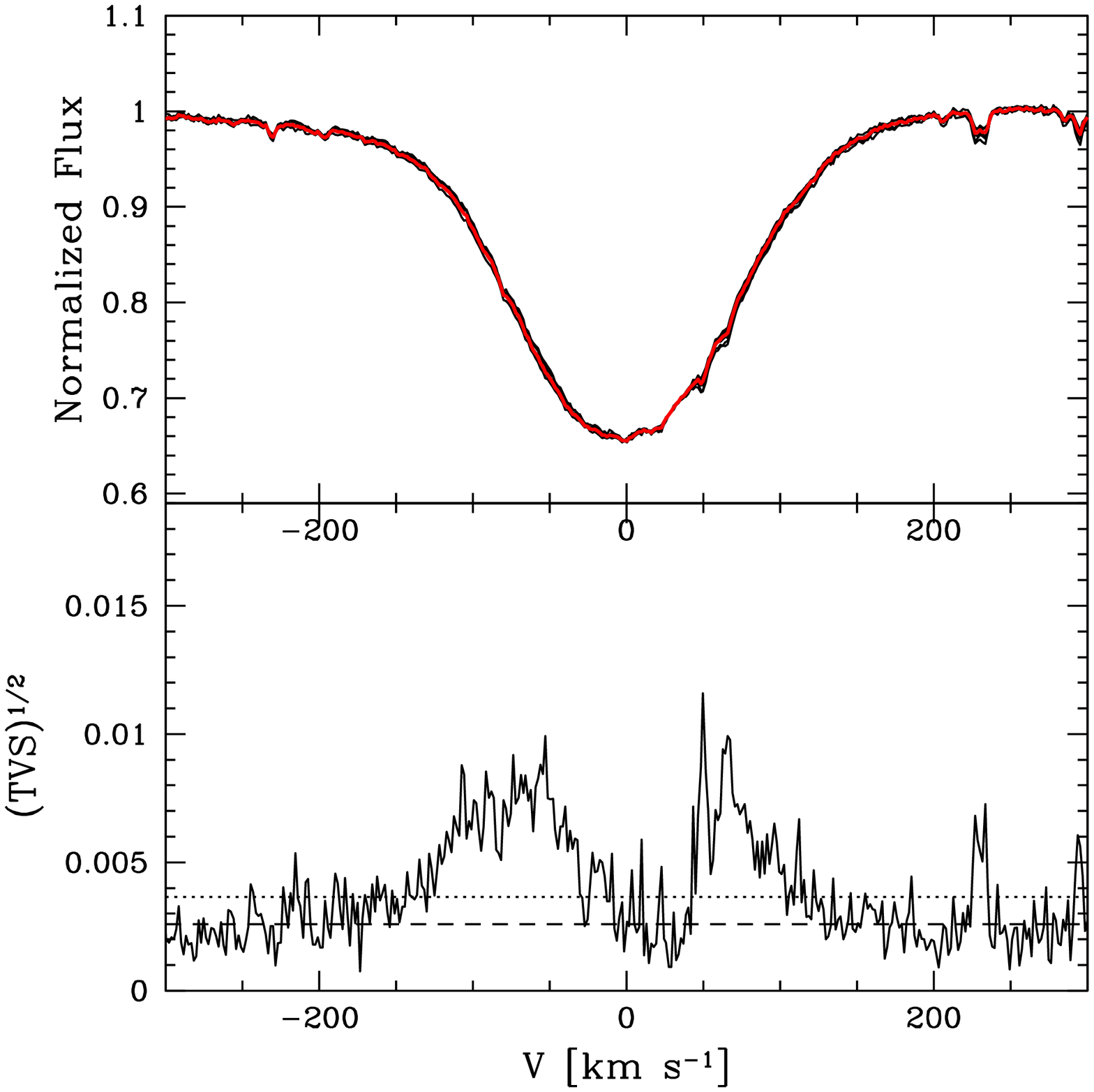}}
     \hspace{0.2cm}
     \subfigure[\ion{He}{II} 4542]{
          \includegraphics[width=.28\textwidth]{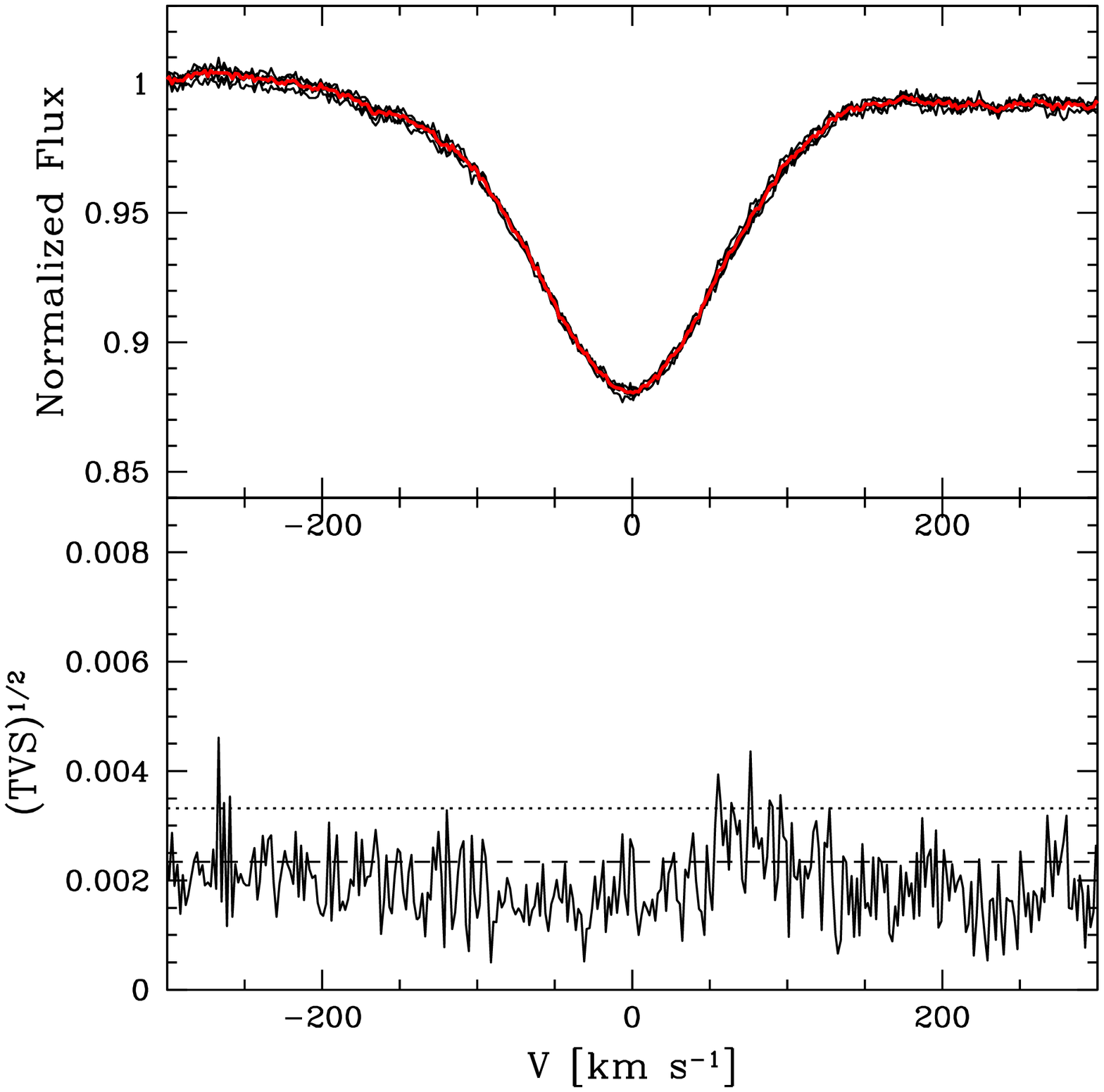}}
     \hspace{0.2cm}
     \subfigure[\ion{He}{II} 4686]{
          \includegraphics[width=.28\textwidth]{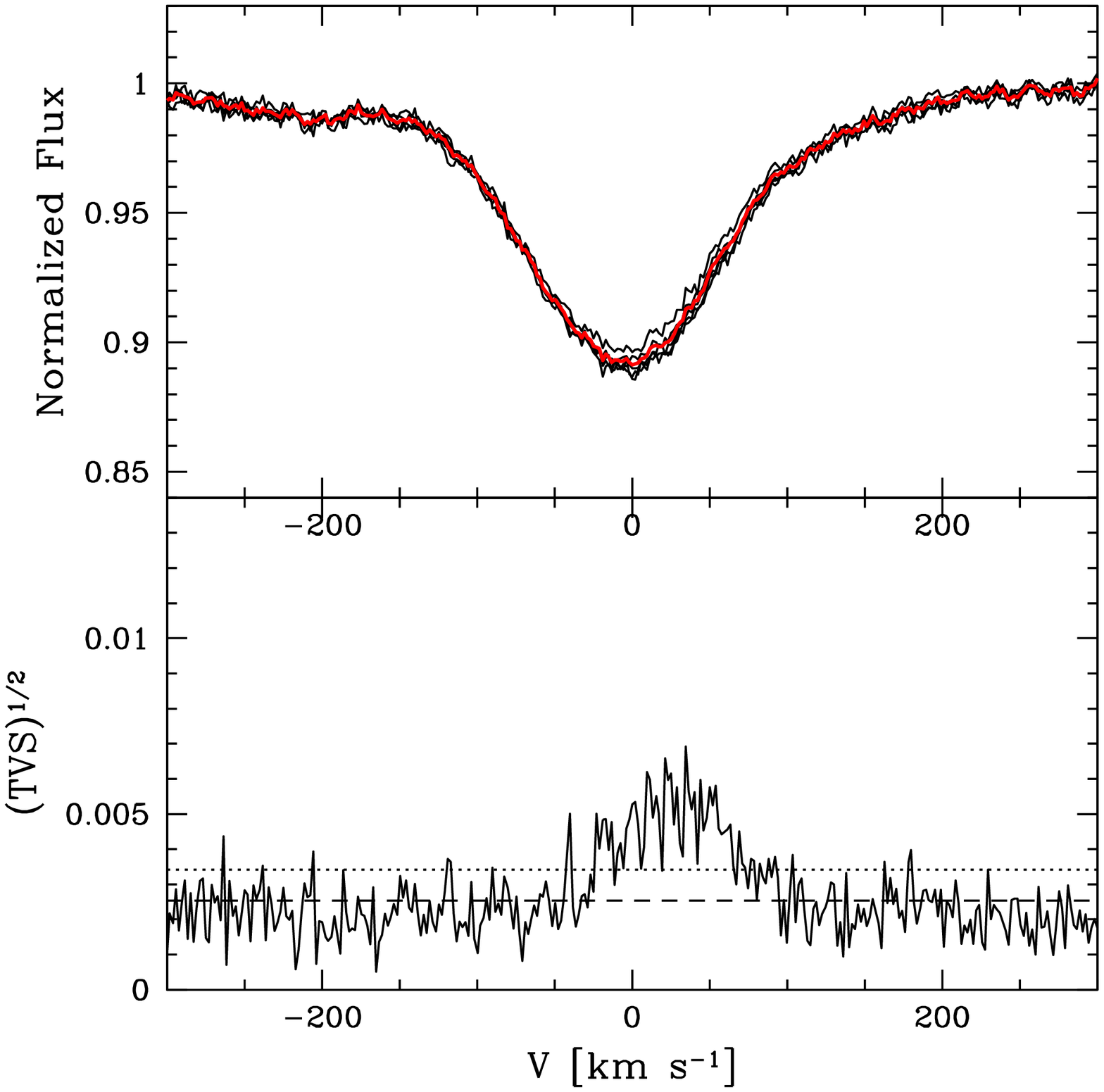}}\\
     \subfigure[\ion{He}{II} 5412]{
          \includegraphics[width=.28\textwidth]{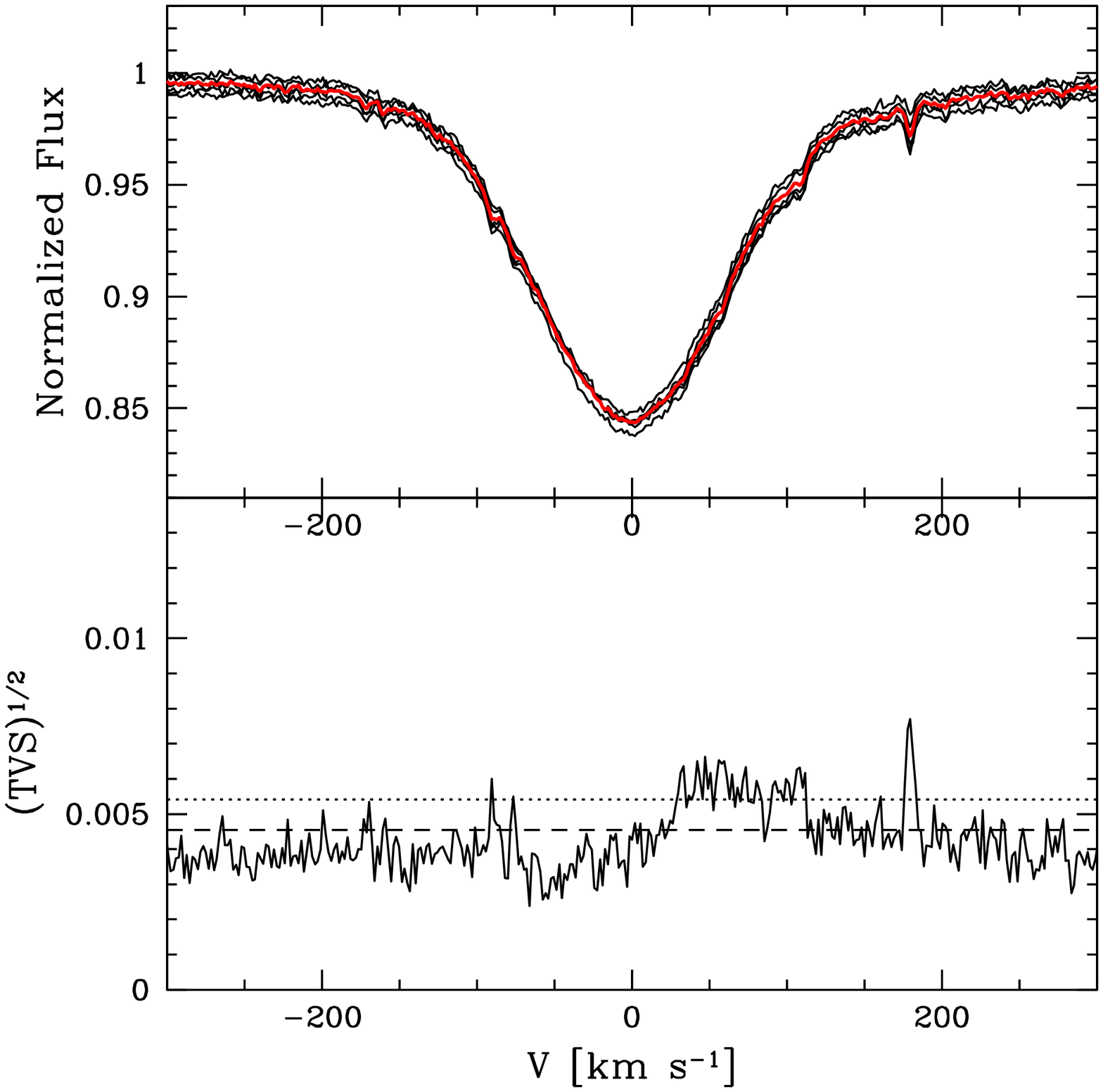}}
     \hspace{0.2cm}
     \subfigure[\ion{O}{III} 5592]{
          \includegraphics[width=.28\textwidth]{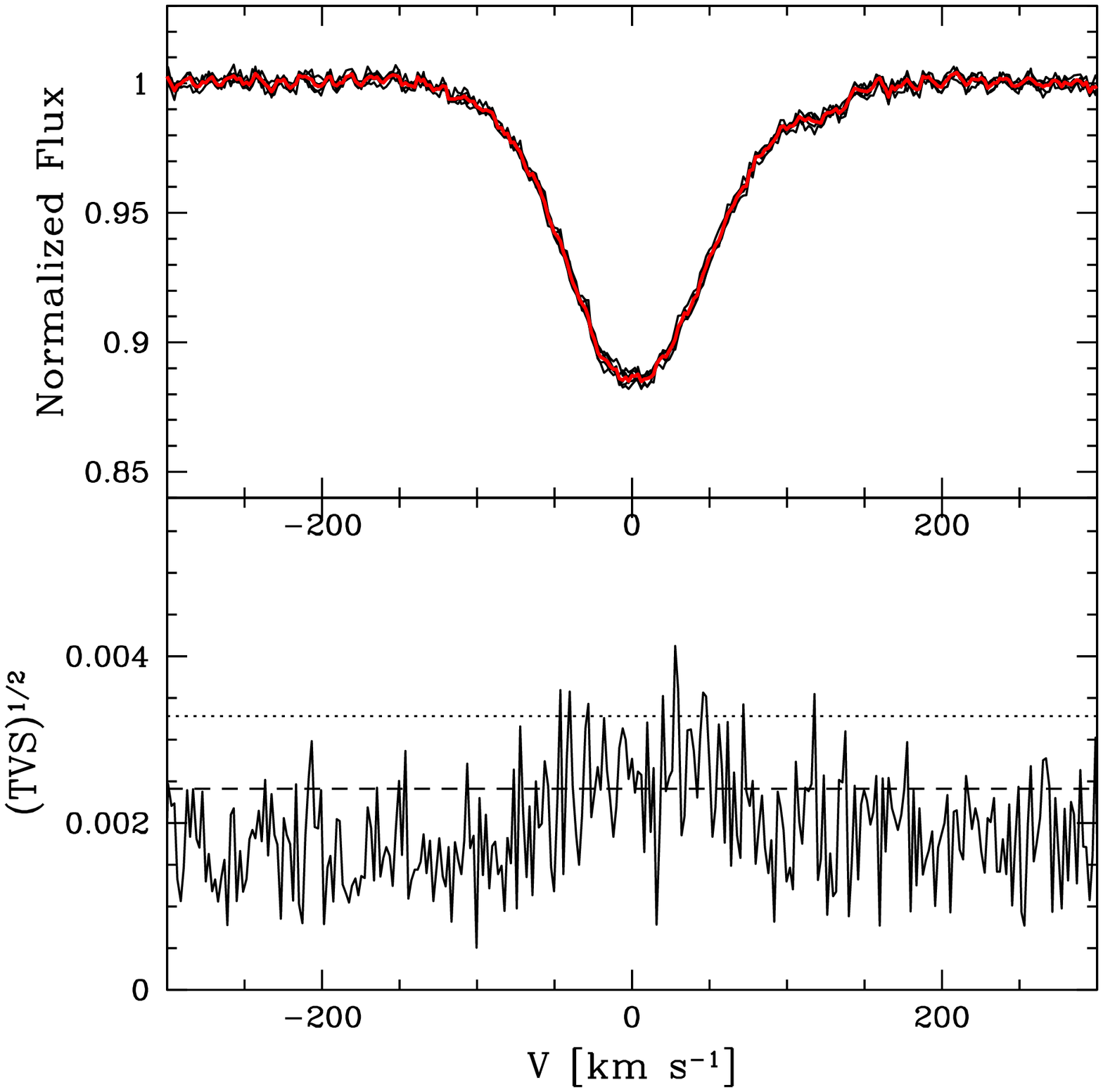}}
     \hspace{0.2cm}
     \subfigure[\ion{C}{IV} 5802]{
          \includegraphics[width=.28\textwidth]{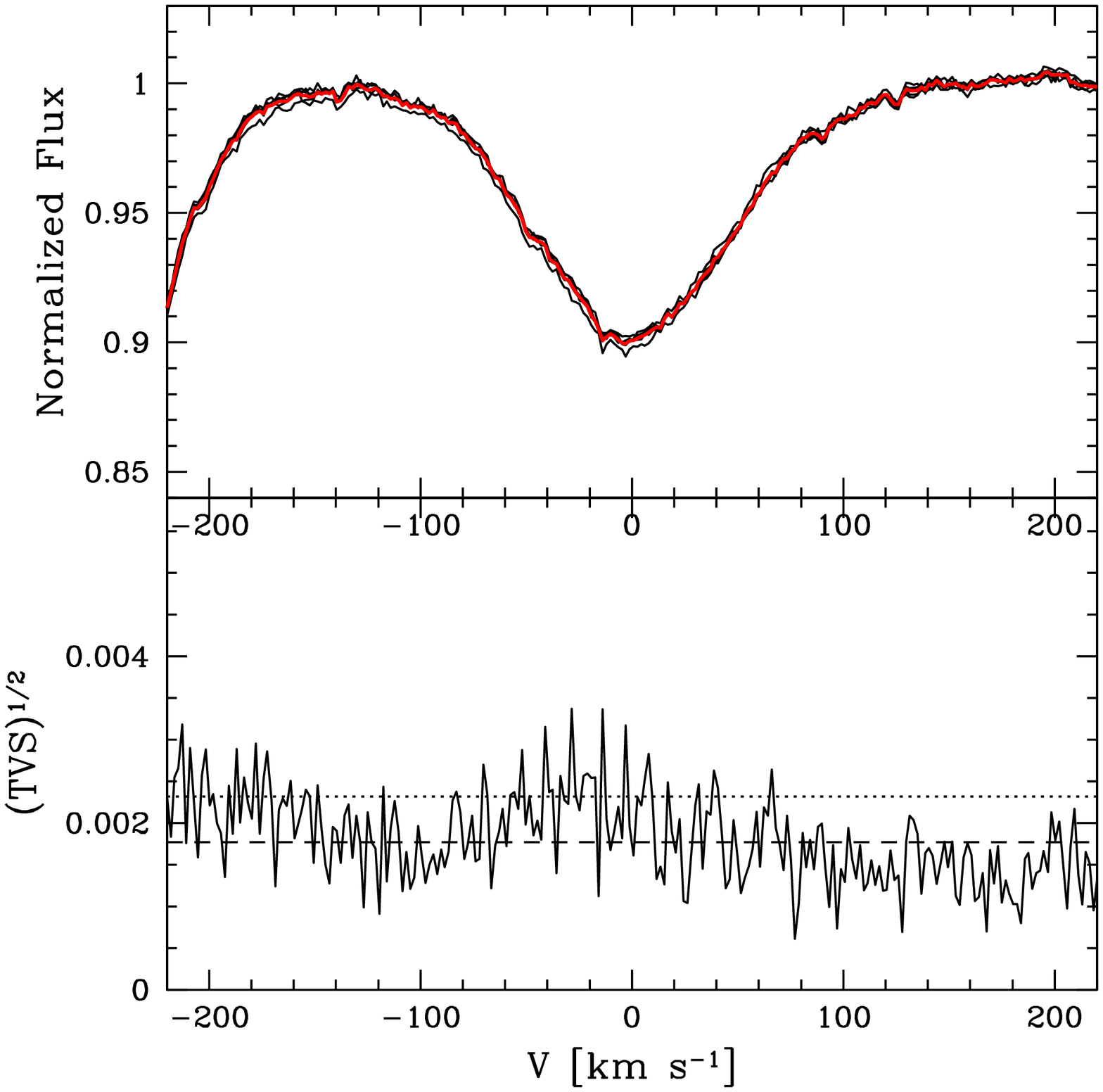}}\\
     \caption{Variability of HD~209975 during the night of June 22$^{nd}$ to 23$^{rd}$ 2008.}
     \label{fig_var_209975_day}
\end{figure*}

\newpage

\section{Best fits to the UV/optical spectra of the sample stars}

\begin{figure*}
\centering
\includegraphics[width=18cm]{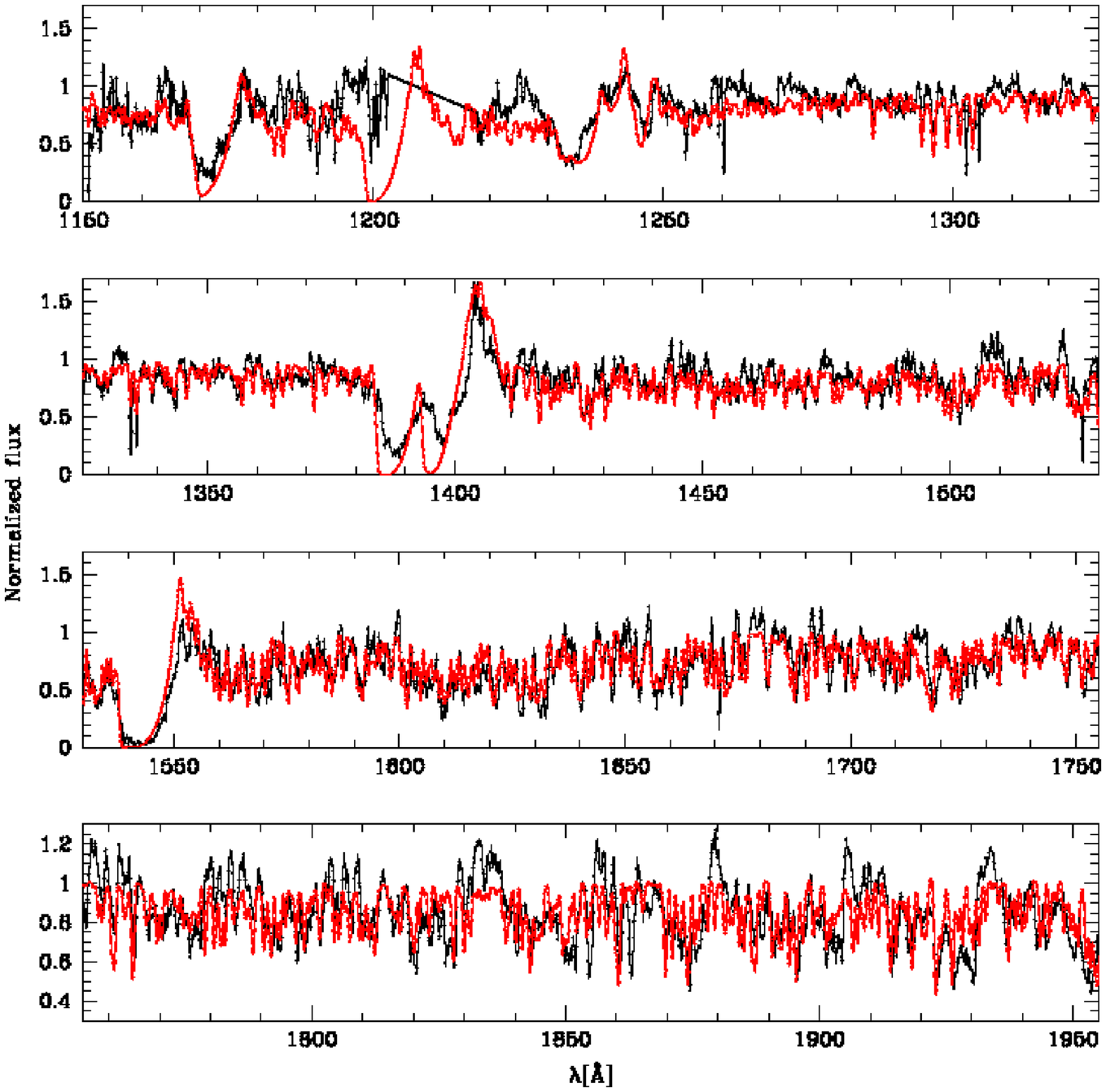}
\caption{Best CMFGEN fit (red solid line) of the UV spectrum (black solid line) of $\epsilon$ Ori.} \label{fit_uv_epsori}
\end{figure*}

\begin{figure*}
\centering
\includegraphics[width=18cm]{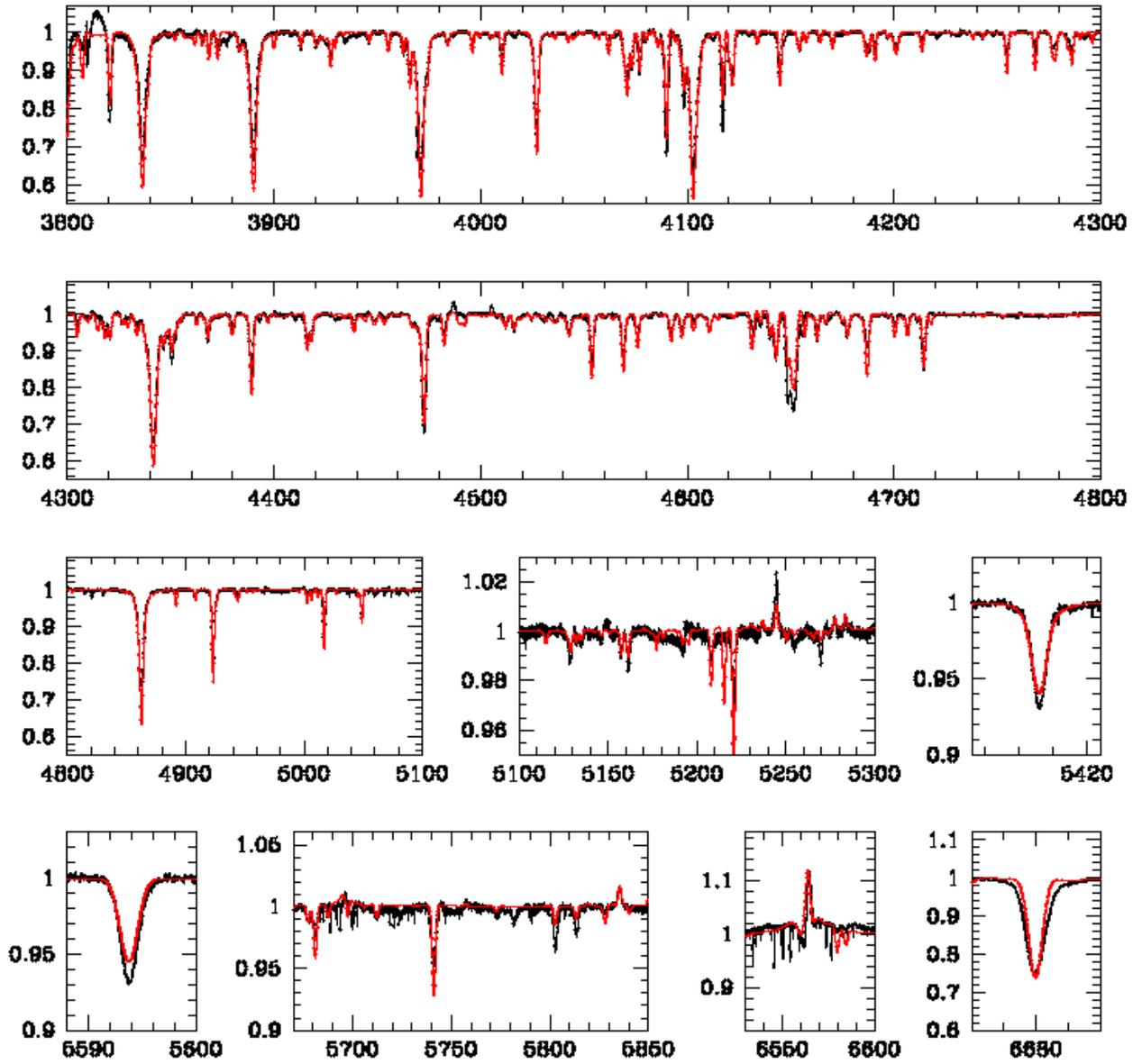}
\caption{Best CMFGEN fit (red solid line) of the optical spectrum (black solid line) of $\epsilon$ Ori.} \label{fit_opt_epsori}
\end{figure*}

\begin{figure*}
\centering
\includegraphics[width=18cm]{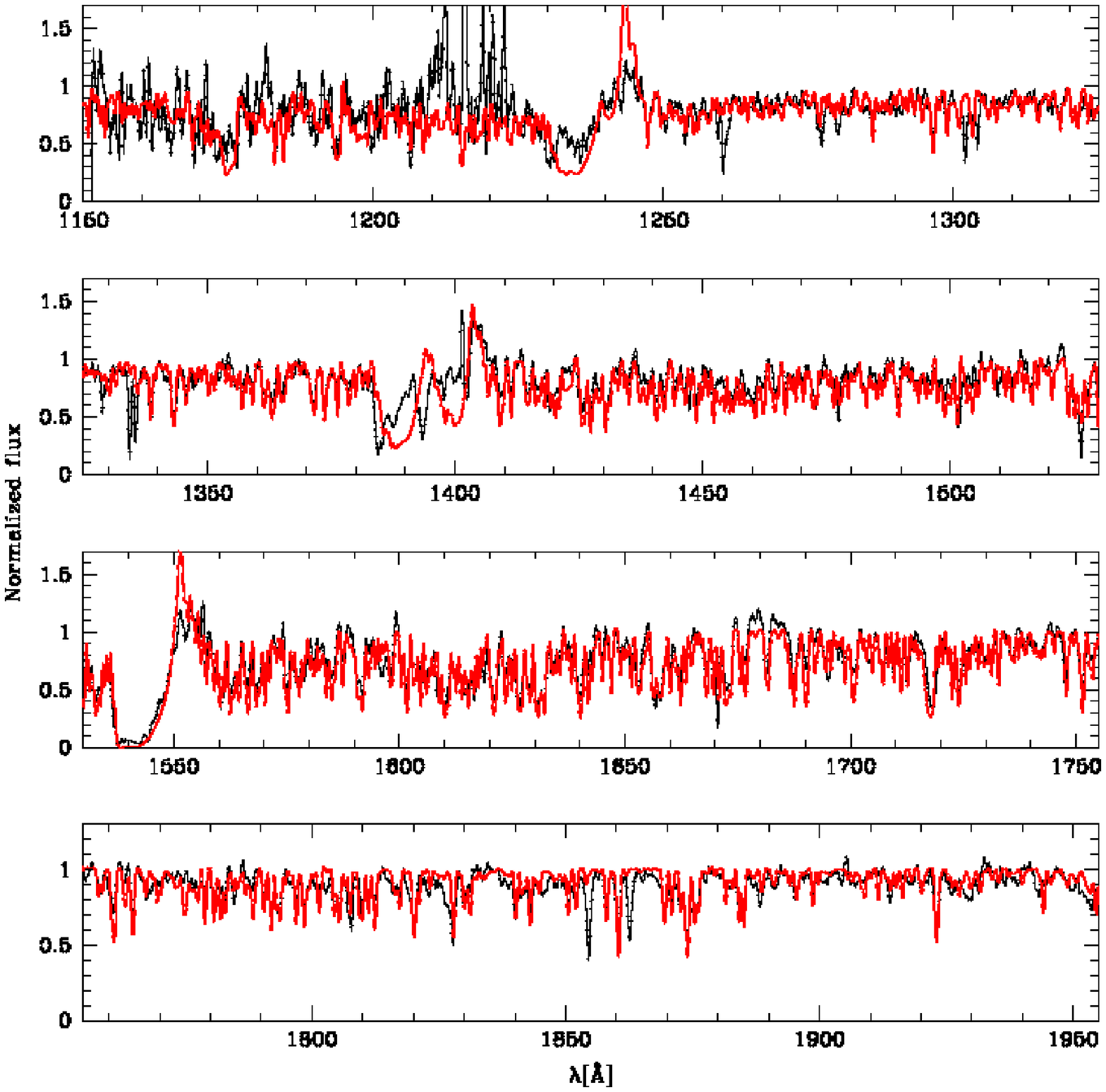}
\caption{Best CMFGEN fit (red solid line) of the UV spectrum (black solid line) of HD~207198.} \label{fit_uv_207198}
\end{figure*}

\begin{figure*}
\centering
\includegraphics[width=18cm]{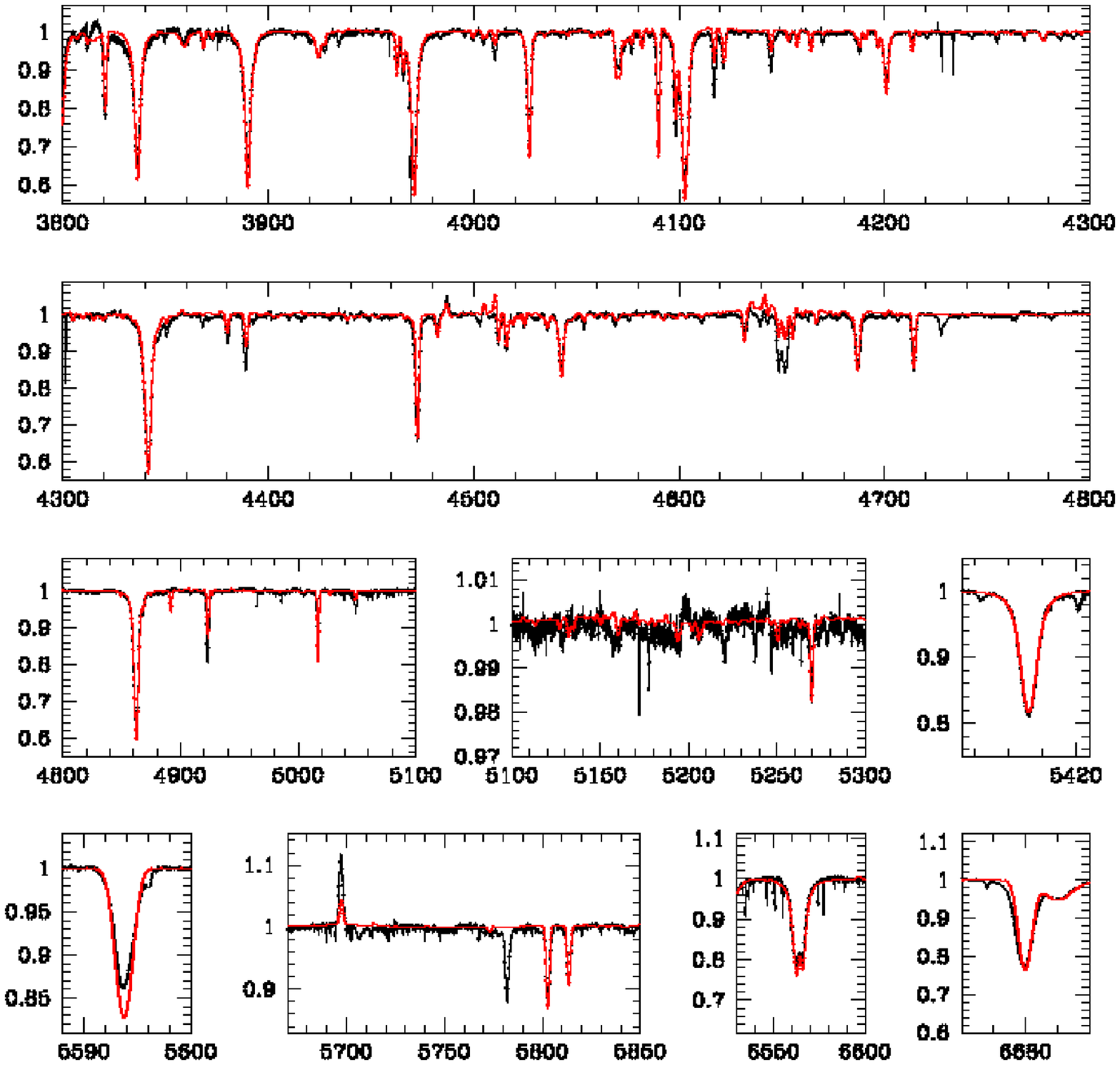}
\caption{Best CMFGEN fit (red solid line) of the optical spectrum (black solid line) of HD~207198.} \label{fit_opt_207198}
\end{figure*}

\begin{figure*}
\centering
\includegraphics[width=18cm]{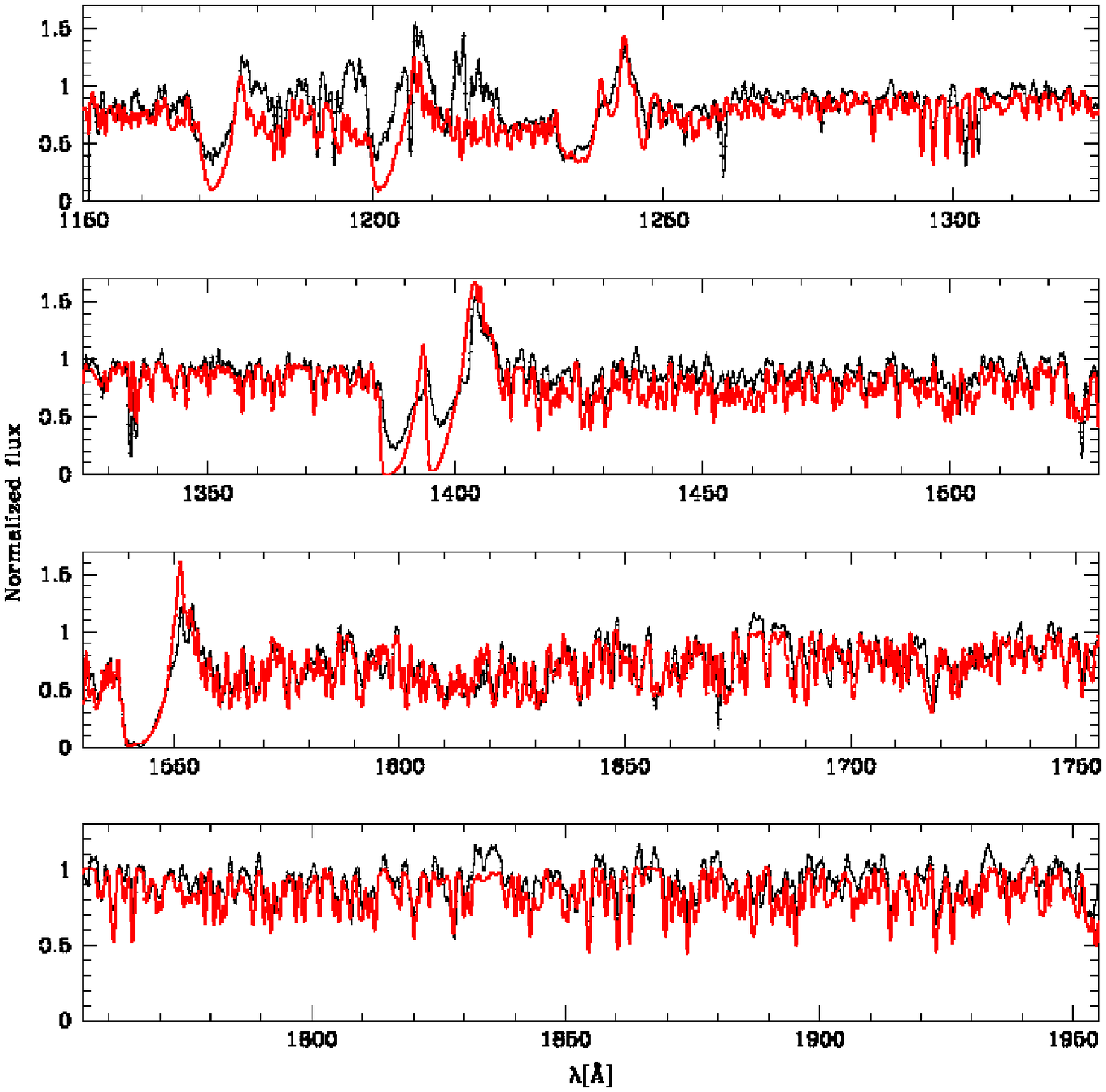}
\caption{Best CMFGEN fit (red solid line) of the UV spectrum (black solid line) of HD~167264.} \label{fit_uv_167264}
\end{figure*}

\begin{figure*}
\centering
\includegraphics[width=18cm]{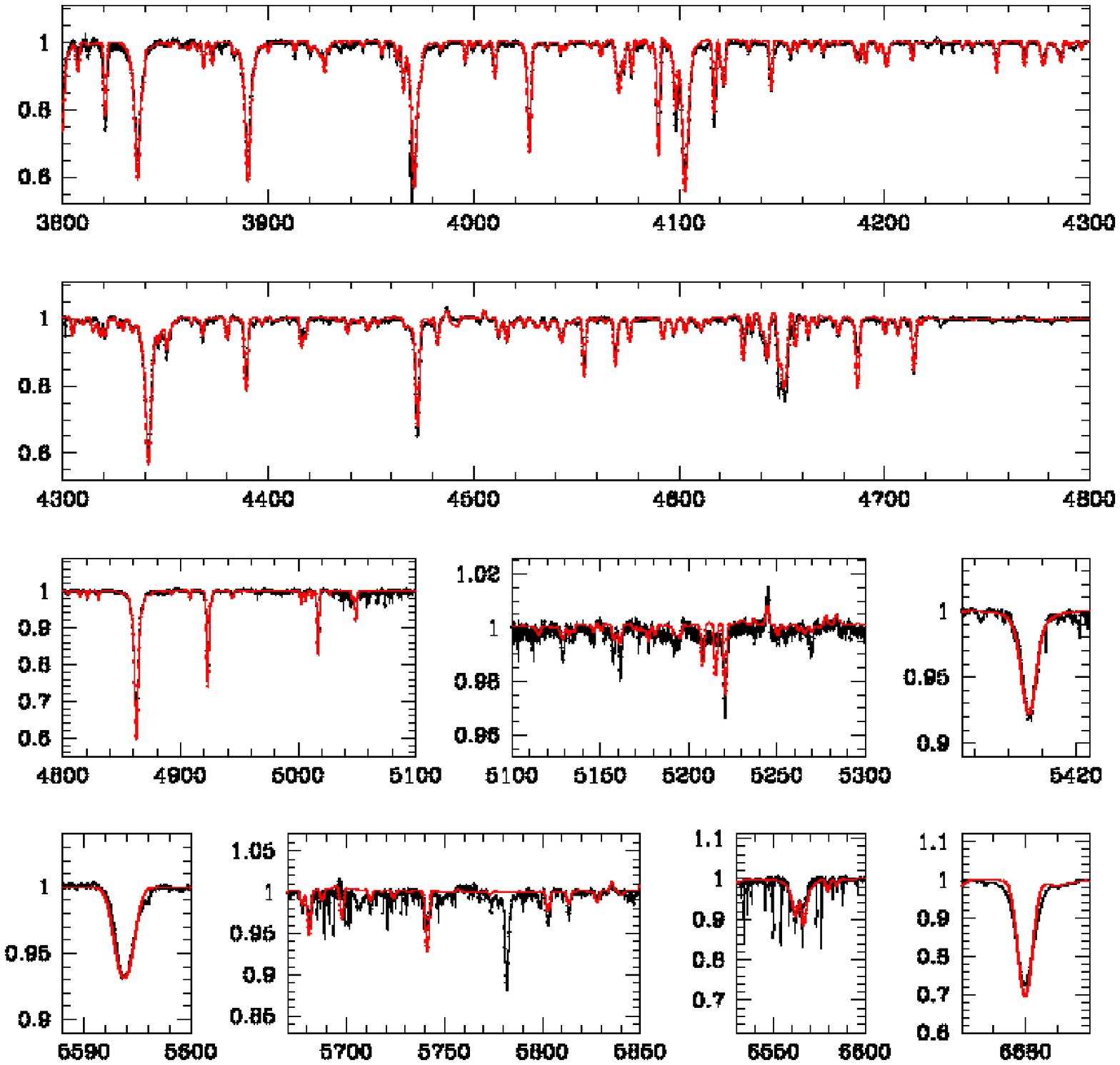}
\caption{Best CMFGEN fit (red solid line) of the optical spectrum (black solid line) of HD~167264.} \label{fit_opt_167264}
\end{figure*}

\pagebreak

\begin{figure*}
\centering
\includegraphics[width=18cm]{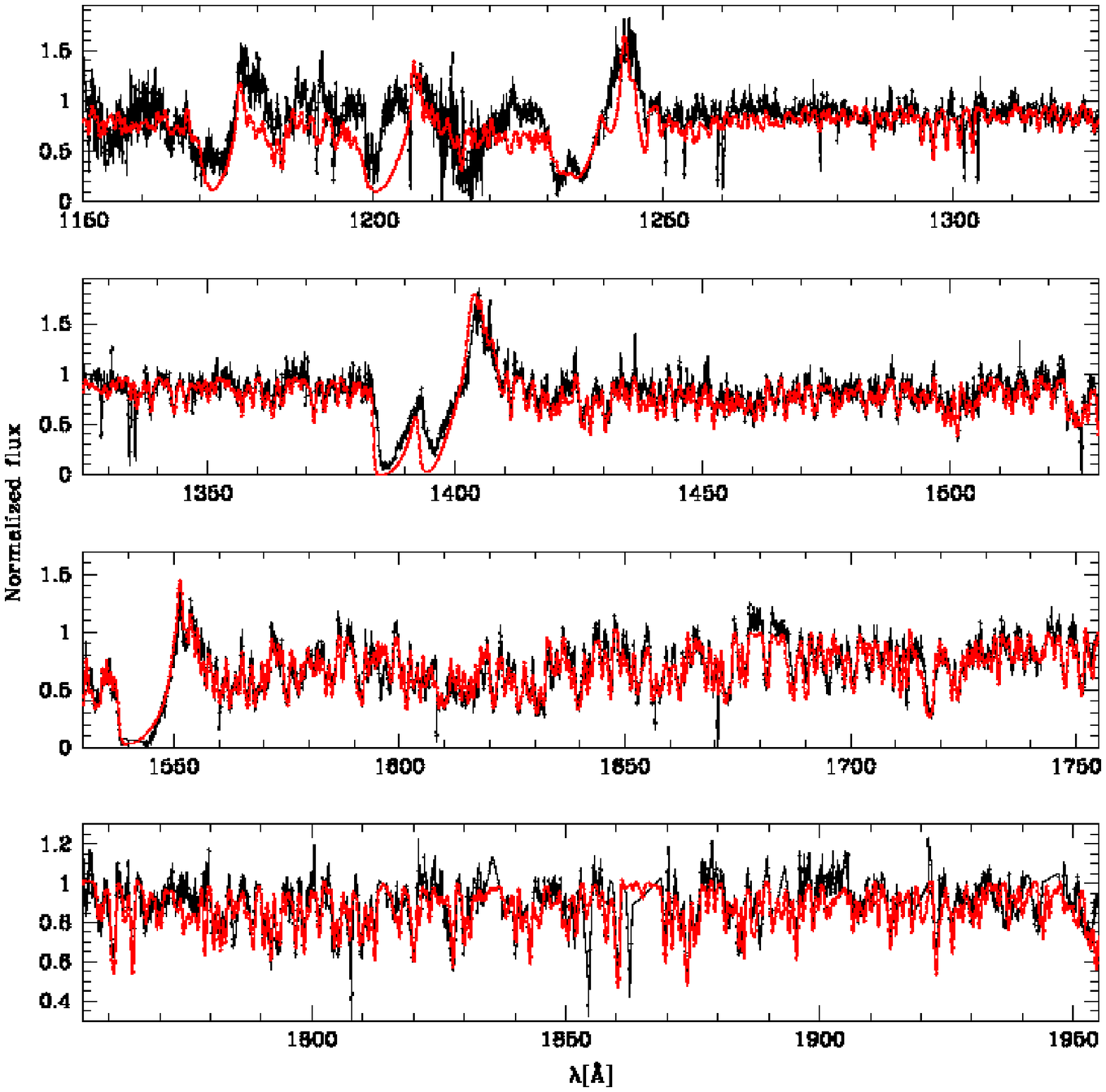}
\caption{Best CMFGEN fit (red solid line) of the UV spectrum (black solid line) of HD~188209.} \label{fit_uv_188209}
\end{figure*}

\begin{figure*}
\centering
\includegraphics[width=18cm]{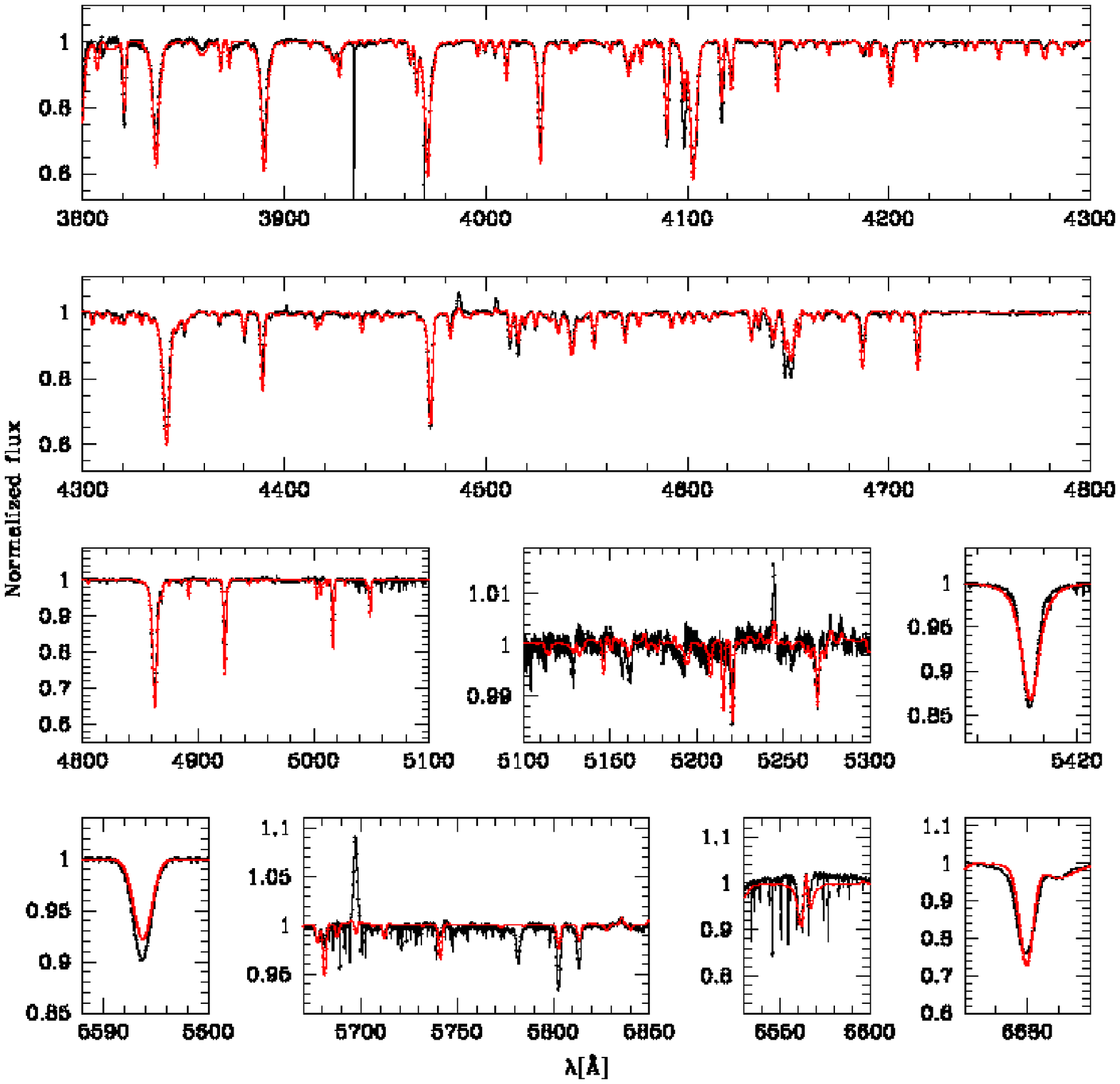}
\caption{Best CMFGEN fit (red solid line) of the optical spectrum (black solid line) of HD~188209.} \label{fit_opt_188209}
\end{figure*}

\pagebreak

\begin{figure*}
\centering
\includegraphics[width=18cm]{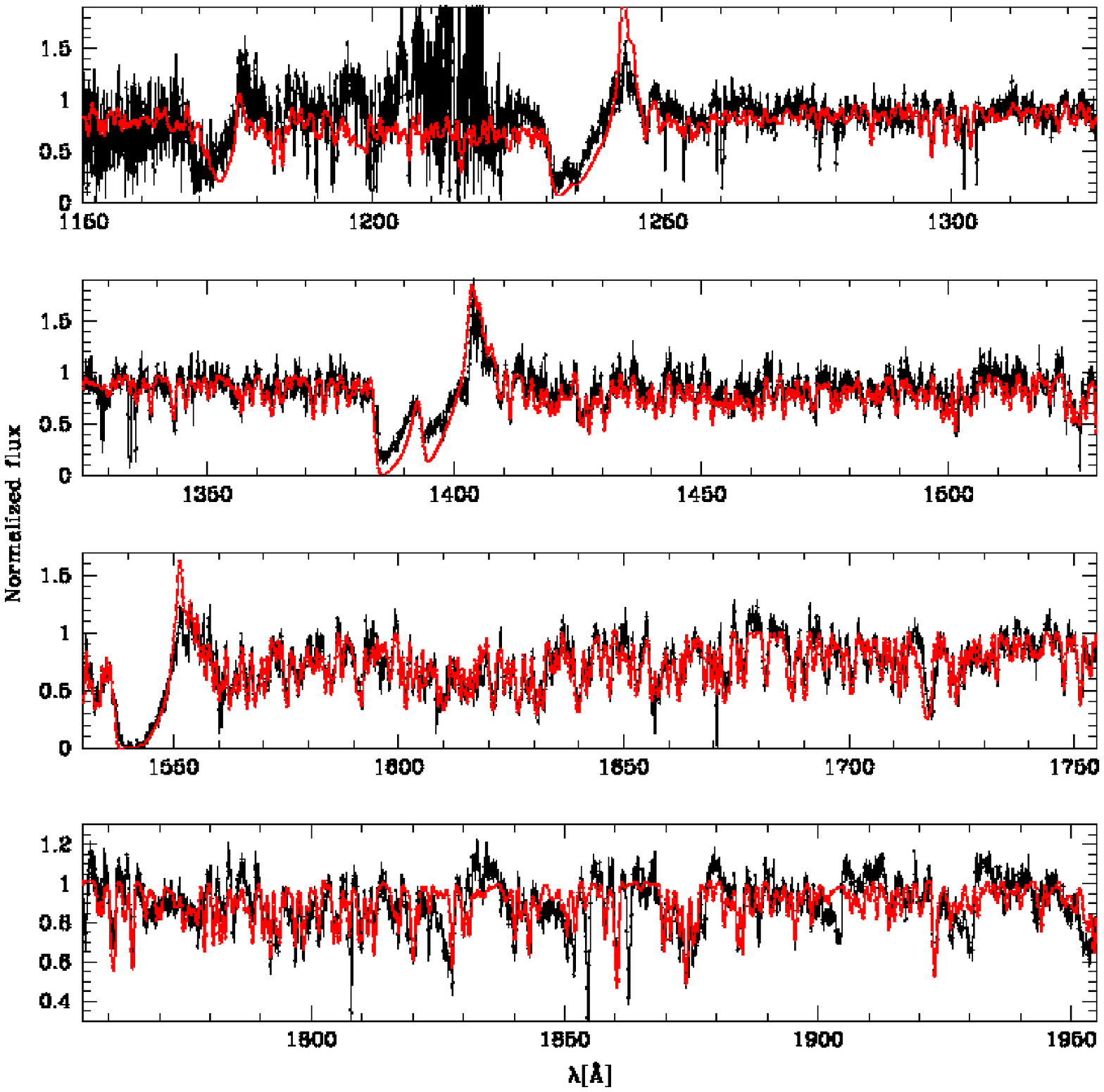}
\caption{Best CMFGEN fit (red solid line) of the UV spectrum (black solid line) of HD~209975.} \label{fit_uv_209975}
\end{figure*}

\begin{figure*}
\centering
\includegraphics[width=18cm]{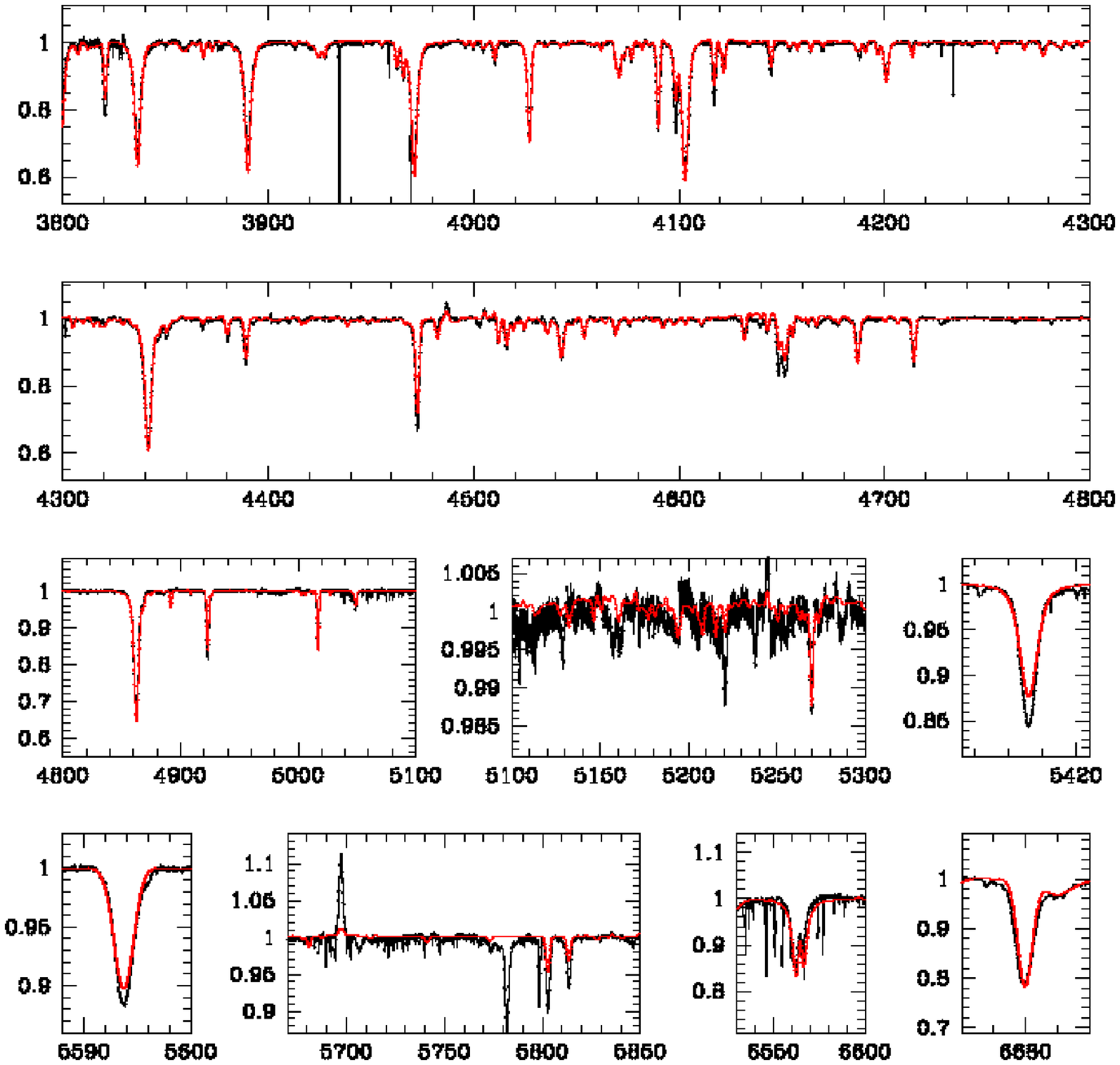}
\caption{Best CMFGEN fit (red solid line) of the optical spectrum (black solid line) of HD~209975.} \label{fit_opt_209975}
\end{figure*}

\pagebreak

\begin{figure*}
\centering
\includegraphics[width=18cm]{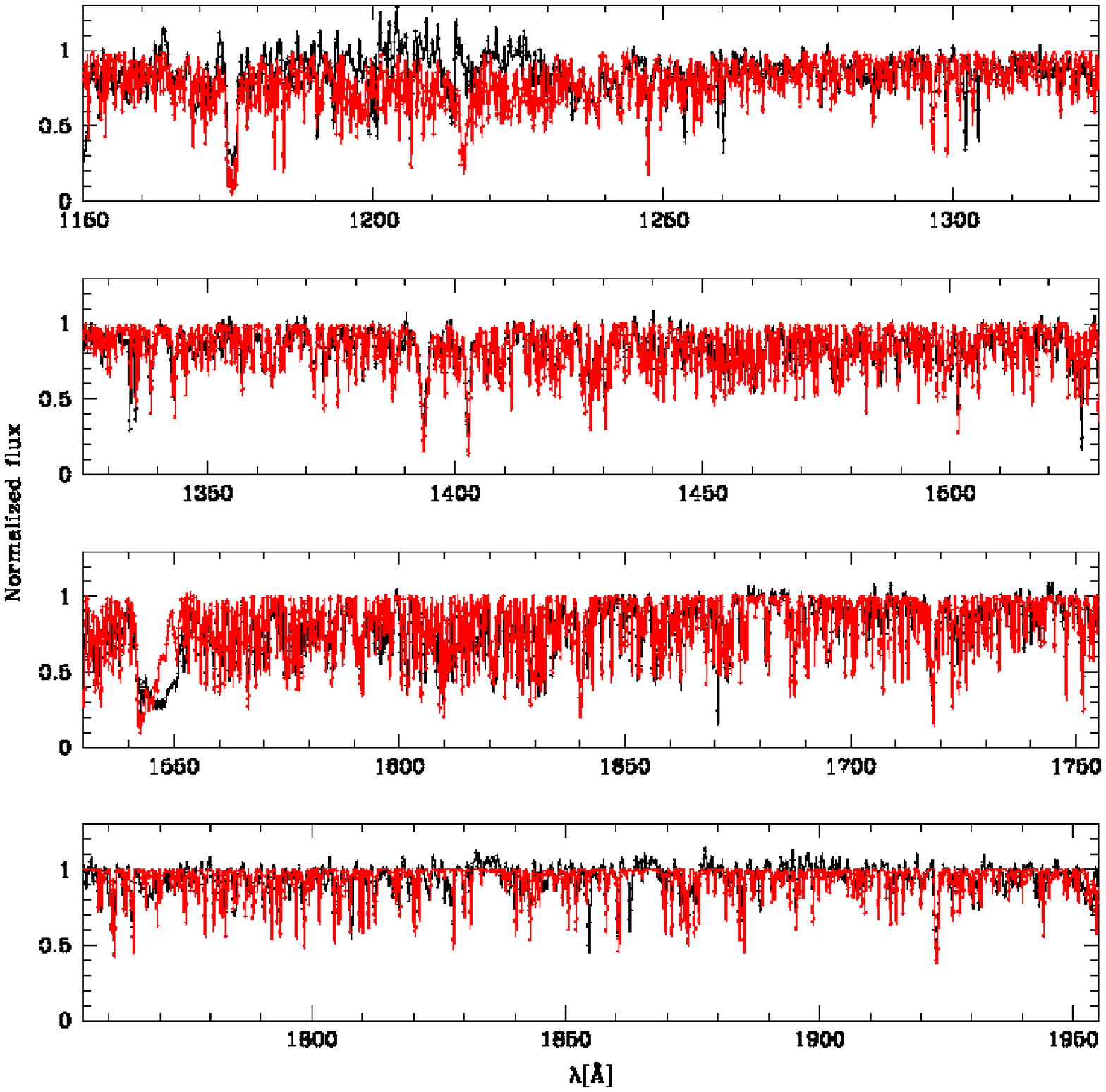}
\caption{Best CMFGEN fit (red solid line) of the UV spectrum (black solid line) of 10~Lac.} \label{fit_uv_10lac}
\end{figure*}

\begin{figure*}
\centering
\includegraphics[width=18cm]{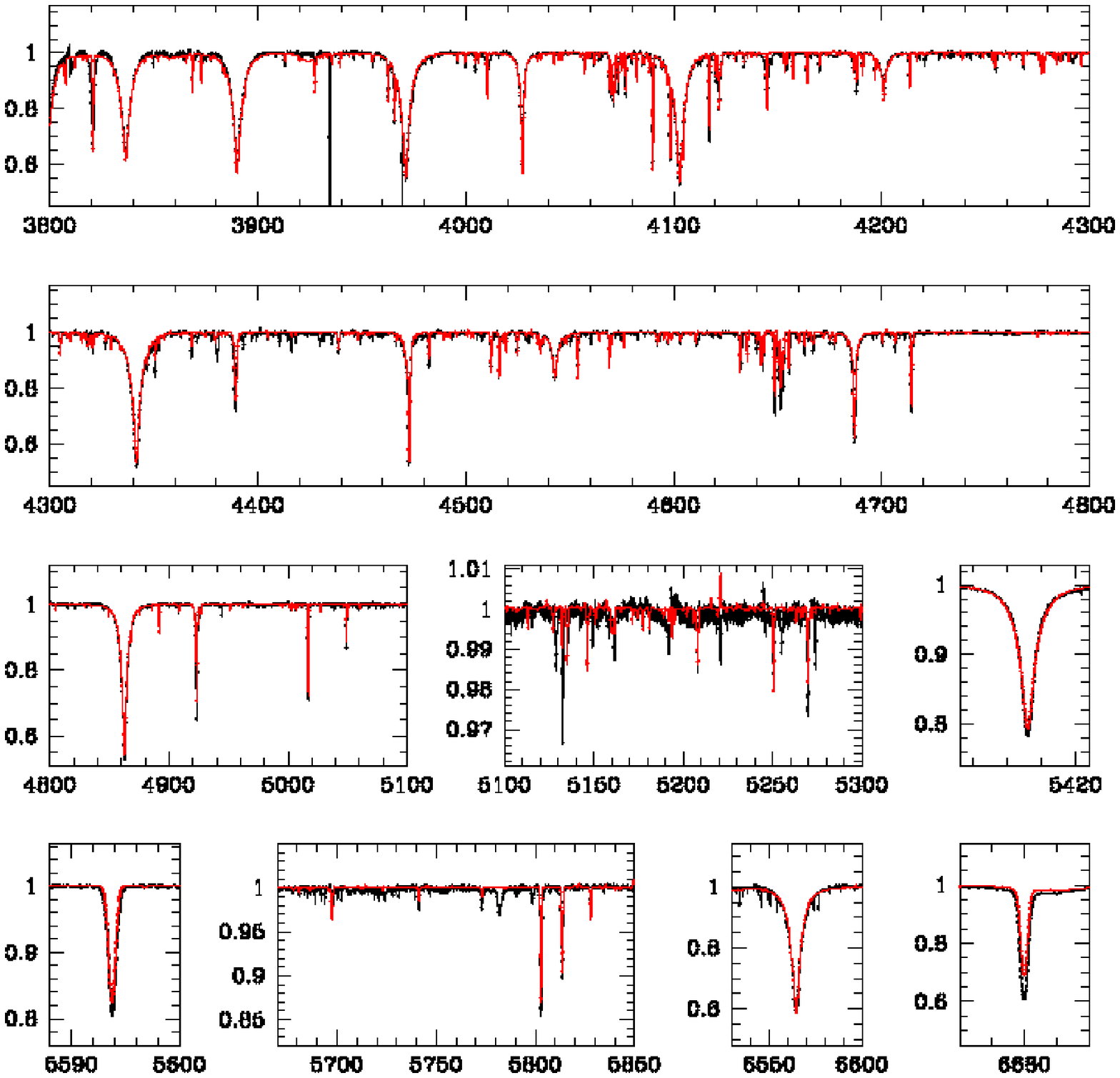}
\caption{Best CMFGEN fit (red solid line) of the optical spectrum (black solid line) of 10~Lac.} \label{fit_opt_10lac}
\end{figure*}

\pagebreak

\begin{figure*}
\centering
\includegraphics[width=18cm]{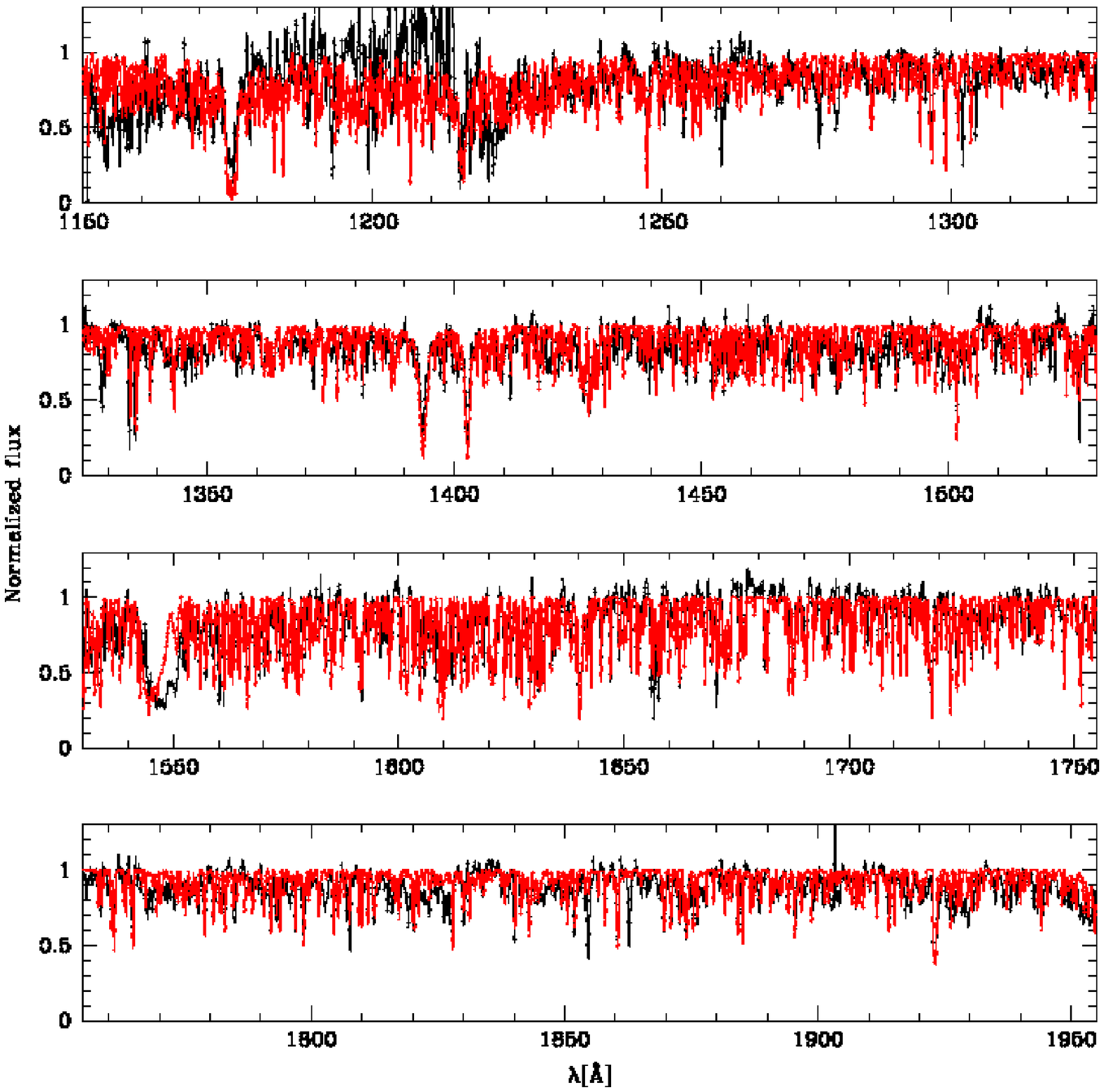}
\caption{Best CMFGEN fit (red solid line) of the UV spectrum (black solid line) of AE~Aur.} \label{fit_uv_aeaur}
\end{figure*}

\begin{figure*}
\centering
\includegraphics[width=18cm]{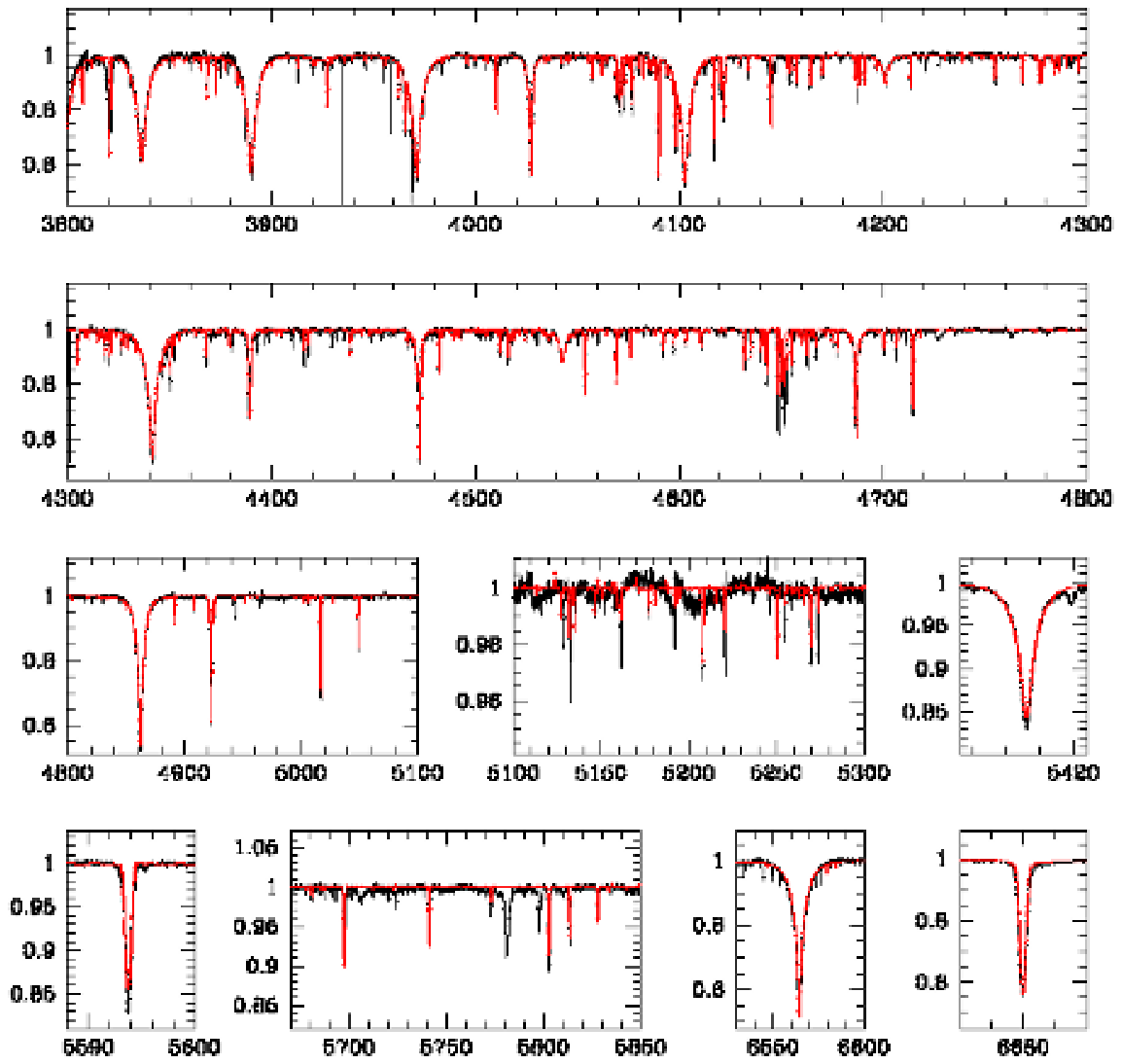}
\caption{Best CMFGEN fit (red solid line) of the optical spectrum (black solid line) of AE~Aur.} \label{fit_opt_aeaur}
\end{figure*}

\end{appendix}

\end{document}